\documentclass[a4paper,twocolumn,11pt,accepted=2026-03-17]{quantumarticle}
\pdfoutput=1

\usepackage[T1]{fontenc}
\usepackage[british]{babel}

\usepackage{amsmath,amssymb,amsfonts,amsthm,mathtools,latexsym}
\usepackage{graphicx}
\usepackage{bbm,dsfont}
\usepackage{array}
\usepackage{float}
\usepackage{colortbl}
\usepackage{xcolor}
\usepackage[normalem]{ulem}
\usepackage[numbers,sort&compress]{natbib}

\usepackage[all,defaultlines=2]{nowidow}

\usepackage{hyperref}
\hypersetup{
  colorlinks,
  linkcolor={red!50!black},
  citecolor={blue!50!black},
  urlcolor={blue!80!black}
}
\usepackage{cleveref}

\usepackage{tikz}
\usetikzlibrary{shapes.geometric,arrows,calc,through,positioning,matrix,graphs,graphs.standard,arrows.meta}

\numberwithin{equation}{section}
\numberwithin{figure}{section}

\newcommand{\kb}[2]{|#1\rangle\langle#2|}
\newcommand{\bk}[2]{\langle#1|#2\rangle}

\newcommand{\ket}[1]{|#1\rangle}
\newcommand{\bra}[1]{\langle#1|}

\newtheorem{equivalence}{Equivalence}[section]
\newtheorem{strategy}{Strategy}[section]

\theoremstyle{definition}
\newtheorem{defn}{Definition}[section]
\newtheorem{problem}{Problem}[section]

\theoremstyle{plain}
\newtheorem{conjecture}{Conjecture}[section]
\newtheorem{prop}{Proposition}[section]
\newtheorem{thm}{Theorem}[section]
\newtheorem{cor}{Corollary}[section]
\newtheorem{lem}{Lemma}[section]

\theoremstyle{remark}
\newtheorem*{claim*}{Claim}

\makeatletter
\renewcommand*\l@section[2]{%
  \ifnumgreater{\c@tocdepth}{\z@}{%
    \addpenalty\@secpenalty
    \addvspace{0.5em}
    \setlength\@tempdima{1.5em}%
    \begingroup
      \parindent \z@
      \rightskip \@pnumwidth
      \parfillskip -\@pnumwidth
      \leavevmode
      \bfseries
      \advance\leftskip\@tempdima
      \hskip -\leftskip
      #1\nobreak\hfil
      \nobreak\hb@xt@\@pnumwidth{\hss #2}\par
    \endgroup
  }{}%
}
\makeatother

\begin{document}

\title{Mutually Unbiased Bases in Composite Dimensions -- A Review}

\author{Daniel McNulty}
\email{daniel.mcnulty@uniba.it}
\affiliation{Dipartimento di Fisica, Università di Bari, Bari, Italy}

\author{Stefan Weigert}
\email{stefan.weigert@york.ac.uk}
\affiliation{Department of Mathematics, University of York, York, UK}

\begin{abstract}
Maximal sets of mutually unbiased bases are useful throughout quantum physics, both in a foundational context and for applications. To date, it remains unknown if complete sets of mutually unbiased bases exist in Hilbert spaces of dimensions different from a prime power, i.e. in composite dimensions such as six or ten. Fourteen mathematically equivalent formulations of the existence problem are presented. We comprehensively summarise analytic, computer-aided and numerical results relevant to the case of composite dimensions. Known modifications of the existence problem are reviewed and potential solution strategies are outlined.
\end{abstract}

\maketitle

\tableofcontents
\newpage

\begin{flushright}
Given a value of $q$,\\
all possible values of $p$ are \emph{equally likely}.\footnote{Originally in German: ``Bei einem gegebenen Wert von $q$ sind alle möglichen Werte von \emph{$p$ gleich wahrscheinlich}.'' \citep[p.~814]{Jordan1927}.}\\
Pascual Jordan -- 1927
\end{flushright}

\section{Introduction}

\subsection{Motivation}

\label{subsec:Motivation}

Given a particle with momentum $p$ moving along a straight line,
what can we say about its position? If we are dealing with a \emph{classical}
point particle, we know that it will occupy a precise location which
a measurement would reveal with certainty. In contrast, the position
$q$ of a \emph{quantum} particle with momentum $p$ cannot be predicted
with certainty. The particle will be found with\emph{ equal} probability
\begin{equation}
\mbox{prob}(q,dq)=\left|\bk pq\right|^{2}dq=\frac{1}{2\pi}dq\,,\label{eq:particle prob}
\end{equation}
irrespective of the chosen interval of length $dq$ about the value
$q$. An analogous statement holds upon exchanging the roles of position
and momentum, with $\mbox{prob}(p,dp)=\left|\bk qp\right|^{2}dp$.

For a free particle on the line, the basis formed by the momentum
eigenstates $\ket p,p\in\mathbb{R}$, is called \emph{unbiased} with
respect to the basis of the position eigenstates $\ket q,q\in\mathbb{R}$,
since the transition probabilities from each state $\ket q$ into
any state $\ket p$ do not depend on the value of $q$. Due to their
symmetry, the bases are, in fact, \emph{mutually unbiased} (MU) since
an equivalent statement applies to the situation in which the roles
of position and momentum are swapped. Any pair of orthonormal bases
with this property—certainty about the value of one observable implies
complete uncertainty about the value of the other observable, and
\emph{vice versa}—is called MU.

\emph{Mutual unbiasedness} captures an important aspect of Bohr's
concept of complementarity \citep{Bohr1928}. It provides a precise
mathematical formulation of what it means for observables such as
position $\hat{q}$ and momentum $\hat{p}$ to be a \emph{complementary}
pair. Importantly, it carries over naturally to bases of \emph{finite-dimensional}
Hilbert spaces $\mathcal{H}\simeq\mathbb{C}^{d}$ which are used to
describe quantum systems with $d$ orthogonal states. Let us now define
the notion central to this review.
\begin{defn}[Pairs of MU bases]
\label{def: MU pair of bases} Two orthonormal bases ${\cal B}$
and ${\cal B}^{\prime}$ of the $d$-dimensional Hilbert space $\mathcal{H}\simeq\mathbb{C}^{d}$
are \emph{mutually unbiased} if the transition probabilities between
any two states $\ket v\in{\cal B}$ and $\ket{v^{\prime}}\in{\cal B}^{\prime}$
are equal, 
\begin{equation}
\left|\bk v{v^{\prime}}\right|{}^{2}=\frac{1}{d}\,,\quad v,v^{\prime}=0,1\ldots d-1\,.\label{eq: def MU pairs}
\end{equation}

The constant value cannot be chosen arbitrarily: completeness of the
basis ${\cal B}$ implies that the sum of the values $\left|\bk v{v^{\prime}}\right|{}^{2}$
over $v$ equals one, forcing the right-hand-side of Eq.~\eqref{eq: def MU pairs}
to take the value $1/d$. The eigenstates of any two Cartesian components
of the spin operator of a spin $1/2$ provide the simplest example
of a pair of mutually unbiased bases: they form a pair of orthonormal
bases such that the squared modulus of the overlap of any two vectors
taken from different bases equals $1/2$. As a relation between states
and bases, mutual unbiasedness is \emph{symmetric} but neither reflexive
nor transitive, hence not an equivalence relation.
\end{defn}
It is natural to associate \emph{complex Hadamard matrices} (see Sec.~\ref{subsec: Hadamard equivalence}
for details) with mutually unbiased bases. To see why, consider the
$d$ orthonormal basis vectors of the bases ${\cal B}$ and ${\cal B}^{\prime}$
as columns of two unitary matrices $V$ and $V^{\prime}$, respectively,
each of size $d$. A suitable unitary transformation maps this pair
to the pair $\left\{ \mathbb{I},H\right\} $, where $\mathbb{I}$
is the identity matrix corresponding to the standard basis in $\mathbb{C}^{d}$,
and $H$ is a unitary matrix the entries of which have modulus $1/\sqrt{d}$,
expressing the fact that its column vectors are mutually unbiased
to the column vectors of the identity matrix.
\begin{defn}[Complex Hadamard matrix]
\label{def: CHM-1} A $\emph{complex Hadamard matrix}$ of order
$d$ is a $d\times d$ unitary matrix $H$ whose entries satisfy $|H_{jk}|=1/\sqrt{d}$
for all $j,k=1\ldots d$.
\end{defn}
Returning to the case of a spin $1/2$, the eigenstates of all three
Cartesian spin components actually provide an example of \emph{three
}pairwise mutually unbiased bases. This observation immediately raises
the question of whether there is a limit on the number of pairwise
MU bases in the Hilbert space $\mathbb{C}^{d}$. The answer is known:
a $d$-dimensional complex Hilbert space supports at most $(d+1)$
pairwise mutually unbiased bases, forming what is known as a complete
or maximal set.
\begin{defn}[Complete sets of MU bases]
 \label{def: complete MU set}A \emph{complete }(or\emph{ maximal})
\emph{set of mutually unbiased bases} in the Hilbert space $\mathbb{C}^{d}$
consists of $(d+1)$ orthonormal bases ${\cal B}_{b}$, $b=0,1,2\ldots d$,
which are pairwise mutually unbiased.
\end{defn}
More explicitly, a complete set of MU bases requires the scalar products
between all its pairs of states to take the values 
\begin{equation}
\left|\bk{v_{b}}{v^{\prime}_{b^{\prime}}}\right|{}^{2}=\begin{cases}
\frac{1}{d} & b\neq b^{\prime}\\
\delta_{vv^{\prime}} & b=b^{\prime}
\end{cases}\,,\quad v,v^{\prime}=0\ldots d-1\,.\label{eq: completeMUsetfor(d+1)}
\end{equation}
If the dimension $d$ of the underlying Hilbert space is given by
a power of a prime number,
\begin{equation}
d=p^{k}\,,\qquad p\in\mathbb{P}\,,\quad k\in\mathbb{N}\,,\label{eq: primepowers}
\end{equation}
then explicit constructions of maximal sets are known, some of which
are summarised in Appendix \ref{sec: complete sets in pp dimensions},
both as a backdrop and for easy reference. It will be useful to denote
the set of prime powers by $\mathbb{PP}$—read: prime power—so that
Eq.~ \eqref{eq: primepowers} would abbreviate to $d\in\mathbb{PP}$.
We will call integer numbers $d$ \emph{composite} if they are \emph{not
equal to the power of a prime}, i.e. $d\notin\mathbb{PP}$. Thus,
our usage of the term differs from the convention in number theory
where all numbers except primes are considered to be composite.

With pairs of mutually unbiased bases capturing the idea of complementarity,
larger sets can be thought of as realising \emph{multiple complementarity.
}Complete sets of mutually unbiased bases, made from states satisfying
the conditions of \eqref{eq: completeMUsetfor(d+1)}, are then required
to achieve ``maximal complementarity'' in a $d$-dimensional state
space. In this way, the traditional concept of complementarity, which
was limited to \emph{pairs }of non-commuting observables such as position
and momentum, opens up in an unexpected direction.

Somewhat surprisingly, it is not known whether a complete set of mutually
unbiased bases exists in state spaces of \emph{composite }dimension,
$d\notin\mathbb{PP}$, i.e. whenever the number $d$ is \emph{not}
equal to a power of a prime or explicitly,
\begin{equation}
d=\prod^{r}_{j=1}p^{n_{j}}_{j}\,,\qquad p_{j}\in\mathbb{P}\,,\quad n_{j}\in\mathbb{N}\,,\quad r\geq2\,,\label{eq: def composite numbers}
\end{equation}
involving at least \emph{two} different prime numbers, $p_{1}\neq p_{2}$.
A list of small integers seems to suggest that composite numbers are
rare: up to $d=11$, for example, we encounter only two, $d=6$ and
$d=10$. However, the opposite is true: composite dimensions abound
since the proportion of prime numbers—and hence prime-powers—among
all integers below a given number $N\in\mathbb{N}$ decreases with
$N$ (see Sec.~\ref{subsec:Composite-dimensions}). 

The upper bound allows for up to seven and eleven MU bases in the
spaces $\mathbb{C}^{6}$ and $\mathbb{C}^{10}$, respectively, but
in both cases, no more than \emph{three }mutually unbiased bases have
been found so far. The construction of these \emph{triples} of mutually
unbiased bases hinges on the lowest factor $p_{1}=2$ in the prime
decomposition \eqref{eq: def composite numbers} of six and ten, as
will be explained in more detail later (cf. in Sec.~\ref{sec:larger_sets_of_MU_bases}).
More generally, the number of MU bases we are able to construct in
a space of dimension $d$ depends on its prime decomposition and is
significantly smaller that the maximum of $(d+1)$ bases. Thus, we
are led to the main question which has been driving the research reviewed
here.
\begin{problem}[\emph{Existence problem in composite dimensions}]
 \label{general existence problem}Do complete sets of mutually unbiased
bases exist in \emph{composite} dimensions $d\notin\mathbb{PP}$,
i.e. whenever $d\neq p^{k}$, $p\in\mathbb{P}$, $k\in\mathbb{N}$?
\end{problem}
For composite dimensions $d\notin\mathbb{PP}$, all known results—be
they in closed form, computer-algebraic or numerical—point to the
\emph{non-existence} of $(d+1)$ MU bases. More specifically, for
dimension six, all findings are compatible with \emph{Zauner's conjecture}
(cf. Sec.~\ref{subsec:Composite-dimensions}) that only three of
the seven potential MU bases exist. Lifting the conjecture from $d=6$
to arbitrary composite dimensions in a sweeping speculative move,
the existence of complete sets in arbitrary composite dimensions is
called into question.
\begin{conjecture}[\emph{Non-existence}]
 \label{conj: Zauner}Complete sets of mutually unbiased bases do
not exist in \emph{composite} dimensions $d\notin\mathbb{PP}$, i.e.
whenever $d\neq p^{k}$, $p\in\mathbb{P}$, $k\in\mathbb{N}$.
\end{conjecture}
A natural attempt to upper-bound the number of MU bases, related to
the prime decomposition of $d\notin\mathbb{PP}$, turns out to be
\emph{false}.
\begin{claim*}[\emph{False upper bound}]
 \label{conj: stronger Zauner}No more than $(p^{n_{1}}_{1}+1)$
mutually unbiased bases exist in the \emph{composite} dimension $d=p^{n_{1}}_{1}p^{n_{2}}_{2}\ldots$,
where $p_{j}\in\mathbb{P}$, $n_{j}\in\mathbb{N}$, and $p^{n_{1}}_{1}$
is the smallest prime-power present in the decomposition of $d$.
\end{claim*}
Thm.~\ref{thm:wocjanMUBs} of Sec.~\ref{subsec:Maximally-entangled-bases}
provides counterexamples to this claim in specific dimensions such
as $d=2^{2}\cdot13^{2}$: a construction using Latin squares leads
to $(p^{n_{1}}_{1}+2)$ MU bases.\emph{ }A universal \emph{lower}
bound on the number of MU bases in composite dimension is stated in
Thm. \ref{thm:reduce_to_primes} of Sec.~\ref{sec:larger_sets_of_MU_bases}.

The existence problem of complete sets of MU bases has made it into
the list of the ``ten most annoying questions in quantum computing''
\citep{Aaronson2014}. It also figures on another long-standing list
of open problems in quantum information \citep{WernerOpenProblems2003},
and a solution wins you a prize \citep{Horodecki2020}. 

\subsection{Goals, scope and outline}

The \emph{goal} of this review is to collect what is known about mutually
unbiased bases in composite dimensions $d\notin\mathbb{PP}$, complementing
earlier surveys \citep{kibler08,durt+10} which were focused on prime-power
dimensions, $d\in\mathbb{PP}$. Accordingly, we can afford to be brief
on topics such as the construction of complete sets (see Appendix
\ref{sec: complete sets in pp dimensions}). 

\begin{figure*}[t]
\begin{centering}
\includegraphics[width=15cm]{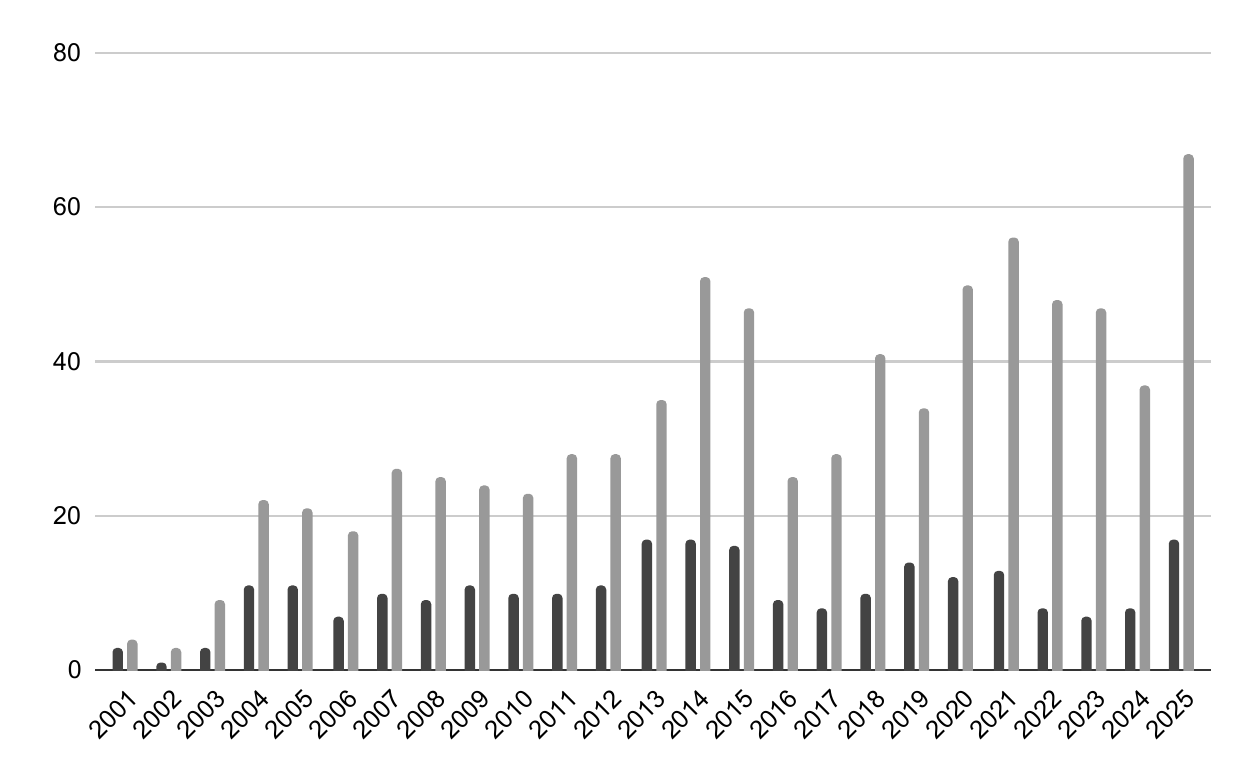}
\par\end{centering}
\caption{Number of preprints uploaded annually to arXiv.org in the sections computer sciences,
mathematics and physics containing the expression \textquoteleft mutually
unbiased\textquoteright{} in the \emph{title} (dark) and in the \emph{abstract}
(light), respectively. \label{fig:Number-of-preprints}}
\end{figure*}

This review will feel much less conclusive than the substantial monograph
by Durt \textit{et al}. \citep{durt+10}, simply because mutually
unbiased bases in composite dimensions have not yet revealed their
secret. With a steady number of papers directly addressing MU bases,
and an increasing interest in using MU bases (see Fig.~\ref{fig:Number-of-preprints}),
it seems desirable to describe the state of play, in spite—or because—of
the limited progress concerning the existence problem. We wish to
provide a reliable snapshot of the current state of the art (summer
2024), as well as an easy-to-navigate resource for future work.

There is much more to say about mutually unbiased bases in composite
dimensions than just stating our inability to construct complete sets.
We will show that considerable ingenuity has led to many results which,
for example, restrict the possible form of a complete set (see the
summary in Sec.~\ref{subsec:Properties-of-a-complete-set}), although
they do not settle the central existence question. In the absence
of an over-arching theory, we can still point to quite a few relevant—if
seemingly disconnected—observations, and to a surprising wealth of
mathematical concepts. They have been published in journals specialising
in pure mathematics, combinatorics, computer science or physics, mostly
quantum information and, hence, are not always easy to track down.
We hope that this review allows researchers wishing to think about
MU bases in composite dimensions to easily access what is known. 

The impatient reader may wish to jump to Secs.~\ref{sec:rigorous_results_any_d}
and \ref{sec:rigorous_results_d=00003D6}, which constitute the core
of the survey containing rigorous results in general composite dimensions
and dimension six, respectively. Sec.~\ref{sec: Numerical_results}
summarises results based on numerical work. MU bases for continuous
variable systems require infinite-dimensional Hilbert spaces and will
be discussed in Sec.~\ref{sec: Modifications-of-the-problem}, along
with other modifications of the original existence problem.

Readers with more time on their hands may wish to explore other parts
of the survey. Let us briefly \emph{outline} its overall structure.
Sec.~\ref{sec: conceptual history} summarises the history of mutually
unbiased bases in a non-technical way, which we hope newcomers to
the field will find useful. Divided into three parts, it tracks down
instances of mutual unbiasedness before the name was coined (Sec.~\ref{subsec: avant-la-letter}),
then describes the emergence of results for the case of prime-power
dimensions (Sec.~\ref{subsec: MU bases prime power dimensions})
and for composite dimensions (Sec.~\ref{subsec:Composite-dimensions}).
Next, in Sec.~\ref{sec: MU-bases-in-QT} we explain why physicists
are interested in mutually unbiased bases, especially within quantum
theory. We cover a broad range of topics including quantum state spaces,
Kochen-Specker type arguments, inequalities saturated by complete
sets of MU bases, and we report experiments in which the states forming
MU bases have been created successfully. Applications of (complete
sets of) MU bases can be found in Sec.~\ref{sec:applications}, containing
both traditional and more recent achievements.

Over time, a variety of alternative characterisations of mutually
unbiased bases have been discovered. They range from specific sets
of coupled polynomial equations and Hadamard matrices to Lie algebras
with specific properties and so-called quantum designs. Fourteen \emph{Equivalences},
which rigorously express complete sets of MU bases in a different
mathematical setting, are described in Sec.~\ref{sec: equivalent_formulations}.
Each equivalence gives rise to a separate conjecture which captures
the existence problem of MU bases in terms of a related mathematical
structure. Proving or disproving any of these conjectures is tantamount
to solving the original problem.

The next two sections bring together rigorous results obtained for
mutually unbiased bases in composite dimensions, obtained either by
analytic or computer-algebraic methods for general composite dimensions
(Sec.~\ref{sec:rigorous_results_any_d}), or specifically dimension
six (Sec.~\ref{sec:rigorous_results_d=00003D6}). It is shown, for
example, that certain types of mutually unbiased bases do not extend
to complete sets. Numerical results are presented in Sec.~\ref{sec: Numerical_results}.
While they provide no proof of the non-existence of maximal sets,
they strongly suggest their absence in the dimensions studied. We
pull together evidence which has been accumulated mainly for the case
$d=6$ but also for some other small composite dimensions.

Since the existence problem has, so far, defied all attempts to be
solved in its current form, we will, in Sec.~\ref{sec: Modifications-of-the-problem},
draw attention to various ways in which it has been modified. The
common underlying strategy is to test how robust the problem is under
a partial change in the relations defining it, and ideally to gain
insights which also apply to the original problem. For example, one
can set the problem in a real instead of a complex Hilbert space,
use a $p$-adic number field, change the constraints on the overlaps
between states, or set the problem in a Hilbert space of countably
infinite dimension.

Readers wishing to actively investigate the open existence problem
might find some inspiration in Sec.~\ref{sec: Summary-and-Conclusions}
where, in order to be as explicit as possible, we first summarise
the properties a complete set of seven MU bases in $\mathbb{C}^{6}$
must have. Then, we list promising strategies to solve the existence
problem which have not yet been fully explored, some of which could
be improved simply by using more powerful computational resources.
Next, the section lists ``stepping stones'', i.e. less general problems
that may be helpful on the way towards a solution of the existence
problem, as well as other open questions related to MU bases. Finally,
we will speculate about future directions of the field.

Appendix~\ref{sec: complete sets in pp dimensions} summarises existing
constructions of complete sets of mutually unbiased bases and discusses
the fact that maximal sets are not always unique. Properties of complex
Hadamard matrices are collected in Appendix~\ref{sec:Complex-Hadamard-matrices},
while Appendix~\ref{subsec:affineplanes_mubs_sics} describes how
MU bases relate to both SIC-POVMs (see Sec.~\ref{subsec:MU-bases-and-SICs})
and affine planes.

\section{Conceptual history of MU bases}

\label{sec: conceptual history}

\subsection{MU bases \emph{avant la lettre} }

\label{subsec: avant-la-letter}

In 1822, Fourier \citep[p. 525]{Fourier1822} described a new method
to represent a function $f(x)$ depending on the variable $x\in\mathbb{R}$,
by expressing it as a linear combination of continuously many wave
trains with real coefficients $f(\alpha)$, 
\begin{equation}
f(x)=\frac{1}{2\pi}\int^{\infty}_{-\infty}d\alpha\,f(\alpha)\int^{\infty}_{-\infty}dp\,\cos\left(px-p\alpha\right).\label{eq: JFourier}
\end{equation}
This mathematical relation, which we now interpret as a definition
of Dirac's delta distribution,
\begin{equation}
\delta\left(x\right)=\frac{1}{2\pi}\int^{\infty}_{-\infty}dp\,e^{ipx}\,,\label{eq: Dirac delta}
\end{equation}
illustrates the concept of complementarity of two conjugate variables,
called $x$ and $p$ here, albeit in a non-quantum setting. In order
to localise a function about a point in $x$-space, we must superpose
infinitely many components in Fourier- or $p$-space, with all coefficients
having \emph{equal} moduli. Since there is no conceptual difference
between the original and the Fourier space, this argument also applies
after swapping the variables. Thus, mutual unbiasedness represents
a mathematical relation, independent of quantum theory. This observation
suggests casting our net wide in Sec.~\ref{sec: equivalent_formulations}
where we will gather alternative formulations of mutual unbiasedness.

Another early appearance of MU bases dates back to an 1844 paper by
Hesse \citep{hesse1844} who discusses the geometric properties of
a configuration of nine points and twelve lines in a projective plane.
This geometric structure showcases an intimate connection between
a collection of nine equiangular vectors—known today as the \emph{Hesse
SIC} discussed in Sec.~\ref{subsec:MU-bases-and-SICs}—and a related
set of twelve vectors which partition into four orthogonal bases of
$\mathbb{C}^{3}$ \citep{bengtsson10}. The overlaps between the vectors
of these four bases satisfy Eqs.~\eqref{eq: completeMUsetfor(d+1)},
resulting in (perhaps) the first instance of a complete set of MU
bases.

Sylvester describes pairs of MU bases in arbitrary finite dimensions
$d$ in a paper\footnote{The introduction of the paper ends by specifying the wide-ranging
audience which Sylvester has in mind: he states that his results would
be ``furnishing interesting food for thought, or substitute for the
want of it, alike to the analyst at his desk and the fine lady in
her boudoir'' (\uline{\mbox{\citep{sylvester67}}}, p. 461).} published in 1867 \citep{sylvester67}: His ``inverse orthogonal
matrices'' effectively satisfy the same conditions as complex Hadamard
matrices. Sylvester conjectures (erroneously) that the construction
provided exhausts all matrices of the desired form. The approach leads
to matrices of order $d$ with the first row and column being real,
i.e. they are given in ``dephased'' form (cf. Appendix \ref{sec:Complex-Hadamard-matrices}). 

In prime dimensions, $d\in\mathbb{P}$, Sylvester obtains the \emph{Fourier
matrices} $F_{d}$ with entries $F_{d,jk}=\omega^{jk}/\sqrt{d}$ in
row $j$ and column $k$, where $j,k=0\ldots d-1$, and $\omega$
is a $d$-th root of unity (see Sec.~\ref{sec:pairs_of_MU_bases_C^d}).
The Fourier matrices $F_{3}$ and $F_{5}$ are displayed explicitly.
Furthermore, for any decomposition of the number $d=d_{1}d_{2}\ldots$,
the construction of inverse orthogonal matrices results in direct
products $F_{d_{1}}\otimes F_{d_{2}}\otimes\ldots$ of Fourier matrices,
some of which are now known to be equivalent to each other. Sylvester
pays particular attention to dimensions which are powers of two, with
explicit expressions given for $d=2,4,8$. In the case of $d=4$,
both $F_{4}$ and $F_{2}\otimes F_{2}$ are displayed. The Fourier
matrices will play an important role throughout the discussion of
mutually unbiased bases (e.g. see Sec.~\ref{sec:fourier_family}).

Nearly thirty years later, Hadamard searches for the maximal value
of matrix determinants given that the moduli of their entries do not
exceed a given value \citep{hadamard93}. Choosing the bound to be
$1$, he shows that Sylvester's ``inverse orthogonal matrices''
produce the desired maximum, describing them as Vandermonde determinants
formed by the roots of unity. Hadamard points out that these are not
the only solutions: In dimension four he adds a collection of matrices
depending on one continuous parameter, known today as the Fourier
family (see Sec.~\ref{sec:pairs_of_MU_bases_C^d} for details). This
set contains the special cases already considered by Sylvester, and
it provides—in modern parlance—the first examples of complex Hadamard
matrices with entries other than roots of unity. The paper concludes
with the question ``for which values of $n$ there exist maximal
determinants with real elements'' (\citep[p. 246]{hadamard93}) which—under
the heading of the \emph{Hadamard conjecture}—has fascinated mathematicians
ever since \citep{Hedayat1978}. Sets of real Hadamard matrices with
all elements $\pm1$ define mutually unbiased bases consisting of
real vectors only; they have been studied in their own right (see
Sec.~\ref{subsec:Real-MU-bases}) without shedding much light on
the existence problem in complex spaces.

One main driving force behind further studies of mutually unbiased
bases has been quantum theory (see Sec.~\ref{sec: MU-bases-in-QT}).
The motto of this review stems from Jordan's fundamental paper \citep{Jordan1927}
in 1927, where he rolls the four variants of quantum theory existing
at the time into a single one; his remark may be one of the first
explicit descriptions of unbiasedness. We are not aware of other authors
stating mutual unbiasedness before 1927.

The interplay between position and momentum measurements also appears
in von Neumann's classic book, published originally in 1932:
\begin{quote}
If $q$ has a very small dispersion in an ensemble, then the $p$
measurement with the accuracy (i.e., dispersion) $\varepsilon$ sets
up a $q$ dispersion of at least $h/4\pi\varepsilon$ {[}$\ldots${]},
i.e., everything is ruined \citep[pp. 305-6 (footnote 162)]{Neumann1955}.
\end{quote}
Note the negative connotation of the word ``ruin'' ('verderben'
in the German version\footnote{``Wenn $q$ in einer Gesamtheit kaum streut, so wird die $p$-Messung
mit der Genauigkeit (d. h. Streuung) $\varepsilon$ die $q$-Streuung
auf mindestens $h/4\pi\varepsilon$ heraufsetzen {[}$\ldots${]}—d.
h. alles verderben.'' \citep[p. 256 (footnote 162)]{von_Neumann_1932}}) used to describe the effect of measuring the observable complementary
to position.

In their well-known paper from 1935, Einstein, Podolski and Rosen
\citep{Einstein1935} use an entangled state of two particles to argue
that quantum theory is incomplete. Their reasoning starts by describing
the impossibility to attribute a definite value of position to a single
quantum particle which resides in a momentum eigenstate, and it invokes
Born's probability interpretation of the squared overlaps of states
\citep{Born1926}. Their discussion closely follows Weyl's 1928 book
on group theory and quantum mechanics \citep{Weyl1928,Weyl1950} in
which he points out that the $x$-component of the momentum of a particle
``\emph{cannot assume a definite value with certainty when} $x$
{[}the $x$-component of its position{]} \emph{does, and conversely}''
(p. 55; original emphasis). This statement also captures mutual unbiasedness
but not as precisely as Jordan's wording. 

An important step in the history of mutually unbiased bases is due
to Schwinger in 1960, building on Weyl's work. The influential study
of unitary operator bases \citep{schwinger60} opens up three new
vistas: (\emph{i}) MU bases are defined and subsequently constructed
in a finite-dimensional space; (\emph{ii}) the generators of a pair
of MU bases in a finite-dimensional Hilbert space ${\cal H}$ can
be used to construct a basis of the unitary operators acting on ${\cal H}$;
(\emph{iii}) mutual unbiasedness is linked to the finite-dimensional
Heisenberg-Weyl group on the space $\mathbb{C}^{d}$. 

Introducing a pair of unitary operators $U$ and $V$ by the requirement
that they cyclically permute their respective eigenstates, Schwinger
derives the commutation relation
\begin{equation}
VU=e^{\frac{2\pi i}{d}}UV\,,\label{eq: HW group commutator}
\end{equation}
which implies that the operators $U$ and $V$ are maximally incompatible,
i.e., their eigenstates form orthonormal bases and come with pairwise
constant overlaps, Eq.~\eqref{eq: completeMUsetfor(d+1)}. When $d$
is a prime number, suitable products of $U$ and $V$ form a basis
of the bounded operators ${\cal B}(\mathbb{C}^{d})$ acting on the
space $\mathbb{C}^{d}$, and their eigenstates define a set of $\left(d+1\right)$
MU bases (see Sec.~\ref{subsec:Maximally-commuting-unitary}). A
finite-dimensional equivalent of the relation Eq.~\eqref{eq: Dirac delta}
exists for the eigenstates of $U$ and $V$, which goes hand in hand
with interpreting them as $d$ eigenstates of finite displacements
and of momentum boosts, respectively. It becomes plain to see that
the Fourier transformation underpins the concept of mutual unbiasedness
in both finite- and infinite-dimensional spaces. 

Also in 1962, Schlögl \citep{schlogl60} points out that the eigenstates
of the Pauli matrices in $\mathbb{C}^{2}$ form a triple of MU bases.
In a finite-dimensional setting, he searches for non-commuting observables
which are pairwise ``informations-fremd'' (``information-agnostic''),
a concept which captures the idea of mutual unbiasedness while avoiding
confusion with the notion of ``complementary'' observables. It is
shown that no more than two observables with this property can exist
in dimensions $d>2$—by allowing \emph{Hermitian} operators only,
he misses the fact that the eigenbases of \emph{unitary }operators
may also form sets of states which are ``informations-fremd''. In
dimension $d=2$, however, he identifies the triple of MU bases associated
with the three observables corresponding to orthogonal spin components
which are, of course, both Hermitian and unitary.

Hadamard's influential paper generated a large body of research limited
to matrices with real elements only, although Sylvester's earlier
work allowed for complex matrix elements of modulus one. Historically
speaking, a more accurate naming convention would be to call ``complex
Hadamard matrices'' \emph{Sylvester matrices}, and to reserve the
label \emph{Hadamard matrices }to those with real entries only.\emph{
}In 1962, Butson \citep{butson62} brought Sylvester's \emph{complex}
Hadamard matrices back to the fold, by initiating the study of complex-valued
Hadamard matrices with entries given by finite roots of unity (cf.
Sec.~\ref{subsec:MU-butson-type}).

It took another 20 years for further contributions relevant for MU
bases to emerge. Then, however, within about two years, we witness
progress in several unrelated fields: \emph{signal processing}, \emph{Lie}
or, more generally, \emph{matrix algebras}, and\emph{ maximal Abelian
subalgebras. }These developments all revolve around the mathematical
structure which underpins mutual unbiasedness. MU bases also resurface
in quantum theory, in the context of \emph{quantum state determination
}(see Sec.~\ref{subsec: MU bases prime power dimensions} for this
topic)\emph{.}

\emph{Correlations of complex signals—}Periodic sequences of complex
numbers and their correlations were studied in great detail in the
1970's, to set up efficient communication protocols with radar signals.
Given sets of sequences of $d$ complex numbers, it is important to
control both the overlap between a sequence and its (shifted) complex
conjugate and between two different sequences within one set. These
quantities are known as \emph{auto}- and \emph{cross}-correlations,
respectively. In 1979, Sarwate \citep{Sarwate1979} shows that a trade-off
relation between good auto-correlations and good cross-correlations
exists. Since ``good'' in this context means ``small'', the two
types of correlations should be minimised simultaneously. This requirement
gives rise to an additive type of uncertainty relation, structurally
similar to the one which applies to the position and momentum variables
of a quantum particle, for example.

For odd prime values of $d$, Sarwate identifies $(d-1)$ equimodular
sequences which minimise the trade-off relation. He points out that,
by adding the standard basis of the space $\mathbb{C}^{d}$, there
are $d$ sets of sequences in total which saturate the trade-off relation.
The conditions satisfied by these sequences are equivalent to those
satisfied by $d$ MU bases in the space $\mathbb{C}^{d}$. Since $d$
MU bases uniquely determine an additional basis MU to the given ones
(see Sec.~\ref{subsec:d_MU_bases_sufficient}), Sarwate had, in fact,
implicitly constructed a complete set of MU bases.

The existence of sequences with desirable \emph{auto}-correlations
in terms of complex exponentials depending on quadratic polynomials
had already been noticed by Chu in 1972 \citep{Chu1972}. Sarwate's
contribution consists of adding a condition on the \emph{cross}-correlations
between different sequences, ensuring the mutual unbiasedness of different
bases.

Alltop \citep{alltop80} constructs $(d+1)$ periodic sequences in
$\mathbb{C}^{d}$ for prime numbers $d$ greater than three (cf. Appendix
\ref{subsec:Alltop's-construction}). His result is based on the properties
of complex third-order polynomials as arguments of exponential functions.
In addition to these \emph{cubic} phase sequences, he also constructs
\emph{quadratic} phase sequences, pointing out that they essentially
reproduce Sarwate's result of $d$ sequences.

\emph{Orthogonal decompositions of Lie algebras}—Still in 1972, Kostrikin
\textit{et al}. \citep{kostrikin+81} initiated a search for orthogonal
decompositions of the Lie groups of type $A_{d}$, having Lie algebras
$sl_{d+1}$. Observations made for values of $d\leq8$ resulted in
the \emph{Winnie-the-Pooh problem }\eqref{WtP problem} and the conjecture
that decompositions would only exist for prime-powers, i.e. when \emph{$d+1=p^{n}$},
for any $p\in\mathbb{P}$, $n\in\mathbb{N}$. In Sec.~\ref{subsec: Orthogonal decomp equivalence}
we describe the problem in detail and explain how it achieved its
name. A direct link between orthogonal decompositions of certain Lie
algebras and complete sets of MU bases was established only in 2005
\citep{boykin+07}.

\emph{Maximal Abelian subalgebras }(MASAs)—In 1983, Popa \citep{popa83}
investigates mutually orthogonal and maximal Abelian $\ast$-subalgebras
(MASAs) of von Neumann algebras. The results pertaining to finite-dimensional
matrix algebras are, with hindsight, statements about the existence
of MU bases (cf. Sec.~\ref{subsec: masa}). Effectively, Popa constructs
$(p+1$) MASAs for complex matrices of order $p\times p$, i.e. in
dimensions $d=p$. When the dimension $d$ is not a prime, at least
$(p_{1}+1)$ MASAs are constructed, with $p_{1}$ being the smallest
prime divisor of $d$. It is stated clearly that there are at least
three MASAs in \emph{all} dimensions greater than two, $d\geq2$,
and that there cannot be more than $(d+1)$ MASAs for dimension $d$.
In today's quantum parlance, Popa's arguments are based on the Heisenberg-Weyl
algebra, using ``clock-and-shift'' operators.

Popa conjectures that in prime dimensions all pairs of MASAs are equivalent
to the standard Fourier pair. This claim was disproved first by de
la Harpe and Jones \citep{Harpe1990} who show that for dimensions
$p\geq13$ and $p\equiv1\mod 4$, other pairs exist. Munemasa and
Watatani \citep{Munemasa1992} extend this result to primes $p$ equal
to or larger than seven if they are of the form $p\equiv3\mod 4$.
Thus, pairs of MU bases inequivalent to the standard Heisenberg-Weyl
pair exist in \emph{all} odd prime dimensions $d\geq7$. Interestingly,
these constructions make use of graph theory—Payley graphs in particular—a
topic which has figured in the context of MU bases only rarely.

A decade later, Haagerup \citep{haagerup97} showed in a \emph{tour
de force }that in dimension $d=5$, all pairs of MU bases are unitarily
equivalent to the standard Fourier pair. This had been the only open
case left after the work in the sequel of Popa's conjecture. Just
as with Sylvester and Kraus \citep{Kraus1987} (cf. below), Haagerup
``dephases'' Hadamard matrices in order to remove trivial equivalences.
As an aside, he also introduced a necessary criterion for the equivalence
of Hadamard matrices of the same order. Given an $n\times n$ Hadamard
matrix $H$, construct its \emph{Haagerup set }$\Lambda(H)$ defined
in Eq.~\eqref{eq:Haagerup_set}, which contains the products of four
matrix elements $H_{jk}$, $j,k=1,\ldots,n$. By construction, this
set of products is invariant under dephasing transformations so that
two Hadamard matrices with different Haagerup sets cannot be equivalent;
there exist, however, inequivalent Hadamard matrices with identical
Haagerup sets (see Appendix \ref{sec:Complex-Hadamard-matrices}).

Haagerup also establishes a close connection between MU bases and
the problem of \emph{cyclic $n$-roots}, i.e., the solutions of a
highly symmetrical set of $n$ equations (see Sec.~\ref{subsec:biunimodular})
introduced by Björck \citep{bjorck85} in 1985. The equations arise
when searching for all \emph{`bi-equimodular' vectors $x\in\mathbb{C}^{n}$,
i.e. all $x$ with coordinates of constant absolute value such that
the Fourier transform of $x$ is a vector with coordinates of constant
absolute value} \citep[p. 331]{bjork+91}, and it was solved for small
values of $n$ using computer-algebraic methods. When $n=5$, for
example, only \textquotedbl classical\textquotedbl{} roots (i.e. of
Fourier-type) exist. In contrast, among the 156 cyclic $6$-roots
given by Björck and Fröberg \citep{bjork+91}, Haagerup identifies
48 which are of modulus one, and thus are relevant in the context
of MASAs or, equivalently, MU bases. Twelve of these roots are the
expected \textquotedbl classical\textquotedbl{} ones and the remaining
36 (which Haagerup spells out explicitly) are of a different type
(see Sec.~\ref{sec:fourier_family} for further analysis).

Kraus, in 1987, considered quantum systems with finite-dimensional
state spaces and, referring back to Schwinger \citep{schwinger60},
used mutual unbiasedness to define pairs of \emph{complementary }observables:
\emph{exact knowledge of the measured value of one observable implies
maximal uncertainty of the measured value of the other} \citep[p. 3070]{Kraus1987}.
He shows that all pairs for $d=2$ and $d=3$, respectively, are equivalent
and explains how to construct complementary pairs of bases in systems
of product dimensions $d=p_{1}p_{2}$, namely by tensoring MU bases
of the factors $\mathbb{C}^{p_{j}}$, $j=1,2$, of the space $\text{\ensuremath{\mathbb{C}^{p_{1}p_{2}}}}$.
However, this construction does not necessarily exhaust all MU pairs,
as he shows for the case $p_{1}=p_{2}=2$. The set of all inequivalent
pairs of MU bases in dimension $d=4$ is presented in the form of
a dephased complex Hadamard matrix (see Sec.~\ref{subsec: Hadamard equivalence})
which depends on \emph{one} free real parameter. Kraus states that
a similar construction for $d=8$ leads to a family of MU pairs depending
on \emph{five} real parameters, while declaring the general classification
to be an (allegedly complicated) open problem. Thus, Kraus provides
early examples both of non-unique complementary \emph{pairs} in product
dimensions and of the existence of continuous families of such pairs
(cf. Sec.~\ref{subsec:Complementarity} for his contribution in the
context of complementarity and Sec.~\ref{subsec:Equivalence-classes}
for the general definition of equivalence classes of sets of MU bases). 

\subsection{Prime-power dimensions }

\label{subsec: MU bases prime power dimensions}

Nearly 30 years after Weyl's and Schwinger's work, the concept underlying
MU bases re-emerged in the 1980's in contributions addressing a fundamental
problem of quantum theory: \emph{quantum state reconstruction,} which
asks for a unique characterisation of arbitrary mixed quantum states
in terms of expectation values. 

In 1981, Ivanović exploited the properties of Gauss sums to construct
states which form $(p+1)$ MU bases in spaces with prime dimensions
$p\in\mathbb{P}$ \citep{ivanovic81}. More specifically, he decomposed
the space of unitary operators into orthogonal hyperplanes which are
associated with sets of commuting operators suitable for quantum state
determination. Ivanović appears to be the first to show that no more
than $(p+1)$ MU bases can exist in the space $\mathbb{C}^{p}$. This
result goes beyond Schwinger who also constructed $(p+1)$ MU bases
but did not show this to be maximal. 

A few years later, Wootters and Fields extended Ivanović's contribution
from Hilbert spaces with prime dimension to spaces of prime-power
dimension $d\in\mathbb{PP}$, both even and odd \citep{wootters+89}.
The authors introduce the notion of \emph{mutual unbiasedness} both
for sets of states forming orthonormal bases and for the associated
measurements, i.e. sets of commuting projections onto one-dimensional
orthogonal subspaces of a Hilbert space $\mathbb{C}^{d}$. The construction
rests on properties of \emph{Galois} (or \emph{finite}) \emph{fields}
with $d$ elements, which exist as long as the dimension of the space
is a power of a prime number. It is also shown that the outlined method
of state reconstruction is \emph{optimal} in the sense that the required
measurements come with the smallest possible statistical errors. Furthermore,
the upper limit of $(d+1)$ MU bases in the space $\mathbb{C}^{d}$
is derived.

Today, many other ways to construct complete sets of MU bases in prime-power
dimensions are known, some of which we present in Appendix \ref{sec: complete sets in pp dimensions}
(see also the review \citep{durt+10}). According to Godsil and Roy
\citep{godsil+09}, nearly all complete sets of MU bases known (in
2009) could, effectively, be obtained from one single construction
presented by Calderbank \emph{et al. }\citep{calderbank+97} in their
study of \emph{symplectic spreads }and $\mathbb{Z}_{4}-$Kerdock codes
used for error correction. Abdukhalikov \citep{Abdukhalikov2015}
introduced an independent construction of complete sets of MU bases
and classified the existing methods.

Are complete sets of MU bases unique for a given prime-power dimension?
Two sets of MU bases are said to be \emph{equivalent} under unitary
transformations if there is a single unitary which maps one set to
the other. For example, all complete sets of MU bases in $\mathbb{C}^{d}$,
with $2\leq d\leq5$, are equivalent \citep{haagerup97,BrierleyWB2010};
in other words, there exists only a single set of $(d+1)$ MU bases
in each of these dimensions (apart from trivial dephased versions). 

The smallest dimension in which \emph{inequivalent} complete sets
of MU bases are known to occur is $d=27$, as pointed out by Kantor
in 2012 \citep{kantor12}. To reach this conclusion, Kantor first
showed that a symplectic spread determines a complete set of MU bases
and, secondly, uses the fact that two spreads are equivalent to each
other if and only if these spreads can be mapped to each other by
a symplectic transformation (cf. \citep{calderbank+97}). Inequivalent
symplectic spreads were identified already in 1983, for example, if
$d=2^{n}$ with an odd number $n>3$, in the context of coding theory
\citep{kantor83}. Explicit examples of inequivalent maximal sets
of MU bases have been given in \citep{Garcia2010}, for $d=p^{n}\in\mathbb{PP}$,
with an odd prime $p$. Some erroneous examples are presented in \citep{sehrawat14}.

The focus of our review is the notable \emph{lack} of MU bases in
\emph{composite }dimensions rather than the structure of MU bases
in prime-power dimensions. Initially barely worth mentioning, the
curious non-existence of complete sets in those dimensions has evolved
into a formidable and long-standing open problem. In the following
section we retrace the steps which led researchers to become aware
of this problem.

\subsection{Composite dimensions $d\protect\notin\mathbb{PP}$}

\label{subsec:Composite-dimensions}

Many early papers on MU bases were focused on the explicit construction
of\emph{ }complete sets in prime and prime-power dimensions, $d\in\mathbb{PP}$.
Checking the first few integers for the relative frequencies of prime
and composite numbers suggests that composite dimensions ($d\not\notin\mathbb{PP}$)
are rare: up to $d=27$, say, five out of six number are primes or
prime-powers. However, this observation is deceptive: fewer than 0.1\%
of the numbers below $10^{1000}$ are primes or prime-powers. In fact,
the probability for a randomly sampled positive integer less than
$N$ to be composite approaches the value one for $N\to\infty$,
\begin{equation}
\mbox{prob}(\text{composite }\mbox{\ensuremath{d<N}})\simeq1-\frac{1}{\log N}\,.\label{eq prob of comp dim}
\end{equation}
This relation follows from Hadamard's and de la Vallée Poussin's \emph{prime-number
theorem} upon ignoring the minute contribution of the prime-powers.
The punchline is that for most quantum system with a large, randomly
chosen dimension $d$ we currently are \emph{not} in a position to
construct a complete set of MU bases. Let us review how the important
insight into this shortcoming emerged and describe some of the attempts
to clarify the unexpected situation.

In the early 1960's, both Schwinger's construction of unitary operator
bases \citep{schwinger60} and Butson's construction of complex Hadamard
matrices \citep{butson62} were set in spaces of arbitrary integer
dimensions $d$. However, it did not occur to them to look for a possible
difference between spaces of prime-power and composite dimensions;
effectively, they were interested in problems which translate into
a search for \emph{pairs }of MU bases known to exist in \emph{all}
dimensions $d\in\mathbb{N}$.

Nearly two decades later and only \emph{en passant}, Sarwate \citep{Sarwate1979}
and Alltop \citep{alltop80} mention the case of complex sequences
with periods given by \emph{composite} numbers, $d\notin\mathbb{PP}$.
They point out that their techniques can still be used to construct
some sequences with the desired properties but only a small number,
limited by the lowest prime factor $p_{1}$ of $d$ if the dimension
is given by $d=p_{1}p_{2}\ldots p_{r}$, with $p_{1}<p_{2}<\ldots<p_{r}$. 

Soon after, Kostrikin \textit{et al}. \citep{kostrikin+81} envisage—apparently
for the first time—that complex Hilbert spaces of composite and prime-power
dimensions, respectively, might differ fundamentally when constructing
specific mathematical objects. If there is no solution to Winnie-the-Pooh's
problem (cf. Sec.~\ref{subsec: Orthogonal decomp equivalence}),
i.e. the desired orthogonal decomposition of the relevant Lie algebra
does not always exist, then the number of MU bases in a Hilbert space
of composite dimension must drop \emph{below} $(d+1)$, the strict
upper bound on their number in any dimension.

In the context of MASAs (cf. Sec.~\ref{subsec: avant-la-letter}),
Popa \citep{popa83} obtained a similar result about MU bases in spaces
of composite dimension $d$: he showed how to construct at least $(p_{1}+1)$
MASAs where $p_{1}$ is the smallest prime divisor of $d$. In addition,
he clearly states in his paper from 1983—possibly for the first time—that
there are at least three MASAs, and hence MU bases, in \emph{all}
dimensions $d\geq2$.

Independently of Sarwate and Alltop, Kraus \citep{Kraus1987} spells
out in 1987 the tensor-product construction of pairs of complementary
observables for bipartite finite quantum systems of dimension $d=p_{1}p_{2}$.
As already mentioned, he was aware of the fact that pairs of MU bases
in $\mathbb{C}^{p_{1}p_{2}\ldots}$ can be found from tensoring MU
bases, but he does not launch a systematic investigation into the
construction of larger sets in composite dimensions.

While Wootters and Fields successfully construct complete sets of
MU bases in all prime-power dimensions \citep{wootters+89}, they
also point out in their 1989 paper that the method will \emph{not}
work in composite dimensions $d\notin\mathbb{PP}$. Since ``there
is no finite field whose number of elements is not a power of a prime'',
any method producing a complete set of MU bases in other dimensions
``must be very different from the procedures used here'' \citep[p. 376]{wootters+89}.
Lacking such a complete set would rule out, for example, the optimal
state reconstruction method presented in their paper. 

In an unpublished diploma thesis completed in 1991, Zauner proposes
a ``Vermutung''—i.e. an educated guess—which states that ``there
are no four non-commuting $6\times6$ matrices, each having six different
eigenvalues, which are independent with respect to\emph{ $\mathbb{I}/6$}''
\citep[p. 75]{Zauner1991}. His notion of independence is an alternative
way to demand that the matrices represent MU bases, with their column
vectors possessing the correct overlaps. The main thrust of Zauner's
study had been to develop the concept of independent observables in
quantum theory in close analogy to the concept of independent observables
in classical probability theory.

Retrospectively—as pointed out by Klappenecker and Rötteler \citep{klappenecker+04}
in 2004—the main novelty in Zauner's work is the explicit conjecture
that in the Hilbert space of composite dimension $d=6$, the number
of MU bases is limited to \emph{three} (and not \emph{seven, }which
the prime-power case would suggest). In his thesis from 1999, the
author surmises that such a limitation exists, using the language
of quantum designs \citep{Zauner1999} (see \citep{Zauner2011} for
an English translation). Here we reproduce the original statement
corresponding to the non-existence conjecture presented in Sec.~\ref{subsec:Motivation}.
\begin{conjecture}[Zauner 1999]
Presumably a complex, affine quantum design with $b=g=6$, $r=1$,
$\lambda=1/6$, and $k=4$ does not exist either. \citep[p. 494]{Zauner2011}
\end{conjecture}
A \emph{quantum design} (see Sec.~\ref{subsec:zauner_conjecture})
consists of $k$ sets of $g$ orthogonal projection matrices in a
$b$-dimensional complex Hilbert space. The rank of each projector
is $r$, and the scalar product of two projectors belonging to different
sets is given by $\lambda$. The case of rank-one projectors, $r=1$,
corresponds to the case of MU bases if $\lambda=1/d$, and the design
is \emph{affine} (or of degree 2) if the scalar products between any
two projectors are equal to either zero or $\lambda$.

Zauner's construction of three MU bases in the space $\mathbb{C}^{6}$
is based on the triple existing in dimension two, i.e. the smallest
factor of six. More generally, this method leads to $(p^{n_{1}}_{1}+1)$
MU bases where $p^{n_{1}}_{1}$ is the smallest prime-power factor
in the decomposition of the dimension $d=p^{n_{1}}_{1}p^{n_{2}}_{2}\ldots p^{n_{r}}_{r}$
\citep{klappenecker+04}. This suggests the (false) \emph{Claim} in
Sec.~\ref{subsec:Motivation} that the upper-bound on the number
of MU bases in a composite dimension is $(p^{n_{1}}_{1}+1)$.

In a long and influential paper published in 1997, Haagerup \citep{haagerup97}
explores the existence of Hadamard matrices—and thus MU bases—in composite
dimensions. He shows that for dimensions $d=pq$ it is possible to
construct families of Hadamard matrices depending on $(p-1)(q-1)$
parameters; the implied two-parameter set for $d=6$ is given explicitly.
Two further Hadamard matrices are constructed which are inequivalent
to the $6\times6$ Fourier matrix. These results are important, of
course, because each known complex Hadamard matrix may represent a
basis in a larger set of MU bases. From 2004, onwards, new $6\times6$
Hadamard matrices have appeared sporadically, including Diţă's one-parameter
family \citep{dita04}, a three-parameter family by Karlsson that
encompasses all earlier non-isolated examples \citep{karlsson11},
and a non-constructive proof by Bondal and Zhdanovskiy that a four-parameter
family exists \citep{bondal16}.

In 2005, Archer \citep{archer05} attempted to systematically transfer
known constructions of complete sets of MU bases in prime-power dimensions
to composite dimensions, without success. This negative outcome explains
the growing interest in alternative ways to search for larger sets
of MU bases, championed mainly by researchers with a background in
quantum theory.

Around the same time, Grassl used \emph{computer-algebraic methods
}(Sec.~\ref{sec:fourier_family}) to prove that no pair of Heisenberg-Weyl-type
MU bases can be extended to more than three bases in $d=6$ \citep{grassl04}.
This important result, confirming the findings obtained for cyclic
$n$-roots (see Sec.~\ref{subsec:biunimodular}), paved the way for
further computer-assisted studies carried out in the quantum community.
In 2007, Bengtsson \textit{et al}. \citep{bengtsson+07} searched
\emph{numerically }for Butson-type Hadamard matrices of order six
by restricting the elements to $d$-th or $(2d)$-th roots of unity,
based on the observation that all known complete sets of MU bases
were of this particular form. This paper also introduced a \emph{distance
}between bases in a Hilbert space, allowing one to frame the search
for MU bases in terms of a global minimum of a scalar function (Sec.~\ref{subsec:nonexistence_of_quadruples}).
Butterly and Hall took up the idea in an attempt to identify \emph{four}
MU bases in dimension six \citep{butterley+07} but failed to find
such a set. A measure of unbiasedness related to the average success
probability of quantum random access codes was shown by Aguilar \textit{et
al}. \citep{aguilar18} to be useful for the same purpose.

The numerical approach was refined soon after by introducing the concept
of \emph{MU} \emph{constellations }\citep{brierley+08}\emph{, }defined
as sets of vectors which are either MU or orthogonal, but not necessarily
forming entire orthonormal bases. The increased flexibility leads
to considerable computational simplifications. In dimension six, for
example, it is much less onerous to numerically search for a single
state that is mutually unbiased to three MU bases than to deal with
a set of four MU bases. The results are among the most extensive numerical
data, strongly supporting the non-existence of constellations containing
more than three MU bases (cf. Sec.~\ref{subsec:nonexistence_of_constellations}).

Some success—in the form of analytic results—has been achieved by
strategies that impose additional constraints on sets of MU bases.
For example, we now have analytic proofs that no more than three \textit{nice}
\citep{aschbacher+07}, \textit{monomal} \citep{boykin+07}, or \textit{product}
\citep{mcnulty+12impossibility} MU bases exist in a Hilbert space
of dimension six (cf Secs.~\ref{subsec:Nice-error-bases}-\ref{subsec:MU-product-bases}).
Bases are ''nice'' or ''monomial'' if the unitary operator basis from
which they are constructed is a nice error basis or a monomial basis,
respectively. Both of these properties are shared by all known complete
sets of MU bases in prime-power dimensions.

It is also possible to express the existence problem as a \emph{semi-definite
program} \citep{brierley+10} (cf. Sec.~\ref{subsec:Solution-strategies}).
But again, no results have been obtained which go beyond a proof of
principle, i.e. the reproduction of known results in low dimensions.

It is instructive to compare the number of \emph{free parameters}
in $(d+1)$ arbitrary bases of the space $\mathbb{C}^{d}$ with the
number of constraints which these $d(d+1)$ states must satisfy to
represent a complete set of MU bases \citep{brierley+08}. In dimension
seven, for example, 56 pure states—which depend on 288 independent
parameters—have to satisfy 1176 constraints. This example shows that
the surprising feature of MU bases is, actually, not the \emph{absence
}of\emph{ }complete sets in composite dimensions but their \emph{existence}
in prime-power dimension, for any $d>2$. The constraints must align
with some fundamental structure prevailing in the space $\mathbb{C}^{7}$
and degenerate in some way, which then allows for the existence of
a complete set. It is likely that the number-theoretic consequences
of $d$ being prime—such as the existence of a suitable Galois field—play
a central role. This perspective explains, in some sense, why it will
be hard to prove Zauner's conjecture: spaces of composite dimensions
do not support structures which, in the case of prime-power dimensions,
allow for the existence of complete sets.

Could it be that the existence of maximal sets of MU bases is an\emph{
undecidable }question?\emph{ }The answer is no: by coarse-graining
the parameter space, Jaming \textit{et al}. \citep{jaming+09,jaming+10}
showed that the existence problem can be expressed in terms of a set
of rigorous inequalities which, in principle, can be checked numerically
to hold (or not) in any composite dimension (cf. the discussion of
Thm.~\ref{thm:Fourierfamilyexclusion}). However, the resources required
to implement this \emph{algorithm} are formidable, even for $d=6$.
As a proof of principle, a restricted case—rather than the general
existence problem in dimension six—has been studied successfully:
it is impossible to extend pairs consisting of the identity and any
member of the Fourier family to a quadruple of MU bases. Other algorithmic
approaches to the existence problem exist which do not rely on numerical
approximations but remain unfeasible from a practical point of view
\citep{Cifuentes23}. 

In view of the overall difficulty to resolve the existence problem,
various authors have introduced\emph{ variations} of the original
problem which will be reviewed in Sec.~\ref{sec: Modifications-of-the-problem}.
The modifications either relax the original constraints or impose
additional\emph{ }ones to define new but structurally similar existence
problems. For example, the problem may be set in a real, quaternionic,
$p$-adic or infinite-dimensional Hilbert space, instead of the standard
\emph{complex} Hilbert space. Alternatively, one can search for measurements
which satisfy an approximate condition of unbiasedness, or non-projective
measurements which satisfy another form of complementarity.

\section{MU bases in quantum theory}

\label{sec: MU-bases-in-QT}

This section will illustrate the role played by mutually unbiased
bases in quantum physics. We describe their relation to other important
concepts, ranging from foundational aspects such as Kochen-Specker's
theorem to experimental realisations of MU bases. Applications of
MU bases, i.e. scenarios in which they are crucial to achieve specific
tasks such as quantum cryptography or quantum state reconstruction,
will be considered separately, in Sec.~\ref{sec:applications}. Nevertheless,
both sections aim at answering the question raised in 2005 \citep{bengtsson05}:
``Given a complete set of MU bases, what can we do with them?''—and
thereby motivating the search for complete sets in composite dimensions.

\subsection{Quantum degrees of freedom}

\label{subsec:Quantum-dofs}

In his paper on the representations of the Heisenberg-Weyl group,
Schwinger\emph{ }\citep{schwinger60}\emph{ }considered Hilbert spaces
of finite dimension $d$, with prime decomposition $d=p_{1}p_{2}\ldots p_{k}$.
He proposed to associate one degree of freedom to each of the factors.
This suggestion is motivated by the idea that a quantum degree of
freedom should be associated with an irreducible representation of
operators satisfying the canonical commutation relations, as the position
and momentum of a quantum particle do. However, in a finite-dimensional
Hilbert space only ``integrated'' versions of these commutation
relations exist, satisfied by operators forming the Heisenberg-Weyl
algebra. Hence, a quantum system with a six-dimensional state space
$\mathbb{C}^{6}$ would have two degrees of freedom.\footnote{Connes is reported to have expressed this idea in the context of quantum
field theory when stating that he ``was immediately led to the idea
that somehow passing from the integers to the primes is very similar
to passing from quantum field theory, as we observe it, to the elementary
particles, whatever they are'' \citep[p. 204]{sabbagh02}.}

While natural, the conceptual advantage of defining degrees of freedom
in this way remains unclear, and it is not obvious whether observable
consequences follow. Some quantum systems may have six orthogonal
levels because they are indeed composed of two distinct physical systems
with two and three orthogonal states, respectively; hence, two degrees
of freedom may be introduced naturally as they are associated with
the subsystems. However, the decomposition into distinct parts may
not be of interest \citep{goryca14} or, in the case of a—yet to be
discovered—elementary particle with spin $s=5/2$, no physical constituents
may exist. 

It is an intriguing idea to associate degrees of freedom to irreducible
representations. However, combined with a hypothetical lack of complete
sets of MU bases in composite dimensions, an awkward \emph{structural}
discrepancy would emerge \citep{brierley+08}, raising physicists'
eyebrows. Consider systems with four and six levels, for example,
both of which would be endowed with two degrees of freedom. Nevertheless,
the collections of observables in these systems possess fundamentally
distinct properties if a complete set of MU bases exists only for
one of them. Such a difference seems acceptable only when the number
of degrees of freedom differs, e.g. when comparing systems of dimensions
five and six.

\subsection{Quantum state-space geometry}

\label{sec:geometric}

In 1981, Ivanović sets up a procedure to determine unknown quantum
states (cf. Sec.~\ref{subsec: Quantum-state-reconstruction}) for
systems with Hilbert space $\mathbb{C}^{d}$ \citep{ivanovic81}.
He points out that a complete set of $(d+1)$ MU bases (if it exists)
induces a highly symmetric, orthogonal decomposition of the quantum
state space, i.e. the density operators of the system. 

To derive this decomposition, the set of $d\times d$ Hermitian matrices
is considered as a real Euclidean space with dimension $d^{2}$, the
scalar product of two Hermitian matrices being defined by the trace
of their product. Non-negative density matrices $\rho$ of a $d$-level
system are a subset of that space satisfying the normalisation condition
$\text{Tr}(\rho\,\mathbb{I}/d)=1/d$; hence, they live in a $(d^{2}-1)$-dimensional
hyperplane of the space. The projectors $P_{b}(v)=\kb{v_{b}}{v_{b}}$
and $P_{b^{\prime}}(v^{\prime})=\kb{v^{\prime}_{b^{\prime}}}{v^{\prime}_{b^{\prime}}}$
on states contained in different MU bases, i.e. for $b\neq b^{\prime}$,
are \emph{not }orthogonal to each other according to Eq.~\eqref{eq: completeMUsetfor(d+1)},
since
\begin{equation}
\text{Tr}\left(P_{b}(v)\,P_{b^{\prime}}(v^{\prime})\right)=\frac{1}{d}\,.\label{eq: non-orthogonality}
\end{equation}
However, the scalar products between their ``shifted'' versions
$\overline{P}_{b}(v)=\kb{v_{b}}{v_{b}}-\mathbb{I}/d$, etc. do vanish,
which means each MU basis is confined to a plane orthogonal to the
plane associated with any other MU basis. Taken together, the $(d+1)$
planes provide—upon adding in the identity $\mathbb{I}$—an orthogonal
decomposition of the state space, explaining the usefulness of MU
bases for quantum state reconstruction.

This geometric picture has been developed further by Bengtsson and
Ericsson \citep{Bengtsson2005,bengtsson05}. By defining the so-called
\emph{complementarity polytope}, the existence problem of MU bases
can be expressed in an entirely geometric setting. A complete set
of MU bases gives rise to this convex polytope in the space of Hermitian
matrices in the following way: construct $(d+1)$ regular simplices
by associating $d$ equidistant vertices to the projectors of each
MU basis and form the convex hull of the $(d^{2}+d)$ vertices characterising
the non-overlapping simplices.

As a geometric object, the \emph{complementarity polytope }may be
rotated in the Euclidean space of Hermitian matrices. In general,
its vertices will move in such a way that they are no longer associated
with rank-one projection operators but other non-positive Hermitian
matrices, except when $d=2$. In other words, the rotated complementarity
polytope will not necessarily sit inside the body of density matrices.
However, every construction of a complete set of MU bases successfully
embeds the complementarity polytope in the set of density matrices,
associating projection operators with its vertices.

Turning this observation around, the existence problem of complete
sets of MU bases can be expressed in the following form: given a complementarity
polytope in the $d^{2}$-dimensional Euclidean space, is it possible
to rotate it in such a way that it will fit inside the body of density
matrices? This geometric picture links the existence problem with
another problem known to be difficult, namely to efficiently characterise
the structure of the convex body of density matrices of quantum systems.
Already for a qutrit, the eight-dimensional analog of the Bloch ball
turns out to be an intricate geometrical object (see e.g. \citep{eltschka21}).

\subsection{Complementarity}

\label{subsec:Complementarity}

Early insights into the properties of pairs of quantum mechanical
observables which give rise to MU bases were mentioned in Sec.~\ref{subsec: avant-la-letter}.
However, the buzzword of the times was \emph{complementarity}, proposed
by Bohr in 1927 as a fundamental property of quantum systems, albeit
expressed in a rather non-technical fashion (see the write-up of his
Como lecture \citep{Bohr1928}, for example). Complementarity is at
the heart of uncertainty relations: if one measures one observable
of a complementary pair—the position of a quantum particle, say—then
the outcomes of a subsequent measurement of the other observable—momentum—are
maximally uncertain. In Bohr's words:
\begin{quote}
The momentum of a particle, on the other hand, can be determined with
any desired degree of accuracy {[}$\ldots${]}, but then the determination
of the space co-ordinates of the particle becomes correspondingly
less accurate.\emph{ }\citep[p. 582]{Bohr1928}
\end{quote}
One way to give meaning to the statement that two observables are
complementary is to say that their eigenstates form a \emph{pair of
mutually unbiased bases} satisfying Eq.~\eqref{eq: def MU pairs}
of Definition \ref{def: MU pair of bases}. This is a neat and satisfactory
approach to complementarity which, however, has not yet made its way
into the average textbook. Many of the traditional ways to think about
complementary observables have been reviewed in \citep{busch+95}.

Weyl and Jordan made early \emph{qualitative} statements about the
complementary observables of a quantum particle (cf. Sec.~\ref{subsec: avant-la-letter}).
In 1960, Schwinger provided a \emph{quantitative} analysis of this
concept by setting it in a state space of finite dimension $d$. He
considers a pair of ``properties $U$ and $V$'' described by unitary
operators satisfying the relation \eqref{eq: HW group commutator},
and shows that the transition probabilities between their eigenstates
are constant,
\begin{equation}
p(u^{\prime},v^{\prime\prime})=\left|\bk{u^{\prime}}{v^{\prime\prime}}\right|^{2}=\frac{1}{d}\,,\label{eq: trans prob 1}
\end{equation}
where $u^{\prime}$ and $v^{\prime\prime}$ label the eigenvalues
of the unitaries $U$ and $V$, respectively. Next, Schwinger considers
the expression 
\begin{equation}
p(u^{\prime},v,u^{\prime\prime})=\sum_{v^{\prime}}p(u^{\prime},v^{\prime})p(v^{\prime},u^{\prime\prime})=\frac{1}{d}\,,\label{eq: trans prob 2}
\end{equation}
i.e. the probability to find a specific value of $u^{\prime}$ given
that the system starts out in another eigenstate of $U$ with label
$u^{\prime\prime}$, say, and assuming an intermediate measurement
in the $V$ basis. The independence of the probability $p(u^{\prime},v,u^{\prime\prime})$
of all its arguments expresses the fact that the intermediate measurement
of $V$ completely erases any information about the initial state.
In Schwinger's own words, Eq.~\eqref{eq: trans prob 2} shows that
``the properties $U$ and $V$ exhibit the maximum degree of incompatibility''
\citep[p. 575]{schwinger60}. Finally, he remarks that this property
is tantamount to ``the attribute of complementarity'' (p. 579, footnote
3).

Observables which either satisfy the canonical commutation relations
or their integrated version \eqref{eq: HW group commutator} are necessarily
complementary in the sense that their eigenbases are mutually unbiased,
i.e. the squared moduli of their overlaps are state independent, as
in Eq.~\eqref{eq: trans prob 1}. Accardi observed in 1984 that the
converse statement is not necessarily true: constant transition probabilities
between all states of two bases can also arise from operators which
do not satisfy the canonical commutation relations \citep{accardi84}.
Accordingly, he suggests classifying all pairs of observables with
this property, in both finite- and infinite-dimensional Hilbert spaces.
This problem is still open since, in the finite-dimensional case,
it means listing all complex Hadamard matrices of order $d$, a task
not yet realised even for dimension six (see \emph{Problem 10.4} in
Sec.~\ref{subsec:Solution-strategies}). In 2002, Cassinelli and
Varadarajan \citep{Cassinelli2002} answered another question raised
by Accardi's question, regarding the uniqueness of complementary projection-valued
measures, both in the finite- and the infinite-dimensional setting.

Schwinger's paper apparently prompted Kraus to systematically classify
pairs of complementary observables in low dimensions \citep{Kraus1987},
without mentioning Accardi. As described at the end of Sec.~\ref{subsec: avant-la-letter},
Kraus established the existence of parameter-dependent, \emph{unitarily}
\emph{inequivalent }families of MU bases for $d=4$ and $d=8$, while
he found pairs of complementary observables to be unique in dimensions
two and three, up to unitary equivalences.

In a paper written on the occasion of Carl Friedrich von Weizsäcker's
90$^{\textrm{th}}$ birthday, Brukner and Zeilinger revisit his idea
of the ``ur'' as an elementary carrier of information, endowed with
a two-dimensional (complex) Hilbert space. They consider ``mutually
complementary measurements'' which are associated with the orthogonal
components of a quantum spin, thereby using the triple of MU bases
for a spin-$\frac{1}{2}$ \citep{Brukner2002}. Their argument aligns
with Weizsäcker's tenet that a fundamental link might exist between
\emph{(i)} us necessarily performing experiments in \emph{three}-dimensional
Euclidean space and \emph{(ii)} a corresponding number of complementary
measurements. 

In a study of amplitude and phase operators, Klimov \textit{et al}.
(2005) introduce the notion of ``multicomplementary observables''
to denote sets of operators with MU eigenbases \citep{Klimov2005}.
Therefore, a Hilbert space of dimension $d$ can support up to $(d+1)$
pairwise complementary observables. \emph{Mutatis mutandis, }this
terminology also applies to a \emph{triple} of pairwise canonically
conjugate observables \citep{weigert+08} used to construct three
MU bases for a quantum particle described by a pair of canonically
conjugate observables (cf. Sec.~\ref{subsec: MUs for CVs} for details
on MU bases for continuous variables). 

To quantify complementarity, the authors of Ref. \citep{bandyopadhyay+13}
introduce a measure of the non-commuta\-ti\-vi\-ty of sharp observables
based on an operational approach, in a setting typical for quantum
key distribution. Their measure $Q(O)$ attains its maximum if the
observables in a set $O=\left\{ O_{1},\ldots,O_{\mu}\right\} $, $\mu\leq d+1$,
are pairwise mutually unbiased, i.e. when they have mutually unbiased
eigenbases. This result could be used to prove non-existence of sets
of MU bases in composite dimensions $d\notin\mathbb{PP}$ by showing
that the bound cannot be saturated. In dimension $d=6$, for example,
it would be sufficient to show that no four observables $O_{1},\ldots,O_{4}$,
exist which achieve this maximum value.

On a conceptual point, the close link between MU bases and complementarity
raises the question of the extent to which these concepts are genuinely
quantum mechanical. As a mathematical property, unbiasedness arises
in non-quantum settings as well, via Fourier transforms and Sarwate's
trade-off relations between auto- and cross-correlations, for example.
These observations notwithstanding, complementarity has not played
an important role in the description of classical systems.

Within quantum theory, \emph{incompatibility} and \emph{entropic uncertainty
relations }have, over time, supplanted the notion of complementarity
as they are more flexible and easier to handle, especially for systems
with finite-dimensional Hilbert spaces. The next two sections explain
how MU bases relate to these concepts.

\subsection{Incompatibility }

\label{subsec:inc}

In quantum theory, certain sets of measurements exhibit incompatibility,
meaning that they cannot be measured jointly on a single device \citep{heinosaari16}.
In particular, incompatible measurements cannot be obtained from the
marginals of a single (parent) positive operator-valued measure (POVM).
For projective measurements this property is equivalent to non-commutativity
but for general measurements, described by POVMs, joint measurability
does not imply commutativity. Interestingly, incompatibility shares
a one-to-one correspondence with Einstein-Podolsky-Rosen (EPR) steering
(cf. Sec.~\ref{subsec:Steering})—a type of uni-directional correlation
between a quantum state split between two parties \citep{wiseman07}—and
has close connections to non-locality, state discrimination and quantum
coherence.

One can quantify the incompatibility of a set of observables in terms
of its \emph{noise robustness}, defined as the minimum amount of uniform
noise needed to ensure the set is jointly measurable. For a pair of
observables $P$ and $Q$, the robustness is given by
\begin{equation}
I(P,Q)
= \inf \bigl\{ \lambda > 0 \;\big|\; (P_{\lambda}, Q_{\lambda}) \text{ are compatible} \bigr\},
\label{eq:robustness}
\end{equation}
where \emph{noisy} versions of the original observables are defined as
\begin{equation}
P_{\lambda}=(1-\lambda)P+\lambda\mathbb{I}/d\,,\label{eq:depolarising}
\end{equation}
and $Q_{\lambda}$, analogously. In general, it is difficult
to evaluate $I(P,Q)$ analytically unless the observables exhibit
a high degree of symmetry, like mutually unbiased bases. Usually,
semidefinite programming techniques are required, but even this is
limited to relatively small dimensions.

The incompatibility robustness of qubit measurements has been treated
exhaustively by Busch in 1986\emph{ }\citep{busch86}; a few decades
later the special case of MU bases connected by a Fourier transformation
(in arbitrary dimensions) was considered \citep{carmeli12}, followed
by a solution for any pair of MU bases as well as complete sets \citep{Designolle2018}.
Finally, the region of joint measurability for a pair of MU bases
with \emph{different} noise parameters, i.e. $P_{\lambda}$ and $Q_{\nu},$
was determined using state discrimination methods and an incompatibility
witness \citep{carmeli19}. The same noise bounds can be derived by
a relation between incompatibility and quantum coherence (cf. Sec.~\ref{subsec:coherence}).
The joint measurement at this noise boundary can be implemented sequentially
by quantum instruments \citep{kiukas20}.

Complete sets of $(d+1)$ MU bases seem to be among the most incompatible
observables with respect to the robustness measure of Eq.~\eqref{eq:robustness}.
When considering fewer measurements, POVMs have been found which exhibit
a stronger degree of incompatibility than MU bases \citep{bavaresco17}.
However, other quantifiers of incompatibility have been introduced
for which pairs of MU bases are found to be maximally incompatible
\citep{designolle19,mordasewicz22,tendick22}.

Incompatibility robustness can also be used to prove the existence
of operationally inequivalent sets of MU bases since the degree of
incompatibility depends on the particular choice of MU bases (for
sets with cardinality greater than two) \citep{Designolle2018}. A
further distinction between equivalence classes of MU bases arises
from the construction of extremal sets of compatible observables,
as not all MU bases lead to such sets \citep{carmeli19extremal}.

\subsection{Entropic uncertainty relations}

\label{subsec:Entropic-inequalities}

Over time, the traditional variance-based approach to uncertainty
relations has been complemented by the introduction of entropic uncertainty
relations. This development is partly driven by the focus on finite-dimensional
quantum systems where no two Hermitian operators exist which satisfy
the equivalent of the relation $i[\hat{p},\hat{q}]=\hbar$, the canonical
commutation relations for particle position and momentum observables.
Nevertheless, systems consisting of quantum particles and qudits,
respectively, are conceptually similar since mutual unbiasedness of
orthonormal bases does not rely on the existence of suitable Hermitian
operators. The unifying feature is the constant (i.e. basis-independent)
overlap between states taken from different orthonormal bases.

To establish an uncertainty relation in a Hilbert space of dimension
$d$, Kraus \citep{Kraus1987} uses the \emph{infor\-ma\-tion-theoretic
}or \emph{Shannon entropy }of an observable $P$,\emph{ }
\begin{equation}
S_{\rho}(P)=-\sum^{d}_{j=1}p_{j}\ln p_{j}\,,\qquad p_{j}=\text{Tr}[P(j)\rho]\,,\label{eq: Shannon entropy}
\end{equation}
where $P(j)$ denotes the projection operator onto the the $j$-th
eigenstate of $P$. The entropy $S_{\rho}(P)$ depends solely on the
probabilities $p_{j}$, $j=1\ldots d$, to obtain the $j$-th eigenvalue
of the observable $P$ when measuring it in a system prepared in state
$\rho$. Improving earlier results \citep{Deutsch1983,Partovi1983},
Kraus conjectures that given any two observables $P$ and $Q$ in
the space $\mathbb{C}^{d}$, their entropies must satisfy the inequality
\begin{equation}
S_{\rho}(P)+S_{\rho}(Q)\geq\ln d\,.\label{eq: entropic UR pairs}
\end{equation}
For $d\leq4$, he found this \emph{entropic uncertainty relation}
to be saturated if the observables $P$ and $Q$ form a complementary
pair (cf. Sec.~\ref{subsec:Complementarity}), i.e. if all outcomes
upon measuring the observable $Q$ are equally likely given that a
system resides in an eigenstate of $P$, and \emph{vice versa}. In
other words, the eigenstates of the operators $P$ and $Q$ are necessarily
MU. For arbitrary dimensions $d$, the relation\emph{ }\eqref{eq: entropic UR pairs}
was proved by Maassen and Uffink \citep{Maassen1988} in 1988 as a
special case of more general inequalities satisfied by probability
distributions with finitely many outcomes.

Ivanović raised the question of whether the entropic uncertainty inequality
\eqref{eq: entropic UR pairs} generalises to more than two MU bases
\citep{Ivanovic1992}. A complete set of MU bases corresponding to
the observables $P_{1},P_{2},\ldots,P_{d+1}$, is shown to obey the
inequality
\begin{equation}
\sum^{d+1}_{n=1}S_{\rho}(P_{n})\geq(d+1)\ln\left(\frac{d+1}{2}\right)\,,\label{eq: entropic UR multi}
\end{equation}
which, for $d\geq4$, strengthens an immediate bound obtained from
Eq.~\eqref{eq: entropic UR pairs} by suitably grouping the terms
in the sum. This result means that the multi-complementarity uncertainty
cannot be reduced to the uncertainty resulting from pairs of MU bases,
an observation which is also valid in the case of a triple of pairwise
complementary observables (see Sec.~\ref{subsec: MUs for CVs}) for
a quantum particle \citep{Kechrimparis2014}. Somewhat unexpectedly,
the bound \eqref{eq: entropic UR multi} distinguishes between even
and odd dimensions since there are no states which attain it for even
dimensions \citep{Sanchez-Ruiz1995}. 

Entropic uncertainty relations for both complete and smaller sets
of MU bases \citep{azarchs04,wu09}, as well as others valid in specific
dimensions, have been surveyed in Ref. \citep{Ballester2007}. A host
of additional state-dependent and state-independent uncertainty relations
were derived for MU bases using the Rényi and Tsallis entropies \citep{Rastegin2013}.

It is important to identify those states that minimise a given uncertainty
relation. For the Rényi entropy (of order 2), the minimal uncertainty
states for a complete set of MU bases include all fiducial vectors
of a Heisenberg-Weyl SIC-POVM (see Sec.~\ref{subsec:MU-bases-and-SICs})
in prime dimensions \citep{appleby14a} and the MUB-balanced states
\citep{wootters07,amburg14,appleby14b}. A MUB-balanced state is a
pure state which has the same outcome probability distribution regardless
of which MU basis from the complete set is measured.

Upper bounds on entropic sums, known as \textit{certainty} relations,
have been derived for arbitrarily many observables and any pure state.
Clearly, a state maximises a certainty relation if it is unbiased
to all measurement bases, therefore certainty relations for extendible
sets of MU bases yield only trivial upper bounds. This result is also
true for an arbitrary \textit{pair} of orthonormal bases due to the
existence of at least one vector mutually unbiased to both bases (see
Sec.~\ref{sec:larger_sets_of_MU_bases}). Examples of non-trivial
certainty relations have been obtained both for complete sets of MU
bases \citep{Sanchez-Ruiz1995} and for sets of orthonormal bases
with cardinality greater than two \citep{puchala15,canturk21}. It
has been conjectured that complete sets of MU bases (if they exist)
achieve the smallest difference between upper and lower bounds for
the average entropy among all sets of $(d+1)$ measurements with $d$
outcomes.

\subsection{Coherence in measurements}

\label{subsec:coherence}

Coherence, typically considered a resource of quantum states \citep{streltsov17,baumgratz14}
(see Sec.~\ref{subsec:coherence-states}), can be treated as a measurement
resource, where the free resources are incoherent measurements (diagonal
in the incoherent basis) \citep{kiukas20,oszmaniec19,baek20}. Measurements
which are \emph{maximally }coherent, i.e. the `{}`most valuable''
from a resource perspective, are those which are mutually unbiased
to the incoherent basis. 

To see this, consider a general measurement described by a POVM (positive
operator-valued measure) $M={\{M(i)\}}_{i\in\Omega}$, i.e. a collection
of positive semidefinite operators indexed by the outcome set $\Omega$
such that $\sum_{i\in\Omega}M(i)=\mathbb{I}$. A measure of coherence
known as the \emph{entry-wise coherence} of a measurement is defined
as
\begin{equation}
\text{coh}_{nm}(M)=\sum_{i\in\Omega}|\bra nM(i)\ket m|\,,
\end{equation}
for each $n,m\in\{1,\ldots,d\}$, where $\{\ket n\}$ is the incoherent
basis. This quantity, which cannot increase with free operations,
and satisfies $0\leq\text{coh}_{nm}(M)\leq1$, implies that a measurement
is \emph{maximally coherent} if the upper bound is saturated for all
$m,n$. 

It was shown in \citep{kiukas20} that a POVM $M$ with $d$ outcomes
is maximally coherent if and only if $M$ is mutually unbiased to
the incoherent basis. For example, the measurement $M(\pm)=\frac{1}{2}(\mathbb{I}\pm\sigma_{x})$
is maximally coherent with respect to the eigenbasis of $\sigma_{z}$,
since $\text{coh}_{nm}(M)=1$ for all $n,m\in\{1,2\}$. In light of
this correspondence, one can reformulate the problem of finding pairs
of MU bases as the problem of finding maximally coherent observables.

Measurement coherence turns out to be intimately connected to incompatibility
(Sec.~\ref{subsec:inc}). In particular, necessary and sufficient
conditions can be derived—with the aid of MU bases—that determine
the amount of coherence needed for pairs of observables to be incompatible
\citep{kiukas20}. To understand the fundamental role of MU bases
in this connection, consider a pair of observables $P$ and $Q$ that
are mutually unbiased (and therefore incompatible in the sense of
Sec.~\ref{subsec:inc}), with $P$ the observable whose eigenstates
define the incoherent basis. The pair can be modified by adding noise
in the form of pure decoherence (which is more general than the convex
mixing approach described in Eq.~(\ref{eq:depolarising})), such
that $P$ remains incoherent, and the coherence of $Q$ is reduced.
The interesting point is that if the noisy observables are jointly
measurable (i.e. the noise has destroyed the incompatibility of the
initial pair), then incompatibility is also lost if we start with
\emph{any }observable in place of the complementary observable $Q$.
Hence, MU bases are both maximally coherent and maximally incompatible
in this setting. Consequently, they provide a means to verify joint
measurability for more general observables.

\subsection{Coherence in states}

\label{subsec:coherence-states}

MU bases also play a role in the resource theory of coherence in states,
where the free resources are states diagonal in the incoherent basis.
Many different resource theories of coherence have been introduced
in the literature, mainly due to the freedom to choose between distinct
classes of free operations which map the set of incoherent states
to itself \citep{streltsov17,baumgratz14}. Examples of these operations
include \emph{maximally incoherent operations} (MIO), which cannot
create coherence, and a proper subset thereof: the \emph{incoherent
operations} (IO) which admit a Kraus decomposition so that each Kraus
operator $K_{i}$ cannot create coherence from the incoherent basis,
i.e. $K_{i}\ket m\sim\ket n.$

Quantifiers of state coherence were first introduced to require monotonicity
under IO, and included the relative entropy of coherence, the $l_{1}$
norm of coherence and the trace norm of coherence. Since coherence
is a basis-dependent quantity, to determine the maximum value of the
coherence attributed to a state requires an optimisation over all
bases. This is equivalent to optimising the coherence $C(\rho)$ over
all unitary transformations of a given quantum state $\rho$, i.e.
\begin{equation}
C_{\text{max}}(\rho)=\sup_{U}C(U\rho U^{\dagger}),\label{eq:maxcoherence}
\end{equation}
defining the maximally coherent state(s) $\rho_{\text{max }}=V\rho V^{\dagger}$,
where $V$ is the unitary maximising Eq.~\eqref{eq:maxcoherence}.

The relative entropy of coherence, which is also an MIO monotone,
is given by
\begin{equation}
C_{R}(\rho)=S(\rho_{I})-S(\rho)\leq\log d-S(\rho),
\end{equation}
where $S$ is the von Neumann entropy and $\rho_{I}$ is the diagonal
part of $\rho$. The upper bound is saturated for bases which are
mutually unbiased to the eigenvectors of $\rho$ \citep{yao16}. This
is also true for other coherence measures, including the robustness
of coherence, the coherence weight and the modified Wigner-Yanase
skew information measure \citep{hu17}. On the other hand, the $l_{1}$-norm
of coherence,
\begin{equation}
C_{l_{1}}(\rho)=\sum_{i\neq j}|\rho_{ij}|,
\end{equation}
is an IO monotone that violates MIO monotonicity, and is not maximised
by MU bases \citep{yao16}.

More generally, for \emph{any} MIO monotone, bases mutually unbiased
to the eigenstates of $\rho$ are always optimal. In particular, among
all states with a fixed spectrum $\{\lambda_{i}\}$, the maximally
coherent state (with respect to any MIO monotone $C$) is given by
$\rho_{\text{max }}=\sum_{i}\lambda_{i}\kb{\phi_{i}}{\phi_{i}}$,
where $\{\ket{\phi_{i}}\}$ is MU to the incoherent basis \citep{streltsov18}.
The unitary which achieves the maximum is given by $V=\sum_{i}\ket{\phi_{i}}\bra{\psi_{i}},$
with $\ket{\psi_{i}}$ the eigenstates of $\rho.$ It follows that
$\rho_{\text{max }}$ is a resource state among all states with the
same spectrum. Somewhat surprisingly, maximally coherent states link
the resource theories of coherence and purity. In particular, for
any MIO monotone, the corresponding purity of a quantum state $\rho$
is the maximal coherence $C_{\text{max}}(\rho)$ \citep{streltsov18}.

\subsection{Quantum channels }

\label{subsec:Quantum-channels}

Any orthonormal basis $\left\{ \ket v,v=0\ldots d-1\right\} $ of
the space $\mathbb{C}^{d}$ defines a \emph{quantum channel} which
acts by projecting a density matrix $\rho$ onto the associated diagonal
matrix,
\begin{equation}
\Psi(\rho)=\sum^{d-1}_{v=0}\bra v\rho\ket v\,\kb vv\,.\label{eq: QC channel}
\end{equation}
Given more than one orthonormal basis of $\mathbb{C}^{d}$, convex
mixtures of the identity map $\mathbb{I}$ and the channels $\Psi_{b}(\rho)$,
$b=1,2,\dots$, of type (\ref{eq: QC channel}) are also channels,
\begin{equation}
\Phi=s\mathbb{I}+\sum_{b}t_{b}\Psi_{b}\,.\label{eq: Pauli channels}
\end{equation}

Using MU bases in this construction, the channel $\Phi$ can be shown
to be \emph{completely positive} and \emph{trace preserving} for suitable
choices of the parameters $s$ and $t$ \citep{nathanson+07}. The
resulting \emph{Pauli diagonal channels constant on axes} leave the
completely mixed state invariant, thus generalising \emph{unital }qubit
channels to higher-dimensional spaces.

Whenever complete sets of MU bases exist, i.e. for channels acting
on states living in a Hilbert space of dimension $d\in\mathbb{PP}$,
a link between quantum channels and orthogonal unitary operator bases
(see Sec.~\ref{subsec:Maximally-commuting-unitary}) emerges, entailing
a generalised Bloch sphere representation of qudits in the space $\mathbb{C}^{d}$.
Further properties and applications of Pauli channels have been studied
in \citep{Siudzinska2020}, and in the references provided there.

\subsection{SIC-POVMs, MU bases and frames}

\label{subsec:MU-bases-and-SICs} 

S\emph{ymmetric informationally-complete }POVMs\emph{ }(or SICs, for
short) are defined by prescribing the moduli of the transition amplitudes
between the pure states forming them, just as for sets of MU bases.
However, the condition that the states $\left\{ \phi_{1},\phi_{2},\ldots,\phi_{d^{2}}\right\} $
give rise to a SIC, i.e.,
\begin{equation}
\left|\bk{\phi_{k}}{\phi_{k^{\prime}}}\right|{}^{2}=\frac{1}{d+1}\,,\label{eq: SIC constraints}
\end{equation}
$k,k^{\prime}=1\ldots d^{2},\:k\neq k'$, exhibits a greater degree
of symmetry than Eqs.~\eqref{eq: completeMUsetfor(d+1)}, since \emph{all
}pairs of states are treated in the same way; there is no subdivision
of the states into orthonormal bases of $\mathbb{C}^{d}$. 

SICs are currently thought to exist in Hilbert spaces of arbitrary
dimension $d$ as they have been found when $d\leq193$ partly by
hand and partly by computer-algebraic methods \citep{scott17,fuchs17,Grassl24},
and for some specific dimensions such as $d=n^{2}+3\in\mathbb{P}$
\citep{appleby22} or $d=4p=n^{2}+3$, $p\in\mathbb{P}$ \citep{bengtsson24}.
Both MU bases and SICs form complex projective 2-designs (see Sec.~\ref{subsec: 2-design equivalence})
containing $d(d+1)$ and $d^{2}$ elements, respectively. In spite
of some surprising connections, the existence problems for these structures
do not seem to illuminate each.

For example, Wootters \citep{wootters06} points out that the geometric
connection between an affine plane and a complete set of MU bases
(cf. Appendix \ref{subsec:affineplanes_mubs_sics}) is mirrored by
a SIC when the roles of the $d^{2}$ points and $d(d+1)$ lines are
swapped. Beneduci \textit{et al.} \citep{Beneduci2013} shed further
light on this connection, providing an operational link between SICs
and MU bases for prime-power dimensions. In particular, by introducing
the notion of \emph{mutually unbiased POVMs }(cf. Sec. \ref{subsec:Mutually-unbiased-POVMs}),
they show that a complete set of MU bases arises when a set of commutative
MU-POVMs is obtained from the marginals of a SIC.

Other connections between these superficially similar sets of states
can be made, unrelated to the existence question. For example, Heisenberg-Weyl
covariant SICs can be created once a so-called \emph{fiducial }vector
has been found, simply by applying $d^{2}$ phase-space displacement
operators. Dang, Appleby and Fuchs \citep{appleby14a} show that,
in prime dimensions $d$, the probabilities that characterise a fiducial
vector with respect to measurements in a complete set of MU bases
satisfy a simple condition within each basis: a fiducial vector must
be a minimum uncertainty state measured in terms of the quadratic
Renyi entropy. However, the argument holds only in one direction:
not every state with minimal uncertainty will be a fiducial vector
of a SIC. If it did, the result could be used to construct SICs in
all prime dimensions. Another surprising connection has been identified
in the space $\mathbb{C}^{4}$ \citep{tavakoli20}, while in $\mathbb{C}^{3}$,
a state-independent Kochen-Specker inequality exists which elegantly
combines the elements of a complete set of four MU bases and the nine
states of a SIC (cf. Sec.~\ref{subsec: contextuality}).

Finite tight frames—a generalisation of orthonormal bases—provide
a framework to study collections of vectors on finite dimensional
Hilbert spaces, often with a high degree of symmetry \citep{waldron18}.
A \emph{finite}\textit{ frame }is a collection of vectors which span
the Hilbert space, extending the notion of a basis to overcomplete
spanning sets of vectors. A frame is \textit{tight} if the projections
onto the frame elements sum to a multiple of the identity. Orthogonal
bases as well as collections of MU bases are examples of tight frames.
A SIC is a special type of frame known as an \emph{equiangular} tight
frame, and contains the maximal number of vectors defining equiangular
lines in the underlying space. MU bases and SICs often appear as illustrative
examples of tight frames, e.g. \citep{waldron18,chien16,waldron17}.
The notion of unbiasedness has been extended to equiangular tight
frames \citep{fickus20}, and generalised further to \emph{MU frames}
\citep{perez2021mutually}, as described in Sec. \ref{subsec:Mutually-unbiased-frames}.

\subsection{Contextuality }

\label{subsec: contextuality}

A state-independent Kochen-Specker (KS) inequality has been constructed
by combining the twelve states of a complete set of MU bases in the
Hilbert space $\mathbb{C}^{3}$ with the nine states of a SIC-POVM
\citep{Bengtsson2012}. By dropping the rules characteristic for value
assignments in the KS setting, one can also obtain a state-independent
\emph{non-contextual }inequality, using the same SIC-POVM and four
MU bases. Since these constructions have no obvious generalisations
to higher dimensions, it remains open whether the interplay between
MU bases and SICs is a dimensional coincidence only. Related higher-dimensional
constructions that build on MU bases—including systematic KS sets
and explicit examples in $d=4,5$—can be found in \citep{navara25}.

Another proof that quantum systems with dimension $d\geq4$ violate
non-contextuality has been formulated in \citep{Leifer_2020}, using
complete sets of MU bases. If these sets also existed in composite
dimensions $d\notin\mathbb{PP}$, the underlying inequality assuming
non-contextual value assignments could be violated in \emph{all} dimensions,
not just those with $d$ being a prime-power. 

\subsection{Quantum correlations}

\label{subsec:Quantum-correlations}

Entanglement in quantum systems can be characterised in a variety
of ways, leading to a complex hierarchy of quantum correlations. Some
entangled states, for example, admit a local hidden-state model and
are therefore unsteerable (cf. Sec. \ref{subsec:Steering}) while
others violate Bell inequalities (Bell non-locality) and are highly
entangled. Methods that detect entangled states (including bound entanglement)
via MU bases are discussed in Sec.~\ref{subsec:entanglement-detection}.

For any dimension $d\geq2$, there are tailor-made Bell inequalities
which are maximally violated by $d$-element MU bases \citep{tavakoli19}.
This property can be used to set up a weak form of device-independent
self-testing for pairs of MU bases as well as a means to certify the
presence of a maximally entangled state. These specific Bell inequalities
also allow for the distribution of a key at an optimal rate, in a
device-independent way, using measurements with $d$ outcomes (see
Sec.~\ref{subsec:Quantum-cryptography} for other approaches to quantum
cryptography based on MU bases). Self-testing via Bell inequalities
is also possible for three and four MU bases in two-qutrit systems
\citep{kaniewski19,borkala22}. Performing random MU bases in low
dimensions has been shown to provide a high probability of achieving
Bell violations \citep{tabia22}.

The existence of quantum correlations in arbitrary bipartite non-product
states can be confirmed using a measure based on MU bases which exploits
their complementarity \citep{Guo_2014}. In another approach, MU product
bases (cf. Sec.~\ref{subsec:product_bases}) for bipartite systems
have been used to quantify quantum correlations and entanglement \citep{maccone15}
(see also Sec.~\ref{subsec:entanglement-detection}). The measures
used were the mutual information, the Pearson coefficient and the
sum of conditional probabilities between the complementary bases.
For two-qutrit systems, an experiment based on four MU bases has simultaneously
certified entanglement, non-locality and steering \citep{huang21}.

\subsection{Steering}

\label{subsec:Steering}

Schrödinger first realised that composite quantum systems can exhibit
surprising non-classical behaviour. The phenomenon of ``steering''
is closely related to entanglement but conceptually distinct from
quantum correlations. In a two-qubit setting, say, the set of steerable
states is found to be smaller than the set of entangled states but
larger than the set of states violating Bell's inequality.

In the steering scenario, a bipartite state $\rho_{AB}$ is shared
between Alice and Bob who then measure observables on each of their
respective subsystems. After a measurement (POVM) by Alice, taken
from the set $\{M_{b}(v)\}$, where $v$ labels the possible outcomes
and $b$ the available measurements, Bob's state is given by $\sigma_{v|b}=\text{Tr}_{A}[M_{b}(v)\otimes\mathbb{I}\rho_{AB}]$.
The assemblage $\{\sigma_{v|b}\}$ of all possible states at Bob's
end (and hence the state $\rho_{AB}$) is said to be \emph{steerable
}by Alice's measurements if $\{\sigma_{v|b}\}$ does not admit a \emph{local
hidden-state model}. An assemblage has a local hidden-state model
capable of simulating the statistics of the quantum states if there
exists a set of positive operators $\{\sigma_{\lambda}\}$ with $\sum_{\lambda}\sigma_{\lambda}=\mathbb{I}$,
such that $\sigma_{v|b}=\sum_{\lambda}D(v|b,\lambda)\sigma_{\lambda},$
for all $v$ and $b,$ where $D(v|b,\lambda)\geq0$ and $\sum_{v}D(v|b,\lambda)=1$
describes a stochastic transformation.

A pure state with full Schmidt rank is steerable if and only if Alice's
measurements are incompatible \citep{quintino14,uola14}. This suggests
that MU bases, which are highly incompatible (see Sec.~\ref{subsec:inc}),
play a significant role in demonstrating steerability. Indeed, many
of the approaches to steering rely on MU bases, as summarised in the
review article by Cavalcanti and Skrzypczyk \citep{cavalcanti16}.

One way to convince Bob that a state is steerable is to construct
inequalities which hold if a local hidden-state model exists. For
example, a recent inequality \citep{Zhu2016} detects if an assemblage
of \emph{any} size is steerable—as well as quantifying the steering
robustness (and hence the incompatibility robustness)—while it exhibits
\emph{maximal }violations only if Alice uses MU bases (under the restriction
of projective measurements). In this scenario, the absence of complete
sets of MU bases in composite dimensions $d\notin\mathbb{PP}$ poses
an intriguing open problem: how many bases are needed to construct
a maximally incompatible (steerable) assemblage in composite dimensions?

Other steerability criteria have been obtained for bipartite qubit-qudit
systems based on \emph{mutually unbiased measurements} \citep{kalev+14},
i.e. collections of\textit{ non-projective} POVMs that satisfy a modified
form of the overlap conditions for MU bases (cf. Sec.~\ref{subsec:Mutually-unbiased-measurements}).
Experimental demonstrations of steering using MU bases include loop-hole
free steering \citep{wittmann12} and higher dimensional steering
\citep{zeng18}.

\subsection{Equivalence classes}

\label{subsec:Equivalence-classes}

Two sets of $\mu$ MU bases in $\mathbb{C}^{d}$ are \emph{equivalent},
\begin{equation}
\{\mathcal{B}_{0},\ldots,\mathcal{B}_{\mu-1}\}\sim\{\mathcal{B}'_{0},\ldots,\mathcal{B}'_{\mu-1}\}\,,\label{eq: classes}
\end{equation}
if any combination of the following operations transform one set to
the other: \emph{(i)} a fixed unitary (or anti-unitary) applied to
all states; \emph{(ii)} permuting the states within a basis; \emph{(iii)}
multiplying each state with an individual phase factor; \emph{(iv)}
swapping any two bases within the set. As a consequence, and without
loss of generality, it is standard practice to assume that the first
basis $\mathcal{B}_{0}$ is the canonical basis so that the remaining
bases can be represented as complex Hadamard matrices (see Secs. \ref{subsec:Motivation}
and \ref{subsec: Hadamard equivalence}). The equivalence relations
imply that the second basis $\mathcal{B}_{1}$ can be written as a
\textit{dephased} Hadamard matrix $H$ with its first column and row
given by $H_{i1}=H_{1i}=1/\sqrt{d}$ for $i=1\ldots d$. A detailed
discussion on the constructions of different equivalence classes is
given in Sec.~\ref{sec:pairs_of_MU_bases_C^d}. 

In the results described so far, unbiasedness of a set of bases—independent
of the choice of equivalence class—is the essential property behind
their practical utility. However, in some cases, the choice of equivalence
class can lead to different outcomes. One discrepancy between inequivalent
sets is their incompatibility content. As already noted in Sec.~\ref{subsec:inc},
inequivalent sets of MU bases can exhibit different degrees of incompatibility
\citep{Designolle2018}. A further discrepancy arises from quantifying
the information extraction capabilities of sets of measurements \citep{zhu22}.
In particular, the estimation fidelity can distinguish inequivalent
MU bases in $d=4$, as has been demonstrated experimentally \citep{yan24}.
Entropies of inequivalent sets of MU bases also do not have to coincide
\citep{serino24}.

Tangible differences in the outcomes of practical tasks can also be
seen in the QRAC protocol (described in Sec.~\ref{subsec:QRAC}):
using inequivalent classes yield different average success probabilities,
although it is conjectured there exists a set of MU bases that always
optimises the average success probability. Similarly, entanglement
witnesses constructed from MU bases (Sec.~\ref{subsec:entanglement-detection})
can depend on the chosen equivalence class \citep{hiesmayr20}. In
some instances, unextendible MU bases (Sec.~\ref{subsec:Unextendible-MU-bases})
detect entanglement more efficiently than extendible ones.

What is more, there are situations in which it is not necessary to
alter the equivalence class to observe different outcomes. When using
MU bases to detect entanglement, permutations of the basis elements
significantly modify the entanglement witness and its ability to detect
bound entangled states \citep{bae22} (see Sec.~\ref{subsec:entanglement-detection}).
A similar phenomenon has been observed for a guessing game described
in \citep{doda20}.

For a discussion on inequivalent complete sets of MU bases, see Appendix
\ref{subsec:other_constructions}.

\subsection{Mathematical topics }

\label{subsec:Mathematical-topics} 

MU bases have found their way into a number of mathematical topics
with no immediate physical motivation. Some of the relations mentioned
in Sec.~\ref{sec: equivalent_formulations} such as Lie theory and
projective $2$-designs are fitting examples. Furthermore, large collections
of MU bases prove useful for probability theory in the context of
random matrices \citep{chan19}, as well as to estimate the value
of the trace of a matrix \citep{fitzsimons16}. They have also inspired
the construction of new arithmetic functions which are multiplicative,
i.e. $f(mn)=f(m)f(n)$ holds whenever the greatest common divisor
of the integers $m$ and $n$ is equal to one \citep{chan20}, and
of maximal orthoplectic fusion frames \citep{bodmann18}. A notion
of MU bases was formulated in the category of sets and relations\textbf{
(Rel}), and a classification was provided via connections to mutually
orthogonal Latin squares \citep{evans09}.

\subsection{Experiments with MU bases}

\label{subsec:Experiments-with-MU-bases}

Quantum states forming MU bases have been created in a number of dimensions
using various physical systems. For $d=2$, pairs and triples of MU
bases were obtained in a quantum optical setting, exploiting the fact
that wave plates cause suitable phase shifts \citep{Hou2015}. D'Ambrosio
\textit{et al}. \citep{ambrosio+13} embedded a photonic quantum system
of dimension $d=6$ in a Hilbert space associated with photon polarisation
coupled to some orbital angular momentum. The 18 product states which
form three MU bases were prepared and certified. Another quantum optical
approach to create and manipulate complete sets of $(d+1)$ MU bases
for dimensions $d<5$ has been implemented by Lukens \textit{et al}.
\citep{lukens18}.

Various tasks at the core of quantum information have been carried
out successfully using MU bases. Two protocols for \emph{quantum key
distributio}n (cf. Sec.~\ref{subsec:Quantum-cryptography}) with
photons based on MU bases were realised experimentally \citep{Bartkiewicz2015};
the security of the protocols is linked to quantum cloning and a temporal
variant of steering. Generating photons with suitable orbital angular
momentum has been verified as a feasible approach, making it possible
to implement higher-dimensional quantum key distribution protocols
\citep{mafu13,mirhosseini15,bouchard18}. A high dimensional implementation
of MU bases $(d>9$) using traverse modes of spatial light has been
applied to verify that ignorance about a certain aspect of the whole
system does not imply ignorance of its parts \citep{kewming20}. Self-testing
of two four-dimensional MU measurements has been demonstrated \citep{farkas2021self}.
A scalable implementation of MU bases is proposed in \citep{ikuta22},
where the number of interferometers scales logarithmically with $d$.
This scheme is relevant to quantum key distribution when $d=4$.

The theoretically appealing idea that MU bases are highly suitable
for \emph{quantum tomography} (cf. Sec.~\ref{subsec: Quantum-state-reconstruction})
has been demonstrated experimentally for two-qubit polarisation states
\citep{adamson10}, with increased fidelity of the reconstructed states
when compared to standard techniques, and for higher-dimensional photonic
qudits \citep{lima11}. Local tomography based on MU bases for the
individual parts of a composite quantum system has led to the successful
reconstruction of the states of bipartite entangled systems \citep{Giovannini2013},
again using the orbital angular momentum of photons. In a similar
spirit, \emph{quantum process tomography} is feasible, either with
complete sets of MU bases for prime dimensions $d$ or using MU bases
obtained from tensoring in the composite dimension $d=6$ \citep{stefano21}
(cf. Sec.~\ref{subsec: Quantum-state-reconstruction}).

MU bases have been used to confirm the first reported quantum \emph{teleportation}
of a qutrit state \citep{luo19} as well as both loop-hole free \emph{steering
}\citep{wittmann12} and higher dimensional steering \citep{zeng18}
(see Sec.~\ref{subsec:Steering}). They have also been instrumental
to experimentally verify bipartite bound entanglement in two-photon
qutrits \citep{hiesmayr13}. A link between pairs of MU bases for
discrete and continuous variables (cf. Sec.~\ref{subsec: MUs for CVs})
was established in \citep{Tasca2018}, along with a quantum optical
implementation of the resulting coarse-grained observables. The findings
have been extended both theoretically and experimentally from two
to three measurements \citep{Paul2018a}.

There is a natural link between MU bases and a discretised model of
plane paraxial geometric optics which may be implemented experimentally.
Hadamard matrices arise as representations of \emph{discrete linear
canonical transforms} that describe the action of an optical system
on $d$-component signals \citep{healy11}.

\section{Applications of complete sets \label{sec:applications} }

We will now summarise applications of complete sets of MU bases in
the field of quantum information: quantum state tomography, quantum
key distribution, secret sharing, the Mean-King problem, entanglement
detection, and quantum random access codes. They provide strong motivation
to search for complete sets in composite dimensions $d\notin\mathbb{PP}$.
In the absence of complete sets, one needs to turn to workarounds,
some of which are discussed when available.

\subsection{Quantum state reconstruction }

\label{subsec: Quantum-state-reconstruction}

A widely known and important application of MU bases is the solution
it provides to the problem of \emph{quantum state determination} /
\emph{reconstruction}, or \emph{quantum tomography}. The idea of complete
sets of MU bases for prime and prime-power dimensions was, in fact,
rediscovered when addressing the problem of optimal quantum state
reconstruction. MU bases play a key role when estimating a given quantum
state since they minimise the statistical error \citep{ivanovic81,wootters+89}.

The reconstruction of a quantum state typically starts by assuming
that there is a large but finite ensemble of identically prepared
systems with $d$ levels. The state of the system is described by
a density matrix $\rho\in\mathcal{S}(\mathcal{H})$, a positive trace
class operator with trace one. To determine the state $\rho$, the
ensemble is divided into $(d+1)$ sub-ensembles of equal sizes. On
each sub-ensemble we repeatedly perform measurements of a different
observable with $d$ outcomes. In the limit of an infinitely large
ensemble, the resulting $(d-1)(d+1)$ independent outcome probabilities
characterise the unknown pre-measurement state $\rho$ unambiguously.
This result conforms with Schrödinger's conception of the wave function
as a `{}`catalogue of expectations'{}' \footnote{Originally in German: ``Die $\psi$-Funktion als Katalog der Erwartung'{}',
in \citep[Sec. 7]{schrodinger35}.}

Any informationally complete set of observables \citep{busch91} will
be sufficient to determine the state $\rho$. However, given a \emph{finite}
ensemble, statistical errors are unavoidable but are minimised by
choosing $(d+1)$ \emph{pair-wise complementary} measurements \citep{wootters+89}.
In this case, measuring the $b$-th observable projects the initial
state onto the elements of an orthonormal basis (associated with $d$
orthogonal projection operators $\{P_{b}(v)\}^{d-1}_{v=0}$) which
is mutually unbiased to the those of the other $d$ measurements.

In the simplest case of a two-dimensional Hilbert space, one wishes
to reconstruct a density matrix $\rho=(\mathbb{I}+\vec{r}\cdot\vec{\sigma})/2$,
with $\mathbb{I}$ being the $2\times2$ identity matrix, while $\vec{\sigma}=(\sigma_{x},\sigma_{y},\sigma_{z})^{T}$
is the operator-valued spin vector constructed from the Pauli matrices,
and the vector $\vec{r}$ runs through all points of the unit ball
in $\mathbb{R}^{3}$. Measuring the complementary spin observables
$\sigma_{x}$, $\sigma_{y}$ and $\sigma_{z}$ separately on three
sub-ensembles, one obtains the components of the vector $\vec{r}$
which determines the unknown state $\rho$ unambiguously. However,
since every measurement comes with some degree of statistical inaccuracy,
the reconstructed value of each component $r_{j},j=x,y,z$, is confined
to a ``fuzzy'' estimate of the exact plane only. The intersection
of all three `slabs' determines the region of the unit ball which
is compatible with the (inaccurate) measurements. As one expects intuitively,
the overall statistical error is minimised if the planes are pair-wise
perpendicular, corresponding to a measurement defined by \emph{mutually
unbiased} bases. The argument carries through for dimensions larger
than $d=2$, underlining the specific role played by MU bases for
quantum state tomography. 

In a bipartite setting, methods of quantum state \emph{verification}—rather
than reconstruction—are more efficient than traditional tomographic
approaches, especially when using complete sets of MU bases. Efficient
protocols for both pure \citep{li19} and maximally entangled states
have been developed \citep{zhu19}.

For quantum process tomography, in which a \emph{quantum} \emph{channel}
is reconstructed, a large number of parameters must be determined
through measurements. Complete sets of MU bases have been used in
different ways to achieve this goal. One may adapt quantum tomographic
methods in prime-power dimensional Hilbert spaces to determine the
parameters characterising the channel in hand \citep{fernandez11}.
Alternatively, one can exploit the fact that they form a $2$-design
(cf. Sec.~\ref{subsec: 2-design equivalence}) \citep{bendersky08}.
This generalises to non-prime-power dimensions \citep{stefano21}
if one tensors complete sets of MU bases. A quantum optical experiment
for quantum process tomography in this scenario has been proposed,
with the quantum state of a $d$-level system encoded as the position
of photons in the transversal direction, by means of apertures with
$d$ slits. A successful implementation of the protocol is reported
for dimension $d=6$.

As noisy intermediate-scale quantum computers advance, it becomes
useful to develop strategies that estimate a system's properties with
minimal measurements, avoiding the need for full-scale quan\-tum
state tomography. For instance, the \emph{classical shadow protocol}
creates a classical approximation of an unknown quantum state, which
can then be used to simultaneous estimate expectation values for non-commuting
observables. The original classical shadow protocol was based on implementing
random Pauli measurements on each qubit \citep{huang20}, i.e. three
MU bases in $\mathbb{C}^{2}$. This was later generalised to $(2^{n}+1)$
MU bases on an $n$-qubit system \citep{wang24}. Another, equally
efficient, strategy to simultaneously estimate expectation values
of relevant observables implements a \textit{joint measurement} (see
Sec.~\ref{subsec:inc}) of noisy versions of three MU bases on each
qubit \citep{mcnulty23}.

\paragraph*{Workaround}

\label{par:QST workaround}

Without knowledge of a complete set, or even if no such set exists,
can we still find an optimal reconstruction procedure for the state
in question? Not surprisingly, alternative approaches for state reconstruction
exist in arbitrary dimensions. The existence of complete sets—while
convenient to have—is not fundamental for optimal quantum tomography.

In the generalisation to arbitrary $d$-level systems, weighted complex
projective $2$-designs (defined in Sec.~\ref{sec:weighted2designs})
play an important role in optimising state reconstruction \citep{roy+07}.
It has been shown that a set of bases which constitutes a weighted
$2$-design forms the orthogonal measurements necessary for optimal
quantum tomography.

Explicit examples of weighted $2$-designs are constructed in \citep{roy+07}
for $d=p^{n}+1$ with $p$ prime, where a set of $(d+2)$ orthonormal
bases is found. This covers dimension six, in which eight orthonormal
bases form a weighted $2$-design. Starting with the standard basis
$\mathcal{B}_{0}=\{\ket 0,\ldots,\ket 5\}$, the remaining orthonormal
bases are given by the states
\begin{equation}
\ket{v_{b}}=\frac{1}{\sqrt{6}}\sum^{5}_{k=0}\omega^{vk}e^{2\pi ib3^{k}/7}\ket k\,,\label{eq:weightedbasesd=00003D6}
\end{equation}
where $b=1\ldots7$, $v=0\ldots5$, and $\omega=e^{2\pi i/6}$. The
overlaps between elements of different bases, given explicitly in
\citep{patra07}, are 
\begin{equation}
|\bk{v_{b}}{v'_{b'}}|^{2}=\left\{ \begin{array}{ll}
\frac{6}{7} & \quad\mbox{if \ensuremath{b\neq b',\,\,v\neq v'}}\,,\\
\frac{1}{36} & \quad\mbox{if \ensuremath{b\neq b',\,\,v=v'}}\,.
\end{array}\right.
\end{equation}
Surprisingly, by performing the measurements associated with the eight
orthonormal bases on the unknown quantum state—the standard basis
is measured in the ratio $7:6$ with respect to each of the remaining
bases—optimal state reconstruction can be achieved. In fact, the same
minimised statistical error is achieved (hypothetically) by implementing
a complete set of seven MU bases.

In higher composite dimensions, when $d\neq p^{n}-1$, the minimum
number of orthonormal bases needed to construct a weighted $2$-design
for optimal state reconstruction is not known explicitly, but an upper
bound of $\frac{3}{4}(d-1)^{2}$ is given in \citep{roy+07}. This
bound was improved in \citep{mcconnell+08}, and weighted $2$-designs
were found to contain roughly $2(d+\sqrt{d})$ bases when $d$ is
odd and $3(d+\sqrt{d})$ for $d$ even.

Another approach to quantum state reconstruction of an arbitrary $d$-level
system is to recast the problem in terms of special types of informationally
complete positive operator value measures (IC-POVM). These are called
\emph{tight} rank-one IC-POVMs \citep{scott06} and are \emph{equivalent}
to complex projective $2$-designs (see Sec.~\ref{subsec: 2-design equivalence}).
Both SIC-POVMs and complete sets of MU bases are examples of tight
rank-one IC-POVMs. It was shown in \citep{scott06} that these POVMs
are optimal for \emph{linear} quantum state tomography. The state
reconstruction is ``linear'' in the sense that it is limited to
a simplified state reconstruction procedure.

\subsection{Entanglement detection \label{subsec:entanglement-detection}}

MU bases provide a simple and efficient criterion for witnessing entanglement
in quantum states. An entanglement witness is a Hermitian operator
$\mathbf{W}$ that satisfies $\text{Tr}[\mathbf{W}\rho_{\mathrm{sep}}]\geq0$
for all separable states $\rho_{\mathrm{sep}}$, and $\text{Tr}[\mathbf{W}\rho]<0$
for at least one entangled state $\rho$. A negative expectation of
the operator $\mathbf{W}$ for a bipartite state indicates the presence
of non-classical correlations between its subsystems.

Let us describe the criterion in the case of a bipartite state with
both subsystems of dimension $d.$ Given a set of $\mu$ MU bases
$\mathcal{B}_{b}=\{\ket{v_{b}}\}$ of the space $\mathbb{C}^{d}$,
it has been shown in \citep{spengler+12} that the operator 
\begin{equation}
\mathbf{B}(\mu):=\sum^{\mu-1}_{b=0}\sum^{d-1}_{v=0}\kb{v_{b}}{v_{b}}\otimes\kb{v^{*}_{b}}{v^{*}_{b}}\,,\label{eq:witness_op}
\end{equation}
satisfies, for any separable state, 
\begin{equation}
\text{Tr}[\mathbf{B}(\mu)\,\rho_{{\rm sep}}]\leq\frac{d+\mu-1}{d}\,.\label{eq:witness_bound}
\end{equation}
Hence, the operator 
\begin{equation}
\mathbf{W}(\mu)=\frac{d+\mu-1}{d}\mathbb{I}_{d}\otimes\mathbb{I}_{d}-\mathbf{B}(\mu)\,,
\end{equation}
is an entanglement witness satisfying $\text{Tr}[\mathbf{W}(\mu)\,\rho_{{\rm sep}}]\geq0$.
If a state $\rho$ violates Eq.~\eqref{eq:witness_bound}, it is
necessarily entangled. For example, Eq.~\eqref{eq:witness_bound}
is violated by all entangled isotropic states—i.e. mixtures of a maximally
mixed and maximally entangled state—when $\mu=d+1$. With fewer measurements,
the witness detects a subset of the entangled isotropic states: a
pair of MU bases, for example, is enough to detect at least half \citep{spengler+12}.

A modified witness, constructed in \citep{bae18}, is given by
\begin{equation}
\mathbf{\widetilde{W}}(\mu)=\mathbf{\widetilde{B}}(\mu)-L_{\mu}\mathbb{I}_{d}\otimes\mathbb{I}_{d}\,,
\end{equation}
for some $L_{\mu}\geq0$, where $\mathbf{\widetilde{B}}(\mu)$ is
identical to Eq.~\eqref{eq:witness_op} but without complex conjugation
in the second system. This leads to a lower bound

\begin{equation}
\text{Tr}[\mathbf{\widetilde{B}}(\mu)\,\rho_{{\rm sep}}]\geq L_{\mu}\,,
\end{equation}
which holds for all separable states. If $\mu=d+1$, then $L_{d+1}=1$
and the bound is violated by all entangled Werner states. On the other
hand, if $\mu<d+1,$ only numerical values of $L_{\mu}$ are known
\citep{bae18}. Interestingly, the bound depends on the choice of
MU bases. For instance, when $d=4$ and $\mu=3$, the ability to detect
certain entangled states depends on the choice of bases from the infinite
family of MU triples \citep{BrierleyWB2010}.

A larger class of entanglement witnesses can be generated by making
orthogonal rotations of the basis states on the first Hilbert space
\citep{chruscinski18}. For example, when the measurement basis is
permuted, the operator

\begin{equation}
\mathbf{B}_{\pi}(\mu)=\sum^{\mu-1}_{b=0}\sum^{d-1}_{v=0}\ket{\pi^{(b)}(v)_{b}}\bra{\pi^{(b)}(v)_{b}}\otimes\ket{v^{*}_{b}}\bra{v^{*}_{b}}\label{eq:probsum-1-1-1}
\end{equation}
generates a new class of entanglement witnesses, where $\pi^{(b)}$
is a permutation of the $d$ elements of the basis $\mathcal{B}_{b}$.
For certain permutations, the witness is non-decomposable and detects
bound entangled states \citep{bae22}. In fact, for $\mu>d/2+1,$
there exist witnesses of this type that are always non-decomposable.

Witnesses have also been derived for MU measurements \citep{chen+14},
defined in Sec.~\ref{subsec:Mutually-unbiased-measurements}, and
more generally 2-designs \citep{kalev13,graydon16}. Continuous variable
systems (cf. Sec.~\ref{subsec: MUs for CVs}) admit structurally
similar entanglement detection methods \citep{spengler+12}, as well
as more general bipartite systems \citep{wang21}.

\subsection{Quantum cryptography$\ldots$ \label{subsec:Quantum-cryptography}}

Measurements on quantum systems tend to disturb the state of the observed
system. This fundamental aspect of quantum mechanics has been the
springboard to applications in quantum cryptography, a \emph{physics-inspired}
means of secret communication \citep{bennett+84,ekert91,bruss98,cerf+02},
fundamentally different from traditional approaches which are based
on the difficulty to solve specific mathematical problems. To establish
a shared \emph{secret key}, necessary for encrypting a message, the
protocols rely on the parties exchanging quantum systems in such a
way that outside entities cannot gain information about it without
leaving traces of their interactions. The resulting modifications
may subsequently be detected by the legitimate parties, only to reveal
that the communication channel is not secure. 

\subsubsection*{$\ldots$ based on pairs of MU bases}

\label{sec:standard_crypto}

The BB84 protocol is one of the earliest examples of quantum key distribution
\citep{bennett+84}. A secret key, usually a \emph{random} sequence
of bits, e.g. 010110, is sent via a series of qubit states to a receiver.
The states are taken from two orthonormal bases, $\mathcal{B}_{z}=\{\ket 0,\ket 1\}$
and $\mathcal{B}_{x}=\{\ket +,\ket -\}$; given by the eigenstates
of the spin operators $\sigma_{z}$ and $\sigma_{x}$, respectively,
this pair of bases is mutually unbiased. Within each basis, the orthogonal
states will represent the bits 0 and 1. The sender \emph{randomly}
chooses a basis and sends off the appropriate state corresponding
to 0 or 1. The state is transmitted using a quantum channel to the
receiver, who then measures the state randomly using either $\sigma_{z}$
or $\sigma_{x}$. If the receiver happens to choose the same basis
as the sender in a particular run—which will occur in half of all
cases—both parties possess the same value for this particular bit. 

Once the measurements have been made, both parties publicly reveal
the bases used by each individual. The, sender and receiver remove
all measurement outcomes from their lists in which their choice of
bases did not agree. The remaining sequence of bits represents the
shared key. To check whether eavesdroppers interfered with the quantum
channel, they publicly compare a subset of the key: if they notice
discrepancies, quantifiable by an \emph{error rate}, a breach must
have occurred. 

Other quantum key distribution protocols have since been developed
and implemented successfully both experimentally and commercially
\citep{scarani+09}. The optimal or most robust protocol is one which
can tolerate large disturbances (errors) and still result in secure
key distribution. The amount of discrepancies caused by the eavesdropper
depends on her strategy; considerable effort is invested towards finding
the optimal method of attack for each protocol. The optimal eavesdropping
strategy for the BB84 protocol is known for \emph{individual} attacks
\citep{cerf+02}. However, the protocol remains secure even if unlimited
resources are available to the eavesdropper. For a correlation-based
cryptographic protocol using pairs of MU bases, see Sec.~\ref{subsec:Quantum-correlations}. 

\subsubsection*{$\ldots$ based on complete sets of MU bases}

Early quantum cryptography protocols based on pairs of MU bases (cf.
Sec.~\ref{sec:standard_crypto}) have been generalised to qutrits
\citep{bechmann00} exploiting all four MU bases and, subsequently,
to $d$-level systems equipped with a complete set of $(d+1)$ MU
bases \citep{cerf+02}. Compared to the approach using a pair of MU
bases, individual attacks by means of a quantum cloning machine lead
to a slightly higher error rate. Nevertheless, protocols using two
bases remain preferable since it is easier to produce longer keys. 

In \citep{brierley09} a protocol for an arbitrary $d$-level systems
is given which improves sensitivity to an eavesdropper with respect
to the error rate. Here, the secret key is constructed from an alphabet
of arbitrary size which is encoded into a basis of the state space
$\mathbb{C}^{d}$. As usual the system is prepared by a sender at
$A$ and transported to a receiver at $B$. The receiver is then set
the task of recovering the initial key by performing measurements
on the system he has received. The protocol aims to maximise the likelihood
of detecting an eavesdropper. Assuming an ``intercept-and-resend''
attack method, one can calculate the error rates which allow the parties
at $A$ and $B$ to detect a breach, with a larger error rate being
beneficial for security. The optimal strategy to share the key is
found to rely on MU bases.

One disadvantage of this method is the relatively large number of
states required to share one bit of information compared to, say,
the six-state protocol \citep{bruss98}. At the same time, the protocol
is much more sensitive to the eavesdropper since the quantum bit error
rate of an attack—proportional to the key elements that contain an
error—is much higher. If one considers higher dimensions and uses
a complete set of MU bases, this protocol reaches a 100\% error rate.
However, in this limit one must use \emph{approximate} MU bases (see
Sec.~\ref{subsec:approximate_mu_bases}). In the absence of complete
sets of MU bases, i.e. for composite dimensions $d\notin\mathbb{PP}$,
the protocol is less efficient.

A hybrid security model has been proposed which offers everlasting
secure-key agreement rather than the stronger \textit{unconditional}
security of standard quantum key distribution methods \citep{vyas20}.
The protocol encodes a single bit using a $d$-dimensional state chosen
from a complete set of $(d+1)$ MU bases. The basis information is
shared between legitimate parties and hidden from Eve under the assumption
of short-term computational security (of one-way functions), until
her quantum memory has decohered. This hybrid approach has benefits
compared to the practicality and implementation issues of traditional
quantum key distribution methods.

Mutually unbiased bases also feature in security proofs of cryptographic
protocols. For example, measuring a complete set of mutually unbiased
bases can ensure statistical security of protocols sharing information
\citep{dupuis22}.

\subsection{Quantum secret sharing}

Quantum secret sharing involves distributing a secret among multiple
parties in a way that access to the secret requires collaboration
between a subset of the participants \citep{hillery99}. A quantum
secret-sharing scheme based on MU bases in a $d$-dimensional system
has been introduced in \citep{yu08}, with perfect security achievable
as $d$ increases. Alternatively, one can set up an efficient protocol
using sequential communication of a single quantum $d$-level system,
with $d\in\mathbb{P}$ a prime number \citep{tavakoli15secret}. The
dimensional restriction is due to its dependence on an algebraic property
known at the time only for MU bases in prime dimensions although a
generalisation to the case of $d=p^{2}$, $p\in\mathbb{P}$, has been
reported \citep{hao19}.

\subsection{Dealing with mean kings \label{subsec: mean kings}}

An interesting—but admittedly slightly artificial—application of MU
bases is their role in the solution to a measurement problem involving
a fictitious ``mean'' king. The problem arises in a scenario where
an observer $A$ prepares a spin-$\frac{1}{2}$ particle in a state
of her choice and then performs a control measurement on the system.
Between the preparation and measurement, a second observer $B$ measures
either $\sigma_{x}$, $\sigma_{y}$ or $\sigma_{z}$ on the particle
state. After the control measurement, observer $A$ is told which
spin component observer $B$ measured and is asked to determine the
corresponding measurement outcome \citep{vaidman+87}.

The generalisation of this problem is usually told by the following
story: a king who lives on a remote island sets a physicist a life
or death challenge. He asks the physicist to prepare a $d$-state
quantum system of her choice and to perform a control measurement
on the system. Before her control measurement, she must hand the state
over to the king while he secretly performs a measurement. After her
control measurement the king reveals his measurement and challenges
the opponent to determine his outcome. In this generalisation, the
choice of measurement made by the king is restricted to a set of pairwise
complementary observables.

The generalised problem was first solved for a system with prime degrees
of freedom in \citep{englert+01} and then extended to include prime-powers
\citep{aravind03}. Crucially, both solutions rely on the existence
of complete sets of MU bases. To resolve the challenge, the physicist
must prepare two $d$-state systems in a maximally entangled state
of the space $\mathbb{C}^{d}\otimes\mathbb{C}^{d}$. The auxiliary
system is kept by the physicist while the king preforms a measurement
of one of $(d+1)$ mutually unbiased bases on the object system.

It was shown subsequently that the existence of a complete set of
MU bases is, in fact, not the essential factor that allows the physicist
to extricate herself \citep{hayashi+05,reimpell07}. Under the above
assumptions, and by restricting the physicist's measurement to a projection-valued
measure (PVM), the mean king's problem for an arbitrary $d$-state
system has a solution only if the maximum number of $(d-1)$ mutually
orthogonal (MO) Latin squares exist. If $d$ is a prime or prime-power,
this maximum is achieved and the solutions agree with those given
in \citep{aravind03,englert+01}. However, in dimension $d=6$ only
three MO Latin squares exist, implying that there is no solution to
the problem for this degree of freedom, regardless of whether a complete
set of seven MU bases exists. Similarly, this is true for $d=10$
since no set of nine MO Latin squares exists \citep{Horibe2005}.
However, in these cases, alternative solutions to the mean-king problem
can be found (cf. the end of this section).

A Wigner function approach to the mean-king problem for qubits has
been developed in \citep{paz05}. The problem has also been transposed
to a setting with continuous variables \citep{botero07} and the associated
MU bases (cf. Sec.~\ref{subsec: MUs for CVs}); again, a phase-space
approach turns out to be instructive. 

\paragraph*{Workaround}

\label{par:Mean-King-other-sol}

If the physicist is allowed to perform POVM measurements on the space
$\mathbb{C}^{d}\otimes\mathbb{C}^{d}$, a complete solution to the
king's problem can be given for arbitrary dimensions $d\in\mathbb{N}$
\citep{kimura+06}. Thus, regardless of whether $(d-1)$ MO Latin
squares or complete sets of MU bases exist, the physicist can always
determine the king's measurement outcome.

\subsection{Quantum random access codes}

\label{subsec:QRAC application} 

A quantum random access code (QRAC), denoted by the symbol $\mu^{d}\rightarrow1$,
is a communication protocol in which the sender Alice encodes $\mu$
classical $d$-level systems ${\bf {x}}=\{x_{1},\ldots,x_{\mu}\}$,
with $x_{j}\in\{1,\ldots d\}$, into a $d$ dimensional quantum state
$\rho,$ and the receiver Bob is asked to correctly identify, via
a set of $\mu$ measurements, one of the $d$ levels $y\in\{x_{1},\ldots,x_{\mu}\}$
chosen randomly. In the most common QRAC protocol a $\mu$-bit message
is encoded in a qubit \citep{wiesner83,ambainis99}, rather than the
$d$-level generalisation considered here \citep{tavakoli15}. For
a given strategy (i.e., a state and measurements) one can find the
average success probability $p(\mu,d)$ of recovering the correct
$d$-level, which is calculated over all inputs ${\bf x}$ and choices
$y$. When $\mu=2$, the \textit{optimal} average success probability
$P(\mu,d)$, maximised over all states and measurements, is given
by
\begin{equation}
P(2,d)=\frac{1}{2}\left(1+\frac{1}{\sqrt{d}}\right),\label{eq: QRAC probability-1}
\end{equation}
and is achieved if and only if Bob measures a pair of MU bases \citep{farkas18,aguilar18}.
Consequently, the $2^{d}\rightarrow1$ QRAC provides a means for self-testing
pairs of MU bases \citep{tavakoli18,farkas18}.

For larger $\mu,$ the optimal strategy does not straightforwardly
generalise to measuring $\mu$ MU bases. For instance, when $\mu=3$
and $d=5$, there exist two inequivalent triples of MU bases and the
average success probability for each triple is different. To find
optimal measurements for a $\mu^{d}\rightarrow1$ QRAC scheme with
$d\geq\mu$, MU bases can be extended to `{}`$\mu$-fold unbiased
bases'{}' \citep{farkas17}. However, this approach imposes an additional
constraint on the bases that seems overly restrictive, and no examples
have been found for $\mu\geq3$ in any dimension $d$.

A variant of the QRAC protocol limits the number of classical $d$
level systems addressed by Bob to any $\nu$ of the total $\mu$ systems.
For $\nu=2\leq\mu$, the resulting `{}`p-QRAC'{}' protocol $(\mu,\nu)^{d}\rightarrow1$
lends itself to characterise MU bases as the measurements of choice
for Bob to maximise his success rate \citep{aguilar18}. The details
are given in Sec.~\ref{subsec:QRAC} where Equivalence \ref{equiv:QRACs}
establishes a rigorous link between the existence of a complete set
of MU bases and the optimal strategy for Bob.

\subsection{Other applications}

The reconstruction of signals from the magnitudes of vectors in real
or complex Hilbert spaces is an important problem in speech recognition.
Complete sets of MU bases have been shown to provide a simple approach
\citep{balan09}, with the 2-design property of the bases (cf. Sec.~\ref{subsec: 2-design equivalence})
a central ingredient. MU bases enable the construction of sequences
with relatively flat Fourier spectra, an important feature of signals
used to efficiently communicate (cf. Sec.~\ref{subsec: avant-la-letter});
a complete set of four MU bases has been used to that effect in~\citep{wu14}. 

\section{Equivalences and conjectures }

\label{sec: equivalent_formulations}

A set of $\mu$ MU bases $\mathcal{B}_{b},b=0\ldots\mu-1$, in the
space $\mathbb{C}^{d}$ is defined by the $d^{2}\mu^{2}$ conditions 
\begin{equation}
|\langle v_{b}|v'_{b'}\rangle|^{2}=\frac{1}{d}(1-\delta_{bb'})+\delta_{vv'}\delta_{bb'}\,,\label{eq:mu_condition}
\end{equation}
many of which are identical due to the symmetries of exchanging $v\leftrightarrow v^{\prime}$
and $b\leftrightarrow b^{\prime}$. Such sets are, as we have seen
in the previous two sections, closely related to a number of concepts
in seemingly disconnected areas of mathematics and physics. We will
now describe many of these and other concepts more rigorously, with
the aim to develop alternative perspectives on the original conditions
defining MU bases. We will, for example, rephrase Eqs.~\eqref{eq:mu_condition}
in terms of Hadamard matrices, unitary operator bases, decompositions
of Lie algebras, quantum designs and random access codes. The final
five reformulations (Secs. \ref{sec:geometric}–\ref{subsec:rigidity})
are slightly weaker in the sense that they hold only when $\mu=d+1$,
i.e. for a complete set of MU bases. For each `{}`representation'{}',
a \emph{conjecture} will be spelled out which expresses the existence
problem in the corresponding mathematical context. Proving or disproving
any of these conjectures will solve all others, including the original
existence problem. 

\subsection{MU bases as global minima\label{subsec:global_min}}

Rather than writing down the large number of constraints \eqref{eq:mu_condition}
on the vectors defining a set of $\mu$ MU bases, $\mu\in\{2,\ldots,d+1\}$,
one can encode them in the global minimum of one \emph{scalar} function
$F_{d}:\mathbb{R}^{p_{\mu}(d)}\to\mathbb{R}$ depending on $p_{\mu}(d)=(d-1)((\mu-1)(d-1)-1)$
free parameters (cf. Sec.~\ref{subsec:nonexistence_of_constellations}
for the parameter count). 
\begin{equivalence}\label{equiv: scalar function for MU conditions}
A set of $\mu$ orthonormal bases $\{\mathcal{B}_{0},\ldots,\mathcal{B}_{\mu-1}\}$
in $\mathbb{C}^{d}$ is mutually unbiased if and only if the non-negative
function
\begin{equation}
F_{d}(\mathcal{B}_{0},\ldots,\mathcal{B}_{\mu-1})\!=\!\!\sum^{\mu-1}_{b,b^{\prime}=0}\sum^{d-1}_{v,v'=0}\left(|\bk{v_{b}}{v'_{b'}}|-\chi^{bb^{\prime}}_{vv^{\prime}}\right)^{2}\label{eq: F encoding complete sets}
\end{equation}
vanishes; here $\chi^{bb^{\prime}}_{vv^{\prime}}=(1-\delta_{bb'})/\sqrt{d}+\delta_{vv'}\delta_{bb'}$
are the positive real square roots of the right-hand-side of Eq.~\eqref{eq:mu_condition}. 
\end{equivalence}

In this representation of the conditions for a set of $(d+1)$ bases
to form a complete set, Conjecture~\ref{conj: Zauner} presented
in Sec.~\ref{subsec:Motivation} turns into a simple statement about
the global minimum of the function $F_{d}(\mathcal{B}_{0},\ldots,\mathcal{B}_{d})$.
\begin{conjecture}
\label{conj: global minima} For $\mu=d+1$, the global minimum of
the function $F_{d}$ over all $(d+1)$-tuples of orthonormal bases
equals zero if and only if $d$ is a prime-power. 
\end{conjecture}
This formulation provides a useful starting point for numerical approaches
to the existence problem in composite dimensions, described in Secs.
\ref{subsec:nonexistence_of_quadruples} and \ref{subsec:nonexistence_of_constellations}.
It is not excluded that the global minimum of the function $F_{d}(\mathcal{B}_{0},\ldots,\mathcal{B}_{d})$
is \emph{degenerate} in prime-power dimensions: complete sets of MU
bases are not necessarily unique (see Appendix~\ref{subsec:other_constructions}).

\subsection{Sets of MU Hadamard matrices}

\label{subsec: Hadamard equivalence}

A complex Hadamard matrix of order $d$ is a generalisation of a real
Hadamard matrix which is a square, unitary matrix with entries consisting
of $\pm1/\sqrt{d}$ (or just $\pm1)$. The generalisation drops the
restriction of real entries to those with modulus $1/\sqrt{d}.$ A
vector is mutually unbiased to the standard basis if all $d$ components
of the vector have modulus $1/\sqrt{d}$ (cf. Eq.~\eqref{eq:mu_poly}).
It then follows that a basis is MU to the canonical basis if and only
if it can be expressed as a complex Hadamard matrix. 

This leads to the important observation that any pair of MU bases
in $\mathbb{C}^{d}$ can be represented by a $d\times d$ complex
Hadamard matrix. To see this we simply apply a unitary transformation—which
maps the orthonormal basis associated with one of the matrices to
the canonical one—to both matrices simultaneously. Since this map
preserves the overlap condition of Eq.~\eqref{eq:mu_condition},
the second matrix maps to a complex Hadamard matrix. Hence, the problem
of searching for pairs of MU bases is identical to the problem of
finding Hadamard matrices. 

A complete classification of Hadamard matrices is known for $d\leq5$
\citep{haagerup97} but it remains an open problem for $d\geq6$,
although a substantial body of knowledge is available. We shall discuss
the current efforts to classify Hadamard matrices of order six in
Sec.~\ref{sec:pairs_of_MU_bases_C^6} and those of higher orders
in Sec.~\ref{sec:pairs_of_MU_bases_C^d}.

The correspondence between pairs of MU bases and Hadamard matrices
extends naturally to an equivalence relation between sets of MU bases
and sets of Hadamard matrices.

\begin{equivalence}\label{equiv:hadamards}
A set of $\mu$ MU bases $\{\mathcal{B}_{0},\ldots,\mathcal{B}_{\mu-1}\}$
in $\mathbb{C}^{d}$ exists if and only if a set of $(\mu-1)$ Hadamard
matrices $\{H_{1},\ldots,H_{\mu-1}\}$ of order $d$ exists such that
the products $H^{\dagger}_{j}H_{j^{\prime}},j\neq j^{\prime}$, are
Hadamard matrices for all values of $j,j^{\prime}=1\ldots\mu-1$.
\end{equivalence}

Thus, finding a complete set of $(d+1)$ MU bases in dimension $d$
corresponds to identifying a collection of $d$ Hadamard matrices
$H_{j}$ with the property that $H^{\dagger}_{j}H_{j^{\prime}}$ is
a Hadamard matrix for all $j\neq j^{\prime}$. This leads to the following
conjecture.
\begin{conjecture}
\label{conj:hadamards}A set of $d$ Hadamard matrices $H_{j}$ of
order $d$ such that $H^{\dagger}_{j}H_{j^{\prime}}$ is Hadamard
for all $j\neq j^{\prime}$ exists if and only if $d$ is a prime-power. 
\end{conjecture}
As well as their relation to MU bases, Hadamard matrices play an important
role in other areas of mathematics and physics. They arise in the
study of quantum groups \citep{banica+07} and have applications in
quantum information, e.g. teleportation and dense coding schemes \citep{werner01}.
Their study has also been motivated by the problem of finding bi-unitary
sequences and cyclic $n$-roots \citep{bjork+91} (see Sec.~\ref{sec:fourier_family})
and they are used in constructing $*$-subalgebras of finite von Neumann
algebras \citep{popa83} and error-correcting codes \citep{agaian85}.

\subsection{Coupled polynomial equations}

\label{subsec: polynomial equivalence} 

Writing out the conditions \eqref{eq:mu_condition} in terms of states
relative to a basis of $\mathbb{C}^{d}$ turns them into a set of
coupled polynomial equations of fourth-order in the expansion coefficients.
Although straightforward, this explicit representation is useful to
implement computational searches for vectors mutually unbiased to
pairs of MU bases (cf. Sec.~\ref{sec:fourier_family}) and to prove
that certain sets are unextendible (cf. Sec.~\ref{subsec: unextendible MU bases}). 

Let us expand all $d\mu$ states $\ket{v_{b}}$ forming a hypothetical
set of $\mu$ MU bases $\mathcal{B}_{b}$, $b=0\ldots\mu-1$ in the
standard basis $\mathcal{B}_{z}=\left\{ \ket j,j=0\ldots d-1\right\} $,
i.e. we write 
\begin{equation}
\ket{v_{b}}=\frac{1}{\sqrt{d}}\sum^{d-1}_{j=0}(v_{b,j}+iv_{b,j+d})\ket j\,,\label{eq:vector_expansion}
\end{equation}
with expansion coefficients 
\begin{equation}
\bk j{v_{b}}\sqrt{d}=v_{b,j}+iv_{b,j+d}\in\mathbb{C}\,,\label{eq: real =000026 im expansion coeff's}
\end{equation}
for $b=0\ldots\mu-1,$ and $v,j=0\ldots d-1,$ where $v_{b,j},v_{b,j+d}\in\mathbb{R}$
are their real and imaginary parts. 

\begin{equivalence}\label{equiv: coupled polynomials}
A set of $\mu$ MU bases $\{\mathcal{B}_{0},\ldots,\mathcal{B}_{\mu-1}\}$
in $\mathbb{C}^{d}$ exists if and only if the equations
\begin{multline}
\sum^{d-1}_{j,k=0}(v_{b,j}+iv_{b,j+d})(v^{\prime}_{b',j}-iv^{\prime}_{b',j+d})\times\\
\qquad\qquad\times(v_{b,k}-iv_{b,k+d})(v^{\prime}_{b',k}+iv^{\prime}_{b',k+d})=d\,,\label{eq:pairwise muness}
\end{multline}
with $b\neq b^{\prime},\,v,v^{\prime}=0\ldots d-1$ and 
\begin{equation}
\sum^{d-1}_{j=0}(v_{b,j}-iv_{b,j+d})(v^{\prime}_{b,j}+iv^{\prime}_{b,j+d})=d\delta_{vv^{\prime}}\,,\label{eq:orthogonality_poly}
\end{equation}
with $v,v^{\prime}=0\ldots d-1$, for all $b$, have a real solution;
the equations are a set of coupled fourth-order polynomial equations
in $2d^{2}\mu$ real variables $v_{b,j},v_{b,j+d}$, $b=0\ldots\mu-1$,
and $v,v^{\prime}=0\ldots d-1$.
\end{equivalence} 

The set of equations can be simplified since the conditions of Eqs.~\eqref{eq:mu_condition}
are invariant under a unitary transformation of the bases $\mathcal{B}_{b}$,
so it is natural to choose $\mathcal{B}_{0}=\mathcal{B}_{z}$, or
$\bk j{v_{0}}=\delta_{jv}$. Then, the $d^{2}(\mu-1)$ complex expansion
coefficients of the remaining $d(\mu-1)$ vectors must have modulus
one, i.e.
\begin{equation}
v^{2}_{b,j}+v^{2}_{b,j+d}=1\,,\label{eq:mu_poly}
\end{equation}
$b=1\ldots\mu-1,$ $j=0\ldots d-1$ and $v=0\ldots d-1$. 

A further reduction of parameters is achieved by \emph{dephasing}
the second basis (see Sec.~\ref{subsec: avant-la-letter} and Appendix
\ref{sec:Complex-Hadamard-matrices}). In particular, for the vector
$\ket{v=0_{b=1}}$, i.e. the first vector of the second basis, we
can fix the coefficients to be $\text{Re}\bk j{0_{1}}=1$ and $\text{Im}\bk j{0_{1}}=0,$
for all $j=0\ldots d-1.$ Furthermore, in all remaining vectors of
the second basis, we can choose $\text{Re}\bk 0{v_{1}}\equiv v_{1,0}=1$
and $\text{Im}\bk 0{v_{1}}\equiv v_{1,d}=0$, $v=1\ldots d-1$. 

The conjecture associated with Equivalence \ref{equiv: coupled polynomials}
reads as follows.
\begin{conjecture}
The fourth-order coupled polynomial equations \eqref{eq:pairwise muness}
and \eqref{eq:orthogonality_poly}, with $\mu=d+1$, have real solutions
if and only if $d$ is a prime-power.
\end{conjecture}
Numerical searches for MU bases and smaller constellations (see Sec.~\ref{subsec:nonexistence_of_constellations})
rely on introducing basis-dependent parameterisations such as Eq.~\eqref{eq:vector_expansion},
especially when formulating the search as an optimisation problem
which may use (variants of) the scalar function $F_{d}(\mathcal{B}_{0},\ldots,\mathcal{B}_{\mu-1})$
in Eq.~\eqref{eq: F encoding complete sets}. 

\subsection{Unitary operator bases}

\label{subsec:Maximally-commuting-unitary}

Unitary operator bases, first introduced by Schwinger \citep{schwinger60},
play an important role in the construction of MU bases. A unitary
operator basis is a set of unitary operators which forms an orthonormal
basis of the Hilbert-Schmidt space consisting of the linear operators
acting on $\mathbb{C}^{d}.$ Let $\mathcal{C}=\{U_{1}=\mathbb{I}_{d},U_{2},\ldots,U_{d^{2}}\}$
be a unitary operator basis for the set of $d\times d$ complex matrices
$\mathbb{M}_{d}(\mathbb{C})$, with any two elements being orthogonal
with respect to the Hilbert-Schmidt inner product, $\text{Tr}[U^{\dagger}_{s}U_{s^{\prime}}]=d\delta_{ss^{\prime}},$
$s,s^{\prime}=1\ldots d^{2}$. The basis $\mathcal{C}$ can be partitioned
into \textit{$(d+1)$ maximally commuting operator classes} if $\mathcal{C}=\mathcal{C}_{0}\cup\ldots\cup\mathcal{C}_{d}$
such that each class $\mathcal{C}_{j}$ contains $(d-1)$ commuting
matrices from $\mathcal{C}$ together with the identity $\mathbb{I}\in\mathcal{C}_{j}$,
for all $j$. This partitioning corresponds directly to a complete
set of $(d+1)$ MU bases in $\mathbb{C}^{d}$, as pointed out in Ref.~\citep{bandyopadhyay+02}.
In particular, such a partitioning exists if and only if a complete
set of MU bases exists. Theorems 3.2 and 3.4 of Ref.~\citep{bandyopadhyay+02}
provide a proof of the underlying equivalence.

\begin{equivalence}\label{equiv:commutingclasses}
A set of $\mu$ MU bases $\{\mathcal{B}_{0},\ldots,\mathcal{B}_{\mu-1}\}$
in $\mathbb{C}^{d}$ exists if and only if there exist $\mu$ classes
$\mathcal{C}_{0},\mathcal{C}_{1},\ldots,\mathcal{C}_{\mu-1},$ each
containing $d$ commuting unitary matrices in $\mathbb{M}_{d}(\mathbb{C})$
such that all matrices in $\mathcal{C}_{0}\cup\mathcal{C}_{1}\cup\ldots\cup\mathcal{C}_{\mu-1}$
are pairwise orthogonal.
\end{equivalence}

The crucial property that establishes this equivalence is the fact
that the vectors of the MU basis $\mathcal{B}_{b}$ are given by the
common eigenstates of the commuting unitary matrices within class
$\mathcal{C}_{b}$. Since each class $\mathcal{C}_{b}$ is maximal
commuting, their common eigenstates are fixed by the requirement of
simultaneous diagonalisation, up to overall phases and ordering. If
a set of $\mu$ MU bases $\mathcal{B}_{b}=\{\ket{v_{b}}\}^{d}_{v=1}$
exists, with $b=0\ldots\mu-1$, one can construct the elements of
each commuting class $\mathcal{C}_{b}=\{U_{b,k}\}$ as 
\begin{equation}
U_{b,k}=\sum^{d-1}_{v=0}e^{2\pi ikv/d}\kb{v_{b}}{v_{b}}\,,
\end{equation}
for $k=0\ldots d-1,$ with $U_{b,0}=\mathbb{I}$. Hence, unitary operator
bases with suitable partitions are as unlikely to exist as complete
sets of MU bases in composite dimensions $d\notin\mathbb{PP}$. 
\begin{conjecture}
\label{conj:max_comm_basis}A partition of a unitary operator basis
of $\mathbb{M}_{d}(\mathbb{C})$ into $d+1$ maximally commuting operator
classes exists if and only if $d$ is a prime-power. 
\end{conjecture}
We will see that Equivalence \ref{equiv:commutingclasses} is useful
in relation to \textit{unextendible} MU bases (Sec.~\ref{subsec:Unextendible-MU-bases}),
\textit{nice} MU bases (Sec.~\ref{subsec:Nice-error-bases}), and
\textit{real} MU bases (Sec.~\ref{subsec:Real-MU-bases}). Further
connections between MU bases and maximally commuting operator classes
are discussed in \citep{thas18}. It is found, for example, that non-isomorphic
sets of $d+1$ maximally commuting operator classes may not yield
inequivalent complete sets of MU bases. It is also shown that a complete
set of MU bases generated from maximally commuting operator classes
under certain restrictions (such as forming a finite and nilpotent
operator group) requires the system size to be a prime-power.

\subsection{Decompositions of Lie algebras}

\label{subsec: Orthogonal decomp equivalence}

The existence problem of complete sets of MU bases is, in fact, equivalent
to a problem related to the classical Lie algebras $sl_{d+1}$ which
are associated to the Lie groups $A_{d}$ of invertible linear transformations
in $(d+1)$ dimensions. It was posed in 1981 \citep{kostrikin+81}
as a response to unsuccessful attempts to construct an `{}`orthogonal
decomposition'{}' (see below) of the algebra $sl_{6}$, in contrast
to the corresponding algebras of other small values of $d$.
\begin{problem}[\emph{Winnie-the-Pooh problem}\footnote{The unconventional name of the problem is explained in Ref.~\citep{kostrikin+94}
as a play on words, incomprehensible to the English reader. The pun
is based on the pronunciation of the Russian word for `{}`again'{}'
being similar to that of `{}`$A_{5}$'{}', leading Kostrikin to
adapt a song from A. A. Milne's `{}`Winnie-the-Pooh'{}' which
revolves around the word `{}`again'{}'.}]
\noindent\looseness=-1\emph{\label{WtP problem}Prove that the Lie algebra of
type $A_{d}$ admits an orthogonal decomposition if and only if $d=p^{n}-1$
for some prime $p$ and for some natural number $n$.} 
\end{problem}
To grasp the \emph{equivalence} of the existence of MU bases for arbitrary
dimension $d$ with orthogonal decompositions shown in \citep{boykin+07},
we need to briefly review a number of concepts from the theory of
Lie algebras. We denote $sl_{d}(\mathbb{C})$ as the algebra of $d\times d$
matrices over $\mathbb{C}$ of trace zero, with multiplication defined
by the commutator $[A,B]=AB-BA$. A \emph{Cartan subalgebra} $H$
of a Lie algebra $L$ is a nilpotent subalgebra which is self-normalising,
i.e. if $[A,B]\in H$ for all $A\in H$, then $B\in H.$ For the algebra
$L=sl_{d}(\mathbb{C})$, a Cartan subalgebra is a maximal Abelian
subalgebra. O\emph{rthogonality} of subalgebras is defined with respect
to the \emph{Killing form,} $K(A,B)=\text{tr}(\text{ad}_{A}\cdot\text{ad}_{B})$,
where $\text{ad}_{A}:sl_{d}\rightarrow sl_{d}$ is the \emph{adjoint
endomorphism,} with $\text{ad}_{A}(C)=[A,C]$ for all $C\in sl_{d}(\mathbb{C})$.
The Killing form, which for $sl_{d}(\mathbb{C})$ is given by $K(A,B)=2d\text{tr}(AB)$,
is non-degenerate on the Lie algebra as well as on the restriction
to any Cartan subalgebra $H$. Thus, two Cartan subalgebras $H_{i}$
and $H_{j}$ are \emph{orthogonal} if $K(H_{i},H_{j})=0$. The algebra
$sl_{d}(\mathbb{C})$ has an orthogonal decomposition if it can be
written as a direct sum of orthogonal Cartan subalgebras, i.e., 
\begin{equation}
sl_{d}(\mathbb{C})=H_{0}\oplus H_{1}\oplus\cdots\oplus H_{d}\,.
\end{equation}

Now we are in a position to spell out the equivalence between sets
of MU bases and orthogonal Cartan subalgebras, first noticed in \citep{boykin+07}.
\begin{equivalence}\label{equiv:OD}
A set of $\mu$ MU bases $\{\mathcal{B}_{0},\ldots,\mathcal{B}_{\mu-1}\}$
of $\mathbb{C}^{d}$ exists if and only if a set of $\mu$ pairwise
orthogonal Cartan subalgebras $\{H_{0},\ldots,H_{\mu-1}\}$ of $sl_{d}(\mathbb{C})$
exists, closed under the adjoint operation.
\end{equivalence}

Here, a Cartan subalgebra is closed under the adjoint operation $\dagger$
(i.e., the conjugate transpose) if $H=H^{\dagger}.$ The equivalence
can be understood as follows. One can construct a Cartan subalgebra
$H$ from an orthonormal basis $\mathcal{B}=\{\ket v\}^{d}_{v=1}$
if $H$ is defined as the linear subspace of $sl_{d}(\mathbb{C})$
consisting of all traceless matrices that are diagonal in $\mathcal{B}$.
Any element $A\in H$ can be written as $A=\sum_{v}a_{v}\kb vv$,
with $\sum_{v}a_{v}=0$. By associating each Cartan subalgebra $H_{i}$
with a mutually unbiased basis $\mathcal{B}_{i}$ in this way, it
is straightforward to show that two Cartan subalgebras $H_{i}$ and
$H_{j}$ are \emph{orthogonal} with respect to the Killing form.

To construct an orthonormal basis $\mathcal{B}$ from a Cartan subalgebra
$H$, one takes the common eigenvectors of all the matrices in $H$
as the elements of $\mathcal{B}$. To show that two bases $\mathcal{B}_{i}$
and $\mathcal{B}_{j}$ which correspond to two orthogonal Cartan subalgebras
$H_{i}$ and $H_{j}$ are mutually unbiased, one simply assumes the
opposite, leading to a contradiction of the orthogonality condition.
Note that since any Cartan subalgebra, closed under adjoint operation,
has a basis of unitary matrices that is orthogonal with respect to
the Killing form, this construction of MU bases is essentially the
same as using maximally commuting operator classes (i.e. Equivalence
\ref{equiv:commutingclasses}).

From Equivalence \ref{equiv:OD} it is clear that a complete set of
$d+1$ MU bases exists if and only if one can find a set of pairwise
orthogonal Cartan subalgebras $H_{0},\ldots,H_{d}$ of the Lie algebra
$sl_{d}(\mathbb{C})$. Turning this into a conjecture on MU bases
is nothing but a reformulation of the Winnie-the-Pooh problem.
\begin{conjecture}
\label{conj:lie_alg}The simple Lie algebra $sl_{d}(\mathbb{C})$
admits an orthogonal decomposition if and only if $d$ is a prime-power. 
\end{conjecture}
In later sections we will see that the correspondence outlined above
has useful consequences in relation to monomial MU bases (Sec.~\ref{sec:monomial_bases})
and the existence of Hadamard matrices (Sec.~\ref{sec:pairs_of_MU_bases_C^6}).
Furthermore, since the automorphism group of a complete set of MU
bases is isomorphic to the automorphism group of the associated orthogonal
decomposition \citep{Abdukhalikov2015}, these can be used to show
inequivalences between complete sets (see Appendix \ref{subsec:other_constructions}).

\subsection{Maximal Abelian subalgebras}

\label{subsec: masa}

The one-to-one correspondence between orthonormal bases in $\mathbb{C}^{d}$
and maximal Abelian subalgebras (MASAs) of $\mathbb{M}_{d}(\mathbb{C})$
mentioned in Sec.~\ref{subsec: avant-la-letter} leads to a natural
reformulation of MU bases in terms of the orthogonality relations
between collections of MASAs. In particular, an orthonormal basis
$\mathcal{B}=\{\ket v\}$ generates a MASA,
\begin{equation}
\mathcal{A}(\mathcal{B})=\left\{ \sum^{d-1}_{v=0}\alpha_{v}\kb vv\text{\,\ensuremath{\bigg|}\,\ensuremath{\alpha_{v}\in\mathbb{C}}}\right\} \,,
\end{equation}
and, conversely, each MASA defines an orthonormal basis. As shown
in \citep{parthasarathy04}, given two bases $\mathcal{B}$ and $\mathcal{B}'$,
and their corresponding MASAs $\mathcal{A}(\mathcal{B})$ and $\mathcal{A}(\mathcal{B}')$,
then $\mathcal{B}$ and $\mathcal{B}'$ are mutually unbiased if and
only if $\mathcal{A}(\mathcal{B})\ominus\mathbb{C}\mathbb{I}$ and
$\mathcal{A}(\mathcal{B}')\ominus\mathbb{C}\mathbb{I}$ are mutually
orthogonal (in the sense of the Hilbert-Schmidt inner product). Here,
$\mathcal{A}(\mathcal{B})\ominus\mathbb{C}\mathbb{I}$ denotes the
orthogonal complement of the subspace $\mathbb{C}\mathbb{I}$ in $\mathcal{A}(\mathcal{B})$,
with $\mathbb{C}\mathbb{I}=\{z\mathbb{I}\,|\,z\in\mathbb{C}\}$. In
recent literature, e.g. \citep{petz07}, the orthogonality relation
between MASAs is usually called \textit{quasi-orthogonality.} Generalising
to larger collections of MASAs, we have the following equivalence
\citep{parthasarathy04}.
\begin{equivalence}\label{equiv:masa}
A set of $\mu$ bases $\{\mathcal{B}_{0},\ldots,\mathcal{B}_{\mu-1}\}$
of $\mathbb{C}^{d}$ are mutually unbiased if and only if the MASAs
$\mathcal{A}(\mathcal{B}_{0}),\ldots,\mathcal{A}(\mathcal{B}_{\mu-1})$
are quasi-orthogonal.
\end{equivalence}

This equivalence leads to a conjecture on the existence of MASAs.
\begin{conjecture}
\label{conj:masa}A set of $(d+1)$ pairwise quasi-orthogonal MASAs
of $\mathbb{M}_{d}(\mathbb{C})$ exists if and only if $d$ is a prime-power.
\end{conjecture}
Studies on MASAs, from as early as 1983, have provided constructions
of complete sets of MU bases in prime dimensions \citep{popa83} (see
Sec.~\ref{subsec: avant-la-letter}) which were later rediscovered
in the language of MU bases, along with several other results. Since
then, MASAs have been relevant for the classification of complex Hadamard
matrices of order $d\leq5$ \citep{haagerup97}, constructions of
strongly unextendible MU bases \citep{szanto16} (Sec.~\ref{subsec:Unextendible-MU-bases}),
as well as to show that $d$ MU bases are sufficient for a complete
set (see Sec.~\ref{subsec:d_MU_bases_sufficient}).

\subsection{$C^{*}$-algebra formulation}

\label{subsec:C*}

Describing a pair of MU bases by their rank-one projectors $P_{1}(v)=\kb{v_{1}}{v_{1}}$
and $P_{2}(v)=\kb{v_{2}}{v_{2}}$ for $v=0\ldots d-1$, it is easy
to check that the following conditions hold: \emph{(i)} $P_{1}(v)P_{1}(v')=\delta_{vv'}P_{1}(v)$;
\emph{(ii)} $\sum_{v}P_{1}(v)=\mathbb{I}$; \emph{(iii)} $P_{1}(v)P_{2}(v')P_{1}(v)=P_{1}(v)/d$;
(iv) $[P_{1}(v)UP_{1}(v),P_{1}(v)VP_{1}(v)]=0$, for all $v,v'=0,\ldots,d-1$
and $U,V\in\mathbb{C}^{d\times d}$. Relaxing the condition that the
projectors are $d\times d$ matrices, a special type of $C^{*}$-algebra
can be defined called a $(d,\mu)$-MUB algebra \citep{navascues12}.
\begin{defn}
A $C^{*}$-algebra $\mathcal{A}$ is called a $(d,\mu)$-MUB algebra
if it contains Hermitian elements $X_{v,b}$ for all $v=0\ldots d-1$
and $b=1\ldots\mu$, which satisfy the following conditions:
\end{defn}
\begin{enumerate}
\item[(i)]  $X_{v,b}X_{v',b}=\delta_{v,v'}X_{v,b}$ for all $v,v'=0\ldots d-1$
and $b=1\ldots\mu$;
\item[(ii)]  $\sum^{d}_{v=1}X_{v,b}=I$ for all $b=1\ldots\mu$;
\item[(iii)]  $X_{v,b}=dX_{v,b}X_{v',b'}X_{v,b}$ for all $v,v'=0\ldots d-1$
and $b,b^{\prime}=1\ldots\mu$ with $b\neq b'$;
\item[(iv)]  $[X_{v,b}UX_{v,b},X_{v,b}VX_{v,b}]=0$ for all $v=0\ldots d-1$
and $b=1\ldots\mu$, and $U,V\in\langle{\bf X}\rangle$, where $\langle{\bf X}\rangle$
denotes the set of monomials in $X_{v,b}$.
\end{enumerate}
If $X_{v,b}$ form the rank-one projectors associated with $\mu$
MU bases of $\mathbb{C}^{d}$, then the conditions (i-iv) are clearly
satisfied, with $U,V\in\mathbb{C}^{d\times d}$. Thus, $\mu$ MU bases
give rise to a $(d,\mu)$-MUB algebra. The converse, proved in \citep{navascues12},
is also true, i.e., the existence of a $(d,\mu)$-MUB algebra with
$I\neq0$, implies the existence of $\mu$ MU bases in $\mathbb{C}^{d}$.
Thus, the existence of MU bases can be reformulated in terms of a
$C^{*}$-algebra \citep{navascues12}.
\begin{equivalence}\label{equiv: C*algebra}
A set of $\mu$ MU bases $\{\mathcal{B}_{0},\ldots,\mathcal{B}_{\mu-1}\}$
in $\mathbb{C}^{d}$ exists if and only if there exist a $(d,\mu)$-MUB
algebra with $I\neq0$.
\end{equivalence}

The equivalence implies a conjecture about MUB-algebras corresponding
to \emph{complete} sets of MU bases.
\begin{conjecture}
A $(d,d+1)$-MUB algebra exists if and only if $d$ is a prime-power.
\end{conjecture}
Formulating the existence question in this manner yields a non-commutative
polynomial optimisation problem \citep{gribling21}. Upon exploiting
existing symmetries, a tractable hierarchy of semidefinite programs
is constructed, leading to numerical sum-of-squares certificates for
the non-existence of $(d+2)$ MU bases for $d\leq8$.

\subsection{Quantum designs}

\label{subsec:zauner_conjecture}

In this section we discuss Zauner's definition of quantum designs
and show how they relate to MU bases and complex projective 2-designs.
A \emph{quantum design} of order $d$ is a set $\mathcal{D}=\{P_{1},\ldots,P_{n}\}$
of complex orthogonal $d\times d$ projection matrices; it is called\textit{
$k$}-coherent if for each unitary $U,$ we have
\begin{equation}
\sum^{n}_{i=1}P^{\otimes k}_{i}=\sum^{n}_{i=1}(UP_{i}U^{-1})^{\otimes k}.\label{eq: k-coherence}
\end{equation}
The set $\mathcal{D}$ is a \emph{quantum $t$-desig}n if it is a
$k$-coherent quantum design for all $1\leq k\leq t$ \citep{Zauner1999}.
In the particular case that $\mathcal{D}$ is a finite subset of the
complex projective space, i.e. $\mathcal{D}\subset\mathbb{C}P^{d-1}$,
as discussed in Sec.~\ref{subsec: 2-design equivalence}, the condition
of \textit{$k$}-coherence is equivalent to the definition of a complex
projective $k$-design, i.e.,
\begin{equation}
\frac{1}{|\mathcal{D}|}\sum_{x\in\mathcal{D}}P(x)^{\otimes k}=\int_{\mathbb{C}P^{d-1}}P(x){}^{\otimes k}d\nu(x)\,,\label{eq: complex k-design}
\end{equation}
where the points $x\in\mathcal{D}$ can be represented as unit vectors
$\ket x\in\mathbb{C}^{d}$ or rank-one projectors $P(x)=\kb xx$,
and $\nu$ is the unitarily invariant Haar measure on $\mathbb{C}P^{d-1}$.

The focus of Zauner's 1999 PhD thesis was on so-called \textit{regular
affine} quantum designs \citep{Zauner1999} (see \citep{Zauner2011}
for an English translation). A \textit{regular} quantum design is
a set $\mathcal{D}=\{P_{1},\ldots,P_{n}\}$ of $d\times d$ projection
matrices such that $\mbox{Tr}[P_{i}]=r$ for all $i=1\ldots n$. The
degree of a quantum design is defined as the number of distinct elements
in the set $\{\mbox{Tr}[P_{i}P_{j}]:1\leq i\neq j\leq n\}$. A quantum
design is called resolvable if the $n$ projections can be partitioned
into $\mu$ subclasses, where each subclass contains $n/\mu$ pairwise
orthogonal projections summing to the identity. An \textit{affine}
quantum design is resolvable and has degree two. When affine designs
are regular, i.e. $r=1$, the elements of $\mathcal{D}$ form a set
of MU bases, and the following equivalence holds \citep{Zauner2011}.
\begin{equivalence}\label{equiv:quantumdesign-1}
A set of $\mu$ MU bases $\{\mathcal{B}_{0},\ldots,\mathcal{B}_{\mu-1}\}$
of $\mathbb{C}^{d}$ exists if and only if an affine quantum design
of order $d$ and $r=1$ exists with $\mu$ orthogonal subclasses.
\end{equivalence}

Unaware of this equivalence, Zauner found that regular affine quantum
designs have $(d+1)$ orthogonal classes when $d$ is a prime or prime-power.
The solution generalises to $r\geq1$, where the existence of $r^{2}(d^{2}-1)(d-r)$
orthogonal classes is shown. He also suggests that ``\emph{Presumably
a complex, affine quantum design with $d=6$, $r=1$ and $\mu=4$
does not exist}'' \citep{Zauner2011}. This appears to be the first
statement conjecturing the non-existence of \textit{four} MU bases
in $\mathbb{C}^{6}$; Winnie-the-Pooh's formulation of the problem
in the Lie algebraic setting (see Sec.~\ref{subsec: Orthogonal decomp equivalence})
was silent with regards to this question. Zauner's statement is easily
generalised to the case of complete sets of MU bases in composite
dimensions $d\notin\mathbb{\ensuremath{PP}}$. 
\begin{conjecture}
\label{conj:quantumdesign} An affine quantum design of order $d$
with $r=1$ and $(d+1)$ orthogonal subclasses exists if and only
if $d$ is a prime-power. 
\end{conjecture}
In Sec.~\ref{sec:triples_of_MU_bases} we describe an infinite family
of triples of MU bases in $\mathbb{C}^{6}$ which were first constructed
as affine quantum designs.

\subsection{Quantum random access codes }

\label{subsec:QRAC}

As already mentioned in Sec.~\ref{subsec:QRAC application}, the
QRAC $\mu^{d}\rightarrow1$ protocol can be modified by giving Alice
an extra piece of information about Bob's actions: Alice is told that
Bob will only be questioned on a certain subset of the $d$ levels.
In particular, let $S_{\nu}$ be the set of all possible subsets of
$\{1,\ldots,\mu\}$ of size $\nu\leq\mu,$ and suppose Alice receives
an input $s\in S_{\nu}$ with the promise that $y\in s$. This setting
defines a `{}`p-QRAC protocol'{}' (with `{}`p'{}' for `{}`promise'{}')
of type $(\mu,\nu)^{d}\rightarrow1$. The average success probability,
$p(\mu,\nu,d),$ is then calculated over all ${\bf x}_{s}$, $y\in s,$
and $s\in S_{\nu},$ where ${\bf x}_{s}=\{x_{j_{1}},\ldots,x_{j_{\nu}}\}$
such that $\{j_{1},\ldots,j_{\nu}\}=s$. When $\nu=2$, the \textit{optimal}
average success probability $P(\mu,2,d)$, maximised over all states
and measurements, is intimately connected to MU bases \citep{aguilar18}.
\begin{equivalence}\label{equiv:QRACs} For a p-QRAC of the type
$(\mu,2)^{d}\rightarrow1,$ the optimal average success probability
satisfies
\begin{equation}
P(\mu,2,d)\leq\frac{1}{2}\left(1+\frac{1}{\sqrt{d}}\right),\label{eq:QRACsucessprob}
\end{equation}
 and achieves equality if and only if Bob's measurement bases are
a set of $\mu$ MU bases $\{\mathcal{B}_{0},\ldots,\mathcal{B}_{\mu-1}\}$
of $\mathbb{C}^{d}$.
\end{equivalence} 

In other words, sets of $\mu$ MU bases in dimension $d$ exist if
and only if the optimal average success probability saturates the
bound in Eq.~(\ref{eq:QRACsucessprob}). It follows that the optimal
average success probability with $\mu=d+1$ cannot reach this bound
in composite dimensions if no complete set of MU bases exists.
\begin{conjecture}
\label{conj:QRAC} A p-QRAC of the type $(d+1,2)^{d}\rightarrow1$
achieves the optimal average success probability given by the RHS
of Eq.~\emph{\eqref{eq:QRACsucessprob}} if and only if $d$ is a
prime-power.
\end{conjecture}
Equivalence \ref{equiv:QRACs} leads naturally to an \emph{operational}
measure of mutual unbiasedness for a set of $\mu$ bases, given by
the average success probability $p(\mu,\nu,d)$ of the $(\mu,2)^{d}\rightarrow1$
p-QRAC protocol, when the $\mu$ bases are measured by Bob. Numerical
calculations of this measure provide additional evidence that only
three MU bases exist in dimension six (cf. Sec.~\ref{subsec:nonexistence_of_quadruples}).

The equivalence relation also provides a strategy to prove the conjecture
that no set of four MU bases exists in $d=6$ (cf. Sec.~\ref{subsec:Solution-strategies}).
Using semidefinite programming methods derived in \citep{navascues15}
to find an upper bound on $P(4,2,d)$ for the $(4,2)^{6}\rightarrow1$
p-QRAC, one can conclude that four MU bases do not exist if the bound
falls below the optimal value in Eq.~\eqref{eq:QRACsucessprob}.

\subsection{Geometry of MU bases}

\label{subsec: Gemometry of MU bases}

A complete set of MU bases for a qubit consists of six states in Hilbert
space $\mathbb{C}^{2}$. They form three orthogonal pairs while states
taken from different pairs have constant overlap of modulus $1/\sqrt{2}$.
The rank-one projectors onto these states can be pictured as points
on the surface of the three-dimensional \emph{Bloch ball} used to
represent both pure and mixed qubit states in the space $\mathbb{R}^{3}$.
The convex combinations associated with the orthogonal quantum states
within a pair form three line segments, each contained in a one-dimensional
subspace of $\mathbb{R}^{3}$. The line segments are pairwise perpendicular
in terms of the standard Euclidian geometry of $\mathbb{R}^{3}$,
and the distance between endpoints belonging to different segments
is constant. Forming the convex hull of the six states of the complete
set, one obtains an \emph{octahedron} contained entirely within the
Bloch ball, known as the \emph{complementarity polytope} associated
with a qubit \citep{Bengtsson2005,bengtsson05}.

Interestingly, polytopes with these characteristics exist in any Euclidean
space of dimension $\mathbb{R}^{d^{2}-1}$, $d\in\mathbb{N}$, used
to represent the state space of a qudit. The one-dimensional line
segments with two endpoints each generalise to \emph{regular} \emph{simplices}
with $(d-1)$ equidistant vertices. In total, $(d+1)$ such simplices
exist in the space $\mathbb{R}^{d^{2}-1}$. They are orthogonal to
each in the Euclidean sense, i.e. they have no point in common except
for the origin, just as the three line segments. Here, the expression
\begin{equation}
D^{2}(A,A^{\prime})=\frac{1}{2}\textrm{Tr}(A-A^{\prime})^{2}\,\label{eq: HD distance}
\end{equation}
defines the (Hilbert-Schmidt) distance in the $(d^{2}-1)$-dimensional
Euclidean space of Hermitian matrices $A,A^{\prime},$ with unit trace.
Using the inner product
\begin{equation}
(A,A^{\prime})=\frac{1}{2}\left(\textrm{Tr}(AA^{\prime})-\frac{1}{d}\right)\label{eq: inner product R^(d^2-1)}
\end{equation}
results in a vector space $\mathbb{R}^{d^{2}-1}$, with the maximally
mixed state $\mathbb{I}/d$ taken as its origin.

The density matrices of a qudit define a convex set of points in $\mathbb{R}^{d^{2}-1}$,
generalising the Bloch ball for a qubit in $\mathbb{R}^{3}$ . While
complementarity polytopes exist for every dimension $d$, they do
not necessarily fit inside the body of density matrices: inscribing
the polytope requires its vertices to correspond to rank-one projectors.
\begin{equivalence} \label{equiv: Complementarity polytope}
The $d(d+1)$ vertices of the complementarity polytope are given by
rank-one projectors if and only if they correspond to the states of
a complete set of MU bases in $\mathbb{C}^{d}$.
\end{equivalence}

In the space $\mathbb{R}^{d^{2}-1}$, it is difficult to geometrically
identify suitable rotations ensuring this property although each construction
of a complete sets of MU bases for $d\in\mathbb{PP}$ guarantees the
existence of such a rotation.
\begin{conjecture}
For a qudit, the $(d^{2}-1)$-dimensional complementarity polytope
fits inside the body of density matrices if and only if $d$ is a
prime-power.
\end{conjecture}

\subsection{Complex projective 2-designs}

\label{subsec: 2-design equivalence}

Loosely speaking, a complex projective $t$-design is a finite set
of points on the unit sphere which can be used to calculate integrals
of functions over the entire unit sphere. More precisely, it is a
finite subset $\mathcal{D}$ of the complex projective space $\mathbb{C}P^{d-1}$
of lines passing through the origin, such that the expectation of
every polynomial of degree at most $t$ over the points (defined as
the intersections of the lines with the unit sphere) is the same irrespective
of the distribution being the Haar measure or the $t$-design.

It is often convenient to represent these points $x\in\mathcal{D}\subset\mathbb{C}P^{d-1}$
as a collection of unit vectors $\ket x\in\mathbb{C}^{d}$ or, alternatively,
by rank-one projectors $P(x)=\kb xx$. For a unit vector $\ket x=\sum^{d}_{j=1}x_{j}\ket j$,
with $x_{j}\in\mathbb{C}$, we define $f:\mathbb{C}P^{d-1}\rightarrow\mathbb{R}$
to be a homogenous polynomial $f(x)=f(x_{1},\ldots,x_{d},x^{*}_{1},\ldots,x^{*}_{d})$
of degree at most $t$ in the coefficients $x_{j}$ , and of degree
at most $t$ in $x^{*}_{j}$. Polynomials of this type are denoted
as $f\in$ Hom$(t,t)$. The identity
\begin{equation}
\frac{1}{|\mathcal{D}|}\sum_{x\in\mathcal{D}}f(x)=\int_{\mathbb{C}P^{d-1}}f(x)d\nu(x)\,,\label{eq:complex_projective_design}
\end{equation}
holds for all $f\in$ Hom$(t,t)$, if and only if $\mathcal{D}$ is
a complex projective $t$-design, where the integral is evaluated
over the unitarily invariant Haar measure $\nu$ on the unit sphere
in $\mathbb{C}^{d}$. This type of identity, which equates an integral
of a function over a domain with a sum of the function values evaluated
at specific points of the domain, is known as a \emph{cubature} (formula).
It represents a convenient starting point to evaluate integrals by
numerical methods. 

The definition of a complex projective $t$-design implies that it
is also a $t'$-design for every $t'\le t$. In particular, any $t$-design
with $t\ge1$ is a projective $1$-design, which in the language of
frame theory is equivalent to a finite unit-norm tight frame. Consequently,
rescaling the rank-one operators $\ket x\bra xd/|\mathcal{D}|$ by
the factor $d/|\mathcal{D}|$ generates a POVM.

Designs of this type are known to exist in every dimension $d$ and
for any $t$ \citep{seymour84}. An example of a 1-design is an orthonormal
basis of the space $\mathbb{C}^{d}.$ The smallest set of vectors
which constitute a 2-design is formed by $d^{2}$ complex equiangular
lines which correspond to the elements of a SIC-POVM \citep{renes04},
as described in Sec.~\ref{subsec:MU-bases-and-SICs}. The set of
qubit stabilizer states form a 3-design \citep{kueng15}.

It can be convenient to define a complex-projective $t$-design (cf.
Eq.~\eqref{eq:complex_projective_design}) via an equivalent expression.
For example, a $t$-design is a set $\mathcal{D}$ such that
\begin{align}
\frac{1}{|\mathcal{D}|}\sum_{x\in\mathcal{D}} P(x)^{\otimes t}
  &= \int_{\mathbb{C}P^{d-1}} P(x)^{\otimes t}\, d\nu(x) \notag\\
  &= \binom{d+t-1}{t}^{-1}\Pi^{(t)}_{\mathrm{sym}}\,,
\label{eq:t-design-symmetric-subspace}
\end{align}
where $\Pi^{(t)}_{\text{sym}}$ is the projector onto the symmetric
subspace of $(\mathbb{C}^{d})^{\otimes t}$ and $P(x)=\kb xx$ is
the projection operator associated with $x\in\mathbb{C}P^{d-1}$.
The equivalence between Eqs.~\eqref{eq:complex_projective_design}
and \eqref{eq:t-design-symmetric-subspace} is shown in Ref.~\citep{renes04}.

Barnum \citep{barnum02} appears to have been among the first to point
out—in the context of information–disturbance tradeoffs—that the uniform
ensemble supported on a complete set of MU bases forms a complex projective
2-design. Klappenecker and Rötteler \citep{klappenecker+05b} subsequently
related Zauner's notion of quantum designs (see Sec.~\ref{subsec:zauner_conjecture})
to MU bases, showing that a set of $(d+1)$ MU bases forms a complex
projective $2$-design containing $d(d+1)$ elements. If the elements
of a design $\mathcal{D}$ are restricted to a set of $\mu$ orthonormal
bases, then $\mathcal{D}$ is a 2-design only when $\mu\geq d+1$,
with equality if and only if the bases are mutually unbiased \citep{roy+07}.
This observation leads to an equivalence between complex projective
2-designs and MU bases \citep{roy+07}. 
\begin{equivalence}\label{equiv:2-designs}
Suppose $\mathcal{D}$ is a set of $(d+1)$ orthonormal bases in $\mathbb{C}^{d}$.
Then $\mathcal{D}$ is a complex projective 2-design if and only if
$\mathcal{D}$ is a set of MU bases.
\end{equivalence}

Thus, any set of $(d+1)$ MU bases $\mathcal{B}_{b}=\{\ket{v_{b}}\}^{d-1}_{v=0}$,
satisfies the condition
\begin{equation}
\sum^{d}_{b=0}\sum^{d-1}_{v=0}P_{b}(v)^{\otimes2}=2\Pi^{(2)}_{\text{sym}}\,,\label{eq:MU-design-property}
\end{equation}
where $P_{b}(v)=\kb{v_{b}}{v_{b}}$. With this equivalence in mind
we formulate a conjecture on the existence of complex projective $2$-designs.
\begin{conjecture}
\label{conj:2-designs} A complex-projective 2-design formed from
$(d+1)$ orthonormal bases in $\mathbb{C}^{d}$ exists if and only
if $d$ is a prime-power. 
\end{conjecture}
We note that an equivalence relation also exists between a projective
toric 2-design \citep{kuperberg06} and a complete set of MU bases,
as shown in Refs. \citep{iosue22,iosue23}. In particular, a set of
$d$ complex Hadamard matrices of order $d$ is mutually unbiased
if and only if their columns form a projective toric 2-design. A projective
toric 2-design is a finite set of points in $P(T^{d})$ that satisfies
a condition similar to Eq.~\eqref{eq:complex_projective_design},
replacing the polynomials with monomials on $T^{d}$ of a specific
degree. Here, $T^{d}$ denotes the $d$-dimensional torus $T=\mathbb{R}/2\pi\mathbb{Z}$, and $P(T^{d})$ consists of the set of points on $T^{d}$ identified
up to a constant additive factor (i.e. without the global phase redundancy).

Interestingly, a projective toric 2-design can be used to construct
a complex projective 2-design with $d(d+1)$ elements that does not
form complete sets of MU bases \citep{iosue23}, disproving a conjecture
by Zhu \citep{zhu15}.

In Secs. \ref{subsec: Quantum-state-reconstruction} and \ref{subsec:entanglement-detection}
we discussed applications of MU bases which included quantum state
tomography and entanglement detection, respectively. However, these
applications are not restricted to MU bases—other complex projective
2-designs \citep{roy+07,bae18} may be used instead. Furthermore,
in composite dimensions $d\notin\mathbb{PP}$, the 2-design property
of the $(d+1)$ MU bases plays an important role in restricting their
entanglement content (cf. Sec.~\ref{subsec:entanglement_content}).
In Sec.~\ref{subsec:groups} we will see that a search for projective
toric 2-designs that form groups reveals a clear distinction between
complete sets in dimension six and those in prime (or prime-powered)
dimensions. Finally, Sec.~\ref{sec: Modifications-of-the-problem}
deals with modifications of Eq.~\eqref{eq:complex_projective_design}
which extend the definition of complex projective 2-designs to\emph{
approximate} (Sec.~\ref{sec:approximate2-designs}), \emph{weighted}
(Sec.~\ref{sec:weighted2designs}), and \textit{conical} (Sec.~\ref{subsec:Mutually-unbiased-measurements})
2-designs.

\subsection{Welch bounds}

\label{subsec: Welch bound equivalence}

Sets of points in the complex projective space $\mathbb{C}P^{d-1}$
are not only related to designs but also come with bounds on the moduli
of overlaps between the states associated with them. The \emph{Welch
bounds }\citep{welch74} 
\begin{equation}
\frac{1}{|\mathcal{X}|^{2}}\sum_{x,y\in\mathcal{X}}|\bk xy|^{2k}\geq{d+k-1 \choose k}^{-1}\,,\label{eq:Welch-bound}
\end{equation}
$k\in\mathbb{N}_{0}$, apply to any finite set $\mathcal{X}=\{x:x\in\mathbb{C}P^{d-1}\}$
with $|\mathcal{X}|>0$ elements. Equality holds for all integer values
of $k$ up to $t$, i.e. $0\leq k\leq t$, if and only if the set
$\mathcal{X}$ is a complex projective $t$-design \citep{klappenecker+05b}.
Thus, if the set $\mathcal{X}$ consists of $(d+1)$ MU bases—which
form a complex projective 2-design according to Sec.~\ref{subsec: 2-design equivalence}—then
we have the two identities
\begin{equation}
\sum_{x,y\in\mathcal{X}}|\bk xy|^{2}=d(d+1)^{2},\label{eq:MUwelchbound1}
\end{equation}
and
\begin{equation}
\sum_{x,y\in\mathcal{X}}|\bk xy|^{4}=2d(d+1)\,,\label{eq:MUwelchbound2}
\end{equation}
for $k=1$ and $k=2$, respectively. The equality for $k=0$ simply
counts the number of ordered pairs of vectors from the set $\mathcal{X}$,
and the $k=1$ identity reduces to the normalisation condition. Hence,
the Welch bounds can be used to reformulate the existence problem
of MU bases. 

\begin{equivalence}\label{equiv:welchbounds}

Suppose that $\mathcal{X}$ is a set of $(d+1)$ orthonormal bases
in $\mathbb{C}^{d}$. Then $\mathcal{X}$ saturates the Welch bounds
in Eq.~\emph{\eqref{eq:Welch-bound}} for $0\leq k\leq2$ if and
only if $\mathcal{X}$ is a set of $(d+1)$ MU bases.

\end{equivalence}
The proof can be found in \citep{roy+07}. It relies on the observation
in \citep{klappenecker+05b} that $\mathcal{X}$ saturates the Welch
bound for $0\leq k\leq2$ if and only if $\mathcal{X}$ is a 2-design.
It then follows from Equivalence \ref{equiv:2-designs} that $\mathcal{X}$
constitutes a complete set of MU bases. Thus, the existence problem
for MU bases in composite dimensions turns into an open problem for
bases saturating Welch bounds.
\begin{conjecture}
\label{conj:welch_bounds} A set of $(d+1)$ orthogonal bases in $\mathbb{C}^{d}$
saturating the Welch bound of Eq.~\emph{\eqref{eq:Welch-bound}}
for $k=0,1,2$ exists if and only if $d$ is a prime-power.
\end{conjecture}
Using Welch bounds, it is also possible recast the existence question
of MU bases as one involving only orthogonal vectors. Suppose that
$A$ and $B$ are matrices of order $d,$ and let $A\circ B$ denote
the Hadamard product, i.e., $(A\circ B)_{jk}=A{}_{jk}\times B{}_{jk}$,
and $A^{(\ell)}$ the $\ell$-th Hadamard power of $A$, with matrix
elements $A^{(\ell)}{}_{jk}=A{}^{\ell}_{jk}.$ For a suitable matrix,
the following proposition demonstrates an equivalence between its
columns saturating a Welch bound and its rows satisfying a set of
orthogonality conditions \citep{belovs+08}.
\begin{prop}
\label{thm:welsh_bounds} Let $B$ be a $d\times n$ matrix and denote
its sets of columns and rows by $\mathcal{C}=\{c_{1},\ldots,c_{n}\}\subset\mathbb{C}^{d}$
and $\mathcal{R=}\{r_{1},r_{2},\ldots,r_{d}\}\subset\mathbb{C}^{n}$,
respectively. Then the column vectors $\mathcal{C}$ saturate the
Welch bound for $k=2$ if and only if all vectors from the set $\{r^{(2)}_{j}\}\cup\{\sqrt{2}r_{j}\circ r_{k}|1\leq j<k\leq d\}$
are pairwise orthogonal and of equal length. 
\end{prop}
If the columns of $B$ are given by the columns of $d$ Hadamard matrices
of order $d$, the row vectors of $B$ form a set of $d$ vectors
in $\mathbb{C}^{d^{2}}$. Subsequently, via a Hadamard product, every
pair of these $d$ row vectors maps to a new vector in $\mathbb{C}^{d^{2}}$.
It follows from the above proposition that the $d$ Hadamard matrices
are MU if and only if the latter vectors in $\mathbb{C}^{d^{2}}$
are pairwise orthogonal. This gives rise to yet another characterisation
of complete sets of MU bases.

\begin{equivalence}\label{equiv:welch_bounds2}
Let $\{H_{1},\ldots,H_{d}\}$ be a set of Hadamard matrices of order
$d$, and let matrix $B$ be the $d\times d^{2}$ matrix formed from
the set of normalised column vectors of each Hadamard matrix. The
set of columns and rows of $B$ are given by $\mathcal{C}=\{c_{1},\ldots,c_{d^{2}}\}\subset\mathbb{C}^{d}$
and $\mathcal{R=}\{r_{1},r_{2},\ldots,r_{d}\}\subset\mathbb{C}^{d^{2}}$,
respectively. Then $\{H_{1},\ldots,H_{d}\}$ are pairwise mutually
unbiased if and only if all vectors from the set $\{r_{j}\circ r_{k}|1\leq j\leq k\leq d\}$
are pairwise orthogonal.
\end{equivalence}

With this equivalence in mind, the existence problem of a complete
set of MU bases leads to a conjecture about the orthogonality of a
set of vectors in $\mathbb{C}^{d^{2}}$.
\begin{conjecture}
\label{conj:welch_bounds2} A set of pairwise orthogonal vectors $\{r_{j}\circ r_{k}|1\leq j\leq k\leq d\}$
in $\mathbb{C}^{d^{2}}$ with the properties defined in Equivalence
\ref{equiv:welch_bounds2} exists if and only if $d$ is a prime-power.
\end{conjecture}

\subsection{Rigidity of MU bases}

\label{subsec:rigidity}

The existence problem of complete sets of MU bases can be reformulated
in terms of the overlap constraints on a set of vectors, without reference
to orthonormal bases \citep{matolcsi21}.

\begin{equivalence}\label{equiv:rigidity}
Let $\mathcal{S}_{d}=\{\ket{\phi_{i}}\:|\,i=1,\ldots,d(d+1)\}$ be
a collection of $d(d+1)$ unit vectors in $\mathbb{C}^{d}$. A set
of $(d+1)$ MU bases in $\mathbb{C}^{d}$ exists if and only if $|\bk{\phi_{i}}{\phi_{j}}|^{2}\in\{0,1/d\}$
for all $i\neq j$.
\end{equivalence}

Clearly, the existence of a complete set of MU bases implies all $d(d+1)$
vectors are either orthogonal or mutually unbiased. The converse is
less obvious: for example, as many as $(d-1)^{2}$ unit vectors in
$\mathbb{C}^{d}$ exist which are pairwise unbiased. The proof of
the equivalence \citep{matolcsi21} relies on graph theoretic methods.
Consequently, we have the following conjecture on the existence of
a collection of vectors that can be pairwise orthogonal or mutually
unbiased.
\begin{conjecture}
\label{conj:rigidity} A set \textup{$\mathcal{S}_{d}$ }of $d(d+1)$
unit vectors \textup{\emph{satisfying the conditions}}\textup{ }$|\bk{\phi_{i}}{\phi_{j}}|^{2}\in\{0,1/d\}$
for all $i\neq j$ exists if and only if d is a prime-power.
\end{conjecture}

\section{Rigorous results: Composite dimensions}

\label{sec:rigorous_results_any_d}

This section collects rigorous existence and non-existence results
relating to MU bases in non-prime-power dimensions. We start with
the problem of identifying all pairs of MU bases, which is equivalent
to listing all complex Hadamard matrices. Moving on to the existence
of larger sets of MU bases, we review various construction techniques
and discuss their limitations. Most constructions (in Sec. \ref{subsec:Nice-error-bases}–\ref{subsec:Maximally-entangled-bases})
impose additional constraints on the bases (e.g. niceness, monomiality,
separability and entanglement) which help simplify the problem, leading
to larger sets of MU bases or proofs of unextendibility (cf. Sec.~\ref{subsec:Unextendible-MU-bases}).
Finally, in Sections \ref{subsec:delsarte} and \ref{subsec:linear-constraints},
we review some mathematical techniques that shed light on the structure
of complete sets, and may be useful in a solution to the existence
problem. In Sec.~\ref{sec:rigorous_results_d=00003D6}, we will make
explicit the consequences of these results for $d=6$ and summarise
additional ones specific to this dimension.

\subsection{Pairs of MU bases}

\label{sec:pairs_of_MU_bases_C^d}

Any pair of mutually unbiased bases $\{\mathcal{B}_{0},\mathcal{B}_{1}\}$
of the space $\mathbb{C}^{d}$ is equivalent to the existence of a
complex Hadamard matrix of order $d$ (see Equivalence \ref{equiv:hadamards}).
Thus, listing all $d\times d$ Hadamard matrices provides a full characterisation
of pairs of MU bases in dimension $d$. While every Hadamard matrix
is known up to and including dimension five \citep{haagerup97,Kraus1987},
there is no exhaustive list for dimensions six and above.

The study of real and complex Hadamard matrices has been an active
research area for several centuries, and continues to be so. Instead
of describing all constructions, which is beyond the scope of this
review, we will summarise results which are relevant in our context.
For standard definitions such as the equivalence between Hadamard
matrices, their defects, isolated Hadamard matrices as well as affine
and non-affine families, see Appendix \ref{sec:Complex-Hadamard-matrices}.

An early construction of a complex Hadamard matrix dates back to Sylvester
(see Sec. \ref{subsec: avant-la-letter}) who described the $d\times d$
Fourier matrices $F_{d}$ in prime dimensions $d\in\mathbb{P}$, with
matrix elements,
\begin{equation}
F_{jk}=\frac{1}{\sqrt{d}}e^{2\pi ijk/d}\:,\qquad j,k=0\ldots d-1\,.
\end{equation}
For every prime-power dimension $d=p^{n},$ with $n>1,$ there exists
an affine Fourier family with the number of free parameters given
by 
\begin{equation}
\text{{def}}(F_{d})=\sum^{d-1}_{n=1}(\text{gcd}(n,d)-1)\,.\label{eq:fourier_defect}
\end{equation}
The value $\text{{def}}(F_{d})$ is known as the \emph{defect} of
the Fourier matrix \citep{tadej+06,barros+12} and, in general, gives
an upper bound on the number of free parameters of any smooth dephased
Hadamard family stemming from $F_{d}$ (see Appendix \ref{sec:Complex-Hadamard-matrices}).
When $d$ is a prime-power, there exist simple constructions of families
saturating this bound. However, for \emph{composite} dimensions $d\in(7,100]$,
the number of free parameters is strictly \emph{less} than Eq.~\eqref{eq:fourier_defect}
\citep{barros+12}.

Hadamard matrices derive their name from a 1893 paper by Hadamard
in which he studies the maximal determinant of matrices with entries
having modulus at most one \citep{hadamard93}. However, already in
1867 Sylvester constructed $2^{k}\times2^{k}$ Hadamard matrices using
tensor products \citep{sylvester67}. In 1933, Paley uses quadratic
residues in $GF(p^{k})$ to construct Hadamard matrices of order $(p^{k}+1)$
for $p^{k}+1\equiv0\mod 4$ \citep{paley33}.

The set of Butson-Hadamard matrices $BH(d,r)$ consists of $d\times d$
matrices, the elements of which are $r$-th roots of unity \citep{butson62}
(see Sec. \ref{subsec:MU-butson-type}). When $r=2$, the class $BH(d,2)$
equates to \emph{real} Hadamard matrices of order $d$, famously conjectured
to exist when $d=4k$, for every $k\in\mathbb{N}$. More generally,
Hadamard matrices with real entries give rise to real MU bases, discussed
in Sec.~\ref{subsec:Real-MU-bases}. Whether matrices of type $BH(d,r)$
exist for arbitrary $r$ is still an open problem. For instance, if
$p\in\mathbb{P}$, there exist $p\times p$ Butson-Hadamard matrices
\textbf{$BH(p,2^{j}p^{k})$} for all\textbf{ $0\leq j\leq k$} \citep{butson62}.\textbf{ }

For many other constructions of $d\times d$ complex Hadamard matrices
we refer the reader to the compendia \citep{agaian85,tadej+06,horadam12}
and the 400-page ``Invitation to Hadamard matrices'' \citep{banica19},
as well as the \emph{online catalogue} \citep{bruzda+12}. In Sec.~\ref{sec:pairs_of_MU_bases_C^6}
all known complex Hadamard matrices for $d=6$ are described. 

\subsection{Larger sets of MU bases}

\label{sec:larger_sets_of_MU_bases}

It is well-known that up to $(d+1)$ MU bases in total may be found
(see Appendix \ref{sec: complete sets in pp dimensions}) when aiming
to go beyond \emph{pairs }of MU bases in a complex vector space $\mathbb{C}^{d}$.
The following theorem by Klappenecker and Rötteler \citep{klappenecker+04}
establishes the minimal number of MU bases that are guaranteed to
exist in any given finite-dimensional Hilbert space.
\begin{thm}
\label{thm:reduce_to_primes} Let $d=p^{n_{1}}_{1}\ldots p^{n_{r}}_{r}$
be a factorisation of $d$ into distinct primes $p_{i}$ and let $N(d)$
denote the number of mutually unbiased bases in dimension $d$. Then
$N(d)\geq\text{min}\{N(p^{n_{1}}_{1}),N(p^{n_{2}}_{2}),\ldots,N(p^{n_{r}}_{r})\}$. 
\end{thm}
To see this, denote the smallest number of MU bases by $\mu=\text{min}_{i}N(p^{n_{i}}_{i})$
and choose for every prime-power factor (i.e. $\mathbb{C}^{p^{n_{i}}_{i}}$)
a set of $\mu$ MU bases labeled by $\mathcal{B}^{(i)}_{1},\ldots,\mathcal{B}^{(i)}_{\mu}$.
A set of $\mu$ MU bases of the space $\mathbb{C}^{d}$ is then given
by the product bases $\mathcal{B}^{(1)}_{k}\otimes\ldots\otimes\mathcal{B}^{(r)}_{k}$,
where $k=1,\ldots,\mu$.

Since we know that every prime-power dimension $p^{n_{i}}_{i}$ has
$(p^{n_{i}}_{i}+1)$ MU bases, it follows that $N(d)\geq3$ for $d\geq2$,
i.e., at least three MU bases exist in any finite dimension. The existence
of triples of MU bases in dimension six is an obvious consequence. 
\begin{cor}
In dimension six there exist at least three MU bases. 
\end{cor}
This statement is contained in Zauner's PhD thesis (cf. Proposition
2.20 of \citep{Zauner2011}), expressed in the language of affine
quantum designs (see Sec.~\ref{subsec:zauner_conjecture}). The triple
in dimension $d=2$ and the $(p+1)$-tuples in prime dimensions were
known since 1960 (cf. Sec.~\ref{subsec: avant-la-letter}). 

\subsection{Sets of $d$ MU bases are sufficient \label{subsec:d_MU_bases_sufficient}}

There is a general result which makes the search for complete sets
a little less challenging \citep{weiner09}.
\begin{thm}
\label{thm:missing_basis} Suppose that $\{\mathcal{B}_{0},\ldots,\mathcal{B}_{d-1}\}$
is a collection of $d$ MU bases in $\mathbb{C}^{d}$. Then there
exists a basis $\mathcal{B}_{d}$ such that $\{\mathcal{B}_{0},\ldots,\mathcal{B}_{d-1},\mathcal{B}_{d}\}$
is a complete collection of $\left(d+1\right)$ MU bases. 
\end{thm}
The proof makes use of the connection between MU bases and maximal
Abelian subalgebras (see Equivalence \ref{equiv:masa}). In particular,
the proof shows that, given a set of $d$ quasi-orthogonal MASAs,
one can always find a MASA quasi-orthogonal to all of them. An alternative
proof was later given in \citep{cariello16}. 

Interestingly, the property observed in Thm.~\ref{thm:missing_basis}
mirrors an analogous result on mutually orthogonal Latin squares,
adding a further connection between the two related structures (cf.
Appendices \ref{sec:affineplanes} and \ref{subsec:affineplanes_mubs_sics}).
In particular, a collection of $(d-2)$ mutually orthogonal Latin
squares can always be completed to $(d-1)$ mutually orthogonal Latin
squares.

Theorem \ref{thm:missing_basis} implies that any collection of six
MU bases in $\mathbb{C}^{6}$ leads to the existence of a full set
of seven MU bases. If two bases are missing from a complete set, however,
then it is not necessarily possible to ``complete'' the set (see
Sec.~\ref{subsec:Unextendible-MU-bases} for an example). Thus, five
MU bases in dimension six would not prove the existence of seven MU
bases in $\mathbb{C}^{6}$.

\subsection{Biunimodular sequences}

\label{subsec:biunimodular}

First considered by Björck in 1985 \citep{bjorck85}, a \emph{biunimodular
sequence} is defined as a vector\footnote{In this section, we will not distinguish between a vector $x$ and
its transpose $x^{T}$, for notational simplicity.} $x=(x_{0},x_{1},\ldots,x_{d-1})\in\mathbb{C}^{d}$ such that $|x_{j}|=|\hat{x}_{j}|=1$
for all $j$, where $\hat{x}=(\hat{x}_{0},\hat{x}_{1},\ldots,\hat{x}_{d-1})$
is the Fourier transform $\hat{x}_{j}=\frac{1}{\sqrt{d}}\sum^{d-1}_{k=0}\omega^{jk}x_{k}$
of the vector $x$, and $\omega=e^{2\pi i/d}$ is a $d$-th root of
unity. In other words, biunimodularity requires both $x$ and $\hat{x}=F_{d}x$
to be unimodular, where $F_{d}$ is the Fourier matrix. 

A unimodular sequence $x$ is biunimodular if and only if its cyclic
translations are all orthogonal. It follows that finding a biunimodular
sequence can be translated into solving the set of equations
\begin{align}
z_{0}+z_{1}+\ldots+z_{d-1}=0 & \,,\nonumber \\
z_{0}z_{1}+z_{1}z_{2}+\ldots z_{d-2}z_{d-1}+z_{d-1}z_{0}=0 & \,,\nonumber \\
z_{0}z_{1}z_{2}+z_{1}z_{2}z_{3}+\ldots z_{d-1}z_{0}z_{1}=0 & \,,\nonumber \\
\vdots\label{eq:cyclic_n_roots}\\
z_{0}z_{1}\ldots z_{d-2}+\ldots+z_{d-1}z_{0}\ldots z_{d-3}=0 & \,,\nonumber \\
z_{0}z_{1}\ldots z_{d-1}=0 & \,,\nonumber 
\end{align}
defining the \emph{cyclic $d$-roots problem} (the mathematical literature
usually speaks of ``cyclic $n$-roots'') \citep{bjork+91,bjorck+95}.
A solution $z=(z_{0},z_{1},\ldots,z_{d-1})$ is known as a cyclic\emph{
$d$-root.} Imposing the additional constraint $|z_{j}|=1$ gives
rise to a biunimodular sequence $x$, where $z_{j}=x_{j+1}/x_{j}.$

Finding all biunimodular sequences, or solving the cyclic $d$-roots
problem, corresponds to finding all vectors mutually unbiased to the
standard and Fourier basis. Solutions are known for $d\leq14$ (see
\citep{fuhr15} for a summary). There are 156 cyclic $6$-roots and
48 biunimodular sequences among them (see Sec.~\ref{sec:fourier_family}
for details). The next non-prime-power dimension contains 34,940 cyclic
10-roots \citep{faugere02}. For prime $p$ there are ${2p-2 \choose p-1}$
cyclic $p$-roots, if counted with multiplicities \citep{haagerup08}.

For $d$ divisible by a square, the number of cyclic $d$-roots is
infinite \citep{backelin89}. The same statement is also true for
biunimodular sequences \citep{bjorck+95}.
\begin{thm}
\label{thm:biunimodular_square}For $d$ divisible by a square, the
number of biunimodular sequences (with leading entry 1) is infinite.
Equivalently, the number of vectors MU to $\{\mathbb{I},F_{d}\}$
is infinite.
\end{thm}
Since all cyclic translations of a biunimodular sequence $x$ are
orthogonal, one can construct a $d\times d$ circulant Hadamard matrix
from $x,$ defined by the matrix elements $H_{jk}=x_{j-k}$ \citep{bjorck+95}.
\begin{thm}
\label{thm:circular-1} A Hadamard matrix $H$ of order $d$ is circulant
if and only if $H_{jk}=x_{j-k}$ for a biunimodular sequence $x=(x_{0},x_{1},\ldots,x_{d-1})$. 
\end{thm}
This result imposes an interesting restriction on the type of basis
mutually unbiased to the Fourier matrix \citep{bjorck+95}.
\begin{cor}
\label{thm:circular} Any Hadamard matrix whose columns are mutually
unbiased to the pair $\{\mathbb{I},F_{d}\}$ is equivalent to a circulant
Hadamard matrix.
\end{cor}
Here, equivalence is understood up to diagonal phase matrices and
permutations of rows and columns, as defined in Appendix \ref{sec:Complex-Hadamard-matrices}.
Combining Thms.~\ref{thm:biunimodular_square} and \ref{thm:circular-1}
with Cor. \ref{thm:circular} implies that the space $\mathbb{C}^{d}$
supports infinitely many triples of MU bases containing both the Fourier
matrix and a circulant Hadamard matrix if the dimension $d$ is divisible
by a square.

A more general type of biunimodular sequence $x$ can be considered
by requiring both $x$ and $Ax$ to be unimodular, with the freedom
to choose $A$ from the set of unitary matrices rather than the Fourier
matrix \citep{fuhr15}. The existence of these sequences, for an arbitrary
unitary $A$, is known indirectly from other results, e.g. \citep{biran04,cho04,idel15}.
\begin{thm}
\label{thm:biunimodular} For any unitary matrix $A$ of order $d$,
there exists a sequence $x=(x_{0},x_{1},\ldots,x_{d-1})$ such that
both x and $Ax$ are unimodular.
\end{thm}
If $A$ is a Hadamard matrix, a biunimodular sequence corresponds
to a vector which is mutually unbiased to both the standard basis
and $A$. Therefore, we have the following observation \citep{lisi11,korzekwa14,idel15,andersson17}.
\begin{cor}
\label{thm:circular-2} There exists at least one vector mutually
unbiased to any pair of MU bases in $\mathbb{C}^{d}$.
\end{cor}

\subsection{Nice error bases }

\label{subsec:Nice-error-bases}

A set of $\mu$ MU bases in the space $\mathbb{C}^{d}$ is equivalent
to partitioning a subset of a unitary operator basis into $\mu$ commuting
classes, as was described in Sec.~\ref{subsec:Maximally-commuting-unitary}.
Here we will require, in addition, that the elements of the unitary
operator basis generate a certain group, resulting in a \emph{nice
error basis }\citep{boykin+07}. Under this assumption there is a
limit on the number of MU bases we can construct from any partitioning
of the basis \citep{aschbacher+07}.
\begin{defn}
Let $\mathcal{G}$ be a group of order $d^{2}$ with identity element
$e$, and let $\mathcal{N}=\{U_{g}:g\in\mathcal{G}\}\subset\mathbb{C}^{d\times d}$
be a set of unitary matrices which are traceless (except for the identity),
i.e. $\mbox{Tr}[U_{g}]=0$ for all $g\in\mathcal{G}\setminus\{e\}$.
The set $\mathcal{N}$ is a \emph{nice error basis }if its elements
satisfy $U_{g}U_{h}=\omega(g,h)U_{gh}$ for all $g,h\in\mathcal{G}$,
where $\omega(g,h)\in\mathbb{C}$ has modulus one. 
\end{defn}
In other words, the unitaries must form a faithful projective (or
ray) representation of the group $\mathcal{G}$. As a simple example,
consider $d=p^{n}\in\mathbb{PP}$ and take the index group as $\mathcal{G}=\mathbb{Z}^{n}_{p}\times\mathbb{Z}^{n}_{p}$
where $\mathbb{Z}_{p}=\{0,\ldots,p-1\}$ such that $(k,\ell)=(k_{1},\ldots,k_{n},\ell_{1},\ldots,\ell_{n})\in\mathbb{Z}^{n}_{p}\times\mathbb{Z}^{n}_{p}$.
A nice error basis $\mathcal{N}$ can be constructed from the Heisenberg-Weyl
operators $X$ and $Z$, defined in Eq.~\eqref{HWoperators}, by
taking 
\begin{equation}
\mathcal{N}=\{U(k,\ell)\:|\,(k,\ell)\in\mathbb{Z}^{n}_{d}\times\mathbb{Z}^{n}_{d}\}\,,\label{eq:HWniceerrorbasis}
\end{equation}
with 
\begin{equation}
U(k,\ell):=X^{k_{1}}Z^{\ell_{1}}\otimes\ldots\otimes X^{k_{n}}Z^{\ell_{n}}.\label{eq:HWniceerrorbasisunitaries}
\end{equation}

\begin{defn}
The MU bases associated with partitioning a subset of a nice error
basis into maximally commuting operator classes (via Equivalence \ref{equiv:commutingclasses})
are called \textit{nice} MU bases.
\end{defn}
Aschbacher \textit{et al.} \citep{aschbacher+07} showed that there
is a limit on the number of nice MU bases one can construct from partitioning
the nice error basis into maximally commuting classes.
\begin{thm}
\label{thm:niceerrorbasis} For dimension $d=p^{n_{1}}_{1}\ldots p^{n_{r}}_{r}\notin\mathbb{PP}$,
where $p_{i}\in\mathbb{P}$ and $n_{i}\in\mathbb{N},$ no more than
$\text{min}_{i}\left(p^{n_{i}}_{i}+1\right)$ nice MU bases exist.
\end{thm}
The proof establishes a connection between the index group $\mathcal{G}$
of a nice error basis $\mathcal{N}$ and Abelian subgroups of $\mathcal{G}$
constructed from the commuting classes of $\mathcal{N}$. Then, the
known upper bound on the number of trivially intersecting Abelian
subgroups of $\mathcal{G}$ provides the bound in Thm.~\ref{thm:niceerrorbasis}.

It is easy to see that the bound is tight for arbitrary $d.$ For
$d=p^{n}\in\mathbb{PP}$, the nice error basis of Eqs.~\eqref{eq:HWniceerrorbasis}
and (\ref{eq:HWniceerrorbasisunitaries}) can be partitioned into
$(d+1)$ maximally commuting classes, leading to a complete set of
$(d+1)$ nice MU bases. An obvious consequence of Thm.~\ref{thm:niceerrorbasis}
for dimension six is the following.
\begin{cor}
In dimension six, no more than three nice MU bases exist.
\end{cor}
This does not rule out the existence of additional bases which are
mutually unbiased to a set of three\emph{ nice} MU bases, although
a different construction would be required.

However, as shown by Nieter \textit{et al.} \citep{nietert20}, a
sufficiently large set of nice MU bases extends to a complete set
in exactly one way. This observation, based on a connection between
MU bases and combinatorial designs called $k$-nets, leads to Thm.~\ref{thm:unext_niceerrorbasis}
(Sec.~\ref{subsec:Unextendible-MU-bases}) which shows that some
sets of nice MU bases do not extend (by \emph{any} methods) to a complete
set.

\subsection{Monomial bases}

\label{sec:monomial_bases}

It was pointed out in Ref.~\citep{godsil+09} that every known complete
set of MU bases is \emph{monomial} in the following sense.
\begin{defn}
A set of MU bases is monomial if the unitary operator basis from which
it is constructed contains only monomial matrices, i.e. matrices that
have only one non-zero element in each row and column.
\end{defn}
Furthermore, every known complete set of MU bases can be obtained
by partitioning a nice error basis into commuting classes. For $d=p^{n}\in\mathbb{PP}$,
every nice error basis that partitions into maximally commuting classes
is equivalent to the monomial basis defined in Eqs. (\ref{eq:HWniceerrorbasis})
and (\ref{eq:HWniceerrorbasisunitaries}) \citep{aschbacher+07,boykin+07}.
Thus, complete sets of nice MU bases are always monomial. In general,
however, MU bases need not be ``nice'' and monomial simultaneously.
For example, the ``incomplete'' sets of MU bases constructed in
square dimensions from Latin squares (Thm.~\ref{thm:wocjanMUBs},
Sec.~\ref{subsec:Maximally-entangled-bases}) are monomial but cannot
be obtained by partitioning a nice error basis.

Given that all known complete sets of MU bases are monomial, the following
theorem provides quite a severe restriction on our ability to find
complete sets in dimension six.
\begin{thm}
\label{thm:monomial} In dimension six, no more than three monomial
MU bases exist. 
\end{thm}
This result, shown in \citep{boykin+07}, is a consequence of the
equivalence of MU bases in $\mathbb{C}^{d}$ and orthogonal decompositions
of $sl_{d}(\mathbb{C})$, described in Sec.~\ref{subsec: Orthogonal decomp equivalence}.
It follows directly from a result proved in \citep{kostrikin+94}
which shows that no more than three monomial and pairwise orthogonal
Cartan subalgebras of $sl_{6}(\mathbb{C})$ exist. It remains unknown
whether complete sets of monomial MU bases exist only in prime-power
dimensions \citep{boykin+07}.

\subsection{MU product bases \label{subsec:MU-product-bases}}

As discussed in Sec.~\ref{sec:larger_sets_of_MU_bases}, product
bases provide a simple yet effective means to construct MU bases in
composite dimensions $d\notin\mathbb{PP}$. It is natural, therefore,
to ask whether an exhaustive classification of MU product bases is
possible, or whether a tight upper bound on their number can be found.
In dimension six both questions have been answered positively (Sec.~\ref{subsec:product_bases}).
For arbitrary multipartite systems, tight upper bounds are also known,
but a classification of MU product bases has been obtained only in
specific cases.

An orthonormal basis $\mathcal{B}$ of the space $\mathbb{C}^{d}$
with dimension $d=d_{1}d_{2}\cdots d_{n}$ is a product basis if each
basis vector takes the form $\ket{v^{1},\ldots,v^{n}}\equiv\ket{v^{1}}\otimes\ket{v^{2}}\otimes\ldots\otimes\ket{v^{n}}\in\mathbb{C}^{d},$
with states $\ket{v^{r}}\in\mathbb{C}^{d_{r}},r=1,\ldots,n.$ The
following result places an upper bound on the maximum number of MU
product bases in arbitrary composite dimensions.
\begin{thm}
\label{thm:productMUBs} If $d=d_{1}d_{2}\cdots d_{n}\notin\mathbb{PP}$
then there exist at most $(d_{m}+1)$ MU product bases in $\mathbb{C}^{d},$
where $d_{m}$ is the dimension of the subsystem with the least number
of MU bases.
\end{thm}
The proof for $d_{1}=2,3$ was given in Ref.~\citep{mcnulty16},
and the general case was solved in \citep{cariello21} using Thm.
\ref{thm:fixed_entanglement_Schmidt}. We note that the bound is tight
by applying the construction from Thm.~\ref{thm:reduce_to_primes}.
For dimensions $d=d_{1}d_{2}\cdots d_{n},$ with $d_{1}=2$ or $d_{1}=3,$
a classification of maximal sets of MU product bases is also possible.
For example, when $d=2^{n}$ and $d=3^{n},$ there exists a unique
triple and a unique quadruple of MU product bases, respectively, up
to equivalence \citep{mcnulty16}. Furthermore, a vector which is
mutually unbiased to a set of $(d_{1}+1)$ MU product bases must have
a particular entanglement structure.
\begin{lem}
\label{lem:maximallyentangledproperty}Let $d=d_{1}\cdots d_{n}$
with $d_{r}=p^{k_{r}}_{r},$ $p_{r}\in\mathbb{P}$ and $k_{r}\in\mathbb{N},$
$r=1\ldots n$, such that $d_{1}\leq\ldots\leq d_{n}.$ A vector,
mutually unbiased to a set of $(d_{1}+1)$ MU product bases (where
the product bases of $\mathbb{C}^{d}$ contain at least one orthogonal
set of $d_{1}$ vectors in the subsystem $\mathbb{C}^{d_{1}}),$ is
maximally entangled across $\mathbb{C}^{d_{1}}\otimes\mathbb{C}^{d_{2}\cdots d_{n}}$.
\end{lem}
For dimensions $d=2d_{2}$ and $d=3d_{2}$, where $d_{2}$ is prime
and $d_{2}\geq d_{1},$ Lemma \ref{lem:maximallyentangledproperty}
implies that any vector mutually unbiased to a set of $(d_{1}+1)$
MU product bases in $\mathbb{C}^{d}$ is maximally entangled. For
the special case $d=d_{1}\cdots d_{n}=p^{n}$, with $d_{r}=p$ and
$r=1\ldots n,$ it follows that any vector MU to a set of $(p+1)$
MU product bases (where the product bases of $\mathbb{C}^{d}$ contain
at least one orthogonal set of $d_{r}$ vectors in each subsystem
$\mathbb{C}^{d_{r}}),$ is maximally entangled across \textit{all}
bipartitions $\mathbb{C}^{p}\otimes\mathbb{C}^{p^{n-1}}.$

The inconvenient requirement in Lemma \ref{lem:maximallyentangledproperty}—namely
that the product bases of $\mathbb{C}^{d}$ contain at least one orthogonal
set of $d_{1}$ vectors in the subsystem $\mathbb{C}^{d_{1}}$—is
an unfortunate consequence of our inability to fully characterise
the structure of product bases. It also limits our ability to classify
smaller sets of MU product bases (except when $d=6)$. The characterisation
or product bases in terms of their orthogonality relations is particularly
difficult and remains an open problem, although it is expected that
the following simple structure holds.
\begin{conjecture}
\label{conj:productbases}If $\mathcal{B}=\{\ket{v^{1},v^{2}}\}^{d-1}_{v=0}$
is an orthonormal product basis of the space $\mathbb{C}^{d}$, with
$d=d_{1}d_{2}$, then the $d$ vectors $\{\ket{v^{1}}\in\mathbb{C}^{d_{1}}\}^{d-1}_{v=0}$
and the $d$ vectors $\{\ket{v^{2}}\in\mathbb{C}^{d_{1}}\}^{d-1}_{v=0}$
can be grouped into $d_{2}$ orthonormal bases $\mathcal{B}^{(d_{1})}_{j},j=1\ldots d_{2},$
and $d_{1}$ orthonormal bases $\mathcal{B}^{(d_{2})}_{k},k=1\ldots d_{1},$
respectively.
\end{conjecture}
A similar problem was considered for $n$-qubit orthogonal product
bases, with their construction reducing to a purely combinatorial
problem \citep{dokovic17}. For the 4-qubit case, 33 multiparameter
families were found such that any orthogonal product basis of four
qubits is equivalent (under local unitaries and qubit permutations)
to a basis in at least one of these families.

We note that MU product bases play a useful role in constructing a
special class of \textit{isolated} Hadmard matrices (see Appendix
\ref{sec:Complex-Hadamard-matrices}), as described in \citep{mcnulty12b}.

\subsection{Entanglement in MU bases }

\label{subsec:entanglement_content}

One way to learn more about the properties of MU bases in composite
dimensions $d\notin\mathbb{PP}$ is to treat the Hilbert space $\mathbb{C}^{d}$
as a tensor product of its factors, e.g. $\mathbb{C}^{d}\cong\mathbb{C}^{d_{1}}\otimes\mathbb{C}^{d_{2}}$
in a bipartite system of dimension $d=d_{1}d_{2}$. As seen in Thm.~\ref{thm:reduce_to_primes},
the product structure provides a convenient way to construct MU bases
in composite dimensions. This approach opens up the possibility of
investigating quantum correlations in MU bases, leading to novel insights
on their entanglement structure. For example, it has been noticed
that a complete set of MU bases in dimension $d=p^{n}\times p^{n}$,
$p\in\mathbb{P}$ and $n\in\mathbb{N}$, can be expressed in terms
of product states and maximally entangled states of the space $\mathbb{C}^{p^{n}}\otimes\mathbb{C}^{p^{n}}$
\citep{gross07}. 

Another observation concerns bipartite quantum systems of dimension
$d=d_{1}d_{2}$, with $d_{1}\leq d_{2}$: a complete set of MU bases
necessarily has a fixed average entanglement (equivalently, a fixed
average purity of the reduced states) \citep{wiesniak+11}.
\begin{thm}
\label{thm:fixed_entanglement} Suppose $\{\ket{\psi_{i}}\}^{d(d+1)}_{i=1}$
is a complete set of $(d+1)$ MU bases in dimension $d=d_{1}d_{2}$
and let $\mathcal{P}(\rho_{i})\equiv\mbox{Tr}[\rho^{2}_{1|i}]$ denote
the purity of the reduced density operator of the state $\rho_{i}=\kb{\psi_{i}}{\psi_{i}}$,
given by $\rho_{1|i}=\mbox{Tr}_{2}[\kb{\psi_{i}}{\psi_{i}}]$. Then
a complete set of $(d+1)$ MU bases contains a fixed total purity
(entanglement) content of 
\begin{equation}
\sum^{d(d+1)}_{i=1}\mathcal{P}(\rho_{i})=d_{1}d_{2}(d_{1}+d_{2})\,.\label{eq:fixedentanglement}
\end{equation}
\end{thm}
The proof follows by first observing that the average purity of a
bipartite state is $(d_{1}+d_{2})/(d+1)$ \citep{lubkin78}. Since
the purity $\mathcal{P}(\rho_{i})$ is a polynomial of degree two,
we can use the 2-design property of MU bases (cf. Sec.~\ref{subsec: 2-design equivalence})
to show that the average purity over all pure states is equal to the
average purity over the 2-design. This implies that the average purity
of a state in the 2-design is $(d_{1}+d_{2})/(d+1)$ and, hence, summing
over all $d(d+1)$ states gives Eq.~\eqref{eq:fixedentanglement}.

The purity of a state achieves its minimum value of $1/d_{1}$ when
the state is maximally entangled, and its maximum of unity when separable.
Consequently, complete sets must include both entangled \emph{and}
product states (provided we take the first basis as the canonical
one). In other words, the fixed average entanglement makes it impossible
for a complete set to contain only maximally entangled states \emph{or}
product states. Related entanglement constraints hold for multipartite
tight informationally complete measurements, which include complete
sets of MU bases as a special case \citep{czartowski18}. Such measurements
cannot consist entirely of separable states or entirely of maximally
entangled states.

The distribution of entanglement within a complete set is arbitrary.
For example, suppose that for a bipartite quantum system of dimension
$d=d_{1}d_{2}$ with $d_{1}\leq d_{2}$, we have a set of $\left(d_{1}+1\right)$
MU product bases. Then the remaining bases of a (hypothetical) complete
set of $\left(d+1\right)$ bases must contain only maximally entangled
states. 

For composite systems of two, three and four qubits, a detailed analysis
of the entanglement structure of a complete set can be found in \citep{romero+05}.
Two-qubit systems are special in the sense that there exists an \emph{iso-entangled
}complete set of five MU bases, i.e. all states come with exactly
the same amount of entanglement \citep{czartowski20}. As mentioned
before, the standard construction of a complete set in the space $\mathbb{C}^{p^{n}}\otimes\mathbb{C}^{p^{n}}$yields
only product bases and maximally entangled bases \citep{gross07}.

In dimension six, where $d_{1}=2$ and $d_{2}=3$, the purity (or
entanglement) content of any complete set equals $30$. For example,
a set of three MU product bases can be constructed from the tensor
product of three MU bases of the space $\mathbb{C}^{2}$ with three
MU bases of the space $\mathbb{C}^{3}$ (see Eq.~\eqref{eq:product_triples}).
However, due to the fixed average entanglement, the remaining four
bases must contain entangled states only. In Sec.~\ref{subsec:product_bases}
we will discuss some stronger results relating to MU product bases
in dimension six.

Thm.~\ref{thm:fixed_entanglement} can be adapted to sets of $\mu$
MU bases, for $\mu\leq d+1$, in bipartite system $d=d_{1}d_{2}$,
with $d_{1}\leq d_{2}$ \citep{cariello21}.
\begin{thm}
\label{thm:fixed_entanglement_Schmidt} Suppose $\{\ket{\psi_{i}}\}^{d\mu}_{i=1}$
is the set of vectors in a collection of $\mu$ MU bases in dimension
$d=d_{1}d_{2}$ and let $\mathcal{P}(\rho_{i})\equiv\mbox{Tr}[\rho^{2}_{1|i}]$
denote the purity of the reduced density operator of the state $\rho_{i}=\kb{\psi_{i}}{\psi_{i}}$,
given by $\rho_{1|i}=\mbox{Tr}_{2}[\kb{\psi_{i}}{\psi_{i}}]$. The
purity content of the set of $\mu$ MU bases satisfies
\begin{equation}
\sum^{d\mu}_{i=1}\mathcal{P}(\rho_{i})\leq(d^{2}_{1}+\mu-1)d_{2}\,.\label{eq:fixedentanglement-1}
\end{equation}
\end{thm}
The proof, in analogy with Thm.~\ref{thm:fixed_entanglement}, relies
on properties of 2-designs. By requiring that the vectors in a set
of $\mu$ MU bases have Schmidt rank less than or equal to $k,$ an
immediate consequence of Thm.~\ref{thm:fixed_entanglement_Schmidt}
is an upper limit on the number of MU bases.
\begin{cor}
\label{cor:schmidt_bound}Suppose there exists a set of $\mu$ MU
bases in $\mathbb{C}^{d}$, with $d=d_{1}d_{2}$, such that the Schmidt
rank of each vector is at most $k,$ and $k<d_{1}\leq d_{2}$. Then
\begin{equation}
\mu\leq k\left(\frac{d^{2}_{1}-1}{d_{1}-k}\right).\label{eq:schmidt_bound}
\end{equation}
\end{cor}
Notice that when $k=1,$ the bases contain only separable states,
and the bound implies at most $\left(d_{1}+1\right)$ MU product bases
exist. Additional results on product bases can be found in Sec.~\ref{subsec:MU-product-bases}.
A bound similar to Eq.~\eqref{eq:schmidt_bound} also applies for
MU bases in a real vector space (see Sec.~\ref{subsec:Real-MU-bases}).

Entangled bases with a fixed Schmidt rank are considered in \citep{guo15}.
In the next section we focus on bases with maximal Schmidt rank, i.e.
maximally entangled bases.

\subsection{Maximally entangled bases }

\label{subsec:Maximally-entangled-bases}

As we have just seen, the simplest construction of MU bases in bipartite
systems $d=d_{1}d_{2}$ starts from $(d_{1}+1)$ MU bases in each
subsystem $\mathbb{C}^{d_{1}}$ and $\mathbb{C}^{d_{2}}$ (provided
they exist). Tensoring pairs of these bases (one from each subsystem)
produces a set of $(d_{1}+1)$ MU product bases. Due to the fixed
average entanglement condition (cf. Sec.~\ref{subsec:entanglement_content}),
the remaining states in a hypothetical complete set of $(d+1)$ MU
bases must be maximally entangled, i.e. of the form $\ket{\psi}=\frac{1}{\sqrt{d}}\sum^{d-1}_{i=0}\ket i\ket{i'}$
for orthonormal bases $\{\ket i\}$ and $\{\ket{i'}\}$ of $\mathbb{C}^{d_{1}}$
and $\mathbb{C}^{d_{2}},$ respectively. 

This motivates the study of maximally entangled bases and, in particular,
to find bounds on the number of maximally entangled MU bases that
exist. In this section we summarise several methods to construct MU
bases consisting of maximally entangled states. 

\emph{Mutually unbiased unitary bases}—To simplify the problem we
initially consider $d_{1}=d_{2}.$ In this case, there is a one-to-one
correspondence between maximally entangled states in $\mathbb{C}^{d}\otimes\mathbb{C}^{d}$
and $d\times d$ unitary matrices. Furthermore, a unitary operator
basis of $\mathbb{M}_{d}(\mathbb{C})$—i.e. a set of $d^{2}$ unitary
matrices $U_{i}$ such that $\mbox{Tr}[U^{\dagger}_{i}U_{j}]=d\delta_{ij}$
(see Sec.~\ref{subsec:Maximally-commuting-unitary})—is equivalent
to a maximally entangled basis. In particular, a basis of orthogonal
unitary operators $\{U^{(i)}\}$, $i=0,\ldots,d^{2}-1$, corresponds
to a set of orthogonal states
\begin{equation}
\ket{U^{(i)}}=\frac{1}{\sqrt{d}}\sum^{d-1}_{j,k=0}U^{(i)}_{jk}\ket j\ket k\,,
\end{equation}
forming a maximally entangled basis, where $U^{(i)}_{jk}$ are the
matrix elements of $U^{(i)}$. Consequently, finding maximally entangled
bases is equivalent to constructing unitary operator bases, which
is a well studied topic \citep{werner01,knill96}. 

Mutually unbiased unitary operator bases—defined by the condition
that the Hilbert-Schmidt inner product between pairs of unitaries
from different bases is constant—were first introduced by Scott \citep{scott08}
in relation to optimal quantum process tomography. Later, Shaari \textit{et
al.} \citep{shaari16} pointed out that these bases translate into
mutually unbiased maximally entangled bases. In particular, any pair
of \emph{mutually unbiased} unitary operator bases of $\mathbb{M}_{d}(\mathbb{C})$
gives rise to a pair of maximally entangled MU bases in $\mathbb{C}^{d}\otimes\mathbb{C}^{d}$.
For arbitrary $d$, the maximal number of pairwise MU unitary operator
bases is $d^{2}-1$, although a construction which saturates this
bound is known only for the primes $d=2,3,5,7,$ and $11$ \citep{scott08}.
Some lower bounds on the number of maximally entangled MU bases (and
hence, lower bounds on the number of MU unitary operator bases) are
presented at the end of this section.

Is a classification of unitary operator bases (and hence maximally
entangled MU bases) possible? Unfortunately, this seems unlikely.
For example, one of the simplest constructions, called the \textit{shift
and multiply }method, involves a Latin square (see Appendix \ref{sec:affineplanes})
and a family of Hadamard matrices. For example, as shown in \citep{bengtsson17,wehner10},
by taking the Latin square
\begin{equation}
\begin{array}{ccc}
0 & 1 & 2\\
1 & 2 & 0\\
2 & 0 & 1
\end{array}
\end{equation}
of order three, we can easily find three orthogonal entangled states
in $\mathbb{C}^{3}\otimes\mathbb{C}^{3},$
\begin{align}
\ket{\psi_{0}} & =\frac{1}{\sqrt{3}}(\ket 0\ket 0+\ket 1\ket 1+\ket 2\ket 2)\,,\\
\ket{\psi_{1}} & =\frac{1}{\sqrt{3}}(\ket 0\ket 1+\ket 1\ket 2+\ket 2\ket 0)\,,\\
\ket{\psi_{2}} & =\frac{1}{\sqrt{3}}(\ket 0\ket 2+\ket 1\ket 0+\ket 2\ket 1)\,.
\end{align}

Now choosing a Hadamard matrix $H$ with matrix elements $H_{ij},$
each of the states $\ket{\psi_{k}}$ can be mapped into three additional
orthogonal states. The first state $\ket{\psi_{0}}$, for example,
becomes
\begin{equation}
\ket{\psi_{0j}}=\frac{1}{\sqrt{3}}(H_{j0}\ket 0\ket 0+H_{j1}\ket 1\ket 1+H_{j2}\ket 2\ket 2)\,,\label{eq:latinbasis}
\end{equation}
$j=0,1,2$. This leads to a maximally entangled basis of states $\{\ket{\psi_{ij}}\}$
for $i,j=0,1,2,$ from a Latin square and Hadamard matrix. More generally
in $\mathbb{C}^{d}\otimes\mathbb{C}^{d}$, a maximally entangled basis
of this form has basis elements
\begin{equation}
\ket{\psi_{ij}}=\frac{1}{\sqrt{d}}\sum^{d-1}_{k=0}H_{jk}\ket k\ket{L_{ik}}\,,
\end{equation}
for $i,j=0,\ldots,d-1,$ where $H_{jk}$ are the elements of a $d\times d$
Hadamard matrix and $L_{ik}$ is the $(i,k)$-th component of a Latin
square $L$ of order $d.$ Note that the corresponding unitary operator
basis of the shift and multiply method is always a monomial basis
(see Sec.~\ref{sec:monomial_bases}).

Even the classification of unitary operator bases from this simple
construction—dependent on a Latin square and a Hadamard matrix—becomes
unfeasible as the dimension increases. Equivalence classes of Hadamard
matrices have been described in Appendix \ref{sec:Complex-Hadamard-matrices}
(see Definition \ref{def:Hadamard-equivalences}), and we refer the
reader to \citep{fisher35} for more details on equivalences between
Latin squares. In dimensions $d\leq5$, the situation is just about
manageable since all Hadamard matrices and Latin squares are known
\citep{haagerup97,fisher35}. For example, in dimension $d=5$ there
is only one Hadamard matrix up to equivalence, and two Latin squares
exist. For $d=6,$ although the number of inequivalent Latin squares
is 22, there are infinitely many Hadamard matrices, and their classification
remains an open problem. In higher dimensions the number of Latin
squares starts to increase significantly, and it becomes ever harder
to list Hadamard matrices.

While there are several known constructions of unitary operator bases
(e.g. nice error bases \citep{knill96}, see Sec.~\ref{subsec:Nice-error-bases}),
two approaches yield maximally entangled MU bases in composite dimensions
other than prime-powers. These are based on Latin and quantum Latin
squares, as summarised below.

\emph{Latin squares}—One construction of maximally entangled MU bases,
which provides a simple and effective way to build MU bases in square
dimensions, is based on Latin squares \citep{wocjan+05}. In particular,
a pair of maximally entangled bases is constructed from two Latin
squares $L$ and $L'$ of order $d$ and a single $d\times d$ Hadamard
matrix $H$. A first set of $d$ vectors stemming from the Latin square
$L$ is defined as
\begin{equation}
\ket{\psi_{0j}}=\frac{1}{\sqrt{d}}\sum^{d-1}_{i,k=0}E^{L}_{ik}(j)\ket i\ket k\,,\label{eq:squareMUBs}
\end{equation}
$k=0,\ldots,d-1$, where $E^{L}_{ik}(j)=1$ if $L_{ik}=j$ and zero
otherwise, and $L_{ik}$ the $(i,k)$-th entry of $L$ \citep{wocjan+05,wehner10}.
Next, each of these vectors is mapped to $d$ vectors by means of
the Hadamard matrix $H,$ using the method from Eq.~\eqref{eq:latinbasis},
which, taken together, form orthonormal bases $\{\ket{\psi_{ij}}\}$,
$i,j=0,\ldots,d-1.$ Repeating this construction with a second Latin
square $L^{\prime}$, orthogonal to $L$, the resulting bases are
mutually unbiased \citep{wocjan+05}. 
\begin{thm}
\label{thm:wocjanMUBs} Given a set of $\mu$ mutually orthogonal
Latin squares of order $d,$ there exists a set of $(\mu+2)$ MU bases
in $\mathbb{C}^{d}\otimes\mathbb{C}^{d}.$
\end{thm}
The additional two MU bases follow from extending the set of orthogonal
Latin squares to an \textit{augmented} set (see Appendix \ref{sec:affineplanes}).

Rather surprisingly, this approach leads—in specific dimensions—to
\emph{more} MU bases than the standard method based on product bases.
The smallest known dimension with an increased number of MU bases
in $\mathbb{C}^{d}$ is $d=26^{2}$. The known number of MO Latin
squares in this case is four, leading to \emph{six} MU bases rather
than the \emph{five} MU bases established in Thm.~\ref{thm:reduce_to_primes}
in Sec.~\ref{sec:larger_sets_of_MU_bases}, which arise from the
factor of four in the prime decomposition of $d=2^{2}\times13^{2}.$
Although this method offers some hope that larger sets of MU bases
exist, it yields at most $(d+1)$ MU bases in the space $\mathbb{C}^{d}\otimes\mathbb{C}^{d}$
since at most $(d-1)$ MO Latin squares of order $d$ exist. While
the maximum number of MO Latin squares exist when $d$ is a prime
or a prime-power, in most dimensions—except in special cases—the number
is unknown. More details on MO Latin squares and their role in constructing
complete sets of MU bases are described in Appendix \ref{sec:affineplanes}.

\emph{Quantum Latin squares}—The shift and multiply approach for constructing
unitary operator bases has been generalised to a\textit{ quantum shift
and multiply} method using \textit{quantum} Latin squares. A quantum
Latin square is a $d\times d$ array of elements in the space $\mathbb{C}^{d}$
such that every row and column is an orthonormal basis \citep{musto15}.
Suppose $Q$ is a quantum Latin square of order $d$ and $H_{j}$
labels a $d\times d$ Hadamard matrix, then
\begin{equation}
\ket{\psi_{ij}}=\frac{1}{\sqrt{d}}\sum^{d-1}_{k=0}\ket k\ket{Q_{kj}}\bra kH_{j}\ket i\,,\label{eq:latin_entangled_bases}
\end{equation}
for $i,j=0,\ldots,d-1,$ forms a set of $d^{2}$ orthogonal maximally
entangled states. Here, $\ket{Q_{kj}}$ is the vector from the $(k,j)-$th
components of $Q.$ Although the quantum shift and multiply method
reproduces all unitary operator bases from the non-quantum version,
it also yields unitary operator bases which are not monomial \citep{musto15}.

A generalisation of the construction of MU bases from Latin squares
to quantum Latin squares follows rather straightforwardly after defining
orthogonality. A pair of quantum Latin squares $Q$ and $Q'$ are
weakly orthogonal if, for all vector entries $\ket{Q_{ij}}$ and $\ket{Q'_{ij}}$,
there exists a fixed $t\in\{0,\ldots,d-1\}$ such that
\begin{equation}
\sum^{d-1}_{k=0}\ket k\bk{Q_{ki}}{Q'_{kj}}=\ket t\,,
\end{equation}
for all $i,j=0,\ldots,d-1.$ A set of maximally entangled MU bases
is then constructed from a set of weakly orthogonal quantum Latin
squares \citep{musto16}.
\begin{thm}
\label{thm:mubs_squares} Let $Q$ and $Q'$ be a pair of weakly orthogonal
quantum Latin squares of order $d,$ and let $H_{i}$ and $H'_{i}$
denote Hadamard matrices of order $d.$ Then the pair of maximally
entangled bases from Eq.~\eqref{eq:latin_entangled_bases} using
$\{Q,H_{i}\}$ and $\{Q',H'_{i}\}$ are mutually unbiased.
\end{thm}
When the quantum Latin squares reduce to Latin squares, this result
reproduces the set of $(\mu+2)$ MU bases of Thm.~\ref{thm:wocjanMUBs}.
In general, very little is known about weakly orthogonal quantum Latin
squares, as well as their potential for constructing larger sets of
MU bases.

\emph{Other related results}—Ad hoc construction methods of maximally
entangled bases in the space $\text{\ensuremath{\mathbb{C}^{d}\otimes\mathbb{C}^{d'}}}$,
and their relation to MU bases, can be found in \citep{tao16,liu16,luo16,xu17,cheng17,zang22},
some of which are reviewed in Ref.~\citep{shi19}. For instance,
in the space $\mathbb{C}^{d}\otimes\mathbb{C}^{kd}$, the set of $kd^{2}$
states
\begin{equation}
\ket{\phi^{j}_{n,m}(U)}=\frac{1}{\sqrt{d}}\sum^{d-1}_{\ell=0}\omega^{n\ell}\ket{\ell\oplus m}\otimes U\ket{\ell+dj}\label{eq:max_ent_basis1}
\end{equation}
with $m,n=0,1,\ldots,d-1,$ and $j=0,1,\ldots,k-1$, forms a maximally
entangled basis for any unitary $U$, and can be used to construct
sets of five and three maximally entangled MU bases in $\mathbb{C}^{2}\otimes\mathbb{C}^{4}$
and $\mathbb{C}^{2}\otimes\mathbb{C}^{6}$, respectively \citep{tao16}.
Here $\ell\oplus m=(\ell+m)\text{ mod }d$ is addition modulo $d$,
and $\omega=e^{2\pi i/d}.$ 

A pair of maximally entangled MU bases in the space $\text{\ensuremath{\mathbb{C}^{2}\otimes\mathbb{C}^{3}} }$is
presented in \citep{shi19}. This generates—by a recursive construction—pairs
of maximally entangled MU bases in $\mathbb{C}^{d}\otimes\mathbb{C}^{d'}$,
for infinitely many $d$ and $d'$, provided $d$ is not a divisor
of $d'$.

A summary of \emph{lower bounds} on the number $N(d,d')$ of maximally
entangled MU in a bi-partite setting with Hilbert space $\mathbb{C}^{d}\otimes\mathbb{C}^{d'}$
is presented in \citep{shi19}. For example, if $d=p^{n_{1}}_{1}\cdots p^{n_{r}}_{r}$
is even, the bases of Eq.~\eqref{eq:max_ent_basis1} provide a lower
bound $N(d,d)\geq$ $\text{min}_{i}(p^{n_{i}}_{i}-1)$ \citep{liu16}.
For odd $d=p^{n_{1}}_{1}\ldots p^{n_{r}}_{r}$, this bound increases
to $N(d,d)\geq\text{min}_{i}2(p^{n_{i}}_{i}-1)$ \citep{cheng17}
while for $d=p^{n}$ with any prime number $p$, one finds $N(d,d)\geq2(d-1)$
\citep{xu17}. Also, when $d=p^{n},$ the canonical construction of
a complete set of $(d^{2}+1)$ MU bases in $\mathbb{C}^{d}\otimes\mathbb{C}^{d}$
yields $(d^{2}-d)$ maximally entangled bases and $(d+1)$ product
bases, hence $N(d,d)\geq d^{2}-d$ \citep{gross07}.

A more exotic basis, namely an \emph{unextendible maximally entangled}
(UME) \emph{basis} has been introduced in \citep{bravyi+11}. This
is an orthogonal basis of $\mathbb{C}^{d}\otimes\mathbb{C}^{d'}$
which contains $n<dd'$ maximally entangled states $\ket{\psi_{i}}\in\mathbb{C}^{d}\otimes\mathbb{C}^{d'}$
such that no additional maximally entangled vector $\ket{\psi}$ satisfies
$\bk{\psi_{i}}{\psi}=0.$ For a two-qubit system, these bases do not
exist. Constructions are known, however, for $\mathbb{C}^{d}\otimes\mathbb{C}^{d}$
when $d=3,4$ \citep{bravyi+11}, and in $\mathbb{C}^{d}\otimes\mathbb{C}^{d'},$
for $\frac{d'}{2}<d<d'$, containing $d^{2}$ maximally entangled
states \citep{chen+13}. For example, in $\mathbb{C}^{2}\otimes\mathbb{C}^{3}$
there exist two UME bases which are mutually unbiased. Similar examples
of mutually unbiased UME bases have been found in \citep{nizamidin+14,nan15,song18,zhao20}.
Furthermore, by taking a single UME basis and a maximally entangled
basis, other pairs of MU bases can be constructed \citep{zhang16}.

Finally, Shi \textit{et al.} \citep{shi19} presented a construction
of MU bases with fixed Schmidt rank, and derived upper bounds on the
maximal number of these bases, complementing the upper bound \eqref{eq:schmidt_bound}
in Cor. \ref{cor:schmidt_bound}.

\subsection{Unextendible sets of MU bases}

\label{subsec:Unextendible-MU-bases}

Given the difficulty to construct complete sets of MU bases in spaces
of composite dimension, it is instructive to investigate the possibility
to extend particular sets of MU bases by individual vectors or bases.
We have already discussed several positive results, including the
observation in Sec.~\ref{subsec:biunimodular} that a vector mutually
unbiased to a given pair of MU bases always exists.

It is also interesting to ponder whether \emph{any} pair of MU bases
extends to a \emph{triple}. The answer is negative: while every MU
pair extends to a triple in dimensions $d\leq5$ , the specific pair
$\{\mathbb{I},S_{6}\}$ in dimension $d=6$ violates this property
\citep{brierley+09} (cf. Sec.~\ref{subsec: unextendible MU bases}).
In fact, numerical searches suggest that most pairs do not extend
to a triple for $d=6$ (see Sec.~\ref{subsec:nonexistence_of_triples}).
On the other hand, no unextendible pair is known for larger $d$ \citep{grassl17}.

More generally, a set of $\mu$ MU bases is called \emph{extendible
}if it can be enlarged by adding at least one further MU basis, and
\emph{unextendible} otherwise. To distinguish between different types
of unextendibility we will use the following terminology.
\begin{defn}
A set of $\mu$ MU bases in $\mathbb{C}^{d}$ is s\emph{trongly unextendible}
if there exists no additional vector unbiased to any of the $\mu$
bases. A set of $\mu$ MU bases in $\mathbb{C}^{d}$, constructed
by partitioning a unitary operator basis $\mathcal{C}$ (see Sec.~\ref{subsec:Maximally-commuting-unitary})
into $\mu$ commuting classes, is \emph{weakly unextendible} if none
of the remaining elements of $\mathcal{C}$ form an additional commuting
class containing $d$ elements. 
\end{defn}
We have already discussed (indirectly) weakly unextendible MU bases
in the form of nice MU bases when $d\notin\mathbb{PP}$ (see Thm.~\ref{thm:niceerrorbasis}
of Sec.~\ref{subsec:Nice-error-bases}). For prime and prime-power
dimensions, several constructions of weakly and strongly unextendible
MU bases are known. For example, the Heisenberg-Weyl group, which
forms a nice error basis (cf. Sec.~\ref{subsec:Nice-error-bases}),
is partitioned by Mandayam \textit{et al. }\citep{mandayam13} (and
later by Garcia and López \citep{garcia21}) into unextendible maximally
commuting classes (i.e., classes for which no other commuting class
can be constructed from the remaining group elements), resulting in
sets of weakly unextendible MU bases for $n$-qubit systems, with
$n=2,3,4,5$. 

More generally, for an even number of $n=2m$ qubits, Grassl \citep{grassl16}
shows there exists a weakly unextendible set of $(2^{m}+1)$ MU bases.
This observation is based on a connection between weakly unextendible
MU bases and maximal symplectic partial spreads \citep{thas16}. Other
examples of maximal symplectic partial spreads, and therefore weakly
unextendible MU bases, for various even and odd prime-power dimensions
are listed in \citep{grassl16,kantor17}. Thas \citep{thas16} uses
symplectic partial spreads to find sets of weakly unextendible MU
bases for all dimensions $d=p^{2},$ with $p\in\mathbb{P}$.

Nietert \textit{et al.} \citep{nietert20} have shown that Thas' set
of weakly unextendible MU bases \citep{thas16} cannot be enlarged
to a complete set of $(d+1)$ MU bases. This follows from a connection
between MU bases and combinatorial designs called $k$-nets \citep{nietert20}
(cf. Sec.~\ref{subsec:Nice-error-bases}):
\begin{thm}
\label{thm:unext_niceerrorbasis} A weakly unextendible set of at
least $(d+1-\sqrt{d})$ nice MU bases in $\mathbb{C}^{d}$ cannot
be enlarged to $(d+1)$ MU bases.
\end{thm}
A construction of $(p^{2}-p+2)$ strongly unextendible MU bases for
$d=p^{2}$ and $p\equiv3\mod 4$, based on complementary decompositions
and MASAs (Sec.~\ref{subsec: masa}), is found by Szántó \citep{szanto16};
the special Galois MU bases in dimensions $d=p^{2},$ with $p=2,3,5,7,11,$
also form a set of $(d-1)$ strongly unextendible bases \citep{szanto16}.
A recent construction in \citep{jedwab16} finds smaller sets of only
$\left(2^{2n-1}+1\right)$ strongly unextendible MU bases for $d=2^{2n}$
and $n\geq1$.

What is the minimum number of MU bases in an unextendible set? Table
\ref{TAB_SUMMARY} provides a summary up to dimension 16 \citep{grassl17}.
While minimal sets are known for $d=2,\ldots,6$ (although for dimensions
$d=3,5$ these are actually complete sets), the upper bound is equal
to three for $d=7,\ldots,16.$ As already mentioned, the smallest
set of unextendible MU bases for $d=6$ has cardinality two, however,
no\emph{ strongly} unextendible pair exists.
\begin{cor}
In finite dimensions $d\geq2,$ no pair of strongly unextendible MU
bases exists.
\end{cor}
This is a direct consequence of Cor. \ref{thm:circular-2} in Sec.~\ref{subsec:biunimodular}:
any pair of orthonormal bases has at least one additional vector unbiased
to both.

In the infinite-dimensional Hilbert space $L^{2}(\mathbb{T}_{1})$
associated with the continuous variables (see Sec.~\ref{subsec: MUs for CVs})
of a particle moving on a one-dimensional torus $\mathbb{T}_{1}$,
however, a pair of \emph{strongly }unextendible MU bases is known:
not even a single quantum state exists which is MU to both the continuous
position and the discrete momentum basis.

\begin{table}[H]
\begin{centering}
\begin{tabular}{|c|c|}
\hline 
dimension & $\aleph(d)$\tabularnewline
\hline 
\hline 
2 & 3\tabularnewline
\hline 
3 & 4\tabularnewline
\hline 
4 & 3\tabularnewline
\hline 
5 & 6\tabularnewline
\hline 
6 & 3\tabularnewline
\hline 
7 to 16 & $\leq3$\tabularnewline
\hline 
$\infty$ & 2\tabularnewline
\hline 
\end{tabular}
\par\end{centering}
\caption{The cardinality $\aleph(d)$ of the smallest set of unextendible MU
bases for $d=2,\ldots,16$, and $d=\infty$. The minimal sets for
$d=3$ and $d=5$ form complete sets of MU bases. For $d=6$ the pair
$\{\mathbb{I},S_{6}\}$ is unextendible, and for dimensions $7\protect\leq d\protect\leq16$
the minimal set contains at most three bases. For $d=\infty$, see
Sec.~\ref{subsec: MUs for CVs}. \label{TAB_SUMMARY}}
\end{table}
 
For non-prime-power dimensions $d=p^{n_{1}}_{1}p^{n_{2}}_{2}\cdots p^{n_{r}}_{r}$,
one can construct $\mu=\text{min}_{i}(p_{i}+1)$ weakly unextendible
MU bases from the eigenbases of the Heisenberg-Weyl operators $Z,$
$X,$ $XZ,$ $XZ^{2},\ldots,XZ^{\mu-2}$ \citep{aschbacher+07,grassl17}.
For even dimensions $d=10,12,14$, the eigenbases of $X,Z$ and $XZ$
are strongly unextendible \citep{grassl17}. For $d=9,15,$ the eigenbases
of $Z,$ $X,$ $XZ,$ and $XZ^{2}$ lead to four unextendible MU bases,
although smaller unextendible sets of cardinality three have also
been found. It is expected that similar results holds for other composite
dimensions, leading to a conjectured systematic behaviour.
\begin{conjecture}
\label{conj:unextendibleHW} For all composite dimensions \textup{$d=p^{n_{1}}_{1}\ldots p^{n_{r}}_{r}$,
and }$\mu=1+\min_{i}p_{i}$, the eigenbases of the Heisenberg-Weyl
operators $Z,X,XZ,XZ^{2},\ldots,XZ^{\mu-2}$ form a set of $\mu$
unextendible MU bases.
\end{conjecture}
Secs.~\ref{sec:fourier_family}–\ref{subsec: unextendible MU bases}
describe additional results on sets of \textit{\emph{unextendible}}
MU bases valid for dimension $d=6.$ 

\subsection{Positive definite functions }

\label{subsec:delsarte}

We now summarise an approach in which certain positive definite functions
provide upper bounds on the maximum cardinality of a set of MU bases
\citep{matolcsi10,kolountzakis18}. The method applies a generalised
version of Delsarte's linear programming bound in the non-commutative
setting \citep{filho14,delsarte72}. Suppose that $G$ is a compact
group with multiplication as the group operation. Let $A=A^{-1}\subset G$
be a symmetric subset, called the ``forbidden'' set, containing
the identity element $e\in A$. The aim is to determine the maximum
cardinality of a set $B=\{b_{1},\ldots,b_{k}\}\subset G$ such that
$b^{-1}_{i}b_{j}\in A^{c}$ for all $i\neq j$, i.e. all pairwise
differences avoid the forbidden set $A$. Here, $A^{c}=G\setminus A$
denotes the complement of $A$. Bounding the cardinality of $B$ involves
finding a positive definite function on $G$ satisfying certain conditions.
\begin{defn}
For a compact group $G$, a continuous function $h:G\rightarrow\mathbb{C}$
is positive definite if for all $t\text{\ensuremath{\geq1},}$ $g_{1},\ldots,g_{t}\in G$
and $c_{1},\ldots,c_{t}\in\mathbb{C}$, then
\begin{equation}
\sum^{t}_{i,j=1}h(g^{-1}_{i}g_{j})c^{*}_{i}c_{j}\geq0\,.\label{def:positivedefinite}
\end{equation}
\end{defn}

The following result specifies a class of positive definite functions
relevant for bounding the cardinality of $B$ \citep{kolountzakis18}.
\begin{thm}
\label{thm:delsarte1} Let $G$ be a compact group and $A=A^{-1}\subset G$
a symmetric subset with identity $e\in A$. Suppose there exists a
positive definite function $h:G\rightarrow\mathbb{R}$ such that $h(x)\leq0$
for all $x\in A^{c}$ and $\int hd\nu>0$, where $\nu$ is the normalised
Haar measure. For any $B=\{b_{1},\ldots,b_{k}\}\subset G$ such that
$b^{-1}_{i}b_{j}\in A^{c}$ for all $i\neq j$, the cardinality of
$B$ is bounded by $|B|\leq h(e)/\int hd\nu.$
\end{thm}
Matolcsi \citep{matolcsi10} first made the connection to MU bases
in the commutative setting by taking $G=\mathbb{T}^{d}$, where $\mathbb{T}$
is the complex unit circle (see Sec. \ref{subsec:linear-constraints}).
Later, once the extension of Delsarte's bound to non-commutative groups
was established, Kolountzakis \textit{et al.} \citep{kolountzakis18}
formalised a simper relation by taking $G=U(d)$ as the group of unitaries,
and $A^{c}=H(d)$ as the set of complex Hadamard matrices. Since the
maximum number of MU bases in $\mathbb{C}^{d}$ is equivalent to the
maximum cardinality of $B=\{U_{1},\ldots,U_{k}\}\subset G$, where
the differences $U^{-1}_{i}U_{j}$ are elements of $H(d)$ (cf. Sec.
\ref{subsec: Hadamard equivalence}), Thm. \ref{thm:delsarte1} leads
to the following result.
\begin{cor}
\label{cor:pos_def}Let $G=U(d)$ and $A^{c}=H(d)$, and let $h:G\rightarrow\mathbb{R}$
be any positive definite function satisfying the conditions of Thm
\ref{thm:delsarte_poly}. The maximum number $\mu$ of MU bases in
$\mathbb{C}^{d}$ satisfies\textup{ $\mu\leq h(e)/\int hd\nu$.}
\end{cor}
Consider, for example, the polynomial
\begin{equation}
h_{0}(U)=-1+\sum^{d}_{i,j=1}|U_{ij}|^{4},\label{eq:positivedefinite}
\end{equation}
where $U=(U_{i,j})^{d}_{i,j=1}\in U(d)$, which is positive definite
on $U(d)$ and takes the value zero whenever $U\in H(d)$. Since $h_{0}(e)=d-1$
and $\int h_{0}d\nu=(d-1)/(d+1)$, Cor. \ref{cor:pos_def} yields
the well known upper bound $\mu\leq d+1$.

The ultimate aim is to construct a positive definite function for
a non-prime-power $d$ to show $\mu<d+1$. While $h_{0}$ is one function
that vanishes on the set of Hadamard matrices, other possible examples
arise for $d=6$ in the following way. Consider,
\begin{equation}
f_{1}(U)=\sum_{\sigma\in S_{6}}\sum^{6}_{j=1}\prod^{3}_{i=1}U_{\sigma(i),j}U^{*}_{\sigma(i+3),j}\,,
\end{equation}
and $f_{2}(U)=f_{1}(U^{*}),$ with $S_{6}$ the permutation group
of the set of 6 elements. Both functions are real-valued and expected
to vanish for every $U\in H(d)$ (see Conjecture \ref{conj:fourier}).
Furthermore, the three functions $f(U)=(f_{1}(U)+f_{2}(U))^{2},$
$f(U)=f^{2}_{1}(U)+f^{2}_{2}(U)$ and $f(U)=f^{2}_{1}(U)f^{2}_{2}(U)$,
satisfy $f(e)=0$ and $\int_{U\in U(d)}f(U)d\nu>0$. If one can show
that for any $\epsilon>0,$ $h_{1}(U)=h_{0}(U)+\epsilon f(U)$ is
positive definite, then the upper bound on the cardinality of $B$
will fall below $(d+1),$ i.e., $|B|\leq\frac{h_{1}(e)}{\int h_{1}d\nu}<d+1$. 

A recent result suggests the method based on positive definite functions
may not be a suitable strategy to tackle the existence problem. By
considering positive definite polynomials in the entries of $U\in U(d)$
and their conjugates, with degree at most $6$, Bandeira \textit{et
al.} \citep{bandeira22} showed the following.
\begin{thm}
\label{thm:delsarte_poly} The method of positive definite polynomials
of degree at most 6 cannot be used to show that fewer than seven MU
bases exist in $\text{\ensuremath{\mathbb{C}^{6}}}$.
\end{thm}
The proof involves a convex duality argument with computer-aided symbolic
calculations. The techniques could also be applied for larger values
of $d$ and polynomials of degree $q$, but the calculations become
computationally intractable beyond $d=q=6$. Instead, the construction
of a dual certificate is presented that, if verified, would prove
the following claim for arbitrary system sizes \citep{bandeira22}.
\begin{conjecture}
The method of positive definite functions cannot be used to show that
fewer than $(d+1)$ MU bases exist in $\text{\ensuremath{\mathbb{C}^{d}}},$
for all $d>1$.
\end{conjecture}

\subsection{Linear constraints}

\label{subsec:linear-constraints} 

By considering the group $G=\mathbb{T}^{d}$, where $\mathbb{T}$
is the complex unit circle, Matolcsi's original application of positive
definite functions in \citep{matolcsi10} applied a more restricted
version of Theorem \ref{thm:delsarte1} to bound the cardinality of
$B\subset G$, where the elements of $B$ are no longer unitary matrices
but the \textit{columns} of mutually unbiased Hadamard matrices. In
related work \citep{matolcsi+12}, Fourier analytic arguments were
applied to reduce the MU conditions to linear constraints, leading
to two important consequences. First, the constraints reveal several
interesting structural features about sets of MU bases. Second, one
can attempt to demonstrate that the constraints do not hold, providing
a strategy to prove that complete sets do not exist in composite dimensions.

The dual group of $G=\mathbb{T}^{d}$ is given by $\hat{G}=\mathbb{Z}^{d}$,
and the action of a character $\gamma=(n_{1},n_{2},\ldots,n_{d})\in\mathbb{Z}^{d}$
on an element ${\bf u}=(u_{1},u_{2},\ldots,u_{d})\in\mathbb{T}^{d}$
is $\gamma({\bf u})={\bf u}^{\gamma}=u^{n_{1}}_{1}u^{n_{2}}_{2}\cdots u^{n_{d}}_{d}$.
For a set $S\subset G$ the Fourier transform is denoted by $\hat{S}=\sum_{{\bf s}\in S}{\bf s}^{\gamma}$.
Thus, given a complete set of MU bases $\{\mathbb{I},H_{1},\ldots,H_{d}\}$,
where $H_{j}\subset G$ is represented as a $d$ element set $\{{\bf c}_{j1},\ldots,{\bf c}_{jd}\}$
of its columns, the Fourier transform of $H_{j}$ is
\begin{equation}
g_{j}(\gamma)\equiv\hat{H}_{j}(\gamma)=\sum^{d}_{k=1}{\bf c}^{\gamma}_{jk}\,,\label{eq:ft1}
\end{equation}
for each $\gamma\in\mathbb{Z}^{d}$. Two functions, $E(\gamma)$ and
$F(\gamma)$, that prove essential in this framework are defined as
\begin{equation}
E(\gamma)\equiv\sum^{d}_{j=1}E_{j}(\gamma)\,,\label{eq:ft2}
\end{equation}
where $E_{j}(\gamma)\equiv|g_{j}(\gamma)|^{2}$, and
\begin{equation}
F(\gamma)\equiv|f(\gamma)|^{2},\label{eq:ft3}
\end{equation}
with $f(\gamma)\equiv\sum^{d}_{j=1}g_{j}(\gamma)$ for each $\gamma\in\mathbb{Z}^{d}$.
The orthogonality and unbiasedness relations can then be expressed
as \emph{linear} constraints on the functions $E$ and $F$. In other
words, the polynomial relations from the orthogonality and unbiasedness
conditions are transformed into linear relations using Fourier transforms,
and one should expect that these constraints are simpler to deal with.

Explicitly, the orthogonality conditions for each basis are
\begin{equation}
\sum^{d}_{r=1}E_{j}(\gamma+\pi_{r})=d^{2}\,,\label{eq:fourier1}
\end{equation}
for each $\ensuremath{\gamma\in\mathbb{Z}^{d}}$, which imply
\begin{equation}
\sum^{d}_{r=1}E(\gamma+\pi_{r})=d^{3}\,,\label{eq:fourier1-1}
\end{equation}
where $\pi_{r}=(0,\ldots,0,1,0,\ldots,0)\in\mathbb{Z}^{d}$ denotes
the vector with its $r$-th coordinate equal to one. The constant
overlap constraint simplifies to
\begin{equation}
dE(\gamma)+\sum_{r\neq t}F(\gamma+\pi_{r}-\pi_{t})=d^{4}.\label{eq:fourier2}
\end{equation}
The functions exhibit additional trivial constraints such as $F(0)=d^{4}$
and $E(0)=d^{3}$, and satisfy inequalities
\begin{equation}
0\leq F(\gamma)\leq d^{4}\,,\qquad0\leq E(\gamma)\leq d^{3},\label{eq:fourier3}
\end{equation}
and
\begin{equation}
F(\gamma)\leq dE(\gamma)\,,\label{eq:fourier3-1}
\end{equation}
for each $\ensuremath{\gamma\in\mathbb{Z}^{d}}$.

For $d\leq5$ the derivations of several known results on MU bases
follow simply from the linear constraints of Eqs.~ (\ref{eq:fourier1}–\ref{eq:fourier3-1}).
For example, Eqs.~(\ref{eq:fourier1}–\ref{eq:fourier3-1}) imply
that the matrix elements of all $d$ Hadamard matrices in a complete
set are $d$-th roots of unity; which in turn provides a simple means
to classify complete sets in these dimensions.

When $d=5$ the constraints imply that any Hadamard matrix of order
five is equivalent to the Fourier matrix, which was first observed
by Haagerup \citep{haagerup97}. To prove this using the constraints
above, the function $E_{1}(\gamma)=|\hat{H}_{1}(\gamma)|^{2},$ where
$H_{1}$ is a Hadamard of order five, is treated as a \textit{variable}
for each $\gamma\in\mathbb{Z}^{5}$. The function $E_{1}(\gamma)$
satisfies Eq.~(\ref{eq:fourier1-1}), $E_{1}(0)=25$ and $0\leq E_{1}(\gamma)\leq25,$
for all $\gamma\in\mathbb{Z}^{5}$. By choosing $\sigma\in\mathbb{Z}^{5}$
as any permutation of ($5,-5,0,0,0)$ one can show via a linear programming
code that $E_{1}(\sigma)=25$. This condition implies that all elements
of $H_{1}$ are fifth roots of unity, whence $H_{1}$ is the Fourier
matrix.

In higher dimensions, if one can show that $F(\rho)=d^{4}$, where
$\rho$ is a permutation of $(d,-d,0,\ldots,0),$ then the set of
$d$ Hadamard matrices contain only $d$-th roots of unity. Imposing
this structure on a set of bases in $d=6$ is incompatible with the
existence of a complete set of MU bases (see Thm. \ref{thm:roots_of_unity}
of Sec. \ref{subsec: unextendible MU bases}), and would prove Zauner's
conjecture. Alas, the linear constraints of Eqs.~(\ref{eq:fourier1-1})–(\ref{eq:fourier3-1})
do not appear to imply $F(\rho)=d^{4}$, and hence the matrix entries
cannot be restricted to roots of unity \citep{matolcsi+12}.

One of the main consequences of the above approach is a limitation
on the number of real Hadamard matrices in a complete set \citep{matolcsi+12}.
\begin{thm}
\label{thm:realhadamards} Let $\{\mathbb{I},H_{1},\ldots,H_{d}\}$
be a complete system of MU bases, in matrix form, and suppose that
$H_{1}$ is a real Hadamard matrix. Then there is no further purely
real column in any of the matrices $H_{2},\ldots,H_{d}$. In particular
it is impossible to have two real Hadamard matrices in a complete
set of MU bases.
\end{thm}
An additional consequence is that no complete set of MU bases in dimension
six contains the pair $\{\mathbb{I},F_{6}\}$, a result we will see
in Sec.~\ref{sec:fourier_family}. The original proof uses a computer
algebraic method, whereas the method here relies only on Fourier analytic
arguments. Furthermore, a stronger result excluding the existence
of a complete set containing the family of pairs $\{\mathbb{I},F^{(2)}_{6}\}$,
namely Thm. \ref{thm:Fourierfamilyexclusion}, can be proved using
similar arguments \emph{if} the following is true \citep{matolcsi+12}:
\begin{conjecture}
\label{conj:fourier} Let $H$ be any complex Hadamard matrix of order
6, not equivalent to the isolated matrix $S_{6}$ and let $\sigma$
be any permutation of the vector $(1,1,1,-1,-1,-1)$. Then the function
$g_{j}(\sigma)$ defined in Eq.~\eqref{eq:ft1} cannot vanish everywhere.
\end{conjecture}
We note, however, that the conjecture is not necessary to prove Thm.
\ref{thm:Fourierfamilyexclusion}. A computer search first confirmed
this result \citep{jaming+09}, and later an analytic proof based
on improvements to the Delstarte-type linear programming bound \citep{matolcsi15}.

Some progress towards proving Conjecture \ref{conj:fourier} was achieved
in \citep{maxwell15}, but only by restricting the class of matrices
to the three-parameter Karlsson family $K^{(3)}_{6}$, described in
Eq.~(\ref{eq:karlsson-block}). It has been predicted that Conjecture
\ref{conj:fourier}, by offering an additional linear constraint on
the function $E$, may be useful in a proof of the non-existence of
a complete set in dimension six.

\section{Rigorous results: Dimension six}

\label{sec:rigorous_results_d=00003D6}

With $d=6$ being the smallest dimension where the existence problem
manifests itself, quite a few studies have focussed on this specific
case, in the hope of conclusive insights. First, we will summarise
all known pairs and triples of MU bases for dimension six, as well
as the properties of sets which contain the Fourier family of $6\times6$
Hadamard matrices. Then, we review a number of results that exploit
the tensor product structure of dimension six. Finally, we describe
non-existence results which show that certain pairs and triples do
not extend to larger collections of MU bases. Numerical results will
be summarised in the subsequent section.

\subsection{Pairs of MU bases }
\label{sec:pairs_of_MU_bases_C^6}
In Sec.~\ref{sec:pairs_of_MU_bases_C^d} some well-known examples
of $d\times d$ complex Hadamard matrices were discussed. We now focus
on \emph{classifying} complex Hadamard matrices of order six since
each such matrix guarantees the existence of a \emph{pair} of MU bases
in $\mathbb{C}^{6}$ when combined with the identity matrix. An early
review was provided in Ref.~\citep{bengtsson+07}, and an up-to-date
summary can be found online \citep{bruzda+12}.

We begin by summarising the currently known $6\times6$ complex Hadamard
matrices in terms of a theorem. 
\begin{thm}
\label{thm:allHadamards}All currently known $6\times6$ complex Hadamard
matrices belong to one of the three classes: (i) an isolated matrix
$S_{6}$; (ii) a three-parameter family $K^{(3)}_{6}$; and (iii)
a four-parameter family $G^{(4)}_{6}$.
\end{thm}
Let us describe what is known about the classes of matrices listed in
the theorem. To simplify notation, we will occasionally suppress the
normalisation factor $1/\sqrt{d}$ in this subsection and the next
one; in other words, the entries of the Hadamard matrices we consider
may have modulus $1$.

\subsubsection*{The isolated matrix $S_{6}$}

A Hadamard matrix is \emph{isolated} if it does not belong to any
continuous family of Hadamard matrices. In particular, a Hadamard
matrix with zero \emph{defect }must be isolated (cf. Appendix \ref{sec:Complex-Hadamard-matrices}).
The only known isolated matrix of order six can be traced back to
a construction by Butson \citep{butson62} in 1962 of order $2p$
matrices which contain only $p$-th roots of unity, when $p$ is prime.
For $p=3$, the construction leads to a matrix consisting of third
roots of unity only, namely
\begin{equation}
S_{6}=\left(\begin{array}{cccccc}
1 & 1 & 1 & 1 & 1 & 1\\
1 & 1 & \omega & \omega & \omega^{2} & \omega^{2}\\
1 & \omega & 1 & \omega^{2} & \omega^{2} & \omega\\
1 & \omega & \omega^{2} & 1 & \omega & \omega^{2}\\
1 & \omega^{2} & \omega^{2} & \omega & 1 & \omega\\
1 & \omega^{2} & \omega & \omega^{2} & \omega & 1
\end{array}\right),\label{eq:isolatedmatrix}
\end{equation}
where $\omega=e^{2\pi i/3}$. The matrix $S_6$ is also know as \emph{Tao's matrix}, displayed explicitly for the first time in a paper from 2004 \citep{tao04}. Other independent derivations
of $S_{6}$ are based on symmetry conditions \citep{matolcsi+08}
and or use product bases \citep{mcnulty12b} (see Sec.~\ref{subsec:product_bases}).

\subsubsection*{The three-parameter family $K^{(3)}_{6}$}

The three-parameter family $K^{(3)}_{6}$ has been discovered by Karlsson
in 2011 \citep{karlsson11}. Before describing its construction, it
is instructive to summarise the derivations of various one- and two-parameter
families which are contained in the larger set $K^{(3)}_{6}$. In
general, constructions have been quite haphazard, with individual
examples found and later extended or connected to one- and two-parameter
families.

The simplest examples of Hadamard matrices are affine families (see
Appendix \ref{sec:Complex-Hadamard-matrices}). The only known affine
families are the two-parameter Fourier family $F^{(2)}_{6}$ (and
its transpose), and the one-parameter Diţă family $D^{(1)}_{6}$ \citep{dita04}.
The Diţă family provides a full characterisation of \textit{regular}
Hadamard matrices of order six \citep{banica09}. The classification
of all \emph{self-adjoint complex} Hadamard matrices of order six
yields a non-affine one-parameter family $B^{(1)}_{6}$ found in \citep{beauchamp+08}.
Another non-affine example, derived in \citep{matolcsi+08}, is the
one-parameter family of symmetric matrices $M^{(1)}_{6}$. More general
non-affine two-parameter families $X^{(2)}_{6}$ and $K^{(2)}_{6}$
were later discovered by Szöllősi \citep{szollosi10} and Karlsson
\citep{karlsson09}, respectively, each containing previously discovered
one-parameter families, with the inclusions $D^{(1)}_{6}\subset X^{(2)}_{6},B^{(1)}_{6}\subset X^{(2)}_{6}$
and $D^{(1)}_{6}\subset K^{(2)}_{6},M^{(1)}_{6}\subset K^{(2)}_{6}$.

The Karlsson family $K^{(3)}_{6}$, which encompasses \textit{all}
previously known one- and two-parameter families, was found by investigating
matrices that are \emph{$H_{2}$-reducible}; a property demanding
that all $2\times2$ blocks of a $6\times6$ matrix be complex Hadamard
matrices themselves \citep{karlsson11h2}. 
\begin{thm}
\label{thm:h2reducible}Every Hadamard matrix of order six is equivalent
to a matrix where \emph{all} or \emph{none} of the nine (non-overlapping)
$2\times2$ blocks are (proportional to) Hadamard matrices. 
\end{thm}
It is simple to check the $H_{2}$-reducibility property since a matrix
of order six is $H_{2}$-reducible if and only if its dephased form
(cf. Appendix \ref{sec:Complex-Hadamard-matrices}) contains an element
equal to $(-1)$.

To construct $K^{(3)}_{6}$, one starts with a general dephased block
matrix of nine $2\times2$ submatrices. Requiring the matrix of order
six to be $H_{2}$-reducible and using the unitary and unimodularity
constraints on its elements, a complete classification of all $H_{2}$-reducible
Hadamard matrices is given by the set 
\begin{equation}
K^{(3)}_{6}\!\equiv\! K_{6}(\theta,\phi,\lambda)\! = \!\left(\begin{array}{ccc}
F_{2} & Z_{1} & Z_{2}\\
Z_{3} & \tfrac{1}{2}Z_{3}AZ_{1} & \tfrac{1}{2}Z_{3}BZ_{2}\\
Z_{4} & \tfrac{1}{2}Z_{4}BZ_{1} & \tfrac{1}{2}Z_{4}AZ_{2}
\end{array}\right)\label{eq:karlssons-3-param-family}
\end{equation}
as described in \citep{karlsson11}. Three parameters enter this expression
in the following way. The $2\times2$ matrix 
\begin{equation}
A=\left(\begin{array}{cc}
A_{11} & A_{12}\\
A_{12} & -A_{11}
\end{array}\right)\,\label{eq:karlsson-block}
\end{equation}
has elements
\begin{align}
A_{11} & =\frac{1}{2}+i\frac{\sqrt{3}}{2}(\cos\theta+e^{-i\phi}\sin\theta),\\
A_{12} & =-\frac{1}{2}+i\frac{\sqrt{3}}{2}(-\cos\theta+e^{i\phi}\sin\theta),
\end{align}
with $\theta,\phi\in[0,\pi)$, while the matrix $B$ is defined as
$B=-F_{2}-A$, with $F_{2}$ the Fourier matrix. The submatrices 
\begin{equation}
Z_{j}=\left(\begin{array}{cc}
1 & 1\\
z_{j} & -z_{j}
\end{array}\right)\,,\qquad Z_{k}=\left(\begin{array}{cc}
1 & z_{k}\\
1 & -z_{k}
\end{array}\right)\,,\label{eq: four Z-matrices}
\end{equation}
with the left and right versions defined for $j=1,2$ and $k=3,4$,
respectively, depend on unimodular parameters, i.e. $|z_{j}|=|z_{k}|=1$,
which are related by Möbius transformations $\mathcal{M}(z)=\frac{\alpha z-\beta}{\bar{\beta}z-\bar{\alpha}}$.
In particular, one has $z^{2}_{3}=\mathcal{M}_{A}(z^{2}_{1})$, $z^{2}_{3}=\mathcal{M}_{B}(z^{2}_{2})$,
$z^{2}_{4}=\mathcal{M}_{A}(z^{2}_{2})$ and $z^{2}_{4}=\mathcal{M}_{B}(z^{2}_{1})$,
where $\alpha_{A}=A^{2}_{12}$, $\beta_{A}=A^{2}_{11}$, $\alpha_{B}=B^{2}_{12}$
and $\beta_{B}=B^{2}_{11}$. By choosing $z_{1}=e^{i\lambda}$, say,
the remaining three $z$-parameters are uniquely determined through
the Möbius transformations, resulting in the three-parameter family
$K_{6}(\theta,\phi,\lambda)$.

The parameterisation used to describe $K^{(3)}_{6}$ differs from
those of the smaller one- and two-parameter families, so connections
between $K^{(3)}_{6}$ and its subfamilies are difficult to spot.
Exceptions occur when $\theta$ and $\phi$ are constant: for example,
$D^{(1)}_{6}$ is recovered when $\theta=\arccos(1/\sqrt{3})$ and
$\phi=\pi/4$. The Fourier family $F^{(2)}_{6}$ is recovered in the
limit of $\theta=0$, taking $z_{1}$ and $\phi$ as the free parameters
\citep{karlsson11}.

\subsubsection*{The four-parameter family $G^{(4)}_{6}$}

Evidence of the existence of a four-parameter family was first provided
by numerical calculations in Ref.~\citep{skinner+09}. Performing
infinitesimal shifts of phases in the Fourier matrix was found to
preserve the unitary condition of the Hadamard matrix while moving
away along four independent directions. This property was, in fact,
expected already in 2005 since the \emph{defect}—an upper bound on
the dimensionality of a set of Hadamard matrices (see Appendix \ref{sec:Complex-Hadamard-matrices})—equals
four for many Hadamard matrices of order six \citep{bengtsson05}.

The construction of a proposed four-parameter family was first presented
by Szöllősi in 2012 \citep{szollosi12}, but a rigorous proof of its
existence appeared only later by Bondal and Zhdanovskiy \citep{bondal16}.
The result, using methods from algebraic geometry, is formulated in
terms of orthogonal pairs of Cartan subalgebras in $sl_{d}(\mathbb{C})$.
\begin{thm}
\label{thm:4-parameter-orthogonal-pair}There exists a four-parameter
family of orthogonal pairs of Cartan subalgebras in $sl_{6}(\mathbb{C})$. 
\end{thm}
Combining this result with Equivalence \ref{equiv:OD} immediately
implies the existence of a four-parameter family of $6\times6$ complex
Hadamard matrices.

The proof of Thm.~\ref{thm:4-parameter-orthogonal-pair} is not constructive;
instead we review the construction of the four-parameter set $G^{(4)}_{6}$
proposed in \citep{szollosi12}. The general technique aims to embed
a ``well-behaved'' $3\times3$ submatrix $E(a,b,c,d)$ into a larger
complex Hadamard matrix, viz.
\begin{equation}
\begin{array}{c}
\hspace{-16em}G_{6}(a,b,c,d)=\\
=\left(\begin{array}{cccccc}
1 & 1 & 1 & 1 & 1 & 1\\
1 & a & b & e & s_{1} & s_{2}\\
1 & c & d & f & s_{3} & s_{4}\\
1 & g & h & * & * & *\\
1 & t_{1} & t_{3} & * & * & *\\
1 & t_{2} & t_{4} & * & * & *
\end{array}\right)\equiv\left(\begin{array}{cc}
E & B\\
C & D
\end{array}\right)\,,
\end{array}\label{eq: G6(4) construction}
\end{equation}
where $B,C$ and $D$ are also $3\times3$ submatrices. First, the
unimodular entries of $B$ and $C$ are determined by the orthogonality
requirements from the first three rows and columns of $G^{(4)}_{6}$.
Once $e,f,g,h,s_{i}$ and $t_{i}$—all of which depend on $a,b,c$
and $d$—have been calculated, the entries of $D$ are then evaluated.
These are fixed by $D=-CE^{\dagger}(B^{-1})^{\dagger}$, where $\dagger$
denotes the conjugate transpose; if the entries are unimodular, $G^{(4)}_{6}$
is a complex Hadamard matrix. It is conjectured that $E$ can be chosen
to ensure a \textit{finite} number of candidate submatrices $B$ and
$C$, so that it is feasible to check whether the resulting matrix
is a Hadamard matrix.

The explicit algorithm which generates $G^{(4)}_{6}$ is presented
in \citep{szollosi12}. The entries of $G^{(4)}_{6}$ are given by
algebraic functions of roots of sextic polynomials, without known
closed expressions. A Mathematica script which provides random matrices
according to the above construction is available online \citep{bruzda+12}.

Unfortunately, the relation between $G^{(4)}_{6}$ and $K^{(3)}_{6}$
is not fully understood. According to \citep[Prop. 2.16]{szollosi12},
if a Hadamard matrix $H$ is not equivalent to either $S_{6}$ or
a member of $K^{(3)}_{6}$, then it contains a well-behaved submatrix
$E$, and consequently there is the possibility to reconstruct $H$
from the above method. On the other hand, it is not known if $K^{(3)}_{6}$
contains a suitable $3\times3$ submatrix that would result in $K^{(3)}_{6}\subset G^{(4)}_{6}$.

Taken together, the sets of $6\times6$ complex Hadamard matrices
listed in Thm. \ref{thm:allHadamards} are expected to exhaust all
such matrices \citep{szollosi12}.
\begin{conjecture}
\label{conj:hadamardclassification} Every complex Hadamard matrix
of order six is equivalent to a member of either $K^{(3)}_{6}$ or
$G^{(4)}_{6}$, or to Tao's matrix $S_{6}$.
\end{conjecture}
If true, this result would open up a feasible approach to decide how
many MU bases the space $\mathbb{C}^{6}$ can accommodate. One could
attempt an exhaustive computer search in analogy to the proof which
excludes both the Fourier family $F^{(2)}_{6}$ and its transpose
from appearing in a hypothetical set of seven MU bases \citep{jaming+09}.

Examples of non-$H_{2}$-reducible $6\times6$ Hadamard matrices—which
one expects to be members of $G^{(4)}_{6}$—appear in \citep{liang24}.
Further properties of Hadamard matrices of order six can be found,
for example, in \citep{liang21,liang24}.

\subsection{Triples of MU bases}

\label{sec:triples_of_MU_bases}

Given the different pairs of MU bases that result from Thm.~\ref{thm:allHadamards},
it is natural to wonder if they extend to triples of MU bases. Does
every pair of MU bases extend to at least one triple? In other words,
given a Hadamard matrix $H$ does there exist another one, say $K$,
such that their product $H^{\dagger}K$ is also a Hadamard matrix? 

An extensive \emph{numerical} search of matrices $K$ unbiased to
$\{\mathbb{I},H\}$, for every affine and non-affine family (known
before 2009), is given in \citep{brierley+09}. For each family, a
search was conducted both at regular intervals along the parameter
space as well as at random. When $H$ belongs to an affine family,
i.e. the Fourier family $F^{(2)}_{6}$ or the Diţă family $D^{(1)}_{6}$,
a triple of the form $\{\mathbb{I},H,K\}$ is \emph{always} found.
In particular, the number of vectors MU to the Diţă family fluctuates
between 48, 72 and 120 vectors, depending on the value of the parameter.
Meanwhile, 48 vector are found to be mutually unbiased to the Fourier
family, for each sampled parameter value. In contrast, the isolated
matrix $S_{6}$ of Eq.~(\ref{eq:isolatedmatrix}) does not extend
to a triple (see Sec.~\ref{subsec: unextendible MU bases}).

When $H$ belongs to a non-affine family, direct calculations become
more complicated, necessitating certain approximations. Results pertaining
to these cases are discussed further in the section on numerics (Sec.~\ref{subsec:nonexistence_of_triples}).
Numerical methods from several studies suggest extensions to MU triples
are not always possible \citep{goyeneche13,Matolcsi24}. The extendibility
properties of all known pairs of MU bases are summarised in Table~\ref{tab:triples}
of Sec.~\ref{subsec: unextendible MU bases}, combining rigorous
results and numerical evidence discussed in Sec.~\ref{subsec:nonexistence_of_triples}.

Finding \emph{closed-form} expressions of MU triples appears to be
difficult: we are, in fact, aware of only three infinite families
in dimension six. A one-parameter family was first presented by Zauner
\citep{Zauner1991,Zauner2011}, and the method was later used to generate
a two-parameter family by Szöllősi \citep{szollosi10}. Another one-parameter
family was found by Jaming \emph{et al}. \citep{jaming+09}, containing
a subset of the family $F^{(2)}_{6}$.

Let us spell out the closed-form expressions of these examples. The
first two use a result obtained by Zauner \citep{Zauner1991} (see
also \citep{szollosi10}).
\begin{thm}
\label{thm:zauner_triples}If T is a $2n\times2n$ Hadamard matrix
with $n\times n$ circulant blocks, then there exist $2n\times2n$
Hadamard matrices $E_{1}$ and $E_{2}$ such that $T=E^{-1}_{1}E_{2}$.
\end{thm}
This theorem implies that a triple of MU bases $\{\mathbb{I},E_{1},E_{2}\}$
exists given a Hadamard matrix $T$ consisting of circulant blocks.
Zauner considers an explicit construction of $T$, $E_{1}$ and $E_{2}$
in the space $\mathbb{C}^{6}$, with $T$ a one-parameter family of
Hadamard matrices. 

Let $T$ be a $6\times6$ matrix of the form 
\begin{equation}
T=\left(\begin{array}{cc}
A_{11} & A_{12}\\
A_{21} & A_{22}
\end{array}\right)\,,\label{DUPLICATE: zauner-family}
\end{equation}
where $A_{ij}$ are circulant matrices of order three. Since $A_{ij}$
are circulant they may be written in the form $A_{ij}=F^{-1}_{3}\bar{A}_{ij}F_{3}$,
where $F_{3}$ is the $3\times3$ Fourier matrix and $\bar{A}_{ij}=\text{diag}(a^{1}_{ij},a^{2}_{ij},a^{3}_{ij})$.
If $T$ is unitary then the matrices 
\begin{equation}
S_{k}=\left(\begin{array}{cc}
a^{k}_{11} & a^{k}_{12}\\
a^{k}_{21} & a^{k}_{22}
\end{array}\right)\,,\quad k=1,2,3,\label{eq:zauner-family-S}
\end{equation}
are unitary. For each of these $2\times2$ unitary
matrices there exist parameters $b^{k}_{\ell}\in[0,2\pi)$, $\ell=1\ldots4$,
such that 
{
\setlength{\arraycolsep}{3pt}
\begin{equation}
S_k\!=\!\tfrac12
\begin{pmatrix}\label{eq:zauner-family-s2}
(e^{ib^k_1}+e^{ib^k_2}) &
e^{ib^k_4}(e^{ib^k_1}-e^{ib^k_2})\\
e^{-ib^k_3}(e^{ib^k_1}-e^{ib^k_2}) &
e^{-ib^k_3}e^{ib^k_4}(e^{ib^k_1}+e^{ib^k_2})
\end{pmatrix}\!.
\end{equation}
}
We can then write $T=E^{-1}_{1}E_{2}$ where 
\begin{equation}
E_{1}=\frac{1}{\sqrt{2}}\left(\begin{array}{cc}
F_{3} & U_{3}F_{3}\\
F_{3} & -U_{3}F_{3}
\end{array}\right)\,\label{zauner-family1}
\end{equation}
and 
\begin{equation}
E_{2}=\frac{1}{\sqrt{2}}\left(\begin{array}{cc}
U_{1}F_{3} & U_{1}U_{4}F_{3}\\
U_{2}F_{3} & -U_{2}U_{4}F_{3}
\end{array}\right)\,,\label{zauner-family2}
\end{equation}
with diagonal unitaries $U_{\ell}=\text{diag}(e^{b^{\ell}_{1}},e^{ib^{\ell}_{2}},e^{ib^{\ell}_{3}})$,
$\ell=1\ldots4$. Now, since $E_{1}$ and $E_{2}$ are Hadamard matrices
and by requiring that $T$ has entries of constant modulus, the columns
of $E_{1}$ and $E_{2}$ form a pair of MU bases. Finally, taking
{\setlength{\arraycolsep}{3pt}
\begin{equation}
T=
\begin{pmatrix}\label{eq:T-matrix}
1 & e^{-ix} & e^{ix} & -1 & i e^{-ix} & i e^{ix}\\
e^{ix} & 1 & -e^{-ix} & i e^{ix} & -1 & i e^{-ix}\\
-e^{-ix} & e^{ix} & 1 & i e^{-ix} & i e^{ix} & -1\\
1 & i e^{-ix} & i e^{ix} & 1 & e^{-ix} & -e^{ix}\\
i e^{ix} & 1 & i e^{-ix} & -e^{ix} & 1 & e^{-ix}\\
i e^{-ix} & i e^{ix} & 1 & e^{-ix} & -e^{ix} & 1
\end{pmatrix}
\end{equation}}with $x\in[0,2\pi)$, the triple $\{\mathbb{I},E_{1},E_{2}\}$ can
be shown to form a set of three MU bases containing one free parameter.

Szöllősi's two-parameter family of Hadamard matrices $X^{(2)}_{6}\equiv X_{6}(\alpha)\equiv X_{6}((x,y),(u,v))$
also contains $3\times3$ circulant blocks for all parameter values
\citep{szollosi10}, and is therefore a suitable candidate for Thm.~\ref{thm:zauner_triples}.
Written in dephased form,
\begin{equation}
X^{(2)}_{6}\!=\!\left(\begin{array}{cccccc}
1 & 1 & 1 & 1 & 1 & 1\\
1 & x^{2}y & xy^{2} & \frac{xy}{uv} & uxy & vxy\\
1 & \frac{x}{y} & x^{2}y & \frac{x}{u} & \frac{x}{v} & uvx\\
1 & uvx & uxy & -1 & -uxy & -uvx\\
1 & \frac{x}{u} & vxy & -\frac{x}{u} & -1 & -vxy\\
1 & \frac{x}{v} & \frac{xy}{uv} & -\frac{xy}{uv} & -\frac{x}{v} & -1
\end{array}\right)\!,\label{eq: X^2_6}
\end{equation}
where $(x,y)$ and $(u,v)$ are determined by the roots of $f_{\alpha}(z)$
and $f_{-\alpha}(z)$, respectively, where $f_{\alpha}(z)=z^{3}-\alpha z^{2}+\alpha^{*}z-1=0$,
and $\alpha\in\mathbb{D}$. Here, $\mathbb{D}$ is a region defined
by the intersection of two deltoids, as described in \citep{szollosi10}.
For this choice of matrix, Thm.~\ref{thm:zauner_triples} leads to
a two-parameter family of MU triples.
\begin{thm}
\label{thm:ferenc_triples}There exists a two-parameter family of
MU triples $\{\mathbb{I},E_{1}(\alpha),E_{2}(\alpha)\}$ in dimension
six, where $X_{6}(\alpha)=E^{-1}_{1}(\alpha)E_{2}(\alpha)$ is the
two-parameter Szöllősi family of Hadamard matrices.
\end{thm}
Recall that the Diţă family $D^{(1)}_{6}$ and the self-adjoint family
$B^{(1)}_{6}$ are both included in the set $X^{(2)}_{6}$. Consequently,
they both extend to sets of three MU bases.

A \emph{one-parameter }family of triples was found by Jaming \emph{et
al.} \citep{jaming+09}. It consists of the standard basis, the Fourier
family $F^{(2)}_{6}\equiv F_{6}(a,b)$, which can be written in dephased
form as
\begin{equation}
F^{(2)}_{6}=\left(\begin{array}{cccccc}
1 & 1 & 1 & 1 & 1 & 1\\
1 & -\omega^{2}x & \omega & -x & \omega^{2} & -\omega x\\
1 & \omega y & \omega^{2} & y & \omega & \omega^{2}y\\
1 & -1 & 1 & -1 & 1 & -1\\
1 & \omega^{2}x & \omega & x & \omega^{2} & \omega x\\
1 & -\omega y & \omega^{2} & -y & \omega & -\omega^{2}y
\end{array}\right),\label{eq: F^2_6}
\end{equation}
 with $x=e^{2\pi ia}$, $y=e^{2\pi ib}$ and $\omega=e^{2\pi i/3}$,
and the matrix $C(t)$ defined in Appendix A of \citep{jaming+09}.
The one-parameter family of three MU bases consists of the standard
basis combined with $F_{6}(0,b(t))$ and $C(t)$, where $\frac{1}{2}\arcsin\frac{\sqrt{5}}{3}\leq t\leq\frac{\pi}{2}-\frac{1}{2}\arcsin\frac{\sqrt{5}}{3}$.

A simple characterisation of the complex Hadamard matrices appearing
in the known families of MU triples can be found in \citep{matszangosz24}.
In particular, if at least three columns of a complex Hadamard matrix
of order six (in dephased form) contains a $-1$ entry, then it belongs
to the transposed Fourier family $(F^{(2)}_{6})^{T}$ or the Szöllősi
family $X^{(2)}_{6}$. Furthermore, these two families can be easily
distinguished by properties of their submatrices \citep{matszangosz24}.

In view of these constructions, together with the unextendibility
results summarised in Table~\ref{tab:triples}, the options to extend
pairs $\{\mathbb{I},H\}$ appear to be limited.

\subsection{Sets containing the Fourier family}

\label{sec:fourier_family}

Pairs of MU bases involving the Fourier family have been widely studied
in dimension six. As discussed in the preceding section, there exists
a one-parameter family of MU triples involving $F_{6}(0,b)$. Furthermore,
the existence of MU triples containing each member of $F_{6}(a,b)$
is supported by strong numerical evidence \citep{jaming+09}, but
no rigorous proof is known. Let us now attempt a chronological description
of \emph{non-existence results} related to the Fourier family, starting
with the cyclic $d$-roots problem, solved in 1991 by Björck and Fröberg
\citep{bjork+91}.

Searching for vectors mutually unbiased to $\{\mathbb{I},F_{6}\}$
is, according to Sec.~\ref{subsec:biunimodular} on biunimodular
sequences, equivalent to finding biunimodular vectors $x\in\mathbb{C}^{6}$.
In turn, finding these biunimodular sequences means solving the cyclic
$6$-roots problem defined by the set of equations (\ref{eq:cyclic_n_roots}),
with the added constraints $|z_{i}|=1$ for each $i$. Without the
unimodular condition on $z$, a computer-aided search found the full
set of 156 cyclic $6$-roots \citep{bjork+91}. In 1997, Haagerup
showed that 48 of these vectors are unimodular, including twelve classical
solutions of the type $(\alpha,\alpha^{3},\alpha^{5},\alpha^{7},\alpha^{9},\alpha^{11})$
where $\alpha$ is a twelfth root of unity \citep{haagerup97}. He
expressed the remaining 36 vectors as
\begin{equation}
z=\left(\omega^{k}z^{(0)}_{j-\ell}\right)_{j\in\mathbb{Z}_{6}}\,,\qquad k,\ell\in\mathbb{Z}_{6}\,,
\end{equation}
where $\omega=e^{2\pi i/6}$,
\begin{equation}
z^{(0)}=(1/a,i,ia,-ia,-i,-i/a)\,,\label{eq:cyclic-6-roots}
\end{equation}
and $a=\frac{1-\sqrt{3}}{2}+i\left(\frac{\sqrt{3}}{2}\right)^{1/2}$.
The corresponding biunimodular sequences take the form
\begin{equation}
x=c\left(\omega^{kj}x^{(0)}_{j-\ell}\right)_{j\in\mathbb{Z}_{6}}\,,\qquad k,\ell\in\mathbb{Z}_{6}\,,c\in\mathbb{T}\,,
\end{equation}
with $x^{(0)}_{j}=(1,i/a,-1/a,-i,-a,ia)$. In 2004, Grassl rediscovered
these 48 biunimodular sequences in the language of MU bases \citep{grassl04}.
Furthermore, he showed that the pair $\{\mathbb{I},F_{6}\}$ does
not extend to a complete set of MU bases.
\begin{thm}
\label{thm:grassl}There exist only 48 vectors mutually unbiased to
the pair $\{\mathbb{I},F_{6}\}$. One can arrange these vectors into
16 different orthonormal bases $\mathcal{B}_{k}$, $k=1\ldots16$,
to produce 16 MU triples $\{\mathbb{I},F_{6},B_{k}\}$, but no remaining
vector is mutually unbiased to any of them. 
\end{thm}
The proof involves solving a set of polynomial equations obtained
by expressing candidate vectors of $\mathbb{C}^{6}$ in the form 
\begin{equation}
\ket{\psi}=\frac{1}{\sqrt{6}}(1,x_{1}+ix_{6},x_{2}+ix_{7},\ldots,x_{5}+ix_{10})^{T},
\end{equation}
where the variables $x_{j}$ are real and $x^{2}_{j}+x^{2}_{j+5}=1$,
$j=1\ldots5$. By requiring that $\ket{\psi}$ is MU to the columns
of $F_{6}$, the computer algebra system MAGMA finds 48 real solutions
for the set of variables $x_{j}$, which are listed explicitly in
the updated preprint of \citep{grassl04}. Later, in 2012, an independent
analytic proof of the unextendibility of $\{\mathbb{I},F_{6}\}$ to
a complete set was found \citep{matolcsi+12}, based on Fourier analytic
techniques (see Sec.~\ref{subsec:linear-constraints}). 

Further analysis of the structure of these 48 vectors and the corresponding
16 orthonormal bases has been carried out in \citep{bengtsson+07}
and later in \citep{goyeneche13}. Of the 16 orthonormal bases, two
are Fourier matrices enphased with 12th roots of unity, two are equivalent
to $F^{T}(1/6,0)$, six are Björck matrices \citep{bjorck+95} and
six are Fourier matrices enphased with Björck's number $a$ defined
in Eq.~\eqref{eq:cyclic-6-roots}.

As described in \citep{goyeneche13}, the vectors group into three
sets, each corresponding to one orbit under the Heisenberg-Weyl group,
whose elements are the displacement operators $D_{jk}=\tau^{jk}X^{j}Z^{k}$
with $\tau=-e^{\pi i/d}$, $j,k=0,\ldots,\bar{d}-1$, expressed in
terms of the shift and phase operators $X,Z$ defined in Eq.~\eqref{HWoperators}.
The order of $\tau$ depends on the parity of $d$: we have $\bar{d}=d$
if $d$ is odd, and $\bar{d}=2d$ if $d$ is even. The three vectors
generating the orbits are $v_{1}=(1,i,\omega^{4},i,1,i\omega^{4})$,
$v_{2}=(1,-i,\omega^{2},-i,1,-i\omega^{2})$ and $v_{3}=(1,ia,a^{2},-ia^{2},-a,-i)$,
with $\omega=e^{2\pi i/6}$. The vectors $v_{1}$ and $v_{2}$ are
eigenvectors of $D_{\ell,\ell}$ and $D_{\ell,5\ell}$, respectively,
for every $\ell=0,\ldots,\bar{d}-1$, resulting in orbits of six elements
each; they are, in fact, product states after a a suitable permutation
of elements. Every eigenvector of $XZ$ is a member of the orbit generated
by $v_{1}$, while the second orbit is obtained from complex conjugation
of the elements of the first orbit. The vector $v_{3}$ is not an
eigenvector of any displacement operator and gives rise to an orbit
of 36 states. Shifting the elements of $v_{3}$ two places to the
right yields the vector $-ax^{(0)}_{j}$, in agreement with the biunimodular
sequences of Björck, Fröberg and Haagerup.

Jaming \emph{et al}. \citep{jaming+09} obtain unextendibility results
which apply to the entire two-parameter Fourier family. 
\begin{thm}
\label{thm:Fourierfamilyexclusion} The family of MU pairs $\{\mathbb{I},F_{6}(a,b)\}$
does not extend to a quadruple of MU bases. 
\end{thm}
The proof of Thm.~\ref{thm:Fourierfamilyexclusion} relies on a discretisation
scheme and a computational search similar to the one described in
Sec.~\ref{subsec:nonexistence_of_quadruples}, but the result is
rigorous due to exact bounds on the error terms. The search for candidate
MU vectors involves finding \emph{approximate} MU vectors by estimating
the phases of the vector components using $N$-th roots of unity.
Each vector component is evaluated at regular intervals of $2\pi j/N$,
with $j=1,\ldots,N$, and a computer-aided search calculates $N^{\nu}$
states, where $\nu$ denotes the number of free variables (phases)
for the candidate MU vectors. By choosing a sufficiently large positive
integer $N$, rigorous bounds of the errors given by the inner products
of the approximated states can be established. If the errors from
these approximated states are too large, no such MU vectors exist.
Importantly, this method can be generalised and could therefore lead
to a proof of the conjectured non-existence of complete sets in dimension
six (or any other composite dimension $d\notin\mathbb{PP}$), even
without an exhaustive classification of $6\times6$ complex Hadamard
matrices \citep{jaming+10}.

The computational search over the parameter values $(a,b)$ finds
the number of vectors MU to the pair $\{\mathbb{I},F_{6}(a,b)\}$
is 48. In most cases these 48 vectors produce 8 orthonormal bases
$C_{1}(a,b),\ldots,C_{8}(a,b)$, but in exceptional cases one can
construct additional orthonormal bases. For example, there exist 16
and 70 orthonormal bases when $(a,b)=(0,0)$ and $(a,b)=(1/6,0)$,
respectively.

A proof of Thm.~\ref{thm:Fourierfamilyexclusion}—not dependent on
a computer search—was later found by Matolcsi and Weiner \citep{matolcsi15}
by improving the Delsarte linear-programming bound (see Secs.~\ref{subsec:delsarte}
and \ref{subsec:linear-constraints}).

\subsection{MU product bases}

\label{subsec:product_bases}

We have discussed MU \emph{product} bases for multipartite systems
in Sec.~\ref{subsec:MU-product-bases} and found that their classification
is, in arbitrary composite dimensions, quite difficult. However, if
we limit ourselves to product bases in $\mathbb{C}^{2}\otimes\mathbb{C}^{3}$,
an exhaustive list of MU product bases can be given. What is more,
we can find tight upper bounds on maximal sets of MU product bases,
and show that any such set is strongly unextendible.

Let us denote product bases of the space $\mathbb{C}^{6}$ by 
\begin{equation}
\Bigl\{\ket{v^{1},v^{2}},\,v=0,\ldots,5\Bigr\}\,\subset\mathbb{C}^{2}\otimes\mathbb{C}^{3},\label{eq:product_basis_d=00003D6}
\end{equation}
with qubit states $\ket{v^{1}}\in\mathbb{C}^{2}$ and qutrit states
$\ket{v^{2}}\in\mathbb{C}^{3}$, respectively. It was shown in \citep{mcnulty+12a},
via a classification of all product bases of the form (\ref{eq:product_basis_d=00003D6}),
that a pair of bases $\mathcal{B}=\{\ket{v^{1},v^{2}}\}^{5}_{v=0}$
and $\mathcal{B}'=\{\ket{u^{1},u^{2}}\}^{5}_{u=0}$ is MU if and only
if $\ket{v^{1}}$ is MU to $\ket{u^{1}}$, and $\ket{v^{2}}$ is MU
to $\ket{u^{2}}$, for all $v,u=0,\ldots,5$. We note that this statement
has not been extended to arbitrary composite dimensions (cf. Sec.~\ref{subsec:MU-product-bases})
due to our inability to classify product bases is general. 

The unbiasedness condition on the states in $\mathbb{C}^{2}$ and
$\mathbb{C}^{3}$ is essential for the classification of all MU product\emph{
}bases in dimension six. In \citep{mcnulty+12a} an exhaustive list
of pairs and triples of MU product bases is constructed, up to local
`{}`equivalence'{}' transformations. We say that sets of MU product
bases are `{}`equivalent'{}' if we can transform one set into
another by some local (anti-) unitary transformations. These include,
for example, a local unitary transformation mapping one set to another,
permutations of states, and complex conjugation operations.

To express the exhaustive list we use complete sets of MU bases in
dimensions two and three. We denote the complete set in dimension
two by $\{\ket{v_{a}}\}$, $a=z,x,y$, where $v=0,1$, indicate the
orthonormal vectors which are eigenstates of the Pauli operators $\sigma_{a}$.
In dimension three, the eigenstates of the four Heisenberg-Weyl operators
$Z$, $X$, $Y=XZ$ and $W=XZ^{2}$ (defined in Appendix \ref{subsec:constructions_on_unitary_operator_bases})
form the four MU bases and are denoted by $\{\ket{V_{b}}\}$, $b=z,x,y,w$,
and $V=0,1,2$.

For example, consider a basis of $\mathbb{C}^{6}$ given by $\{\ket{0_{z},V_{z}},\ket{1_{z},V_{y}}\}$:
it consists of three states from the tensor product of $\ket{0_{z}}\in\mathbb{C}^{2}$
with the eigenstates of the Heisenberg-Weyl operator $Z$, and three
states resulting from tensoring the states $\ket{1_{z}}\in\mathbb{C}^{2}$
with the eigenstates of $Y$. The states $\ket{0_{z}}$ and $\ket{1_{z}}$
form the canonical basis in dimension two. This construction is an
example of an \emph{indirect product }basis: all the states are product
states of the bi-partite system but they are obtained by tensoring
the elements of \emph{three} bases rather than two, namely the standard
basis of the space $\mathbb{C}^{2}$ and a \emph{pair} of different
bases, $\{\ket{V_{y}}\}$ and $\{\ket{V_{z}}\}$, of $\mathbb{C}^{3}$. 

The complete list of MU product pairs comes in four different flavours
according to the following theorem.
\begin{thm}
\label{thm:All-pairs-d=00003D6} Any pair of MU product bases in the
space $\mathbb{C}^{2}\otimes\mathbb{C}^{3}$ is equivalent to a member
of the families 
\begin{align}
\mathcal{P}_{0} & =\{\ket{v_{z},V_{z}};\,\ket{v_{x},V_{x}}\}\,,\nonumber \\
\mathcal{P}_{1} & =\{\ket{v_{z},V_{z}};\,\ket{0_{x},V_{x}},\ket{1_{x},\hat{R}_{\xi,\eta}V_{x}}\}\,,\nonumber \\
\mathcal{P}_{2} & =\{\ket{0_{z},V_{z}},\ket{1_{z},V_{y}};\,\ket{0_{x},V_{x}},\ket{1_{x},V_{w}}\}\,,\nonumber \\
\mathcal{P}_{3} & =\{\ket{0_{z},V_{z}},\ket{1_{z},\hat{S}_{\zeta,\chi}V_{z}};\nonumber \\
 & \,\quad\quad\quad\,\,\ket{v_{x},0_{x}},\ket{\hat{r}_{\sigma}v_{x},1_{x}},\ket{\hat{r}_{\tau}v_{x},2_{x}}\}\,,
\end{align}
with $v=0,1$, and $V=0,1,2$. The unitary operator $\hat{R}_{\xi,\eta}$
is defined as $\hat{R}_{\xi,\eta}=\kb{0_{z}}{0_{z}}+e^{i\xi}\kb{1_{z}}{1_{z}}+e^{i\eta}\kb{2_{z}}{2_{z}}\,,$
for $\eta,\xi\in[0,2\pi)$, and $\hat{S}_{\zeta,\chi}$ is defined
analogously with respect to the $x$-basis; the unitary operators
$\hat{r}_{\sigma}$ and $\hat{r}_{\tau}$ act on the basis $\{\ket{v_{x}}\}\equiv\{\ket{\pm}\}$
according to $\hat{r}_{\sigma}\ket{v_{x}}=(\ket{0_{z}}\pm e^{i\sigma}\ket{1_{z}})/\sqrt{2}$
for $\sigma\in(0,\pi)$, etc. 
\end{thm}
The pairs $\mathcal{P}_{0}$ and $\mathcal{P}_{2}$ have no parameter
dependence, the pair $\mathcal{P}_{1}$ depends on two parameters,
while $\mathcal{P}_{3}$ is a four-parameter family. The ranges of
the parameters are assumed to be such that no MU product pair occurs
more than once in the list.

If we consider \emph{triples} of MU product bases, there exists only
one example in addition to the expected Heisenberg-Weyl triple:
\begin{thm}
\label{thm:two-triples-d=00003D6} Any triple of MU product bases
in the space $\mathbb{C}^{2}\otimes\mathbb{C}^{3}$ is equivalent
to either 
\begin{align}
\mathcal{T}_{0} & =\{\ket{v_{z},V_{z}};\,\ket{v_{x},V_{x}};\,\ket{v_{y},V_{y}}\}\,,\nonumber \\
\mbox{or }\mathcal{T}_{1} & =\{\ket{v_{z},V_{z}};\,\ket{v_{x},V_{x}};\,\ket{0_{y},V_{y}},\ket{1_{y},V_{w}}\}\,.\label{eq:product_triples}
\end{align}
\end{thm}
A consequence of exhaustively enumerating MU product bases in dimension
six is a bound on their number in a hypothetical complete set. It
is straightforward to see that no single \emph{product} state can
be MU to the two triples $\mathcal{T}_{0}$ and $\mathcal{T}_{1}$.
However, a stronger result is within reach: it is impossible to complement
either $\mathcal{T}_{0}$ or $\mathcal{T}_{1}$ by \emph{any} MU vector
\citep{mcnulty+12impossibility}. Thus, \emph{a complete set of MU
bases in dimension six cannot contain a product triple}. This is in
marked contrast to the prime-power dimension $p^{2}$ where a complete
set of MU bases includes $(p+1)$ MU product bases constructed from
the tensor products of Heisenberg-Weyl operators \citep{lawrence11}.

By further investigating the set of MU product \emph{pairs} given
in Thm.~\ref{thm:All-pairs-d=00003D6}, a stronger statement on the
non-existence of MU product bases can be derived \citep{mcnulty+12c}.
\begin{thm}
\label{thm:noproductbases} If a complete set of seven MU bases in
dimension six exists, it contains at most one product basis. 
\end{thm}
In other words, six of the seven MU bases contain entangled states.
The proof of Thm.~\ref{thm:noproductbases} relies on Theorems \ref{thm:Fourierfamilyexclusion},
\ref{thm:All-pairs-d=00003D6} and \ref{thm:unextendible_pairs}.
It is shown in \citep{mcnulty+12c} that the four pairs in Thm.~\ref{thm:All-pairs-d=00003D6}
are equivalent, under non-local equivalence transformations, to either
$\{\mathbb{I},S_{6}\}$ or $\{\mathbb{I},F_{6}(a,b)\}$. But since
neither pair extends to a complete set (Theorems \ref{thm:Fourierfamilyexclusion}
and \ref{thm:unextendible_pairs}), no pair of product bases appears
in a complete set. It is clear from the proof that the following result
is also true.
\begin{cor}
\label{cor:noproductbases}If a set of four MU bases in dimension
six exists, it contains at most one product basis.
\end{cor}
The refined concept of MU product \emph{constellations} (cf. Sec.~\ref{subsec:nonexistence_of_constellations})
rather than entire bases has been considered to find further limitations
on the number of product states in a complete set \citep{chen17}.
\begin{thm}
\label{thm:product_constellations}Let $H$ be a $6\times6$ Hadamard
matrix in which three of its columns are product vectors of $\mathbb{C}^{2}\otimes\mathbb{C}^{3}.$
Then the pair $\{\mathbb{I},H\}$ does not extend to a set of four
MU bases.
\end{thm}
The proof is based on considering all possible orthogonality relations
between the three product vectors which form the product columns of
$H$, and on showing that $H$ is equivalent to a matrix containing
a $3\times3$ submatrix proportional to a unitary. Then the following
lemma can be applied \citep{chen17}.
\begin{lem}
\label{thm:subunitary}Let $H$ be a $6\times6$ Hadamard matrix containing
a $3\times3$ submatrix $U,$ which is proportional to a unitary matrix.
Then the pair $\{\mathbb{I},H\}$ does not extend to a set of four
MU bases.
\end{lem}
To prove this result, one rewrites the matrix $H$ in block form as
\begin{equation}
H'=\left(\begin{array}{cc}
U & A\\
B & C
\end{array}\right)\,,\label{subunitary}
\end{equation}
where the rows and columns of $H$ have been permuted such that $U$
is now in the top left block of $H'$. One can easily check that if
$H'$ is unitary, then both $\sqrt{2}U$ and $\sqrt{2}B$ are unitaries.
Thus, the pair $\{\mathbb{I},H'\}$ can be mapped to
\begin{equation}
\left\{ \sqrt{2}\left(\begin{array}{cc}
U^{\dagger} & 0\\
0 & B^{\dagger}
\end{array}\right),\sqrt{2}\left(\begin{array}{cc}
\mathbb{I}/2 & U^{\dagger}A\\
\mathbb{I}/2 & B^{\dagger}C
\end{array}\right)\right\} \,,\label{subunitary-map}
\end{equation}
and, since the latter matrix is unitary, we also have $U^{\dagger}A=-B^{\dagger}C$.
Therefore, the columns of both matrices are product vectors in the
space $\mathbb{C}^{2}\otimes\mathbb{C}^{3}.$ Corollary \ref{cor:noproductbases}
is then applied to show that these bases cannot exist in a set of
four MU bases.

We note that a result similar to Thm.~\ref{thm:product_constellations}
has been derived in \citep{chen18} which places a limit on the number
of product states in a set of four MU bases in $\mathbb{C}^{2}\otimes\mathbb{C}^{3}$,
by assuming that the first basis is an arbitrary product basis rather
than the canonical one.

Finally, let us mention an unextendibility result for Hadamard matrices
of order six and low Schmidt rank. Given an operator $P$ acting on
the space $\mathbb{C}^{d_{1}}\otimes\mathbb{C}^{d_{2}}$, its operator-Schmidt
decomposition reads $P=\sum_{j}s_{j}A_{j}\otimes B_{j},$ where $A_{j}$
and $B_{j}$ are operators acting on $\mathbb{C}^{d_{1}}$ and $\mathbb{C}^{d_{2}}$
respectively, and $s_{j}\geq0.$ The \emph{Schmidt rank} of $P$ is
the number of non-zero coefficients $s_{j}$ in this decomposition.
Investigating this quantity for Hadamard matrices of order six leads
to the following result \citep{chen17}.
\begin{thm}
\label{thm:schmidt_rank}Let H be a $6\times6$ Hadamard matrix with
Schmidt rank $r\leq2.$ Then the pair $\{\mathbb{I},H\}$ does not
extend to a set of four MU bases.
\end{thm}
The proof relies on an exhaustive classification of $6\times6$ Hadamard
matrices with Schmidt rank one and two. Any pair $\{\mathbb{I},H\}$,
where $H$ has Schmidt rank at most two, is equivalent to a pair of
MU product bases. Since no set of four MU bases contains a pair of
product bases (see Corollary \ref{cor:noproductbases}) the proof
is complete.

Although a classification of Hadamard matrices with Schmidt rank three
is known \citep{hu20}, a statement analogous to Thm. \ref{thm:schmidt_rank}
is not. Further properties of such matrices are discussed in \citep{chen22}.

\subsection{Group theory of complete sets\label{subsec:groups}}

Some restrictions on the structure of a hypothetical complete set
are known which relate to group theory. We have seen in Sec.~\ref{subsec:Nice-error-bases}
that no complete set in dimension six exists that consists entirely
of nice MU bases. In particular, if seven MU bases exist, they cannot
be constructed from partitioning a nice error basis into maximally
commuting classes. Instead, the MU bases must come from a unitary
operator basis without the appropriate group structure (cf. Thm.~\ref{thm:niceerrorbasis}).

Another restriction arises from the connection between MU bases and
projective toric 2-designs. As described in Sec.~\ref{subsec: 2-design equivalence},
a set of $d$ complex Hadamard matrices of order $d$ is mutually
unbiased if and only if the columns form a projective toric 2-design
of $P(T^{d})$. For $d=6$, an exhaustive computational search confirmed
that such a design, formed from $d$ Hadamard matrices, cannot be
a subgroup of $P(T^{6})$ \citep{iosue23}.
\begin{thm}
\label{thm:toric_designs}A set of six Hadamard matrices of order
six are pairwise mutually unbiased if and only if their columns form
a projective toric 2-design which is not a subgroup of $P(T^{6})$.
\end{thm}
This is markedly distinct from prime and prime-power dimensions, where
all (known) complete sets relate to projective toric 2-designs that
are groups \citep{iosue23}. If, as predicted, a complete set for
$d=6$ does not exist, Thm.~\ref{thm:toric_designs} implies that
toric non-group designs do not exist. More generally, it is conjectured
that for any $d$, complete sets of MU bases originate only from group
projective toric 2-designs \citep{iosue23}.

\subsection{Unextendible MU bases}
\label{subsec: unextendible MU bases}

\begin{table*}[t]
\centering
\resizebox{\textwidth}{!}{%
\begin{tabular}{|c|c|c||c|c||c|c|}
\hline
$\{\mathbb{I},\cdot\}$ & type & defined in & triples exists? & source & no quadruple & source\tabularnewline
\hline
\hline
$S_{6}$ & isolated & \citep{tao04} & no & \citep{brierley+09} & $\checkmark$ & implied\tabularnewline
$F_{6}^{(2)}$ & affine & \citep{Kraus1987} & $(\checkmark)$ & \citep{jaming+09} & $\checkmark$ & \citep{jaming+09,matolcsi15}\tabularnewline
$D_{6}^{(1)}$ & affine & \citep{dita04} & $\checkmark$ & \citep{szollosi10} & ($\checkmark$) & \citep{brierley+09}\tabularnewline
$B_{6}^{(1)}$ & not affine & \citep{beauchamp+08} & $\checkmark$ & \citep{szollosi10} & ? & \tabularnewline
$M_{6}^{(1)}$ & not affine & \citep{matolcsi+08} & (some) & \citep{brierley+09,goyeneche13} & ? & \tabularnewline
$X_{6}^{(2)}$ & not affine & \citep{szollosi10} & $\checkmark$ & \citep{szollosi10} & ? & \tabularnewline
$K_{6}^{(2)}$ & not affine & \citep{karlsson09} & (some) & \citep{goyeneche13} & ($\checkmark$) & \citep{goyeneche13,Matolcsi24}\tabularnewline
$K_{6}^{(3)}$ & not affine & \citep{karlsson11} & (some) & \citep{goyeneche13} & ($\checkmark$) & \citep{goyeneche13,Matolcsi24}\tabularnewline
$G_{6}^{(4)}$ & not affine & \citep{bondal16,szollosi12} & ? &  & ? & \tabularnewline
\hline
\end{tabular}%
}
\caption{\label{tab:triples}Extendability of pairs of MU bases $\{\mathbb{I},\cdot\}$
in dimension six to triples and quadruples, collecting results described
in Secs.~\ref{sec:triples_of_MU_bases}, \ref{subsec: unextendible MU bases}
and \ref{subsec:nonexistence_of_triples}. Checkmarks ``$\checkmark$''
either confirm that the pairs in question \emph{do} extend to an MU
triple, or that they \emph{do not} extend to quadruples; brackets
``($\cdot$)'' indicate the statement is based on numerical evidence.
The rigorous (non-) existence proofs of \textit{triples} follow from
Thms.~\ref{thm:ferenc_triples} and \ref{thm:unextendible_pairs},
while the non-existence result of \textit{quadruples} is a consequence
of Thm.~\ref{thm:Fourierfamilyexclusion}. Numerical evidence in
\citep{jaming+09} supports the existence of a triple containing
$\{\mathbb{I},F_{6}(a,b)\}$ for all $a$ and $b$, but a rigorous proof
is only known for the pairs $\{\mathbb{I},F_{6}(0,b)\}$
(cf. Sec.~\ref{sec:triples_of_MU_bases}).}
\end{table*}

According to Thm.~\ref{thm:Fourierfamilyexclusion}, any triple of
MU bases that includes the Fourier family $F_{6}(a,b)$ is \emph{unextendible},
as is any triple which contains multiple product bases (Corollary
\ref{cor:noproductbases}). In this section, we summarise further
results on (un-) extendibility of pairs of MU bases valid in dimension
$d=6$ (cf. Table~\ref{tab:triples}). The table also contains numerical
results on unextendibility discussed in Sec.~\ref{sec: Numerical_results},
including evidence that certain families of MU pairs do not extend
to triples (Sec.~\ref{subsec:nonexistence_of_triples}) or quadruples
(Sec.~\ref{subsec:nonexistence_of_quadruples}).

Generalising Grassl's computer-algebraic calculation for the Heisenberg-Weyl
pair (Thm.~\ref{thm:grassl}) to pairs of the form $\{\mathbb{I},H\}$,
with a $6\times6$ complex Hadamard matrix $H$, was considered in
Ref.~\citep{brierley+09}. Using Buchberger's algorithm \citep{buchberger76},
the original set of polynomials was mapped to another, easier to solve
`{}`tri-diagonal'{}' set of coupled polynomial equations, the
so-called Gröbner basis. In this way, all vectors mutually unbiased
to a particular pair $\{\mathbb{I},H\}$ were constructed.

The number of vectors MU to $\{\mathbb{I},H\}$, where $H$ is the
Diţă-matrix $D_{6}(0)$, the circulant matrix $C_{6}$ \citep{bjorck+95}
(which is a member of the self-adjoint family $B^{(1)}_{6}$) and
the isolated matrix $S_{6}$ equal 120, 56 and 90, respectively. Ten
triples exist containing the pair $\{\mathbb{I},D_{6}(0)\}$, but
none extend to four MU bases. Similarly, no four MU bases exist containing
$\{\mathbb{I},C_{6}\}$, and no orthonormal basis can be constructed
from the 90 vectors MU to $\{\mathbb{I},S_{6}\}$.
\begin{thm}
\label{thm:unextendible_pairs}The pairs $\{\mathbb{I},C_{6}\}$ and
$\{\mathbb{I},D_{6}(0)\}$, which have an additional 56 and 120 unbiased
vectors respectively, do not extend to four MU bases. The pair $\{\mathbb{I},S_{6}\}$
has 90 unbiased vectors and does not extend to an MU triple.
\end{thm}
The calculations were also carried out for regularly spaced values
of the parameter $x$ in the Diţă family $D_{6}(x)$ and a grid of
points of the two-parameter Fourier family $F_{6}(a,b)$. For the
pair $\{\mathbb{I},D_{6}(x)\}$, the number of MU vectors appears
to be piecewise constant, dropping from 120 to 72 and then to 48 at
the end points of the parameter range. For the Fourier family, there
exist 48 MU vectors at each of the tested parameter values. In both
cases, no set of four MU bases can be formed. This result was later
shown to be valid for all members of the Fourier family, as stated
in Thm.~\ref{thm:Fourierfamilyexclusion}.

While these results provide rigorous limits on the number of vectors
MU to pairs $\{\mathbb{I},H\}$, no definite results were obtained
for non-affine complex Hadamard matrices $H$. For the symmetric,
Hermitian and Szöllősi non-affine families, i.e. $M^{(1)}_{6}$, $B^{(1)}_{6}$
and $X^{(2)}_{6}$, the available computational memory was insufficient
to produce the relevant Gröbner basis, making certain approximations
necessary. Thus, no rigorous conclusion regarding the existence of
a fourth MU basis containing these families could be drawn.

In Hadamard form, all known complete sets of MU bases (cf. Appendix
\ref{sec: complete sets in pp dimensions}) are of Butson-type, with
matrix elements consisting of either $d$ or $2d$ roots of unity
depending on whether the dimension of the Hilbert space $d$ is odd
or even, respectively. Motivated by this observation, a search was
carried out in Ref. \citep{bengtsson+07} for mutually unbiased Butson-type
Hadamard matrices $B(6,12)$ (cf. Sec.~\ref{subsec:MU-butson-type}),
whose entries consist only of twelfth roots of unity. The search confirmed
that a Butson-type matrix $B(6,12)$ does not appear in any MU quadruple.
\begin{thm}
\label{thm:roots_of_unity}If a set of four MU bases in dimension
six exists, it contains no Butson-type Hadamard matrix $BH(6,12)$.
\end{thm}
Non-existence theorems of this type, i.e. for specific \emph{quadruples}
of MU bases, are of considerable interest since they would help rule
out complete sets. A series of papers \citep{liang19,liang19b,chen21}
leads to the statement that no set of four MU bases in dimension six
contains a pair of type $\{\mathbb{I},D^{(1)}_{6}\}$, $\{\mathbb{I},B^{(1)}_{6}\}$,
$\{\mathbb{I},M^{(1)}_{6}\}$, or $\{\mathbb{I},X^{(2)}_{6}\}$. Unfortunately,
the derivation of this result builds on a lemma in Ref.~\citep{chen17}
which is erroneous \citep{mcnulty24}. Thus, numerical evidence (cf.
Sec.~\ref{subsec:nonexistence_of_triples}) remains the main argument
against the presence of these families in quadruples of MU bases,
as shown in Table~\ref{tab:triples}.

\section{Numerical results: Dimension six}

\label{sec: Numerical_results}

This section reports numerical evidence which overwhelmingly points
towards the non-existence of four mutually unbiased bases in dimension
six. Even the existence of a single vector mutually unbiased to any
triple of MU bases seems unlikely.

\subsection{Triples of MU bases}

\label{subsec:nonexistence_of_triples}

According to Thm.~\ref{thm:unextendible_pairs}, the pair $\{\mathbb{I},S_{6}\}$,
defined by the \emph{isolated }Hadamard matrix of Eq.~(\ref{eq:isolatedmatrix}),
does not extend to an MU triple. We now describe numerical evidence
(see Table~\ref{tab:triples} in Sec.~\ref{subsec: unextendible MU bases})
suggesting that many other pairs of MU bases do not extend to triples.

Goyeneche \citep{goyeneche13} carried out a comprehensive computational
search for MU triples based on an iterative procedure which constructs
approximations of vectors MU to pairs $\{\mathbb{I},H\}$, where $H$
is a $6\times6$ complex Hadamard matrix. The solutions (MU vectors)
are attractive fixed points of a ``physical imposition operator''.
The operator is used to transform a state $\ket{\Psi}$, chosen at
random, into $\ket{\Psi'}$, which solves the set of equations $|\bk{\psi_{i}}{\Psi'}|=|\bk{\phi_{j}}{\Psi'}|=1/d$,
where $\ket{\psi_{i}}$ and $\ket{\phi_{j}}$ are vectors from the
MU pair $\{\mathbb{I},H\}$. 

The method successfully confirmed the results of several previously
studied cases (e.g. Thms.~\ref{thm:ferenc_triples} and \ref{thm:unextendible_pairs}).
It outputs, for example, all 90 vectors MU to the pair $\{\mathbb{I},S_{6}\}$,
all 48 vectors MU to the pair $\{\mathbb{I},F_{6}\}$, and a triple
containing $\{\mathbb{I},F_{6}(a,b)\}$ for all sampled $a$ and $b$.
It does not find a single vector MU to a triple containing the pair
$\{\mathbb{I},F_{6}(a,b)\}$. Furthermore, the iterative method confirms
the existence of an MU triple for \emph{each} sampled element of the
non-affine family $B^{(1)}_{6}$, in agreement with the search in
\citep{brierley+09} and Thm.~\ref{thm:ferenc_triples}. The theorem
confirms that any matrix from $D^{(1)}_{6}$ or $B^{(1)}_{6}$ is
contained in an MU triple because both one-parameter families are
subsets of $X^{(2)}_{6}$.

The search \citep{goyeneche13} finds MU triples containing a member
of the non-affine family $M^{(1)}_{6}\equiv M_{6}(t)$ for some but
not all values of the parameter space. In particular, pairs $\{\mathbb{I},M_{6}(t)\}$
were not found to extend to triples in small regions around the values
$t=\pi/2,\pi,3\pi/2,2\pi$. ''Approximate'' triples containing $M^{(1)}_{6}$
were found in an earlier search \citep{brierley+09}.

For the Karlsson families $K^{(2)}_{6}$ and $K^{(3)}_{6}$, the search
identified triples in only a limited region of the parameter space.
An analysis of the numerical results reveals that certain `{}`symmetries'{}'
exist within these families. The Fourier matrix $F_{6}$ and the Diţă
family $D^{(1)}_{6}$ appear as centres of symmetry in the parameter
space of $K^{(2)}_{6}$. Reflection symmetries are observed in $K^{(3)}_{6}$,
as illustrated pictorially in Ref.~\citep{goyeneche13}.

An independent numerical search provides further evidence of widespread
unextendibility among pairs of MU bases \citep{Matolcsi24}. The idea
is to search for three unitary matrices $U_{1},U_{2},$ and $U_{3}$
such that their transition matrices $U^{\dagger}_{i}U_{j}$ are Hadamard
matrices, ensuring that the triple gives rise to three MU bases. The
numerical evidence suggests that the transition matrices must belong
to the Fourier family $F^{(2)}_{6}$, its transpose, or the Szöllősi
family $X^{(2)}_{6}$, leading to a restrictive conjecture concerning
the existence of MU triples.
\begin{conjecture}
\label{conj:matolcsi} In dimension six, a pair \textup{$\{\mathbb{I},H\}$}
can be extended to an MU triple if and only if $H$ is equivalent
to a member of the families of Hadamard matrices $F^{(2)}_{6}$, $(F^{(2)}_{6})^{T}$
or $X^{(2)}_{6}$.
\end{conjecture}
If true, the construction from Thm.~\ref{thm:ferenc_triples} is
likely to describe all MU triples \citep{Matolcsi24}. It is important
to realise that a proof of Conjecture~\ref{conj:matolcsi} would
be a major step towards confirming Zauner's conjecture that a set
of four MU bases does not exist. Since Thm.~\ref{thm:Fourierfamilyexclusion}
states that an MU quadruple cannot contain any member of the families
$F^{(2)}_{6}$or $(F^{(2)}_{6})^{T}$, it would be sufficient to prove
that no member of $X^{(2)}_{6}$ is contained in a quadruple.

The findings that predict Conjecture~\ref{conj:matolcsi} are consistent
with the evidence in Table~\ref{tab:triples} due to known relations
between families of Hadamard matrices discussed in Sec.~\ref{sec:pairs_of_MU_bases_C^6},
such as the inclusions $D^{(1)}_{6}\subset X^{(2)}_{6}$ and $B^{(1)}_{6}\subset X^{(2)}_{6}$.
In addition, it seems necessary that the family $K^{(2)}_{6}$ must
overlap in yet unknown ways with $X^{(2)}_{6}$; in particular one
expects at least a subset of $M^{(1)}_{6}$ to be contained in the
set $X^{(2)}_{6}$. The numerical evidence resulting in Conjecture~\ref{conj:matolcsi}
also rules out the possibility that \emph{all} members of the three-parameter
families $K^{(3)}_{6}$ and $G^{(4)}_{6}$ figure in an MU triple.

\subsection{Non-existence of MU quadruples}

\label{subsec:nonexistence_of_quadruples}

The search for MU bases can be recast as an optimisation problem of
a function over a set of orthonormal bases whose maximum (or minimum)
is achieved if and only if the bases are mutually unbiased (see Sec.~\ref{subsec:global_min}).
A slightly modified version of the function considered in Eq.~\eqref{eq: F encoding complete sets}
was derived in \citep{bengtsson+07} as a measure of the squared distance
between pairs of orthonormal bases,
\begin{equation}
D_{b,b'}=1-\frac{1}{d-1}\sum^{d-1}_{v',v=0}\left(|\bk{v_{b}}{v'_{b'}}|^{2}-\frac{1}{d}\right)^{2},
\end{equation}
which takes the maximum value of $D_{b,b'}=1$ if and only if the
bases \textbf{$\mathcal{B}_{b}$ }and $\mathcal{B}_{b'}$ are mutually
unbiased. This measure of ``unbiasedness'' is quite natural when
viewing the basis vectors of the space $\mathbb{C}^{d}$ as density
matrices in a real vector space of dimension $d^{2}-1$. In this representation,
a $d$-dimensional Hilbert space spans a $(d-1)$-plane in a real
vector space of dimension $d^{2}-1$, and two MU bases correspond
to two orthogonal $(d-1)$-planes.

The degree of unbiasedness for a set of $\mu$ bases can be quantified
by the average squared distance between all pairs of bases, i.e., 
\begin{equation}
\overline{D}_{\mu}=\frac{2}{\mu(\mu-1)}\sum_{0\leq b<b'\leq\mu-1}D_{b,b'}\,,\label{eq:ADS}
\end{equation}
which achieves the value $\overline{D}_{\mu}=1$ if and only if all
$\mu$ bases $\{\mathcal{B}_{0},\ldots,\mathcal{B}_{\mu-1}\}$ are
mutually unbiased. In \citep{butterley+07}, a numerical search for
the minimum of the expression $1-\overline{D}_{\mu}$ was carried
out for dimensions $d\leq7$, producing strong evidence in support
of the conjecture that only three MU bases exist in the space $\mathbb{\mathbb{C}}^{6}$.
The minimisation, which is a non-linear least squares problem, was
approached by means of the Levenberg-Marquadt algorithm. While this
finds only \emph{local} minima, the algorithm was run repeatedly from
different starting points, in an attempt to identify the global minimum.
In all dimension below $d=6$, almost all test runs (i.e. $\geq99.98\%$)
converge to $\overline{D}_{d+1}=1$, implying the existence of a complete
set of MU bases. However, in dimension $d=6,$ searching over sets
of four and seven orthonormal bases found maximal values of only $\overline{D}_{4}=0.9982917$
and $\overline{D}_{7}=0.9849098$, respectively. In the case $d=7,$
complete sets of MU bases were found, but only in 3 of the 250 test
runs.

Further analysis in \citep{raynal+11}, using the steepest ascent
method to maximise the average distance square function of Eq.~\eqref{eq:ADS},
also supports the earlier numerical result that only three MU bases
exist in dimension six. Furthermore, a set of four ``most distant''
bases with $\overline{D}_{4}=0.9982917$ is identified. In particular,
a two-parameter family of three orthonormal bases is explicitly derived
which, together with the canonical basis, achieve the numerically
determined maximum of $\overline{D}_{4}$ for certain parameter values.
Of the four bases, three are equidistant and the remaining basis is
mutually unbiased to the others. Thus, the set can be written as the
identity matrix together with three complex Hadamard matrices containing
two parameters. All three Hadamard matrices have the same determinant
and are members of the transposed Fourier family. For their explicit
parameterisation we refer the reader to the original paper \citep{raynal+11}.

A relation between MU bases and the average success probability of
quantum random access codes (QRACs), as described in Equivalence \ref{equiv:QRACs}
of Sec.~\ref{subsec:QRAC}, provides an alternative way to measure
the unbiasedness of a set of $\mu$ orthonormal bases \citep{aguilar18}.
For $\mu$ bases $\mathcal{B}_{b}=\{\ket{v_{b}}\}^{d-1}_{v=0}$ in
dimension $d,$ the maximum attainable average success probability
of a $(\mu,2)^{d}\rightarrow1$ p-QRAC is given by
\begin{equation}
\overline{P}_{\mu}=\frac{2}{\mu(\mu-1)}\!\!\sum_{\{b,b'\}\in S_{2}}\!\!\left(\frac{1}{2}+\frac{1}{2d^{2}}\!\sum^{d-1}_{v,v'=0}\!|\bk{v_{b}}{v'_{b'}}|\right)\!,
\end{equation}
where the bracketed quantity is averaged over all pairs of bases $\mathcal{B}_{b}$
and $\mathcal{B}_{b'}$ from the set of $\mu$ bases, and $S_{2}$
is the set of all possible subsets of $\{1,\ldots,\mu\}$ of size
2. As shown in Sec.~\ref{subsec:QRAC}, this saturates the upper
bound in Eq.~(\ref{eq:QRACsucessprob}) if and only if the bases
are mutually unbiased. Normalising this function provides a new measure
of unbiasedness such that the upper bound of one is reached for MU
bases and the zero lower bound when the bases are identical.

While semidefinite programming techniques may be used to find this
upper bound—and possibly prove non-existence by finding an upper bound
smaller than Eq.~\eqref{eq:QRACsucessprob}—so far only indirect
numerical calculations have been applied \citep{aguilar18}. In particular,
the $(4,2)^{6}\rightarrow1$ protocol is optimised using the see-saw
method, involving an iterative process which alternates between optimising
Alice's state and Bob's measurement to find the global maximum. Using
these techniques the $(4,2)^{6}\rightarrow1$ protocol yields a maximum
guessing probability of $\overline{P}_{4}=0.703888$, with an average
squared distance $\overline{D}_{4}=0.9982839.$ This falls just short
of the optimal value of $\overline{D}_{4}=0.9982917$ calculated in
\citep{butterley+07}. If one applies the $(4,2)^{6}\rightarrow1$
pQRAC to the four bases which are maximally unbiased with respect
to $\overline{D}$, the protocol has an optimal average success probability
of $\overline{P}_{4}=0.703887$. This implies that the two measures
of unbiasedness are inequivalent and have different partial orderings
on the set of $\mu$ bases.

Another numerical search relies on Bell inequalities (see Sec.~\ref{subsec:Quantum-correlations})
which are maximally violated if and only if a set of $\mu$ MU bases
exists in $\mathbb{C}^{d}$ \citep{colomer22}. In analogy with the
methods described above, i.e. quantifying unbiasedness and QRACs,
the search can be recast as an optimisation of the function that gives
rise to the Bell inequality. Numerical approaches to the optimisation
problem included a see-saw optimisation, a non-linear SDP and Monte
Carlo techniques. Each approach failed to find a fourth MU basis in
dimension six, and the bases which optimised the function are close
to the four ``most distant'' bases given in \citep{raynal+11}.
The Monte Carlo approach also suggests that no fourth basis exists
in dimension ten.

\subsection{Non-existence of MU constellations}

\label{subsec:nonexistence_of_constellations}

Extensive numerical searches for mutually unbiased \emph{constellations},
fleetingly mentioned in Sec.~\ref{subsec:product_bases}, were carried
out in Ref.~\citep{brierley+08}. The findings provide further substantial
evidence that any three MU bases in dimension six are \emph{strongly}
unextendible (cf. Sec.~\ref{subsec:Unextendible-MU-bases}): not
even a single vector is MU to three MU bases. 

A mutually unbiased\emph{ constellation} is a set of vectors, partitioned
into sets of orthonormal states, such that the states within each
set are MU to all others not in the set. Constellations are denoted
by $\{x_{1},\ldots,x_{n}\}_{d}$, where $x_{j}$ is the number of
orthonormal vectors in the $j$-th set, and $d$ is the dimension
of the vector space. The constellation $\{6,6,4\}_{6}$, for example,
contains two sets of six orthogonal states, i.e. two MU bases, and
a set of four orthogonal states MU to all members of the first two
bases. We will abbreviate this notation to $\{6^{2},4\}_{6}$ where
$6^{2}$ denotes the two sets of six orthonormal states. Since $(d-1)$
orthonormal states already determine an orthonormal basis of the space
$\mathbb{C}^{d}$ uniquely, one can leave out one vector from each
basis, hence denoting the MU constellation $\{6^{2},4\}_{6}$ by $\{5^{2},4\}_{6}$
instead. 

The idea underlying a search for constellations is simple. If a complete
set of seven MU bases exists, any MU constellation also exists by
suppressing a suitable set of states. Hence, the non-existence of
\emph{any} MU constellation between a triple and a complete set implies
the non-existence of any larger constellation, including the complete
set. The obvious advantage of using constellations rather than entire
sets of MU bases is a substantial reduction in the number of free
parameters. 

\begin{table*}[t]
\begin{centering}
\begin{tabular}{c|ccccc|r@{\extracolsep{0pt}.}lr@{\extracolsep{0pt}.}lr@{\extracolsep{0pt}.}lr@{\extracolsep{0pt}.}lr@{\extracolsep{0pt}.}l}
\hline 
$d=6$ & \multicolumn{5}{c||}{%
\mbox{%
Parameters $p_{\mathcal{C}}$%
}} & \multicolumn{10}{c}{%
\mbox{%
Success rate%
}}\tabularnewline
\hline 
$x,y$ & \multicolumn{5}{c||}{$z$} & \multicolumn{10}{c}{$z$}\tabularnewline
 & 1 & 2 & 3 & 4 & 5 & \multicolumn{2}{c}{1} & \multicolumn{2}{c}{2} & \multicolumn{2}{c}{3} & \multicolumn{2}{c}{4} & \multicolumn{2}{c}{5}\tabularnewline
\hline 
\hline 
1,1 & 10 &  &  &  &  & 100&00 & \multicolumn{2}{c}{} & \multicolumn{2}{c}{} & \multicolumn{2}{c}{} & \multicolumn{2}{c}{}\tabularnewline
\hline 
2,1 & 15 &  &  &  &  & 100&00 & \multicolumn{2}{c}{} & \multicolumn{2}{c}{} & \multicolumn{2}{c}{} & \multicolumn{2}{c}{}\tabularnewline
2,2 & \textbf{20} & 25 &  &  &  & 100&00 & 100&00 & \multicolumn{2}{c}{} & \multicolumn{2}{c}{} & \multicolumn{2}{c}{}\tabularnewline
\hline 
3,1 & \textbf{20} &  &  &  &  & 100&00 & \multicolumn{2}{c}{} & \multicolumn{2}{c}{} & \multicolumn{2}{c}{} & \multicolumn{2}{c}{}\tabularnewline
3,2 & 25 & 30 &  &  &  & 99&95 & 100&00 & \multicolumn{2}{c}{} & \multicolumn{2}{c}{} & \multicolumn{2}{c}{}\tabularnewline
3,3 & 30 & 35 & 40 &  &  & 99&42 & 39&03 & 0&00 & \multicolumn{2}{c}{} & \multicolumn{2}{c}{}\tabularnewline
\hline 
4,1 & 25 &  &  &  &  & 100&00 & \multicolumn{2}{c}{} & \multicolumn{2}{c}{} & \multicolumn{2}{c}{} & \multicolumn{2}{c}{}\tabularnewline
4,2 & 30 & 35 &  &  &  & 92&92 & 44&84 & \multicolumn{2}{c}{} & \multicolumn{2}{c}{} & \multicolumn{2}{c}{}\tabularnewline
4,3 & 35 & 40 & 45 &  &  & 12&97 & 0&00 & 0&00 & \multicolumn{2}{c}{} & \multicolumn{2}{c}{}\tabularnewline
4,4 & 40 & 45 & 50 & 55 &  & 0&74 & 0&00 & 0&00 & 0&00 & \multicolumn{2}{c}{}\tabularnewline
\hline 
5,1 & 30 &  &  &  &  & 95&40 & \multicolumn{2}{c}{} & \multicolumn{2}{c}{} & \multicolumn{2}{c}{} & \multicolumn{2}{c}{}\tabularnewline
5,2 & 35 & 40 &  &  &  & 76&71 & 10&96 & \multicolumn{2}{c}{} & \multicolumn{2}{c}{} & \multicolumn{2}{c}{}\tabularnewline
5,3 & 40 & 45 & 50 &  &  & 1&47 & 0&00 & 0&00 & \multicolumn{2}{c}{} & \multicolumn{2}{c}{}\tabularnewline
5,4 & 45 & 50 & 55 & 60 &  & 0&00 & 0&00 & 0&00 & 0&00 & \multicolumn{2}{c}{}\tabularnewline
5,5 & 50 & 55 & 60 & 65 & 70 & 0&00 & 0&00 & 0&00 & 0&00 & 0&00\tabularnewline
\hline 
\end{tabular}
\par\end{centering}
\caption{Success rates for searches of MU constellations $\mathcal{C}=\{5,x,y,z\}_{6}$
in dimension six, each based on 10,000 randomly chosen initial points.
The number of free parameters $p_{\mathcal{C}}$ equals the number
of constraints only for the two constellations $\left\{ 5,2^{2},1\right\} _{6}$
and $\left\{ 5,3,1^{2}\right\} _{6}$ shown in bold; all larger constellations
are overdetermined. (Reproduced with permission from Ref.~\citep{brierley+08},
correcting an erroneous count of the number of constraints pointed
out by M. Matolcsi.) \label{tab:Success-rates-d=00003D6}}
\label{dim6}
\end{table*}

When searching for MU constellations, it is useful to write them in
a form containing as few free parameters as possible. Exploiting the
dephased form of the associated Hadamard bases, one finds that all
candidates for a pair of MU bases $\{(d-1)^{2}\}_{d}$ in dimension
$d$, for example, depend on $p_{2}(d)=(d-2)(d-1)$ real phases. The
first Hadamard matrix can be taken as the identity, and dephasing
the second matrix leaves $(d-2)$ column vectors with $(d-1)$ free
phase factors each (note that one of the columns can be suppressed
since $(d-1)$ orthonormal vectors in $\mathbb{C}^{d}$ fix the last
vector). For larger constellations, every additional Hadamard matrix
brings another $(d-1)^{2}$ free phases which means that the candidates
$\{(d-1)^{\mu}\}_{d}$, $\mu\in\left\{ 2\ldots d+1\right\} $, for
MU constellations depend on $p_{\mu}(d)=(d-1)((\mu-1)(d-1)-1)$ free
parameters. Similar arguments lead to the numbers $p_{\mathcal{C}}$
of parameters for the constellations $\mathcal{C}=\{5,x,y,z\}_{6}$
listed in Table \ref{dim6}. 

The numerical searches for MU constellations given in \citep{brierley+08}
use the same optimisation approach as \citep{butterley+07}. They
were carried out for subsets of four bases of the form $\left\{ d-1,x,y,z\right\} _{d}$,
with $d=5,6,7$, and $x,y,z\in[0,d-1]$. In dimension $d=5$ \emph{all
}of these constellations were identified with significant success
rates. For dimension $d=7$, only 46 out of 56 constellations were
found directly, although with small success rates; however, \emph{all
}constellations were found \emph{indirectly }(the existence of the
MU constellation $\{6,4^{2},2\}_{7}$, for example, was inferred from
identifying a larger one such as $\{6^{2},5,2\}_{7}$ containing it). 

The success rates (out of 10000 runs for each case) to find constellations
$\{5,x,y,z\}_{6}$ in the space $\mathbb{C}^{6}$ are reproduced in
Table \ref{tab:Success-rates-d=00003D6}. The \emph{largest} MU constellations
found were $\{5,4^{2},1\}_{6}$ and $\{5^{2},3,1\}_{6}$, consisting
of 15 and 16 states, respectively. The \emph{smallest} MU constellations
which the search failed to find were $\{5,3^{3}\}_{6}$ and $\{5,4,3,2\}_{6}$
each containing 15 states. The 19-state MU constellation $\{5^{3},1\}_{6}$
was never found, implying the non-existence of a set of three MU bases
together with an additional MU state. 

\section{Modifying the problem}

\label{sec: Modifications-of-the-problem}

The difficulty to construct complete sets of MU bases in arbitrary
dimensions has led to modifications of the existence problem. The
study of closely related structures may provide insights into the
original problem, as well as workarounds in applications such as optimal
state reconstruction. In this section we review these approaches.
A natural extension consists of replacing the field of complex numbers
over which the problem is originally defined, by real numbers, roots
of unity, quaternions or finite fields (Secs.~\ref{subsec:Real-MU-bases}–\ref{subsec:MU-bases-over-finite-fields}).
Another approach is to modify, in a consistent way, the overlaps defining
MU bases and to investigate the resulting structures (Secs.~\ref{subsec:Weak-MU-bases}–\ref{subsec:approximate_mu_bases}).
One can generalise the design property of MU bases to include approximate,
weighted and conical designs, as described in Secs.~\ref{sec:approximate2-designs}–\ref{subsec:Mutually-unbiased-measurements}.
More general measurements may also be considered, e.g., POVMs or projectors
with rank larger than one (Sec.~\ref{subsec:Mutually-unbiased-POVMs}–\ref{subsec:dimension-independent}).
Finally, infinite-dimensional counterparts of MU bases have been studied,
corresponding to quantum systems with continuous variables over the
real line, a circle $S_{1}$, or over the $p$-adic numbers, with
associated Hilbert spaces $L^{2}(\mathbb{R})$, $L^{2}(S_{1})$ and
$L^{2}(\mathbb{Q}_{p})$, respectively (Secs.~ \ref{subsec: MUs for CVs}–\ref{subsec: p-adic-MU-bases}). 

\subsection{Real MU bases}

\label{subsec:Real-MU-bases}

The difficulty of the MU existence problem is related to the large
number of parameters needed to parameterise $d(d+1)$ vectors of $\mathbb{C}^{d}$.
Setting the problem in a real Hilbert space $\mathbb{R}^{d}$ significantly
reduces the number of parameters \citep{myrheim99,stueckelberg60}.
The modified problem turns the existence question into a search for
sets of lines in $\mathbb{R}^{d}$ which \emph{(i)} are orthogonal
within each set and \emph{(ii)} otherwise intersect at a specific,
fixed angle which depends on the dimension $d$. These sets of `{}`equiangular'{}'
lines in $\mathbb{R}^{d}$ have been studied in their own right for
a long time. Geometric pictures of real MU bases existing in the spaces
$\mathbb{R}^{3}$ and $\mathbb{R}^{4}$ have been developed in Ref.
\citep{bengtsson07}.

An upper limit on the number of MU bases in the space $\mathbb{R}^{d}$
has been derived by Boykin\textit{ et al.} \citep{boykin+05real}:
for any $d\geq2$, at most $(d/2+1$) real MU bases exist (also see
\citep{kantor12}). The proof relies on the link between MU bases
and maximally commuting classes of a unitary operator basis (see Equivalence
\ref{equiv:commutingclasses} of Sec.~\ref{subsec:Maximally-commuting-unitary}).
In particular, a set of $\mu$ real MU bases $\{\mathcal{B}_{0},\ldots,\mathcal{B}_{\mu-1}\}$
in $\mathbb{R}^{d}$ exists if and only if there exists $\mu$ classes
$\mathcal{C}_{0},\ldots,\mathcal{C}_{\mu-1}$, of \textit{real symmetric}
unitary matrices in $\mathbb{M}_{d}(\mathbb{R})$, each containing
$d$ commuting matrices (including the identity), such that all matrices
in $\mathcal{C}_{0}\cup\ldots\cup\mathcal{C}_{\mu-1}$ are pairwise
orthogonal. The restriction to \textit{real symmetric} matrices ensures
their eigenstates are real and thus generate MU bases of $\mathbb{R}^{d}$,
not $\mathbb{C}^{d}$. Comparing the span of this subset of matrices
to the space of all real symmetric matrices (which has dimension $d(d+1)/2$),
one finds that $\mu\leq d/2+1$. Earlier proofs of this result in
different contexts can be found in \citep{delsarte+75,calderbank+97}.

Stricter bounds can be placed on the number of real MU bases if more
is known about the dimension $d$. If $d\neq4m$, for $m\in\mathbb{N}$
and $d\text{>2}$, then \emph{no} real Hadamard matrix of order $d$
exists, and therefore no \emph{pair} of real MU bases, in striking
contrast to the complex case. Hadamard famously conjectured that if
$d$ is a multiple of four then a (real) Hadamard matrix exists \citep{paley33},
leading to the expectation that at least two real MU bases exist for
$d=4m$. However, general constructions of Hadamard matrices have
been found only in certain dimensions, e.g. \citep{sylvester67,hadamard93},
and the smallest matrix for which the conjecture remains unverified
is of order $668$.

When $d$ is a \textit{non-square} dimension of the form $d=4m$,
$m\in\mathbb{N}$, then at most two real MU bases exist \citep{boykin+05real}.
Clearly, such pairs exist if and only if a real Hadamard matrix of
order $d$ exists. Now suppose that $d$ is a square dimension divisible
by four. If $d=4s^{2}$ for any positive odd integer $s$, then a
simple proof based on lattice lines shows that there exist at most
three real MU bases. If $d=4^{k}s^{2}$ with $s$ any positive odd
integer, and $k>1$, then at least $(\ell+2)$ real MU bases exist
provided a real Hadamard of order $2^{k}s$ exists. Here, $\ell$
is the maximum number of MO Latin squares of order $2^{k}s$. This
construction, found in \citep{wocjan+05}, is discussed further in
Sec.~\ref{subsec:Maximally-entangled-bases}. Finally, for $d=4^{k}$,
one can apply connections to algebraic coding theory to construct
the maximum number of $(d/2+1)$ real MU bases \citep{calderbank+97,cameron73}.
It remains unknown if maximal numbers of MU bases in $\mathbb{R}^{d}$
exist when the dimension is not a power of four. For example, when
$d=144$, the construction from Ref.~\citep{wocjan+05} finds only
seven bases as opposed to the 73 that could hypothetically exist.

It is possible to establish an equivalence between MU bases in $\mathbb{R}^{d}$
and specific so-called `{}`4-class cometric association schemes'{}'
as explained in Refs.~\citep{lecompte10,abdukhalikov09}. This observation
leads to sets of $(d/2+1)$ real MU bases for $d=4^{k}$, which are
identical to those in \citep{calderbank+97,cameron73}. Furthermore,
a relation between real MU bases and representations of finite groups
of odd order is shown in \citep{gow17}. This yields a set of $(d/q+1)$
MU bases in $\mathbb{R}^{d}$, when $d=q^{2r},$ with $q$ a power
of 2 and $r$ a positive integer, by finding a real orthogonal matrix
of order $q^{2r}$, with multiplicative order $(d/q+1)$, whose powers
define the set of real MU bases. When $q=2$ the upper bound of $(d/2+1)$
bases is reached. A proof is also given for the existence of a set
of $(2k+1)$ MU bases in $\mathbb{R}^{d}$ when $d=2^{2^{k-1}},$
by taking the real representation of an arbitrary group $G$ of odd
order $(2k+1$) for $k\geq3$.

Other studies consider special classes of real Hadamard matrices,
e.g. the Bush-type matrices (see Sec.~\ref{subsec:Real-MU-bases})
yield up to $(4n-1)$ unbiased real Hadamard matrices when $d=4n^{2}$,
with even $n$ \citep{holzmann10,kharaghani14}. For $d>2,$ it is
always possible to incorporate free parameters into any collection
of real MU bases \citep{goyeneche15}. For a set of $\mu$ MU bases
in $\mathbb{R}^{d}$, one can introduce no fewer than $(\mu-1)d/2$
free parameters.

We conclude with an upper bound on the number $\mu$ of MU bases in
a bipartite real vector space, which holds if every vector has a limited
Schmidt rank. Corollary \ref{cor:schmidt_bound} in Sec.~\ref{subsec:entanglement_content}
provides a bound for bases of a complex Hilbert space $\mathbb{C}^{d_{1}}\otimes\mathbb{C}^{d_{2}}$
assuming the Schmidt rank of every vector is at most $k,$ and $k<d_{1}\leq d_{2}$.
Under the same assumption on the ranks of the vectors in $\mathbb{R}^{d_{1}}\otimes\mathbb{R}^{d_{2}}$,
a different bound on the number of MU bases is obtained \citep{cariello21},
namely
\begin{equation}
\mu\leq\frac{k}{2}\left(\frac{d_{1}(d_{1}+1)-2}{d_{1}-k}\right).\label{eq:schmidt_bound_real}
\end{equation}
The result follows from a modified version of Thm.~\ref{thm:fixed_entanglement_Schmidt},
which limits the purity of a set of MU bases in $\mathbb{R}^{d}$
rather than $\mathbb{C}^{d}.$ If $k=1$, all states in each basis
are separable, and the bound implies that at most $(d_{1}+2)/2$ real
MU product bases exist.

\subsection{Butson-type MU bases}

\label{subsec:MU-butson-type}

As discussed in Sec.~\ref{sec:pairs_of_MU_bases_C^d}, a special
class of $d\times d$ Hadamard matrices known as Butson-Hadamard matrices
$BH(d,r)$ have matrix elements which are $r$-th roots of unity \citep{butson62}.
When $r=2$, the matrices $BH(d,2)$ are the real Hadamard matrices
of Sec.~\ref{subsec:Real-MU-bases}; in all other cases, some matrix
elements will have non-zero imaginary parts. As all matrix elements
are taken from a finite set of complex numbers, searching for MU bases
associated with Butson-Hadamard matrices is less complicated than
for general complex Hadamard matrices. 

Well-studied examples include the Butson-Hadamard matrices $BH(d,4)$
containing elements from the set $\{\pm1,\pm i\}$, first considered
in \citep{wallis73,turyn1970}. Constructions appear in orders $d$
where symmetric conference matrices exist, which require dimensions
$d=1+a^{2}+b^{2}$, with integers $a$ and $b$. However, it is conjectured
that they exist for every even $d.$ If $d$ is an odd integer, at
most two unbiased matrices $BH(2d,4)$ exist, while it is conjectured
that pairs exist for all odd integers $d$ when $2d$ is the sum of
two squares. A computer search confirms this for $BH(10,4)$ and $BH(18,4)$
matrices \citep{best10}. An early computer search of all MU Butson-Hadamard
matrices $BH(6,12$) with twelfth roots of unity was carried out in
\citep{bengtsson+07} (see Thm.~\ref{thm:roots_of_unity} of Sec.~\ref{subsec: unextendible MU bases}).

\emph{Bush-type Butson-Hadamard matrices }$BH(d^{2},r)$ of square
order can be divided into $d$ blocks of order $d$, such that every
block is either a matrix consisting of all ones or a matrix with each
row and column summing to zero. It was shown that for every prime
power $d\in\mathbb{PP}$, there exists an unextendible set of $d$
MU Bush-type Butson-Hadamard matrices $BH(d^{2},r)$ \citep{best10}.

\subsection{Quaternionic MU bases}

\label{Quaternionic-MU-bases}

Conceptually, the generalisation from the complex inner product space
$\mathbb{C}^{d}$ to the quaternionic case $\mathbb{H}^{d}$ is straightforward.
An upper limit on the number of MU bases is given by $(2d+1)$, as
shown by Chterental and Ðoković \citep{Chterental_2008} who adapt
the proof based on the intersection of hyperplanes in the complex
case described in Sec.~\ref{sec:geometric} (cf. Refs. \citep{ivanovic81,Bengtsson2005}).
For dimensions $d=2^{n-2}$, an upper bound of $(2d+1)d$ lines in
$\mathbb{H}^{d}$ which are either orthogonal or intersect at an angle
$\cos^{-1}(1/\sqrt{d})$ had already been found in 1982 \citep{hoggar82,levenshtein82}.
Examples of lines saturating this bound for $\mathbb{H}^{2}$ and
$\mathbb{H}^{4}$, which correspond to complete sets of five and nine
quaternionic MU bases, respectively, can be found in \citep{hoggar82}.

The five members of the complete set in $\mathbb{H}^{2}$ consist
of the $2\times2$ identity $\mathbb{I}$ and four structurally simple
matrices,
\begin{equation}
\left(\begin{array}{cc}
1 & 1\\
1 & -1
\end{array}\right)\quad\text{and \ensuremath{\quad}}\left(\begin{array}{cc}
1 & 1\\
\mathbf{{\bf a}} & -\mathbf{{\bf a}}
\end{array}\right),
\end{equation}
where ${\bf a}=\mathbf{i},\mathbf{j},\mathbf{k}$ are quaternions
satisfying $\mathbf{i}^{2}=\mathbf{j}^{2}=\mathbf{k}^{2}=\mathbf{i}\mathbf{j}\mathbf{k}=-1$;
the common factor of $1/\sqrt{2}$ has been dropped. This complete
set is unique up to equivalence \citep{Chterental_2008}.

In the space $\mathbb{H}^{3}$, a complete set consists of seven MU
bases. All \emph{triples} of MU bases in $\mathbb{H}^{3}$ have been
identified in \citep{Chterental_2008}; they necessarily contain (an
equivalent of) the standard complex-valued Fourier matrix. The third
basis can be chosen either from one three-parameter family or from
one of five families depending on two parameters each. Furthermore,
a \emph{quadruple }of quaternionic MU bases depending on three real
parameters has been identified. It is unextendible (see Sec.~\ref{subsec:Unextendible-MU-bases})
and contains a one-parameter family which, in turn, has the complete
set of four MU bases in $\mathbb{C}^{3}$ as a subset. No sets with
more than four elements have been constructed, meaning that the (expected)
complete set of seven MU bases for $d=3$ remains elusive.

The existence of complete sets in general quaternionic spaces $\mathbb{H}^{d}$
with arbitrary dimension $d\in\mathbb{N}$, is, to our knowledge,
an open question. However, in 1995, Kantor presented quaternionic
line-sets in spaces of dimension $d=2^{n-2}$, with $n$ even, whose
angles are either $90^{\circ}$ or $\cos^{-1}(2^{-(n-2)/2})$ \citep{kantor95}.
These sets translate into complete sets of $(2d+1)$ MU bases in $\mathbb{H}^{d}$.
This approach, in analogy with constructions of complete sets described
in Appendix \ref{subsec:other_constructions} for $\mathbb{C}^{d}$,
utilises symplectic spreads.

More recently, constructions of quaternionic Hadamard matrices of
small orders—each such matrix giving rise to a pair of quaternionic
MU bases—as well as infinite families of larger orders, have been
found in \citep{higginbotham21} and \citep{barrera24}, respectively.
Quaternionic Hadamard matrices prove useful in connection to generalised
unbiasedness (see Sec.~\ref{subsec:dimension-independent}).

Quaternionic extensions of tight frames (cf. Sec.~\ref{subsec:MU-bases-and-SICs}),
equiangular lines, and $t$-designs, have also been been studied \citep{waldron20tight,Waldron2020}.
A maximal set of equiangular lines in $\mathbb{C}^{d}$ consists of
$d^{2}$ vectors and defines a SIC, while in $\mathbb{H}^{d}$ there
are at most $(2d^{2}-d)$ equiangular lines. Thus, for $d=2$, six
such lines may exist for $\mathbb{H}^{2}$ that, in fact, have been
described explicitly \citep{khatirinejad08}. The allowed maximum
number of equiangular lines in $\mathbb{H}^{d}$ is conjectured to
exist for every finite $d$ \citep{Waldron2020}, in analogy with
SICs in $\mathbb{C}^{d}$.

It is worth pointing out that, obviously, each set of MU bases in
$\mathbb{C}^{d}$ also can be thought of as a set in the space $\mathbb{H}^{d}$.
Consequently, the known complete sets of $(d+1)$ MU bases in prime-power
dimensions define sets in $\mathbb{H}^{d}$ which are, however, still
$d$ bases away from a complete set.

\subsection{MU bases over finite fields}

\label{subsec:MU-bases-over-finite-fields}

The definition of mutually unbiased bases extends naturally to vector
spaces over finite fields, as shown by McConnell \textit{et al.} \citep{mcconnell21}.
Let $F^{d}$ denote a $d$-dimensional vector space over a finite
field $F$ with a characteristic not dividing $2d$. For $u=(u_{j}),v=(v_{j})\in F^{d}$,
a Hermitian form is defined as $\bk uv=\sum^{d}_{j=1}u^{\sigma}_{j}v_{j}$,
where $\sigma$ denotes the automorphic involution of $F$ which leaves
invariant a subfield $K$ contained in $F$. Pairs of (normalised)
vectors $u,v\in F^{d}$ are mutually unbiased if the identity $\bk uv\bk uv^{\sigma}=1/d$
holds. As in the case of real or complex number fields, orthonormal
bases are mutually unbiased if every pair of vectors taken from different
bases satisfy the identity just given. The infinite-dimensional equivalent
of this generalisation corresponds to $p$-adic MU bases \citep{vandam+11}
considered in Sec. ~\ref{subsec: p-adic-MU-bases}.

It is possible to bound the cardinality of the largest set of MU bases
in $F^{d}$. For any quadratic extension of $F/K$ and $d>2$ coprime
to the characteristic $p$ of $K$, the maximum number $\mu$ of MU
bases in $F^{d}$ satisfies $\text{1}\leq\mu\leq d+1$. Interestingly,
three MU bases in $F^{d}$ do not always exist, in contrast to the
case of $\mathbb{C}^{d}$. However, for prime powers $d\in\mathbb{PP}$
, complete sets of $(d+1)$ MU bases in $F^{d}$ exist for many fields
$F$, as described in \citep{mcconnell21}.

The properties of MU bases over finite fields in a vector space of
dimension $d=6$ can be considered as partial evidence both \emph{for}
and \emph{against }Conjecture \ref{conj: Zauner}. For prime powers
$q=p^{r}\in[5,41]$, there exist at most three MU bases in $F^{6}$
when $F=\mathbb{F}_{q^{2}}$. However, for some values $q\equiv5\mod 12$
there exist three MU bases in $F^{6}$ together with four orthonormal
vectors that are mutually unbiased to the triple \citep{mcconnell21}.
In contrast, all known MU triples in $\mathbb{C}^{6}$ appear to be
strongly unextendible, i.e. no vector unbiased to three MU bases has
been found (cf. Sec. \ref{sec:rigorous_results_d=00003D6}). This
situation may represent a ''shadow'' \citep{mcconnell21} of an original
(unknown) set of MU bases in $\mathbb{C}^{6}$, mirroring a phenomenon
that has been observed for SIC-POVMs.

Extensions to finite fields have also been considered for other discrete
structures in Hilbert space. For example, Greaves \textit{et al.}
\citep{greaves22} study equiangular lines as well as frames (see
Sec.~\ref{subsec:MU-bases-and-SICs} and Sec.~\ref{subsec:Mutually-unbiased-frames})
over finite fields.

\subsection{Weak MU bases}

\label{subsec:Weak-MU-bases}

A attempt to allow for overlaps other than the standard ones between
vectors in Def. \ref{def: MU pair of bases} has been proposed for
composite dimensions $d\notin\mathbb{PP}$ in Ref.~\citep{shalaby+12}.
Suppose $d=d_{1}\ldots d_{r}$ is a particular factorisation of the
Hilbert space dimension $d.$ Then, a pair of orthonormal bases $\mathcal{B}_{b}=\{\ket{v_{b}}\}$
and $\mathcal{B}_{b'}=\{\ket{v'_{b'}}\}$ are weak MU bases if either
$|\bk{v_{b}}{v'_{b'}}|^{2}=\frac{1}{d}$ for all $v$ and $v'$, \emph{or}
$|\bk{v_{b}}{v'_{b'}}|^{2}\in\{\frac{1}{d_{j}},0\}$ for a given $d_{j}$
$(\neq1,d)$ and all $v$ and $v'$.

For an explicit example of the modification, consider the case when
$d=p_{1}p_{2}$ with both $p_{1}$ and $p_{2}$ prime. Then a pair
of bases $\mathcal{B}_{b}=\{\ket{v_{b}}\}$ and $\mathcal{B}_{b'}=\{\ket{v'_{b'}}\}$,
with $v,v'=1,\ldots,d$, of the space $\mathbb{C}^{p_{1}}\otimes\mathbb{C}^{p_{2}}$
are \textit{weakly} MU if their overlap is either $(i)$ standard:
$|\bk{v_{b}}{v'_{b'}}|^{2}=1/d\quad\forall v,v'$, or one of the conditions
\begin{equation} \emph{(ii)}\quad|\bk{v_{b}}{v'_{b'}}|^{2}=\left\{ \begin{array}{ll} 1/p_{1} & v\equiv v'\,\,\text{mod}\,\,p_{2}\,,\\ 0 & \text{otherwise}\,; \end{array}\right. \end{equation} \begin{equation} \emph{(iii)}\quad|\bk{v_{b}}{v'_{b'}}|^{2}=\left\{ \begin{array}{ll} 1/p_{2} & v\equiv v'\,\,\text{mod}\,\,p_{1}\,,\\ 0 & \text{otherwise} \end{array}\right. \end{equation}holds.
A set of bases $\mathcal{B}_{0},\ldots,\mathcal{B}_{\mu-1}$, are
weak MU bases if every pair satisfies one of the constraints $(i)$–$(iii)$.
For example, suppose $\mathcal{B}^{(2)}_{0},\mathcal{B}^{(2)}_{1},\mathcal{B}^{(2)}_{2}$
and $\mathcal{B}^{(3)}_{0},\mathcal{B}^{(3)}_{1},\mathcal{B}^{(3)}_{2},\mathcal{B}^{(3)}_{3}$
are complete sets of MU bases in $\mathbb{C}^{2}$ and $\mathbb{C}^{3},$
respectively. We can construct 12 weak MU product bases $\mathcal{B}^{(2)}_{j_{1}}\otimes\mathcal{B}^{(3)}_{j_{2}}$
by using all combinations of the labels $j_{1}=0,1,2$, and $j_{2}=0,1,2,3$. 

Thm.~IV.2 of \citep{shalaby+12} states that any set of weak MU bases
of dimension $d=d_{1}d_{2}$ take the form $\mathcal{B}^{(d_{1})}_{j_{1}}\otimes\mathcal{B}^{(d_{2})}_{j_{2}}$,
where $\mathcal{B}^{(d_{1})}_{j_{1}}$ and $\mathcal{B}^{(d_{2})}_{j_{2}}$
are MU bases in $\mathbb{C}^{d_{1}}$ and $\mathbb{C}^{d_{2}}$, respectively,
and that the maximal number of weak MU bases is $(d_{1}+1)(d_{2}+1)$.
The proof makes the assumption that all product bases in composite
dimensions can be written as a direct product. However, \emph{indirect}
product bases (cf. Sec.~\ref{subsec:product_bases}) can also be
used to construct weak MU bases: for $d=6$ the pair $\mathcal{B}^{(2)}_{0}\otimes\mathcal{B}^{(3)}_{0}$
and $\{\ket 0\otimes\mathcal{B}^{(3)}_{1},\ket 1\otimes\mathcal{B}^{(3)}_{2}\}$
are weak MU bases but not covered by Thm.~IV.2; here $\mathcal{B}^{(2)}_{0}$
and $\mathcal{B}^{(3)}_{0}$ are the standard bases of $\mathbb{C}^{2}$
and $\mathbb{C}^{3},$ respectively. Thus, the general structure of
weak MU bases and their maximal number is still unknown. 

\subsection{Approximately MU bases}

\label{subsec:approximate_mu_bases} 

Another approach to introduce bases with overlaps similar to MU bases
are the so-called \emph{approximately} mutually unbiased bases, defined
by the asymptotic behaviour of their inner products. The idea, introduced
by Klappenecker \textit{et al.} \citep{klappenecker+05}, is to construct
a system of $(d+1)$ orthonormal bases of $\mathbb{C}^{d}$ which
satisfies the inequality
\begin{equation}
|\bk{v_{b}}{v'_{b'}}|^{2}\leq\frac{1+o(1)}{d}\,,\label{eq:amubs}
\end{equation}
for all $v,v'$ and $b\neq b'.$ The \textit{o}-notation $f(d)=o(g(d))$
implies that for \textit{any} positive constant $c$, there exists
a constant $d_{0}>0$ such that $0<f(d)<cg(d)$ for all $d\geq d_{0}$,
i.e., the ratio $f(d)/g(d)$ approaches zero in the limit of increasing
dimension, $d\rightarrow\infty$.

A construction of sets of bases which exhibit asymptotic bounds of
this type is provided in \citep{klappenecker+05}. To begin, for arbitrary
dimension $d$, and any positive integer $n$, they construct a set
of $(d^{n}+1)$ orthonormal bases from exponential sums, where the
constant overlap property of Def. \ref{def: MU pair of bases} is
modified such that the inner product of any two vectors from different
bases becomes $O(d^{-1/4})$. When $n=1$, the upper bound can be
improved to $O(d^{-1/3}).$ Here, the \textit{O}-notation $f(d)=O(g(d))$
implies that there exist positive constants $c$ and $d_{0}$ such
that $0<f(d)\leq cg(d)$ for all $d\geq d_{0}$, hence there exists
$c>0$ such that the inner product $|\bk{v_{b}}{v'_{b'}}|\leq cd^{-1/3}$
for all $d\geq d_{0}$. Note that the two asymptotic bounds can differ
since the \textit{O}-notation may or may not lead to asymptotically
tight bounds while the \textit{o}-notation is compatible with an upper
bound that is not tight.

Explicitly, the $(d+1)$ asymptotically MU bases consist of the standard
basis $\mathcal{B}_{0}$ together with the bases $\mathcal{B}_{b}=\{\ket{v_{b}}\}^{d-1}_{v=0}$, $b=1,\ldots,d$, with orthonormal vectors
\begin{equation}
\ket{v_{b}}=\frac{1}{\sqrt{d}}\sum^{d-1}_{k=0}e^{{2\pi ibk^{2}/p}}e^{2\pi ivk/d}\ket k\,.\label{eq:AMUbases-1}
\end{equation}
In a related paper it is shown that the bound $O(d^{-1/3})$ for the
inner products may be strengthened to $O(d^{-1/2}(\log d)^{1/2})$
\citep{shparlinski+06}.

When $p$ is the smallest prime such that $p\equiv1$ mod $d$ for
a given dimension $d$, the construction described in Ref.~\citep{klappenecker+05}
can be improved by using multiplicative and mixed character sums,
to provide a set of $(d^{n}+1)$ orthonormal bases satisfying 
\begin{equation}
|\bk{v_{b}}{v'_{b'}}|^{2}\leq\frac{np^{1/2}}{d}\,.\label{eq:AMUbound}
\end{equation}
If $p=d+1$, this construction gives a set of $(d+1)$ approximately
MU bases satisfying Eq.~\eqref{eq:amubs}. In fact, the bound given
in Eq.~\eqref{eq:AMUbound} improves to $O(d^{-1/2}\log d)$ \emph{if}
one relies on a widely believed conjecture on the distribution of
primes in arithmetic progressions.

A different construction with an overlap bound of $O(d^{-1/2})$ was
given in \citep{shparlinski+06} for almost all $d$, based on elliptic
curves. An extension to all dimensions $d$ exists if one assumes
Cramér's conjecture which provides an estimate for the gap size between
consecutive prime numbers. Two further constructions based on mixed
character sums of functions over finite fields yield sets of $(d+1)$
and $(d+2)$ approximately MU bases when $d=p-1,$ for $p$ a prime-power
\citep{wang18}.

A construction of up to $\sqrt{d}+1$ approximately MU bases for $d=q^{2}$,
where $q$ is a positive integer, is presented in \citep{kumar21}.
If $q\equiv0\mod 4$, and a $q\times q$ real Hadamard matrix exists,
the construction gives rise to approximately real MU bases (cf. Sec.~\ref{subsec:Real-MU-bases}).

\textit{Almost perfect} MU bases \citep{kumar24} represent another
modification of \emph{approximately} MU bases. Given a set of vectors,
a restriction is placed on the magnitudes of all inner products: they
are allowed to take only two values. The resulting vectors contain
a large number of components equal to zero, and every non-zero component
is of equal magnitude. 

\subsection{Approximate 2-designs}

\label{sec:approximate2-designs} 

Let us now turn to a modification of mutual unbiasedness by Ambainis
and Emerson \citep{ambainis+07} which involves revising the design
property of complete sets explained in Sec.~\ref{subsec: 2-design equivalence}.
A\textit{pproximate} complex-projective $t$-designs are defined in
the following way \citep{ambainis+07}:
\begin{defn}
Let $\mathcal{D}=\{x:x\in\mathbb{C}P^{d-1}\}$ be a finite subset
of $\mathbb{C}P^{d-1}$ and a 1-design. Then $\mathcal{D}$ is an
$\epsilon$-approximate $t$-design if for all $f\in$ Hom$(t,t)$,
\begin{align}
(1-\epsilon)\!\!\int_{\mathbb{C}P^{d-1}}\!\!f(x)d\nu(x)
&\le \frac{1}{|\mathcal{D}|}\sum_{x\in\mathcal{D}} f(x)
\label{eq: eps-approx t-design}\\
&\le (1+\epsilon)\!\!\int_{\mathbb{C}P^{d-1}}\!\!\!f(x) d\nu(x).
\nonumber
\end{align}
where the integral is evaluated over the unitarily invariant Haar
measure $\nu$ on the unit sphere in $\mathbb{C}^{d}$.
\end{defn}
The above inequalities clearly relax the identity \eqref{eq:complex_projective_design}
valid for standard $t$-designs. It is known that for $d\geq2t$,
there exists an $O(d^{-1/3})$-approximate $t$-design such that $|\mathcal{D}|=O(d^{3t})$
\citep{ambainis+07} (see Sec.~\ref{subsec:approximate_mu_bases}
for the definition of the $O$-notation). Hence, there exists an $O(d^{-1/3})$-approximate
2-design containing $O(d^{6})$ elements for $d\geq4$.

Since we are interested in finding bases, it makes sense for the $\epsilon$-approximate
2-design $\mathcal{D}\subset\mathbb{C}P^{d-1}$ to be the union of
a set of orthonormal bases $\mathcal{B}_{b}=\{\ket{v_{b}}\}^{d-1}_{v=0}\subset\mathbb{C}^{d}$,
with $b=0,\ldots,\mu-1$. Approximate 2-designs of this type were
considered in relation to cyclic 2-designs \citep{avella24}. These
designs, composed of sets of bases, generalise cyclic MU bases—i.e.
MU bases generated from the powers of a single unitary (cf. Appendix
\ref{subsec:other_constructions})—which appear only to exist in even
prime-power dimensions. 

To our knowledge, the construction of minimal sets of bases which
form an approximate 2-design, for arbitrary dimension $d,$ remains
an open problem.

\subsection{Weighted 2-designs}

\label{sec:weighted2designs} 

Complex projective $t$-designs (defined in Sec.~\ref{subsec: 2-design equivalence})
can be generalised by adding a normalised positive-valued weight function
to the finite set $\mathcal{D\subset\mathbb{C}}P^{d-1}$.
\begin{defn}
Let $\mathcal{D}=\{x:x\in\mathbb{C}P^{d-1}\}$ be a finite subset
of $\mathbb{C}P^{d-1}$, and $w:\mathcal{D}\rightarrow[0,1]$ a normalised
weight function. Then $(\mathcal{D},w)$ is a weighted complex projective
$t$-design if for all $f\in$ Hom$(t,t)$, 
\begin{equation}
\sum_{x\in\mathcal{D}}w(x)f(x)=\int_{\mathbb{C}P^{d-1}}f(x)d\nu(x)\,,\label{eq:weighted-complex-projective-t-design}
\end{equation}
where the integral is evaluated over the unitarily invariant Haar
measure $\nu$ on the unit sphere in $\mathbb{C}^{d}$.
\end{defn}
This relation generalises Eq.~\eqref{eq:complex_projective_design}.
Equivalently, and in analogy to Eqs.~(\ref{eq:t-design-symmetric-subspace}),
a weighted $t$-design $(\mathcal{D},w)$ is a subset of $\mathbb{C}P^{d-1}$
which satisfies
\begin{align}
\sum_{x\in\mathcal{D}} w(x) P(x)^{\otimes t}
&= \int_{\mathbb{C}P^{d-1}} P(x)^{\otimes t}\, d\nu(x) \notag\\
&= \binom{d+t-1}{t}^{-1}\Pi^{(t)}_{\mathrm{sym}}\,,
\label{eq: weighted t-design}
\end{align}
where $P(x)=\kb xx$ is the projection operator associated with $x\in\mathbb{C}P^{d-1}$,
and $\Pi^{(t)}_{\text{sym}}$ is the projector onto the symmetric
subspace of $(\mathbb{C}^{d})^{\otimes t}$. As seen in Sec.~\ref{subsec: 2-design equivalence},
(unweighted) $t$-designs exist for all choices of $t$ and $d$.

Since a complete set of MU bases forms a complex projective 2-design
consisting of $(d+1)$ orthogonal bases (each contributing equally),
one way to generalise sets of MU bases is to consider a \emph{weighted}
2-design formed from the union of a set of orthonormal bases in $\mathbb{C}^{d}$.
Suppose we have a set of $\mu$ orthonormal bases $\mathcal{B}_{0},\ldots,\mathcal{B}_{\mu-1}\subset\mathbb{C}P^{d-1}$,
such that $\mathcal{B}_{b}=\{\ket{v_{b}}\}^{d-1}_{v=0}$, and an associated
weight $w_{b}$ assigned to each basis. Then the weight function together
with the set $\mathcal{D}=\cup_{b}\mathcal{B}_{b}$ forms a 2-design
if 
\begin{equation}
\sum^{\mu-1}_{b=0}w_{b}\sum^{d-1}_{v=0}P_{b}(v)^{\otimes2}={d+1 \choose 2}^{-1}\Pi^{(2)}_{\text{sym}}\,,\label{eq:weighted2design}
\end{equation}
where $P_{b}(v)=\kb{v_{b}}{v_{b}}$. This holds only if $\mu\geq d+1$,
with equality when the bases are mutually unbiased \citep{roy+07}. 

Provided $\mu$ is sufficiently large, weighted 2-designs from bases
exist in all dimensions $d$ \citep{seymour84,roy+07}. The construction
is based on highly non-linear functions on Abelian groups. For prime-power
dimensions $d=p^{n}$, there exist maximally nonlinear functions which
generate a 2-design containing the minimal $\mu=d+1$ orthonormal
bases \citep{roy+07}. When $d=p^{n}-1$, this technique is used to
generate a 2-design with $\mu=d+2$. Let $\mathcal{B}_{0}$ be the
standard basis of $\mathbb{C}^{d}$ with corresponding weight 
\begin{equation}
w_{0}=\frac{1}{d(d+1)}\,.
\end{equation}
For the remaining $(\mu-1)$ bases, the $v$-th vector of basis $\mathcal{B}_{b}$,
$b=1\ldots d+1$, is defined as 
\begin{equation}
\ket{v_{b}}=\frac{1}{\sqrt{d}}\sum^{d-1}_{k=0}e^{2\pi ivk/d}e^{2\pi i\textrm{Tr}[by^{k}]/p}\ket k\,,\label{eq:weightedbases}
\end{equation}
where the trace function is $\textrm{Tr}[x]=x+x^{p}+\ldots+x^{p^{n-1}}$,
and $y$ is a primitive element of $\mathbb{F}_{d+1}$. The weight
assigned to the basis $\mathcal{B}_{b}$ is given by 
\begin{equation}
w_{b}=\frac{1}{(\mu-1)(d+1)}\,,\qquad b=1\ldots d+1\,,
\end{equation}
and the overlaps between elements of different bases read
\begin{equation}
|\bk{v_{b}}{v'_{b'}}|^{2}=\left\{ \begin{array}{ll}
\frac{1}{d^{2}} & \quad\mbox{if \ensuremath{b\neq b',\,\,v\neq v'}}\,,\\
\frac{d+1}{d^{2}} & \quad\mbox{if \ensuremath{b\neq b',\,\,v=v'}}\,.
\end{array}\right.\label{eq:weightedbasesoverlap}
\end{equation}

Weighted $2$-designs of this type provide a good generalisation for
MU bases in quantum tomography as they yield an optimal approach for
quantum state reconstruction (see Sec.~\ref{subsec: Quantum-state-reconstruction}).

\subsection{MU measurements and conical 2-designs }

\label{subsec:Mutually-unbiased-measurements}

Another natural generalisation of MU bases relaxes the rank-one restriction
on the projection operators $P_{b}(v)=\ket{v_{b}}\bra{v_{b}}$ from
the bases $\mathcal{B}_{b}=\{\ket{v_{b}}\}$. Suppose that for each
$b,$ $M_{b}=\{M_{b}(v):v=0\ldots d-1\}$ is a $d$-outcome POVM on
the space $\mathbb{C}^{d}$ such that $\mbox{Tr}[M_{b}(v)]=1,$ for
all \textbf{$v.$} A suitable modification of the fundamental conditions
\eqref{eq: completeMUsetfor(d+1)} leads to one-parameter-families
of sets of \emph{MU measurements}.
\begin{defn}
A collection of $\mu$ POVMs $M_{b},$ with $b=0,\ldots,\mu-1$, forms
a set of $\mu$ \emph{MU} \textit{measurements} if $\mbox{Tr }[M_{b}(v)M_{b'}(v')]=1/d$
for $b\neq b'$, and 
\begin{equation}
\mbox{Tr }[M_{b}(v)M_{b'}(v')]=\kappa\delta_{v,v'}+(1-\delta_{v,v'})\frac{1-\kappa}{d-1}\,,\label{eq: MU measurements}
\end{equation}
for $\mbox{\ensuremath{b=b'}}$, with $v,v'=0\ldots d-1$ and $1/d<\kappa\leq1$.
\end{defn}
When $\kappa=1$ the condition is equivalent to the definition of
MU bases since the operators $M_{b}(v)$ become rank-one projectors.
However, for $\kappa<1$ we obtain a new set of measurements such
that the inner product between any two elements from a single POVM
depends on the parameter $\kappa,$ while the inner product between
any elements from a pair of POVMs have the overlap $1/d$. The value
$\kappa$ determines how close the POVM elements are to rank-one projectors. 

A set of $(d+1)$ MU measurements $M_{b}=\{M_{b}(v):v=0\ldots d-1\}$
has been constructed in Ref.~\citep{kalev+14}. Let $F_{v,b}$ be
the $d(d+1)$ elements of the generalised Gell-Mann operator basis
of bounded operators on the Hilbert space $\mathbb{C}^{d}$, and define
\begin{equation}
F_{b}(v)=\left\{ \begin{array}{ll}
(1+\sqrt{d})F_{b} & \text{for}\,\,v=d\,,\\
F_{b}-(d+\sqrt{d})F_{v,b} & \text{otherwise}\,,
\end{array}\right.
\end{equation}
where $F_{b}=\sum^{d-1}_{v=1}F_{v,b}$. The operators $M_{b}(v)=\mathbb{I}/d+tF_{b}(v)$,
with $t$ chosen such that $M_{b}(v)\geq0$, define a set of MU measurements
for arbitrary $d$ when $\kappa=1/d+2/d^{2}$. Note that while MU
measurements exist for any $d,$ if one initially fixes $\kappa,$
their existence is not guaranteed. 

Since MU measurements share many features with MU bases, they are
useful for applications such as entanglement detection \citep{chen+14,rastegin15,li19b,salehi21},
especially in composite dimensions $d\notin\mathbb{PP}$ where complete
sets are not known.

Mutually unbiased measurements are examples of conical 2-designs \citep{graydon16}
which generalise complex projective 2-designs by dropping the restriction
on the rank of the projection operators. As we have seen in Sec.~\ref{subsec: 2-design equivalence},
a complex projective 2-design $\mathcal{D}$ is a finite set of rank-one
projection operators $P(x)=\ket x\bra x$ such that $\sum_{x\in\mathcal{D}}P(x)\otimes P(x)$
is proportional to the symmetric projection $\Pi_{\text{sym}}$ onto
$\mathbb{C}^{d}\otimes\mathbb{C}^{d}.$ Allowing for asymmetric projections
leads to the definition of conical 2-designs.
\begin{defn}
A \emph{conical 2-design} is a finite set $\mathcal{D}=\{M(x):x=1,\ldots,n\}$
of positive semi-definite operators such that
\begin{equation}
\sum^{n}_{x=1}M(x)\otimes M(x)=k_{s}\Pi_{\text{sym}}+k_{a}\Pi_{\text{asym}}\,,\label{eq:weighted-complex-projective-t-design-1}
\end{equation}
where $k_{s}>k_{a}\geq0$, and $\Pi_{\text{sym}}$, $\Pi_{\text{asym }}$
are the symmetric and antisymmetric projectors onto $\mathbb{C}^{d}\otimes\mathbb{C}^{d}.$
\end{defn}
Both conical and complex projective 2-designs are sets of projection
operators such that their linear combinations $\sum_{x}M(x)\otimes M(x)$
commute with $U\otimes U$ for all unitary operators $U.$ The weighted
complex projective 2-designs discussed in Sec.~\ref{sec:weighted2designs}
form a subset of conical 2-designs if the parameter $k_{a}$ takes
the value zero. In fact, $k_{a}=0$ holds if and only if all projectors
are rank-one. In addition to mutually unbiased measurements, arbitrary-rank
symmetric informationally complete measurements also form conical
2-designs \citep{appleby07}. Another class of generalised designs,
called \emph{mixed $2$-designs} were introduced in Ref.~\citep{brandsen16}
and they also contain MU measurements as a subclass. Mixed designs
of finite cardinality form a subset of the homogeneous conical 2 -designs
\citep{graydon16}.

\subsection{MU POVMs}

\label{subsec:Mutually-unbiased-POVMs}

Beneduci \textit{et al.} \citep{Beneduci2013} generalised the definition
of MU bases to POVMs. This notion of unbiasedness is more general
than the definition of MU measurements from Sec.~\ref{subsec:Mutually-unbiased-measurements}.
\begin{defn}
A pair of $d$-outcome POVMs $M=\{M(i)\}^{d}_{i=1}$ and $N=\{N(j)\}^{d}_{j=1}$
acting on a $d$ dimensional Hilbert space $\mathcal{H}$ are called
\emph{mutually unbiased} if
\begin{equation}
\mbox{Tr }[M(i)N(j)]=1/d\,,\qquad i,j=1,\ldots,d\,.\label{eq:mu_povm}
\end{equation}

Mutually unbiased POVMs shed some light on the connection between
SICs (cf. Sec~\ref{subsec:MU-bases-and-SICs}) and MU bases. For
instance, MU POVMs arise as marginals of a SIC (cf. Sec.~\ref{subsec:inc}),
and any SIC is, in fact, a joint measurement of at least three and
at most $(d+1)$ mutually unbiased POVMs \citep{Beneduci2013}. Given
a set of $(d-1)$ pairwise mutually orthogonal Latin squares of order
$d$ (cf. Appendix~\ref{subsec:affineplanes_mubs_sics}), a SIC yields
$(d+1)$ mutually unbiased POVMs as marginals. If the resulting POVMs
are \emph{commutative}, e.g. the elements of $M(i)$ pairwise commute,
it is possible to construct a system of MU bases. In particular, mutually
unbiased POVMs are smearings of MU bases.
\end{defn}

\subsection{MU frames}

\label{subsec:Mutually-unbiased-frames}

Pérez \textit{et al.} \citep{perez2021mutually} generalised the concept
of unbiasedness to finite frames (cf. Sec.~\ref{subsec:MU-bases-and-SICs}).
\begin{defn}
A pair of frames $\{\ket{\phi_{i}}\}^{m_{1}}_{i=1}$ and $\{\ket{\psi_{j}}\}^{m_{2}}_{j=1}$
in a $d$ dimensional Hilbert space $\mathcal{H}$ are called \emph{mutually
unbiased} \emph{frames} if there exists $c>0$ such that
\begin{equation}
|\bk{\phi_{i}}{\psi_{j}}|^{2}=c\,,\label{eq:mu_frames}
\end{equation}
 for $i=1,\ldots,m_{1}$ and $j=1,\ldots,m_{2}$.

Sets of pairwise mutually unbiased frames include MU bases and SICs,
as well as mutually unbiased rank-one POVMs. If at least one of the
frames is tight, i.e. the projections onto the frame elements sum
to a multiple of the identity, the overlap takes the value $c=1/d$.
Hence, mutually unbiased tight frames satisfy Eq. (\ref{eq:mu_povm})
and correspond to the MU-POVMs of Sec.~\ref{subsec:Mutually-unbiased-POVMs}.

The notion of MU frames covers other extensions of unbiasedness, including
MU equiangular tight frames \citep{fickus20} and MU regular simplices
\citep{fickus20a}. An equiangular tight frame in the space $\mathbb{C}^{d}$
is a collection of $m\geq d$ vectors which have constant overlap
(e.g. a SIC), and a regular simplex is an equiangular tight frame
with $m=d+1$. For equiangular tight frames with $m$ vectors in $\mathbb{C}^{d}$,
the number $\text{\ensuremath{\mu}}$ of pairwise mutually unbiased
equiangular tight frames satisfies
\begin{equation}
\mu\leq\left\lfloor \frac{d^{2}-1}{m-1}\right\rfloor \,.\label{eq:mu_equiangular_frames}
\end{equation}
This reproduces previously derived bounds for MU bases and MU regular
simplices when $m=d$ and $m=d+1,$ respectively. MU equiangular tight
frames have also been applied to construct equiangular tight frames
in composite dimensions \citep{fickus20}.
\end{defn}

\subsection{Generalised unbiasedness \label{subsec:dimension-independent}}

In the context of Bell inequalities (see Sec.~\ref{subsec:Quantum-correlations}),
a notion of unbiasedness was introduced which imposes no constraints
on the dimension of the Hilbert space and fixes only the number of
measurement outcomes \citep{tavakoli19}. Although this definition
differs from the one given in Sec.~\ref{subsec:Mutually-unbiased-measurements},
both share the same name.
\begin{defn}
Two $d$-outcome POVMs $M=\{M(i)\}_{i=1}^{d}$ and $N=\{N(j)\}_{j=1}^{d}$,
acting on a Hilbert space $\mathcal{H}$ (not necessarily of dimension
$d$), are called \emph{mutually unbiased measurements} (MUMs) if
\begin{align}
M(i) &= d\,M(i)N(j)M(i), \notag\\
N(j) &= d\,N(j)M(i)N(j),
\label{eq:dim_indep}
\end{align}
for all $i,j=1,\ldots,d$.
\end{defn}
Here, the dimension of the Hilbert space $\mathcal{H}$ is undefined,
but the notion of complementarity remains a key feature. The formulation
is linked to a class of tailor-made Bell inequalities that are maximally
violated by MU bases. Introducing MUMs ensures that the inequalities
are \textit{only} maximally violated by this class of measurements
and, hence, provides a means of self-testing. A similar modification—removing
the fixed dimension of the Hilbert space and only requiring $d^{2}$
outcomes—has been made for SIC-POVMs \citep{tavakoli19}.

The structure and properties of MUMs as well as their construction
have been studied extensively in \citep{farkas2022mutually}. It is
clear that both MU bases and direct sums of MU bases provide examples
of MUMs. Other distinct examples have been constructed, including
a method based on quaternionic Hadamard matrices. A fundamental difference
between MU bases and MUMs is the existence of an unbounded number
of pairwise $d$-outcome measurements satisfying Eq.~(\ref{eq:dim_indep}).

\subsection{MU bases for continuous variables}

\label{subsec: MUs for CVs}

Mutually unbiased bases in an infinite-dimensional Hilbert space emerge
naturally for position and momentum observables of quantum particles
(cf. Sec.~\ref{subsec:Motivation}) which are known as (a pair of)
`{}`continuous variables'{}'\emph{ }\citep{Braunstein_2005}.
Only a few results have been obtained for composite systems; they
will be described after discussing the case of a single continuous
variable. 

Consider two different bases ${\cal B}_{r}=\{\ket r,r\in I\}$ and
${\cal B}_{s}=\{\ket s,s\in I\}$ of the Hilbert space ${\cal H}=L^{2}(I)$,
with some interval $I\subseteq\mathbb{R}$. The bases consist of (generalised)
states which satisfy orthonormality conditions expressed in terms
of Dirac delta functions, i.e. $\langle r|r^{\prime}\rangle\propto\delta(r-r')$,
$r,r^{\prime}\in I$, and $\langle s|s^{\prime}\rangle\propto\delta(s-s')$,
$s,s^{\prime}\in I$, must hold. The labels of the states vary over
a \emph{continuous} range, and the interval $I\subseteq\mathbb{R}$
may be infinitely large. The bases are mutually unbiased if the \emph{transition
probability} \emph{densities} for all pairs of vectors $\ket r\in{\cal B}_{r}$
and $\ket s\in{\cal B}_{s}$ equal some constant, i.e. $|\langle r|s\rangle|^{2}=\kappa>0$.
The best-known examples of such sets of states are the generalised
eigenstates of position or momentum for the space, with $I=\mathbb{R}$.
According to Eq.~\eqref{eq:particle prob}, the constant overlaps
guarantee that the probabilities $\mbox{prob}(r,dr)$ and $\mbox{prob}(s,ds)$
to find outcomes near $r$ and s, initially given $\ket s$ and $\ket r$,
respectively, are all equal.

MU bases can be defined either directly in the setting of an infinite-dimensional
Hilbert space or by constructing them through a limiting procedure
of a $d$-dimensional Hilbert space and taking the limit of $d\to\infty$.
The second approach has been carried out in considerable detail in
\citep{durt+10}. 

The state space $L^{2}(\mathbb{R})$ of a quantum particle hosts,
in fact, not only pairs of MU bases associated with the position operator
$\hat{q}$ and the momentum operator $\hat{p}$. All rotated position
observables $\hat{q}_{\theta}=\hat{q}\cos\theta+\hat{p}\sin\theta$,
$\theta\in[0,\pi)$, possess complete sets of (generalised) eigenstates.
The overlap of two states $\ket{q_{\theta}}$ with $\ket{q_{\theta'}}$
associated with the bases ${\cal B}_{\theta}$ and ${\cal B}_{\theta^{\prime}}$,
respectively, can be calculated from their Wigner functions \citep{weigert+08}, 
\begin{equation}
|\langle q_{\theta}|q_{\theta'}\rangle|^{2}=\frac{1}{2\pi\hbar|\sin(\theta-\theta')|}\,.\label{eq: CV overlap}
\end{equation}
Since the overlap only depends on the angles $\theta$ and $\theta'$
(i.e. the chosen pair of bases) but not on the values $q_{\theta}$
and $q_{\theta^{\prime}}$, the bases ${\cal B}_{\theta}$ and ${\cal B}_{\theta'}$
are mutually unbiased. This property of the rotated position bases
has been noticed previously, for example in \citep{botero07}, where
the mean-king problem (cf. Sec.~\ref{subsec: mean kings}) is formulated
and solved in the continuous-variable setting. The fact that the overlap
can vary as a function of the difference $(\theta-\theta')$ , i.e.
the pair of bases considered, represents a new feature of the continuous-variable
case. In finite-dimensional spaces, the overlap of MU bases cannot
vary from one pair to the next since, for given dimension $d$, a
specific fixed value is required for consistency. 

As an immediate consequence of Eq.~\eqref{eq: CV overlap} we are
now presented with two possibilities when looking for \emph{sets}
of MU bases; both choices have their merits. If we allow for \emph{basis-dependent
}overlaps, then a \emph{continuous }family of MU bases ${\cal B}_{\theta}$,
$\theta\in[0,\pi)$ in the space $L^{2}(\mathbb{R})$ \citep{weigert+08}
exists. This situation is natural in the sense that the number of
MU bases in prime-power dimensions increases without bound. In addition,
the MU bases of Heisenberg-Weyl type in finite-dimensional systems,
for ever larger dimensions $d$, have been shown to approach the MU
bases ${\cal B}_{\theta}$, $\theta\in[0,\pi)$ for a particle moving
on a line \citep{durt+10}. Taking the limit is, however, a rather
subtle affair since the density of prime-powers among all integers
approaches zero for for $d\to\infty$ according to Eq.~\eqref{eq prob of comp dim}.

Alternatively, one may decide to only consider sets of MU bases with
\emph{basis-independent} pairwise overlaps, just as in the finite-dimensional
case where this property is the only option. In other words, we require
all overlaps to take a single value only. This stronger condition
leaves us with a maximum of \emph{three }MU bases among the set ${\cal B}_{\theta}$,
$\theta\in[0,\pi)$ (cf. below). In both cases, the link between sets
of MU bases and non-redundant quantum tomography seems to be broken:
three MU bases are not tomographically complete for arbitrary quantum
states while measurements associated with a continuous family of MU
bases will be overcomplete, i.e. they contain redundant information.
A suitably chosen countable subset, however, might represent the most
appropriate equivalent of a complete set of MU bases in a finite-dimensional
space. 

A \emph{symmetric} triple of MU bases \citep{weigert+08} is given
by the set $\{{\cal B}_{0},{\cal B}_{2\pi/3},{\cal B}_{4\pi/3}\}$
where ${\cal B}_{0}$ is the basis of position eigenstates $\ket q$,
$q\in\mathbb{R}$, and two other bases associated with the observables
$\hat{q}_{\pm}=\hat{q}\cos(2\pi/3)$ $\pm\hat{p}\sin(2\pi/3)$, obtained
by three-fold rotations of the position operator. Denoting their eigenstates
by $\ket{q_{\pm}}$, $q_{\pm}\in\mathbb{R}$, the second basis is
given by ${\cal B}_{2\pi/3}\equiv\mathcal{B}_{+}=\left\{ \ket{q_{+}},q_{+}\in\mathbb{R}\right\} $,
and we define ${\cal B}_{4\pi/3}\equiv\mathcal{B}_{-}$ in a similar
way. The observables $\hat{p}$ and $\hat{q}_{\pm}$ are (non-unitarily)
equivalent to the \emph{asymmetric} set of operators $\hat{q}$, $\hat{p}$
and $\hat{r}=-\hat{q}-\hat{p}$ which have the advantage of being
`{}`equi-commutant'{}' \citep{Kechrimparis2014}, in the sense
that 
\begin{equation}
[\hat{p},\hat{q}]=[\hat{q},\hat{r}]=[\hat{r},\hat{p}]=\frac{\hbar}{i}\,.\label{eq: three equicommutant ops}
\end{equation}
The \emph{Schrödinger triple} $(\hat{p},\hat{q},\hat{r})$ is a \emph{maximal} set of equi-commutant observables and is \emph{unique} (up to unitary
transformations). First, assume that another observable $\hat{r}^{\prime}\neq\hat{r}$
exists which equi-commutes with both $\hat{p}$ and $\hat{q}$, i.e.
we have
\begin{equation}
[\hat{q},\hat{r}^{\prime}]=[\hat{r}^{\prime},\hat{p}]=\frac{\hbar}{i}\,,\label{eq: CCRs for operator r'}
\end{equation}
in analogy to Eq.~\eqref{eq: three equicommutant ops}. It follows
immediately that the operator $(\hat{r}^{\prime}-\hat{r})$ commutes
with both generators of the Heisenberg algebra but the only Hermitian
operators with this property are multiples of the identity so that
\begin{equation}
\hat{r}^{\prime}=\hat{r}+c\hat{I}\,,\quad c\in\mathbb{R}\,.\label{eq: relation between r' and r}
\end{equation}
Thus, discounting the irrelevant constant shift, there is no observable
other than $\hat{r}$ which equi-commutes to $\hbar/i$ with both
position and momentum. 

The \emph{uniqueness} of the Schrödinger triple $(\hat{p},\hat{q},\hat{r})$
follows from the fact that all irreducible representations of equi-commutant
triples are unitarily equivalent. This result is a consequence of
the theorem by \emph{Stone} and \emph{von Neumann} stating that (under
some mild conditions) all irreducible representations of the canonical
commutation relation $[\hat{p},\hat{q}]=\hbar/i$ are unitarily equivalent
\citep{vonneumann31}. Now assume that we have three observables $(\hat{P},\hat{Q},\hat{R})$
satisfying the relations \eqref{eq: three equicommutant ops} and
that an irreducible representation of these relations on the space
$L^{2}(\mathbb{R})$ is given. Consequently, we have an irreducible
representation of two observables with a canonical commutator, $[\hat{P},\hat{Q}]=\hbar/i$.
However, due to the Stone-von Neumann theorem, there exists a unitary
operator $\hat{U}$ which maps this representation to the one we have
used to initially represent the particle's position and momentum,
i.e. we conclude
\begin{equation}
\hat{P}=\hat{U}\hat{p}\hat{U}^{\dagger}\,,\mbox{ and }\hat{Q}=\hat{U}\hat{q}\hat{U}^{\dagger}\,.\label{eq: unitary equiv of Pp and Qq}
\end{equation}
This relation leads to
\begin{equation}
[\hat{p},\hat{q}]=[\hat{q},\hat{R}^{\prime}]=[\hat{R}^{\prime},\hat{p}]=\frac{\hbar}{i}\,,\label{eq: CCRs for triple pqR'}
\end{equation}
where $\hat{R}^{\prime}\equiv\hat{U}^{\dagger}\hat{R}\hat{U}$. However,
as shown before, the only observable equi-commutant to $\hbar/i$
with position $\hat{q}$ and momentum $\hat{p}$ is given by $\hat{R}^{\prime}\equiv\hat{r}$,
leading to
\begin{equation}
\hat{R}=\hat{U}\hat{r}\hat{U}^{\dagger}\,.\label{eq: unitary equiv of Rr}
\end{equation}
Combining this result with Eqs.~\eqref{eq: unitary equiv of Pp and Qq}
establishes the unitary equivalence of the triples $(\hat{P},\hat{Q},\hat{R})$
and $(\hat{p},\hat{q},\hat{r})$, i.e. the essential uniqueness of
triples equi-commutant to $\hbar/i$. Therefore, no more than three
MU bases can indeed be associated with a set of three equi-commutant
observables, and these triples are all unitarily equivalent. 

The argument just given is not strong enough to limit the number of
MU bases for continuous variables with basis-independent overlaps
to three; other constructions may exist, unrelated to equi-commutant
operators. Interestingly, the situation changes if one replaces the
real numbers $\mathbb{R}$ by the $p$-adic numbers $\mathbb{Q}_{p}$
since the space $L(\mathbb{Q}_{p})$ is known to support $(p+1)$
MU bases (see Sec.~\ref{subsec: p-adic-MU-bases}).

MU bases for other continuous degrees of freedom have been introduced
as well. For both motion on a half-line and a finite line segment,
continuously many pairs of MU bases exist \citep{durt+10}, each containing
subsets of three bases with basis-independent overlaps.

The quantum mechanical rotor, i.e. motion on a circle, poses an unexpected
challenge. The basis associated with a shift of the discretised angular
momentum states (which obey a Kronecker-delta type normalisation)
and the basis associated with the continuous rotations of the azimuthal
position (with a Dirac-delta type normalisation) are mutually unbiased.
However, this pair of bases is \emph{strongly} unextendible (cf. Sec.~\ref{subsec:Unextendible-MU-bases}):
not a single state MU to both of them can be constructed, seemingly
due to the different normalisation conditions \citep{durt+10}. In
stark contrast, no pair of MU bases in finite dimensions is strongly
unextendible: one can always find at least one vector MU to any two
bases (Cor. \ref{thm:circular-2}). Guided by the fact that Hilbert
spaces with countably infinite dimensions are isomorphic, two different
continuous sets of triples of MU bases have, however, been identified
for the quantum rotor \citep{lu12}.

In composite systems with $N>1$ continuous variables, it is straightforward
to construct infinitely large families of MU bases with basis-dependent
overlaps, simply by tensoring copies of the bases ${\cal B}_{\theta}$,
$\theta\in[0,\pi)$. The overlaps of the product states $\ket{q_{\theta_{1}}\ldots q_{\theta_{N}}}$
follow from Eq.~\eqref{eq: CV overlap},
\begin{equation}
|\langle q_{\theta_{1}}\ldots q_{\theta_{N}}|q_{\theta^{\prime}_{1}}\ldots q_{\theta^{\prime}_{N}}\rangle|^{2}=\kappa\prod^{N}_{n=1}\frac{1}{|\sin(\theta_{n}-\theta^{\prime}_{n})|}\,,\label{eq: product CV overlap}
\end{equation}
where $\kappa=\left(2\pi\hbar\right)^{-N}$. This construction is
expected to also work for the other types of continuous degrees of
freedom discussed in \citep{durt+10}. Further MU bases may be found
if the restriction to product states is dropped. 

Asking for\emph{ basis-independent} overlaps drastically reduces the
number of MU bases. For two continuous variables, five MU product
bases with unit overlap have been constructed explicitly \citep{weigert+08}.
They are characterised by pairs of vectors with two components all
of which are elements of a Galois field obtained by a quadratic extension
of the integer numbers. Having three and five MU bases in both the
two- and the infinite-dimensional Hilbert space, respectively, suggests
an analogy which has not been formalised but speculated about \citep{blume08}.
However, the analogy is unlikely to be straightforward since a complete
set of MU bases for two qubits with state space $\mathbb{C}^{2}\otimes\mathbb{C}^{2}$
contains both product states and entangles states, while the five
bases of the space $L^{2}(\mathbb{R})\otimes L^{2}(\mathbb{R})$ contain
product states only. In addition, a one-parameter family of inequivalent
sets of five MU bases exists, and one can construct yet other pentuples
\citep{beales} taking inspiration from the continuous-variable analogue
of indirect product bases described in Sec.~\ref{subsec:product_bases}.

For $N\geq2$, conditions for the existence of specific types of MU
bases, not necessarily of product form, have been expressed in terms
of metaplectic operators \citep{weigert+08}. So far, no solutions
have been found. 

The position and momentum variables of a quantum particle can be `{}`coarse-grained'{}'
in a periodic fashion to introduce a situation which sits between
the finite- and the infinite-dimensional case \citep{Tasca2018}.
The new observables are constructed by means of a set of infinite
periodic square waves which, when taken together, cover all of $\mathbb{R}$.
Effectively, one lumps together all those projectors onto (generalised)
position eigenstates which have labels in regions where the square
waves are non-zero. The construction ensures that quantum states contained
entirely in one of the position bins, say, is evenly spread out across
all momentum bins \citep{silva22}.

The construction can be generalised by using arbitrary straight lines
in phase space as a starting point \citep{Paul2018a}. It has been
shown that this approach cannot lead to more than three MU bases with
basis-independent overlaps. Both of the papers just mentioned report
quantum optical experiments which implement the required measurements
and confirm that they are unbiased. Composite systems with two or
more (pairs of) continuous variables have not been considered.

A conceptual link between Feynman's path integral and mutually unbiased
bases for continuous variables has been described in Refs. \citep{Tolar_2009,svetlichny2007}.
Furthermore, it is pointed out that the basis of position eigenstates
at time $t$ and its image under the free-particle dynamics at time
$t^{\prime}$ are mutually unbiased since we have 
\begin{equation}
\bra{x^{\prime},t^{\prime}}x,t\rangle=\left(\frac{m}{2\pi i(t^{\prime}-t)}\right)^{\frac{1}{2}}\exp\left[\frac{im\left(x^{\prime}-x\right)^{2}}{\hbar t}\right].\label{eq: free particle MU-ness}
\end{equation}
This relation effectively defines a continuous family of MU bases
for a continuous variable with basis-dependent overlaps, parameterised
by the variable $t$. In contrast to the three rotated position bases
satisfying Eq.~\eqref{eq: CV overlap}, only two bases with identical
overlap exist. 

\subsection{$p$-adic MU bases}

\label{subsec: p-adic-MU-bases}

van Dam and Russell \citep{vandam+11} studied the properties of MU
bases in the infinite-dimensional Hilbert space $L^{2}(\mathbb{Q}_{p})$,
where $\mathbb{Q}_{p}$ is the field of $p$-adic numbers. The $p$-adic
numbers, which pervade the field of number theory, as well as other
areas of mathematics \citep{gouvea97}, have led to $p$-adic variants
of much of theoretical physics, including quantum mechanics and quantum
field theory \citep{vladimirov94}. 

In a non-relativistic setting, states take the form $\ket{\psi}=\int_{x\in\mathbb{Q}_{p}}\psi(x)\ket xd\nu(x)\in L^{2}(\mathbb{Q}_{p})$,
with $\nu$ being the Haar measure on $\mathbb{Q}_{p}$ \citep{vandam+11}.
At least $(p+1)$ MU bases in $L^{2}(\mathbb{Q}_{p})$ can be found
for $p>2$. The construction is a generalisation of Wootters' approach
to obtaining $(d+1)$ MU bases in $\mathbb{C}^{d}$ for prime-powers
$d$, as described in Appendix \ref{subsec:Odd-prime-powers}. The
original method relies on properties of quadratic Gauss sums over
finite fields, while the $p$-adic construction relies on similar
results for quadratic Gauss integrals over $\mathbb{Q}_{p}$. More
explicitly, suppose that $p$ orthonormal bases $\mathcal{B}_{b}=\{\ket{v_{b}}\}$
of $L^{2}(\mathbb{Q}_{p})$ are given, with $b\in\{0,\ldots,p-1\}$
and $v\in\mathbb{Q}_{p}$. Each basis is a collection of states
\begin{equation}
\ket{v_{b}}=\int_{x\in\mathbb{Q}_{p}}e^{bx^{2}+vx}\ket xd\nu(x)\,,\label{eq: p-adic states}
\end{equation}
with the function $e:\mathbb{Q}_{p}\rightarrow\mathbb{C}$ defined
in the $p$-adic sense as $e(x)=\exp(2\pi i\{x\})$ where $\{x\}$
is the fractional part of $x\in\mathbb{Q}_{p}$. In addition, there
is a basis $\mathcal{B}_{\infty}=\{\ket{v_{\infty}}\}$ defined by
the Fourier transform of the states in $\mathcal{B}_{0}$. Taken together,
one obtains a collection of $(p+1)$ MU bases in $L^{2}(\mathbb{Q}_{p})$. 

A similar construction for the Hilbert space $L^{2}(\mathbb{R})$
over the field of \emph{real} numbers rather than $\mathbb{Q}_{p}$
produces only three MU bases with basis-independent overlaps (cf.
Sec.~\ref{subsec: MUs for CVs}). The greater abundance of MU bases
in $L^{2}(\mathbb{Q}_{p})$ is due to a `{}`coarser'{}' $p$-adic
norm than the one for real numbers, as explained in Ref.~\citep{vandam+11}.
There are more possibilities to create numerically identical `{}`overlaps'{}'
in the $p$-adic equivalent of Eq.~\eqref{eq: CV overlap}. It is
likely that products of the MU bases of $L^{2}(\mathbb{Q}_{p})$ will
lead to MU bases in product spaces such as $L^{2}(\mathbb{Q}_{p})\otimes L^{2}(\mathbb{Q}_{p})$
but we are not aware of any published results. Relations between coherent
states over a field of $p$-adic numbers, canonical commutation relations
and MU bases have been explored in Ref.~\citep{zelenov23}.

\section{Summary and outlook}

\label{sec: Summary-and-Conclusions} 

This section summarises what we know about MU bases in composite dimensions
$d\notin\mathbb{PP}$ by making explicit the \emph{properties} which
a complete set in dimension $d=6$ would need to possess. We will
also collect \emph{strategies} to solve the existence problem which
have not yet been fully exhausted. Then, we formulate a number of
open problems related to the existence of complete sets. The problems
are interesting in themselves but do not necessarily contribute directly
to the existence problem in non-prime-power dimensions $d\notin\mathbb{PP}$.
We conclude with a few remarks and an outlook.

\subsection{Constraints on complete sets in $\mathbb{C}^{6}$}

\label{subsec:Properties-of-a-complete-set}

Sections \ref{sec: equivalent_formulations} and \ref{sec:rigorous_results_any_d}
describe structural features of complete sets of MU bases $\mathcal{M}_{d}=\{\mathcal{B}_{0},\ldots,\mathcal{B}_{d}\}$
for arbitrary dimensions $d\geq2$. We spell out some of these properties
for the case of $d=6$ and combine them with results which apply only
to bases in composite dimension (cf. Sec.~\ref{sec:rigorous_results_any_d})
or dimension six (cf. Sec.~\ref{sec:rigorous_results_d=00003D6}).
Hence, if a complete set of seven MU bases ${\cal M}_{6}=\{\mathcal{B}_{0},\ldots,\mathcal{B}_{6}\}$
were to exist, it would necessarily have the properties listed in
the corollary.
\begin{cor}
\label{cor:list}A complete set of seven MU bases $\mathcal{M}_{6}=\{\mathcal{B}_{0},\ldots,\mathcal{B}_{6}\}$
in the space $\mathbb{C}^{6}$ has the following properties:
\end{cor}
\begin{enumerate}
\item The seven bases $\mathcal{B}_{0},\ldots,\mathcal{B}_{6}$ can be written
in terms of \emph{six mutually unbiased Hadamard matrices} $\mathcal{M}_{6}=\{\mathbb{I}_{6}\}\cup\{H_{1},\ldots,H_{6}\}$,
i.e. the union of the identity and six complex Hadamard matrices $H_{i}$
of order six such that the product $H^{\dagger}_{i}H_{j}$ of any
two of them is another Hadamard matrix (see Equivalence \ref{equiv:hadamards}).
\item The matrix $H_{1}$ can be written in \emph{dephased} form with its
first row and column entries given by $H_{1j}=H_{i1}=1/\sqrt{6}$
for $i,j=1,\ldots,6$ (see Sec.~\ref{subsec:Equivalence-classes}).
\item Given \emph{five }mutually unbiased Hadamard matrices one can construct
a sixth one to obtain the complete set $\mathcal{M}_{6}$ (Thm.~\ref{thm:missing_basis}).
\item The set $\mathcal{M}_{6}$ defines a \emph{unitary operator basis}
of the space $\mathbb{C}^{6}$ which partitions into $d+1$ maximally
commuting classes (Equivalence \ref{equiv:commutingclasses}).
\item The bases in $\mathcal{M}_{6}$ provide a set of seven \emph{pairwise
orthogonal Cartan subalgebras} such that the simple Lie algebra $sl_{6}(\mathbb{C})$
admits an orthogonal decomposition (Equivalence \ref{equiv:OD}).
\item The vectors in $\mathcal{M}_{6}$ saturate the \emph{Welch bounds}
(\ref{eq:MUwelchbound1}) and (\ref{eq:MUwelchbound2}) (Equivalence
\ref{equiv:welchbounds}).
\item The vectors in $\mathcal{M}_{6}$ form a \emph{complex projective
2-design} and hence satisfy Eq.~\eqref{eq:MU-design-property}.
\item The average \emph{entanglement content} over the states in $\mathcal{M}_{6}$
is fixed by Thm.~\ref{thm:fixed_entanglement}.
\item The set $\mathcal{M}_{6}$ contains at most one product basis (Thm.~\ref{thm:noproductbases}).
\item No Hadamard matrix in $\mathcal{M}_{6}$ contains three or more columns
which are product states (Thm.~\ref{thm:product_constellations}).
\item The set $\mathcal{M}_{6}$ contains no Butson-type Hadamard matrices
$BH(6,12)$ (Thm.~\ref{thm:roots_of_unity}).
\item The six Hadamard matrices in $\mathcal{M}_{6}$ form a projective
toric 2-design of $P(T^{6})$ that is not a group (Thm.~\ref{thm:toric_designs}).
\item No more than three bases in $\mathcal{M}_{6}$ are monomial (Thm.~\ref{thm:monomial}).
\item Each Hadamard matrix in $\mathcal{M}_{6}$ has Schmidt rank $r>2$
(Thm.~\ref{thm:schmidt_rank}).
\item No Hadamard matrix in $\mathcal{M}_{6}$ contains a $3\times3$ submatrix
proportional to a unitary (Lem.~\ref{thm:subunitary}).
\item Each Hadamard matrix in ${\cal M}_{6}$ is equivalent to a Hadamard
in which all or none of the nine $2\times2$ blocks are\emph{ unitary}
(Thm.~\ref{thm:h2reducible}).
\item The set $\mathcal{M}_{6}$ does not include the isolated matrix $S_{6}$
(Thm.~\ref{thm:unextendible_pairs}) or the Fourier family $F^{(6)}_{6}$
(Thm.~\ref{thm:Fourierfamilyexclusion}).
\end{enumerate}
The first seven statements are generic in the sense that complete
sets of MU bases in any dimension have analogous properties. \emph{Mutatis}
\emph{mutandis}, property 8 holds for other composite dimensions,
as well as a weaker version of property 9 (Thm.~\ref{thm:productMUBs}).
The remaining eight properties have been derived for the case of $d=6$;
their generalizations to other dimensions may exist but are not known.
Furthermore, in dimensions where real Hadamard matrices exist, we
also know that a complete set will contain at most one of those (Thm.~\ref{thm:realhadamards}).

At this point, musings by Sylvester spring to mind when, in 1887,
he pondered the existence of odd perfect numbers:
\begin{quote}
$\ldots$ a prolonged meditation on the subject has satisfied me that
the existence of any one such {[}odd perfect number{]} – its escape,
so to say from the complex web of conditions which hem it in on all
sides – would be little short of a miracle \citep[pp. 152-3]{sylvester1887}.
\end{quote}
The problem is still open today. The existence problem for complete
sets of MU bases puts us in a similar situation, although perhaps
not (yet) as dire as Sylvester's case. A viable strategy to prove
their non-existence is to further ``hem them in on all sides'' so
that their construction becomes impossible. The following section
summarises some alternative proof strategies.

\subsection{Solution strategies}

\label{subsec:Solution-strategies}

\subsubsection*{Arbitrary composite dimension}

We now summarise several strategies which could be used to prove non-existence
of complete sets in any composite dimension, i.e. when $d\notin\mathbb{PP}$.

\begin{strategy}Equivalent formulations \end{strategy}

If true, each of the fourteen conjectures listed in Sec.~\ref{sec: equivalent_formulations}
ensures the non-existence of complete MU bases in any composite dimension.
For instance, one could attempt to show that no suitable partitioning
of a unitary operator basis (as defined in Sec.~\ref{subsec:Maximally-commuting-unitary})
exists, or that $sl_{d}(\mathbb{C})$ has no orthogonal decomposition
(cf. Sec.~\ref{subsec: Orthogonal decomp equivalence}). Alternatively,
a solution would follow if the Welch bounds were found not to be tight,
or if the optimal success probability of a p-QRAC is unachievable
(see Secs.~\ref{subsec: Welch bound equivalence} and \ref{subsec:QRAC},
respectively). One popular (numerical) strategy is to search for the
global minimum (or minima) of the function defined in Eq.~(\ref{eq: F encoding complete sets})
of Sec.~\ref{subsec:global_min}, with the main findings of this
approach summarized in Secs. \ref{subsec:nonexistence_of_quadruples}
and \ref{subsec:nonexistence_of_constellations}. Similar methods
can be applied to other functions optimised by MU bases, e.g., the
measure of non-commutativity introduced by Bandyopadhyay and Mandayam
\citep{bandyopadhyay+13} (cf. Sec.~\ref{subsec:Complementarity}).
While some of these techniques rely on numerics, others can be formulated
in terms of semidefinite programming, allowing for the possibility
of a rigorous non-existence proof.

\begin{strategy}Semidefinite programming\end{strategy}

Expressing the search for MU bases as a semi-definite program is a
promising computational approach to rigorously prove Conjecture~\ref{conj: Zauner}.
There are different methods to do so, depending on the type of optimisation,
and whether commutative or non-commutative polynomial optimisations
are considered.

One technique involves solving a system of coupled polynomials $p_{i}(\alpha)$
which represent the constraints for the MU basis vectors \citep{brierley+10,thiang10}
(see Sec.~\ref{subsec: polynomial equivalence}). The real \emph{commutative}
variables ${\alpha}\equiv(\alpha_{1},\ldots,\alpha_{n})$, with $\alpha_{i}\in[0,2\pi)$,
parameterise the candidate MU vectors. The idea is to minimise one
polynomial, $(p_{k}(\alpha))^{2}$, subject to the condition $p_{i}(\alpha)=0$
for all $i\neq k$. Since the polynomials are non-convex, Lasserre's
hierarchy of semidefinite programs can be applied \citep{lasserre01},
and for each level of relaxation the existence of MU bases is ruled
out or the result is inconclusive. If a positive global bound is obtained
by this minimisation, the coupled polynomial equations have no solution
and, therefore, no set of MU bases exists. This method successfully
confirms the non-existence of an MU triple containing the pair $\{\mathbb{I},S_{6}\}$,
where $S_{6}$ is the isolated Hadamard matrix, but has been unsuccessful
for the constellation $\{5^{3},1\}$ (see Sec.~\ref{subsec:nonexistence_of_constellations}
for the definition of constellations), due to the increase in computational
complexity \citep{brierley+10}.

One can also formulate the existence question in terms of a certain
$C^{*}$-algebra (Sec.~\ref{subsec:C*}) to provide a \emph{non-commutative}
polynomial optimisation strategy, leading to semidefinite programming
relaxations \citep{gribling21}. Proving infeasibility of these relaxations
provides a strategy to rule out the existence of $(d+1$) MU bases
in dimension $d$. In \citep{gribling21}, by exploiting symmetries
of the problem, this technique was used to show that $d+2$ MU bases
do not exist if $d\leq8$.

Another approach, which also involves a non-commutative polynomial
optimisation, uses the equivalence between QRACs and MU bases, as
described in Sec.~\ref{subsec:QRAC}. The idea is to apply a specific
semidefinite-programming hierarchy \citep{navascues15} to find an
upper bound for the optimal success probability of the $(4,2)^{6}\rightarrow1$
p-QRAC. If the bound falls below the optimal value defined in Eq.~\eqref{eq:QRACsucessprob}
one can conclude that four MU bases do not exist. As with the commutative
case, the method has been unsuccessful due to insufficient computational
resources. For example, the $k$-th level of the hierarchy for the
$(4,2)^{6}\rightarrow1$ p-QRAC requires roughly $2^{32k}$ bits of
memory, although a quadratic reduction is feasible by exploiting certain
symmetries of the QRAC \citep{aguilar18}.

\begin{strategy}Discretisation of the parameter space \end{strategy}

Thm.~\ref{thm:Fourierfamilyexclusion} of Sec.~\ref{sec:fourier_family}
states that any triple of MU bases containing the two-parameter Fourier
family is unextendible. The proof, which relies on a discretisation
of the parameter space and a computational search over a finite set
of vectors, can also be applied as a strategy to disprove the existence
of any four MU bases in $d=6$, as well as generalising to larger
dimensions. Suppose that four MU bases $\{\mathbb{I},H_{1},H_{2},H_{3}\}$
exist which can be parameterised in terms of $m$ phases, $\alpha\equiv(\alpha_{1},\ldots,\alpha_{m}),$
where $\alpha_{j}\in[0,2\pi)$. The discretisation approximates the
original MU bases by the set $\{\mathbb{I},\widetilde{H}_{1},\widetilde{H}_{2},\widetilde{H}_{3}\}$
where the phases are restricted to $N$-th roots of unity. For instance,
if the phase $\alpha_{j}\in[0,2\pi)$ initially lies in the interval
$[(2k-1)\pi/N,(2k+1)\pi/N)$, then we approximate $\alpha_{j}$ by
$\tilde{\alpha}_{j}=2\pi k/N.$ If we choose a sufficiently large
positive integer $N$, rigorous bounds of the errors given by the
inner products of the approximated states can be established. If the
errors lie outside these bounds, the original bases cannot be mutually
unbiased. The efficiency of this approach can be improved by considering
suitable MU constellations which will depend on fewer parameters (see
Sec.~\ref{subsec:nonexistence_of_constellations}).

\begin{strategy}Positive definite functions \end{strategy}

An approach exploiting positive definite functions (Sec.~\ref{subsec:delsarte})
on the group $G=U(d)$ of unitary matrices may provide a means to
upper-bound the cardinality of the maximal set of MU bases for a given
dimension. Introduced in \citep{matolcsi10,kolountzakis18}, the cardinality
of the maximal set of MU bases is bounded from above by $h(e)/\int hd\nu$
where $h:G\rightarrow\mathbb{R}$ is a positive definite function
with the properties \emph{(i)} $h(H)\leq0$ for all $d\times d$ Hadamard
matrices $H$; \emph{(ii)} $h(e)=0$; and \emph{(iii)} $\int hd\nu>0$,
where $\nu$ is the normalised Haar measure. 

The function given in Eq.~(\ref{eq:positivedefinite}) yields a tight
upper bound when the dimension is a prime power. The aim is to construct
a positive definite function which would lower this bound for composite
dimensions $d\notin\mathbb{PP}$.

\begin{strategy}Linear constraints \end{strategy}

Treating the columns of a set of mutually unbiased Hadamard matrices
as elements of the group $\mathbb{T}^{d},$ the usual polynomial conditions
of orthogonality and unbiasedness for a set of MU bases can be transformed
into linear constraints. The constraints, which are derived via Fourier
analytic methods, apply to the functions $E(\gamma)$ and $F(\gamma)$
(see Eqs.~(\ref{eq:fourier1-1})–(\ref{eq:fourier3-1}) of Sec.~\ref{subsec:linear-constraints})
and are linear by treating each function $E(\gamma)$ and $F(\gamma)$
as a variable for each value of $\gamma\in\mathbb{Z}^{d}$.

There are two ways to apply these constraints to prove that complete
sets do not exist. First, one can attempt to find some structural
results on the entries of the Hadamard matrices. A contradiction would
follow if the additional structure imposes a restriction on the complete
set that cannot hold. For instance, if $F(\rho)=d^{4}$ holds for
all permutations of $\rho=(d,-d,0,\ldots,0)\in\mathbb{Z}^{d},$ then
the matrix entries of a set of MU bases are all $d$-th roots of unity.
While this constraint has been used to classify all MU bases for $d\leq5$,
it is not expected to hold when $d=6$ \citep{matolcsi+12}.

Secondly, one can attempt to prove directly the non-existence of a
complete set by applying a linear programming code to show that the
constraints in Eqs.~(\ref{eq:fourier1-1})–(\ref{eq:fourier3-1})
do not hold. For instance, if maximising the variable $E(\gamma)$
for a given $\gamma\in\mathbb{Z}^{d}$ yields $E(\gamma)\geq d^{3},$
then a contradiction and ultimately a proof would follow. It is surmised
that a proof of Conjecture \ref{conj:fourier} will be an important
ingredient in this approach since it establishes an additional linear
constraint which is expected to hold for almost all $6\times6$ Hadamard
matrices. While the conjecture is true for the three-parameter Karlsson
family $K^{(3)}_{6},$ a linear program is unable to capitalise on
the additional constraint and fails to rule out the existence of a
complete set containing only Hadamard matrices from $K^{(3)}_{6}$
\citep{maxwell15}.

\subsubsection*{Dimension six}

The ultimate goal, to prove that a complete set of MU bases in $\mathbb{C}^{d}$
exists if and only if the dimension is a prime-power, would simultaneously
prove the conjectures of Sec.~\ref{sec: equivalent_formulations}.
A slightly more feasible challenge is to focus on dimension six rather
than arbitrary composite dimensions $d\notin\mathbb{PP}$. 
\begin{problem}
Show that no complete set of seven MU bases exists when $d=6$.
\end{problem}
Simplifying further, one could aim for a proof of Zauner's conjecture
given in Sec.~\ref{subsec:Composite-dimensions}.
\begin{problem}
Show that no set of four MU bases exists when $d=6$.
\end{problem}
This approach is, of course, sufficient to solve the existence problem
for complete sets in $\mathbb{C}^{6}$. Instead, it might be easier
to solve a pared-down version of this question by showing that any
three MU bases are strongly unextendible when $d=6.$ 
\begin{problem}
Show that no single vector is mutually unbiased to any triple of three
MU bases when $d=6$.
\end{problem}
An exhaustive classification of all MU triples would probably be necessary
for this method. To make progress, an exhaustive classification of
all pairs of MU bases in $d=6$ (Sec.~\ref{sec:pairs_of_MU_bases_C^6})
should be given.
\begin{problem}
Classify all complex Hadamard matrices of order six.
\end{problem}
This problem was addressed in Sec.~\ref{sec:pairs_of_MU_bases_C^6}
where all known Hadamard matrices of order six were reported: an isolated
matrix $S_{6}$, a three-parameter family $K^{(3)}_{6}$ and a four-parameter
family $G^{(4)}_{6}$. It is expected that any Hadamard matrix is
equivalent to a member of these families, but a proof that this list
is exhaustive remains elusive (Conjecture~\ref{conj:hadamardclassification}).
The structure of the four-parameter family and its relation to $K^{(3)}_{6}$
is not well understood; for instance, does the general construction
presented in \citep{szollosi12} include $K^{(3)}_{6}$ as a subfamily?
Even simpler problems remain open, such as the question of whether
$F_{6}$ is a member of a four-parameter family (although substantial
evidence points towards this being true; cf. Appendix~\ref{sec:Complex-Hadamard-matrices}).

Given an exhaustive list, all candidates for the six MU Hadamard matrices
would be known. Still, a non-trivial task remains, namely to identify
all triples of MU bases in dimension six or, equivalently, to identify
all pairs of Hadamard matrices $H_{1}$ and $H_{2}$ such that $H^{\dagger}_{1}H_{2}$
is another Hadamard matrix (see Sec.~\ref{sec:triples_of_MU_bases}).
\begin{problem}
Find all pairs of $6\times6$ complex Hadamard matrices $H_{1}$ and
$H_{2}$ such that their product $H^{\dagger}_{1}H_{2}$ is another
Hadamard matrix.
\end{problem}
Conjecture \ref{conj:matolcsi} suggests a solution to this problem,
with the Fourier family $F^{(2)}_{6}$, its transpose, and Szöllősi's
family $X^{(2)}_{6}$ proposed as the only viable candidates. 

If these problems are still too difficult, one could perhaps try to
complete Table \ref{tab:triples} by providing a rigorous proof confirming
the numerical evidence that certain families are unextendible.
\begin{problem}
Show that a given pair of MU bases in $\mathbb{C}^{6}$ does not extend
to a triple (or quadruple) of MU bases.
\end{problem}
Examples would include a proof that, for certain parameter values,
$\{\mathbb{I},M^{(1)}_{6}\}$, $\{\mathbb{I},K^{(2)}_{6}\}$ or $\{\mathbb{I},K^{(3)}_{6}\}$
do not extend to a triple, in analogy with the isolated case $\{\mathbb{I},S_{6}\}$.
A general analytic proof similar to Thm.~\ref{thm:Fourierfamilyexclusion}
would represent a major step forward.

\subsection{Related open problems}

For general composite dimensions, and for the case of continuous variables,
there are several open problems related indirectly to the existence
of MU bases. Some are mathematical in nature, others concern the physics
behind MU bases.

For even dimensions such as $d=2\times p$ with $p$ an odd prime,
Thm.~\ref{thm:reduce_to_primes} yields only three MU bases, even
when the dimension $d$ is arbitrarily large. Do more effective construction
methods exist in this situation, even if they do not lead to complete
sets?
\begin{problem}
Can one improve the lower bound provided in Thm.~\ref{thm:reduce_to_primes},
to find larger sets of MU bases in composite dimensions?
\end{problem}
Theorem \ref{thm:productMUBs} states that the lower bound cannot
be improved if one considers product bases only. However, a method
based on Latin squares and maximally entangled bases successfully
improves the bound in certain square dimensions, as discussed in Sec.~\ref{subsec:Maximally-entangled-bases}
and Thm.~\ref{thm:wocjanMUBs}.

One can also attempt to extend the minimal set of MU bases used to
derive the lower bound in Thm.~\ref{thm:reduce_to_primes}.
\begin{problem}
Is the set of MU product bases from Thm.~\ref{thm:reduce_to_primes}
unextendible?
\end{problem}
The solution is known for dimension six: any set of three MU product
bases is \emph{strongly unextendible} (Sec.~\ref{subsec:product_bases}).
We may also ask a similar question for the Latin square construction
in square dimensions.
\begin{problem}
Is the set of $(\mu+2)$ MU bases from Thm.~\ref{thm:wocjanMUBs}
unextendible?
\end{problem}
One construction method of MU bases proceeds by partitioning a subset
of the elements of a nice error basis into maximally commuting classes
(cf. Sec.~\ref{subsec:Nice-error-bases}). According to Thm.~\ref{thm:niceerrorbasis},
a set of $\min{}_{i}(p^{n_{i}}_{i}+1)$ weakly unextendible ``nice''
MU bases results for $d=p^{n_{1}}_{1}\ldots p^{n_{r}}_{r}\notin\mathbb{PP}$.
\begin{problem}
For $d=p^{n_{1}}_{1}\ldots p^{n_{r}}_{r}\notin\mathbb{PP}$, is a
set of $\min_{i}\left(p^{n_{i}}_{i}+1\right)$ nice MU bases \emph{strongly}
unextendible?
\end{problem}
This relates to another open problem, namely Conjecture~\ref{conj:unextendibleHW},
which predicts that the eigenbases of a smaller set of Heisenberg-Weyl
operators (defined globally rather than as a tensor product of each
system) are unextendible. The conjecture has been confirmed for $d\leq15$
(Sec.~\ref{subsec:Unextendible-MU-bases}). Recall, from Sec.~\ref{subsec:Unextendible-MU-bases},
that if a set of weakly unextendible nice MU bases is sufficiently
large, it cannot form a complete set.

MU bases also give rise to an interpretational question which asks
about the consequences of inequivalent pairs of complementary observables
for the description or the properties of a physical system (see Sec.~\ref{subsec:Equivalence-classes}).
\begin{problem}
What does it mean for a physical system if inequivalent complementary
pairs of observables exist? 
\end{problem}
Since a continuous family of inequivalent Hadamard matrices exists
already for $d=4$, this question can be addressed in full for a quantum
system composed of two qubits. So far, inequivalent \emph{pairs} are
not known to exhibit operational differences. For example, methods
of self-testing (cf. Sec.~\ref{subsec:Quantum-correlations}) are
unable to distinguish between complementary pairs. However, for \emph{sets
of $\mu$ }MU\emph{ bases} in $\mathbb{C}^{d}$, with $2<\mu<d$,
there are some known consequences. Different sets affect the success
of a p-QRAC protocol (Sec.~\ref{subsec:QRAC application}) and the
ability to detect (bound) entangled states (Sec.~\ref{subsec:entanglement-detection}).
Furthermore, certain sets exhibit more measurement incompatibility
than others (Sec.~\ref{subsec:inc}). Curiously, these discrepancies
no longer hold for complete sets, which suggests another question.
\begin{problem}
What are the consequences for a physical system if inequivalent complete
sets of MU bases exist?
\end{problem}
So far, the existence of inequivalent complete sets (see Appendix~\ref{subsec:other_constructions})
is known for \emph{prime-power} dimensions $d=p^{n}$ with $p\in\mathbb{P}$
and $n>1$. No examples have been found when the dimension is a \emph{prime}
number, $d\in\mathbb{P}$.
\begin{problem}
Do inequivalent complete sets of MU bases exist in prime dimensions?
\end{problem}
MU bases for continuous variables bring their own existence problems.
For example, the maximal number of MU bases with identical overlaps
(see Sec.~\ref{subsec: MUs for CVs}) is not known even in the simplest
case of a single continuous-variable pair. 
\begin{problem}
For a quantum particle with one degree of freedom, do more than three
MU bases with basis-independent overlaps exist? 
\end{problem}
A somewhat less general question concerns the extendability of the
``standard triple'' of MU bases by a single state.
\begin{problem}
For a quantum particle with one degree of freedom, is the Heisenberg-Weyl
type triple of MU bases with basis-independent overlaps strongly unextendible?
\end{problem}
A positive answer would establish a desirable similarity between the
three MU bases of Heisenberg-Weyl type for a discrete and for a continuous
degree of freedom, respectively.

\subsection{Conclusions}

The concept of mutual unbiasedness in linear spaces with an inner
product turns out to be \emph{structurally rigid}: the definition
of MU bases makes sense for real, complex, quaternionic and even $p$-adic
numbers (cf. Sec.~\ref{sec: Modifications-of-the-problem}). In the
standard setting of the space $\mathbb{C}^{d}$, a parameter count
reveals that the existence of complete MU sets in prime-power dimensions
is, in fact, a remarkable surprise: the number of constraints exceeds
the number of free parameters for $d>2$, both in composite \emph{and}
prime-power dimensions (see Sec.~\ref{subsec:Composite-dimensions}).
To understand the (im-) possibility to solve the coupled polynomial
equations defining the existence problem (cf. Equivalence \ref{equiv: coupled polynomials})
is probably at the heart of the matter. The number-theoretical identities
which entail solutions in prime-power dimension are absent in composite
dimensions. At the same time, the existence problem of complete sets
of MU bases in a given composite dimension $d\notin\mathbb{PP}$ is
\emph{decidable} since an algorithm detecting MU bases (or their absence)
is known (cf. the end of Sec.~\ref{subsec:Composite-dimensions}).

Being set in a complex inner product space, quantum theory is a natural
home for mutually unbiased bases. However, it is important to acknowledge
that they also arise when studying complex-valued \emph{classical}
signals or within Fourier theory (cf. Sec.~\ref{subsec:Motivation}).
Nevertheless, the existence problem is particularly important in quantum
theory and quantum information since complete sets of MU bases often
represent ``extreme'' cases: they maximise or minimise uncertainty
relations or entropies, they exhibit maximal non-classicality in the
form of measurement incompatibility and coherence, and they optimise
information processing protocols making them suitable for benchmarking
(cf. Sec.~\ref{sec: MU-bases-in-QT}).

The existence problem of complete sets clearly represents yet another
instance where quantum theory meets number theory (see e.g. \citep{shor94,berry99,mack02}).
For discrete structures such as SICs (cf. Sec.~\ref{subsec:MU-bases-and-SICs}),
the link is even more explicit and possibly ``mathematically deeper''
(cf. \citep{Appleby_2018,appleby22}). The problem is also linked
with the tensor-product structure used to describe composite quantum
systems. There are, in fact, alternatives to the commonly used tensor
product \citep{Kl_y_1987} such as the ``maximal'' or ``minimal''
tensor product, which differ from the standard one used in quantum
theory that sits between the extreme cases.

Mapping the search for complete sets of MU bases onto equivalent problems
(cf. Sec.~\ref{sec: equivalent_formulations}) has led to \emph{new
perspectives} and confirms the \emph{genuine difficulty} of the open
question. For example, rephrasing the existence problem in terms of
orthogonal decompositions of $sl_{d}(\mathbb{C})$ has revealed the
existence of a single family of mutually unbiased bases which—although
it has not yet been fully characterised analytically—is conjectured
to contain all previously known examples, except for isolated ones.
The equivalence of a complete set of MU bases with a complex projective
2-design implies that their entanglement content is restricted in
any composite dimension. Furthermore, research into MU bases has (re-)
established interest in \emph{complex} Hadamard matrices which have
become a mathematical topic of its own.

This review collects a considerable number of properties of mutually
unbiased bases in composite dimensions. Nevertheless, no overarching
or underlying structure emerges, not even in dimension six (cf. Sec.~\ref{subsec:Properties-of-a-complete-set}),
the composite dimension for which most is known. A definite \emph{lack
of recognisable patterns} reflects our lack of understanding. Attempts
to simplify or modify the main existence problem—by focussing on a
particular dimension or by restricting the candidate states of MU
bases to product form, for example—have led to a plethora of apparently
unrelated observations, without shedding much light on the original
existence problem.\footnote{``There are thing{[}s{]} we know that we know. There are known unknowns.
That is to say there are things that we now know we don't know. But
there are also unknown unknowns. There are things we don't know we
don't know'' \citep{Rumsfeld}.} We continue to divide but do not conquer. From a mathematical point
of view, our current fragmented knowledge about mutually unbiased
bases makes one think of the problem of radicals before Galois' solution
by introducing entirely new mathematical concepts \citep{Giroud2019}.

On the basis of the results gathered in these pages, we expect the
generalisation of \emph{Zauner's Conjecture} to be true: complete
sets of mutually unbiased bases do not exist in dimension six or any
other composite dimensions. While the search for a mathematical proof
of Zauner's conjecture continues, we suggest pondering the relevance
of MU bases for the description of nature: would existence or non-existence
of complete sets have far-reaching consequences in quantum theory,
or in some other context? Are complete sets of MU bases ``luxury
items'' which are nice to have in prime-power dimensions but not
really essential otherwise?

\section*{Acknowledgements}

The authors would like to thank M. Grassl and M. Matolcsi for their
valuable feedback on a draft of the manuscript, and F. Szöllősi for
insightful comments regarding Hadamard matrices. D.M. has received
funding from the European Union's \emph{Horizon 2020} research and
innovation programme under the Marie Skłodowská-Curie grant agreement
No 663830, and PNRR MUR Project No PE0000023-NQSTI.

\bibliographystyle{quantum-alpha}
\bibliography{ReferencesForD=6Review}

\appendix

\section{Constructions of complete sets}

\label{sec: complete sets in pp dimensions}

Early constructions of complete sets of MU bases covering prime dimensions
$d=p\in\mathbb{P}$ were provided in the early eighties by Alltop
and Ivanović \citep{alltop80,ivanovic81}, at a time when MU bases
had not been formally defined. Alltop's implicit construction for
prime dimensions $p\geq5$ was given in terms of complex periodic
sequences (see Sec.~\ref{sec: conceptual history}). A year later
Ivanović discovered these bases in the context of optimal quantum
state tomography, as described in Sec.~\ref{subsec: Quantum-state-reconstruction}.
They can be written in the form $\mathcal{B}_{b}=\{\ket{v_{b}}\},b,v\in\mathbb{Z}_{p}$,
with basis vectors,
\begin{equation}
\ket{v_{b}}=\frac{1}{\sqrt{p}}\sum^{p-1}_{k=0}e^{(2\pi i/p)(bk^{2}+vk)}\ket k\,.
\end{equation}
Here the label $b$ denotes different bases, $v$ the vectors within
each basis, and $\ket k$ the elements of the computational basis.
Together with the standard basis they form a set of $(p+1)$ MU bases.
Throughout this appendix, $\mathcal{B}_{0}$ will not necessarily
denote the standard basis.

Ivanović's construction was extended by Wootters and Fields \citep{wootters+89}
to include (both odd and even) powers of primes, $d=p^{n}\in\mathbb{PP}$,
using field extensions $\mathbb{F}_{p^{n}}$. For odd prime-powers
we will follow the method of Wootters and Fields\emph{,} but for even
prime-powers we will review a simpler construction by Klappenecker
and Rötteler \citep{klappenecker+04} based on Galois rings.

We will summarise several other construction methods, roughly in chronological
order, and describe why they fail in non-prime-power dimensions $d\notin\mathbb{PP}$.
Our aim is to give a flavour of the diverse (but not exhaustive) methods
used to construct MU bases.

\subsection{Odd prime-powers}

\label{subsec:Odd-prime-powers}

Since the construction of Wootters and Fields \citep{wootters+89}
relies on finite fields $\mathbb{F}_{p^{n}}$, we will briefly describe
some of their basic properties. To construct the field $\mathbb{F}_{p^{n}}$
from the field $\mathbb{F}_{p}$ one starts by adding an element $\alpha$
which is a root of an \emph{irreducible} polynomial $f(x)$ of degree
$n$, with coefficients from the field $\mathbb{F}_{p}$. A polynomial
of order $n$ is irreducible if it cannot be expressed as the product
of two polynomials over $\mathbb{F}_{p}$ of smaller degree. The remaining
elements of $\mathbb{F}_{p^{n}}$ are then given by the set of all
polynomials of the form $g(\alpha)=g_{0}+g_{1}\alpha+\ldots+g_{n}\alpha^{n}$
were $g_{i}\in\mathbb{F}_{p}$. In this representation we say that
$\{\alpha,\alpha^{2},\ldots,\alpha^{n}\}$ is a \emph{basis} of $\mathbb{F}_{p^{n}}$.

Addition and multiplication within the field are defined by the usual
addition and multiplication of polynomials \emph{modulo} the irreducible
polynomial. In the construction of the extension $\mathbb{F}_{p^{n}}$
one is free to choose any irreducible polynomial of order $n$ and
any root $\alpha$ of the polynomial as they result in the same field
up to isomorphism. If, however, one chooses the irreducible polynomial
to be a \emph{primitive} polynomial $p(x)$, then all non-zero elements
of the field can be generated through powers of $\gamma$, i.e. $\mathbb{F}_{p^{n}}\setminus\{0\}=\{\gamma^{0},\gamma,\ldots,\gamma^{p^{n}-2}\}$,
where $p(\gamma)=0$. A polynomial $p(x)$ of degree $n$ is called
primitive if the smallest positive integer $m$ for which $p(x)$
divides $(x^{m}-1)$ is $m=p^{n}-1$.

For powers of odd primes the construction leads to a set of $d=p^{n}$
bases $\mathcal{B}_{b}=\{\ket{v_{b}}\}$, $b,v\in\mathbb{F}_{p^{n}}$,
consisting of the the vectors 
\begin{equation}
\ket{v_{b}}=\frac{1}{\sqrt{p^{n}}}\sum_{k\in\mathbb{F}_{p^{n}}}e^{(2\pi i/p)\textrm{Tr}[bk^{2}+vk]}\ket k\,.\label{eq:MUvectorsd=00003Dpn}
\end{equation}
The trace of $\alpha\in\mathbb{F}_{p^{n}}$ is defined as $\mbox{Tr }\alpha=\alpha+\alpha^{p}+\ldots+\alpha^{p^{n-1}}$,
and $\ket k$ again denotes elements of the standard basis. The bases
are pairwise mutually unbiased due to the quadratic Gaussian sum over
finite fields taking values independent of both $b$ and $v$, i.e.
\begin{equation}
\left|\sum_{k\in\mathbb{F}_{p^{n}}}e^{(2\pi i/p)\textrm{Tr}[bk^{2}+vk]}\right|=\sqrt{p^{n}}\,,\label{eq:numbertheoryresult}
\end{equation}
if $p$ is an odd prime and $b\neq0$. Thus, together with the standard
basis, the vectors in Eq.~\eqref{eq:MUvectorsd=00003Dpn} form a
complete set of $\left(p^{n}+1\right)$ MU bases.

This approach fails for $p=2$ since Eq.~\eqref{eq:numbertheoryresult}
does not hold. The bases defined through Eq.~\eqref{eq:MUvectorsd=00003Dpn}
can also be written in a form without explicitly referring to the
field elements, as shown in Ref. \citep{wootters+89}. This specific
representation can then be used to write out explicitly the complete
set of MU bases for even dimensions $d=2^{n}$. Galois rings provide
an alternative way to construct complete sets in even prime-powers.

\subsection{Even prime-powers}

\label{subsec:Even-prime-powers}

For spaces $\mathbb{C}^{d}$ with even prime-power dimensions $d=2^{n}$,
a construction of complete sets of MU bases using Galois rings was
found by Klappenecker and Rötteler \citep{klappenecker+04}. In particular,
they employ properties of exponential sums over $GR(4,n)$ (the Galois
ring of degree $n$ over $\mathbb{Z}/4\mathbb{Z}$) to obtain sets
consisting of $(2^{n}+1)$ MU bases.

The Galois ring $GR(4,n)$ is the unique Galois extension of $\mathbb{Z}/4\mathbb{Z}$
(the residue class ring of integers modulo 4) of degree $n$. The
$4^{n}$ elements of $GR(4,n)$ can be written in the form $r=b+2v$
where $b$ and $v$ are elements of the Teichmüller set $\mathcal{T}_{n}$
of order $(2^{n}-1)$. For $b\in\mathcal{T}_{n}$, elements of $\mathcal{B}_{b}$
are given by
\begin{equation}
\ket{v_{b}}=\sum_{k\in\mathcal{T}_{n}}\frac{1}{\sqrt{2^{n}}}e^{(2\pi i/4)\textrm{Tr}[(b+2v)k]}\ket k\,.\label{eq:klapp_bases}
\end{equation}
Together with the standard basis they provide the desired set of $(2^{n}+1)$
MU bases of $\mathbb{C}^{2^{n}}$. The trace map $\mbox{Tr}:GR(4,n)\rightarrow\mathbb{Z}/4\mathbb{Z}$
is defined as $\mbox{Tr }x=\sum^{n-1}_{j=0}\sigma^{j}(x)$ where $\sigma$
is the automorphism $\sigma(b+2v)=b^{2}+2v^{2}$. The proof follows
straightforwardly by replacing Eq.~\eqref{eq:numbertheoryresult}
with the identity 
\begin{equation}
\left|\sum_{x\in\mathcal{T}_{n}}e^{(2\pi i/4)\textrm{Tr}[rx]}\right|=\left\{ \begin{array}{lll}
0 & r\in2\mathcal{T}_{n},\,\,r\neq0;\\
2^{n} & r=0;\\
\sqrt{2^{n}} & \text{otherwise}.
\end{array}\right.\label{eq:numbertheoryresult2}
\end{equation}

The relations \eqref{eq:numbertheoryresult} and (\ref{eq:numbertheoryresult2})
have no equivalents in composite cases $d\notin\mathbb{PP}$. Moreover,
it has been shown that a generalisation of Eqs.~\eqref{eq:MUvectorsd=00003Dpn}
and \eqref{eq:klapp_bases} to any finite ring leads to no more than
$(p^{n_{1}}_{1}+1)$ MU bases in $\mathbb{C}^{d}$, where $p^{n_{1}}_{1}$
is the smallest prime-power factor of a given composite number $d$
\citep{archer05}.

\subsection{Generalised Alltop construction}

\label{subsec:Alltop's-construction}

Alltop's construction of complete sets in terms of complex periodic
sequences for prime dimensions $p\geq5$ (mentioned in Appendix~\ref{subsec:Odd-prime-powers})
has been generalised by Klappenecker and Rötteler in \citep{klappenecker+04}
to prime-power dimensions $d=p^{n}$ with $p\geq5$. For a finite
field $\mathbb{F}_{p^{n}}$ of characteristic $p\geq5$, the bases
$\mathcal{B}_{b}=\{\ket{v_{b}}:v\in\mathbb{F}_{p^{n}}\}$ are expressed
as 
\begin{equation}
\ket{v_{b}}=\frac{1}{\sqrt{p^{n}}}\sum_{k\in\mathbb{F}_{p^{n}}}e^{(2\pi i/p)\textrm{Tr}[(k+b)^{3}+v(k+b)]}\ket k\,,\label{alltop_bases}
\end{equation}
where $b\in\mathbb{F}_{p^{n}}$ and $\ket k$ are standard basis elements
of $\mathbb{C}^{p^{n}}$. Together with the standard basis they form
a set of $(p^{n}+1)$ MU bases. The proof follows from a result on
Weyl sums over finite fields, which is applicable only when $p$ is
odd.

\subsection{Unitary operator bases}

\label{subsec:constructions_on_unitary_operator_bases}

We now exploit the link between maximally commuting operator classes
and MU bases (cf. Equivalence \ref{equiv:commutingclasses} in Sec.~\ref{subsec:Maximally-commuting-unitary})
to construct complete sets using the Heisenberg-Weyl group, following
Bandyopadhyay \emph{et al.'s }approach \citep{bandyopadhyay+02}.

To recap, we require a unitary operator basis of $\mathbb{M}_{d}(\mathbb{C})$
with $d^{2}$ elements that partition into $(d+1)$ subsets, each
containing a maximal set of commuting unitary matrices. One such candidate
is the Heisenberg-Weyl group generated from the cyclic\emph{ shift}
(modulo $d$) and \emph{phase} operators $X$ and $Z$, respectively,
defined as 
\begin{equation}
X\ket k=\ket{k+1}\qquad\mbox{and}\qquad Z\ket k=\omega^{k}\ket k\,,\label{HWoperators}
\end{equation}
where $\omega=e^{2\pi i/d}$ is a $d$-th root of unity and $\{\ket k\}$
is the standard basis with $k=0,\ldots,d-1$. An overview of the role
of this group in the context of MU bases can be found in Ref. \citep{bengtsson17}.

For prime dimensions $d=p$, the group elements $X^{k}Z^{l}$ can
be split into $(p+1)$ cyclic subgroups generated by the operators
$Z,X,XZ,\ldots,XZ^{p-1}$, which form the commuting subclasses required.
The eigenstates of these operators provide the $(p+1)$ MU bases.
Each basis can be written in a compact form, with the eigenstates
of $Z$ giving the standard basis $\{\ket 0,\ldots,\ket{p-1}\}$ and
the remaining bases $\mathcal{B}_{b}=\{\ket{v_{b}}\}$ consisting
of the eigenstates of $X(Z)^{b}$, for $0\leq b,v\leq p-1$, as 
\begin{equation}
\ket{v_{b}}=\frac{1}{\sqrt{p}}\sum^{p-1}_{k=0}(\omega^{v})^{p-k}(\omega^{-b})^{s_{k}}\ket k\,,
\end{equation}
where $s_{k}=k+\ldots+(p-1)$.

The generalisation to the case of prime powers, $d=p^{n}$, uses Heisenberg-Weyl
unitary operators acting on the Hilbert space $\mathbb{C}^{p}\otimes\ldots\otimes\mathbb{C}^{p}$,
which are of the form 
\begin{equation}
X(k_{1}\ldots k_{n})Z(l_{1}\ldots l_{n})\equiv X^{k_{1}}Z^{l_{1}}\otimes\ldots\otimes X^{k_{n}}Z^{l_{n}},\label{pauli_group_elements}
\end{equation}
where $k_{j},l_{j}\in\mathbb{F}_{p}$. The elements $\omega^{j}X(k_{1}\ldots k_{n})Z(l_{1}\ldots l_{n})$
form the Heisenberg-Weyl group (or generalised Pauli group $\mathbb{P}_{j}(p,n)$
) of unitary operators on $\mathbb{C}^{p}\otimes\ldots\otimes\mathbb{C}^{p}$
for some integer $j>0$.

One can form a unitary operator basis of the set of $p^{n}\times p^{n}$
complex matrices $\mathbb{M}_{p^{n}}(\mathbb{C})$ from the elements
of (\ref{pauli_group_elements}). These operators split into commuting
classes $\mathcal{C}_{1},\ldots,\mathcal{C}_{d}$, and we associate
with each class a $(n\times n)$ matrix $A_{b}$ over $\mathbb{Z}_{p}$
with $b=1,\ldots,p^{n}$. By extending each matrix to a matrix $(\mathbb{I}_{n},A_{b})$
of size $(n\times2n)$, with $\mathbb{I}_{n}$ the identity, the rows
of $(\mathbb{I}_{n},A_{b})$ are chosen to represent the exponents
$(k_{1},\ldots,k_{n},l_{1},\ldots,l_{n})$ of the operators (\ref{pauli_group_elements})
in class $\mathcal{C}_{b}$. Thus, each class contains $n$ operators.

The MU bases are built by making use of the following result: let
$A_{b}$ and $A_{b'}$ be a symmetric pair of $(n\times n)$ matrices
over $\mathbb{Z}_{p}$ such that $\det(A_{b}-A_{b'})\neq0$ mod $p$,
then $A_{b}$ and $A_{b'}$ correspond to a pair of MU bases in $\mathbb{C}^{p^{n}}$.
Thus, by including the standard basis, a set of matrices $\{A_{1},\ldots,A_{p^{n}}\}$
satisfying these properties corresponds to a complete set of MU bases.
To construct these matrices we require $n$ symmetric nonsingular
matrices $B_{1},\ldots,B_{n}\in\mathbb{M}_{n}(\mathbb{C})$ such that
the matrix $\sum^{n}_{j=1}b_{j}B_{j}$ is nonsingular for every nonzero
vector $(b_{1},\ldots,b_{n})\in\mathbb{Z}^{n}_{p}$. Then, the $p^{n}$
matrices $A_{b}=\sum b_{j}B_{j}$, $(b_{1},\ldots,b_{n})\in\mathbb{Z}^{n}_{p}$,
form a set of matrices $(0_{n},\mathbb{I}_{n})$, $(\mathbb{I}_{n},A_{1})$,
$\ldots$, $(\mathbb{I}_{n},A_{p^{n}})$ which yield the $\left(p^{n}+1\right)$
MU bases.

A general method to find the symmetric nonsingular matrices $B_{1},\ldots,B_{n}$
has been given by Wootters in \citep{wootters+89}. In particular,
let $\gamma_{1},\ldots,\gamma_{n}$ be a basis of $\mathbb{Z}^{n}_{p}$
as a vector space over $\mathbb{Z}_{p}$ (i.e. any element $c\in\mathbb{Z}^{n}_{p}$
can be expressed as $c=\sum_{i}c_{i}\gamma_{i}$ where $c_{i}\in\mathbb{Z}_{p}$).
Then, any element $\gamma_{i}\gamma_{j}\in\mathbb{Z}^{n}_{p}$ can
be written uniquely as $\gamma_{i}\gamma_{j}=\sum^{n}_{l=1}b^{l}_{ij}\gamma_{l}$,
with $b^{l}_{ij}$ the $ij$-th element of $B_{l}$.

Unsurprisingly, this approach fails for dimensions $d\notin\mathbb{PP}$,
as some of the required properties no longer hold. For example, symmetric
matrices satisfying $\det(A_{b}-A_{b'})\neq0$ mod $p$ no longer
correspond to pairs of MU bases in $\mathbb{C}^{p^{n}}$. At best,
one can construct $(p^{n_{1}}_{1}+1)$ MU bases of $\mathbb{C}^{d}$
in this way, where $p^{n_{1}}_{1}$ is the smallest prime-power factor
of $d$. In Ref. \citep{bengtsson17}, the Heisenberg-Weyl group and
its commuting subclasses are described in terms of flowers and their
petals.

\subsection{Discrete geometries}

\label{sec:affineplanes}

Affine planes display structural similarities to sets of MU bases.
These discrete geometric structures consist of points and lines that
satisfy the following three axioms: (\emph{i}) any two points have
exactly one line in common; (\emph{ii}) for any line and additional
point there is a unique line through this point and disjoint (parallel)
from the given line; and finally: (\emph{iii}) there exist at least
three non-collinear points. This being the case, an affine plane of
order $d$ contains $d^{2}$ points and $d(d+1)$ lines, with each
line containing $d$ points. The lines of an affine plane can be partitioned
into $(d+1)$ sets, called \emph{striations}, each containing $d$
parallel lines. Any two non-parallel lines intersect at only one point.

An affine plane can also be represented by a set of orthogonal Latin
squares. A Latin square of order $d$ is an array of $d\times d$
integers ranging from $0$ to $(d-1)$ such that each number appears
exactly once in each row and column. Two Latin squares $L$ and $L^{\prime}$
are orthogonal if all ordered pairs of elements $(L_{ij},L^{\prime}_{ij})$
are distinct. \emph{Pairs} of orthogonal Latin squares exist for all
$d>2$ and $d\neq6.$ For every prime and prime-power $d$, there
exist $(d-1$) mutually orthogonal (MO) Latin squares, a consequence
of the existence of affine planes \citep{dembowski97}. When the order
$d$ of an affine plane equals either $(1\mod 4)$ or $(2\mod 4)$,
the Bruck-Ryser theorem imposes the restriction that $d$ must be
the sum of two squares \citep{bruck+49}. In this way, the existence
of affine planes in an infinite number of composite dimensions, including
orders 6 and 14, can be ruled out. Furthermore, a computation-based
proof has shown there is no affine plane of order 10 \citep{lam+89,lam91}.

A set of $\ell$ mutually orthogonal Latin squares can be extended
to an \emph{augmented} set of MO Latin squares which include two additional
(non-Latin) squares $A$ and $B$ defined via $A_{ij}=i$ and $B_{ij}=j$.
An example of an augmented set of MO Latin squares in $d=3$ is given
by 
\begin{equation}
\begin{array}{ccc}
0 & 1 & 2\\
0 & 1 & 2\\
0 & 1 & 2
\end{array}\quad\begin{array}{ccc}
0 & 0 & 0\\
1 & 1 & 1\\
2 & 2 & 2
\end{array}\quad\begin{array}{ccc}
0 & 1 & 2\\
2 & 0 & 1\\
1 & 2 & 0
\end{array}\quad\begin{array}{ccc}
0 & 1 & 2\\
1 & 2 & 0\\
2 & 0 & 1
\end{array}
\end{equation}
where the last two squares are Latin and all four are mutually orthogonal.
Each square from an augmented set of MO Latin squares corresponds
to a striation of an affine plane, with the points on a line corresponding
to distinct integers. 

An augmented set of orthogonal Latin squares of size $(\ell+2)$ is
equivalent to a combinatorial design known as a \emph{net}. An algorithm
which translates an augmented set of $(\ell+2)$ MO Latin squares
to a \emph{net design} is given in \citep{paterek+09}. The resulting
net design can be written as a table with $(\ell+2)$ rows containing
$d^{2}$ integers, split into $d$ cells of size $d$. The numbers
contained in one cell of a given row are distributed evenly among
all cells of any other row. Then, a correspondence between MU bases
and net designs emerges from linking the cells and rows of the net
to the exponents of the Heisenberg-Weyl group elements generated by
the shift and phase operators $X$ and $Z$, respectively, defined
in Eq.~\eqref{HWoperators}. Each row of the net corresponds to a
subclass of commuting operators from the Heisenberg-Weyl group, and
the common eigenstates associated with each commuting set form $(d+1)$
MU bases, as described in Appendix \ref{subsec:constructions_on_unitary_operator_bases}. 

Since affine planes also exist in prime-power dimensions, one can
generalise the algorithm to construct MU bases for these cases too
\citep{paterek+09}. However, it was pointed out in \citep{hall11}
that this approach only works for \emph{some} sets of MO Latin squares,
and the construction is ultimately based on Galois fields. In particular,
it reproduces the method based on Wigner functions described by Gibbons
\emph{et al.} \citep{gibbons+04}. Thus, evidence suggests the link
between MU bases and affine planes is due to their association with
Galois fields rather than any other, possibly deeper, underlying connection.

The process of constructing MO Latin squares from sets of MU bases
can also be reversed, an idea is developed in \citep{hall+10}, where
complete sets of MU bases for odd prime-powers are used to generate
complete sets of MO Latin squares. Two methods are highlighted, one
using linear combinations of vectors from MU bases to construct MO
Latin squares, while the other one relies on so-called \emph{planar}
functions. More similarities between MU bases and affine planes are
discussed briefly in Appendix \ref{subsec:affineplanes_mubs_sics}.
Latin and \emph{Quantum} Latin squares, which are related to maximally
entangled MU bases in square dimensions \citep{musto15,musto16},
are considered in Sec.~\ref{subsec:Maximally-entangled-bases}.

\subsection{Relative difference sets}

\label{subsec:Relative-difference-sets}

\emph{Relative difference sets }(defined below) are a tool from discrete
geometry which can be used to construct MU bases. We will summarise
results by Godsil and Roy \citep{godsil+09} using relative $(d,\mu,d,\lambda)$-difference
sets to obtain $(\mu+1)$ MU bases in $\mathbb{C}^{d}$.

A difference set $D$ of a group $G$ is a subset in which every element
of $G$ (excluding the identity) can be expressed as a difference
$d_{1}-d_{2}$ of elements of $D$ in exactly $\lambda$ ways. A \textit{relative}
difference set is defined as follows.
\begin{defn}
Let $G$ be a group, $N$ a normal subgroup of $G$, and $R$ a subset
of $G$ such that $|G|=\mu d$, $|N|=\mu$ and $|R|=r$. Then $R$
is a \emph{relative} $(d,\mu,r,\lambda)$\emph{-difference set} if
there exists $\lambda$ such that 
\[
[\{r_{1,2}\in R:\,r_{1}-r_{2}=b\}|=\left\{ \begin{array}{lll}
d\mu, &  & b=0\\
0, &  & b\in N\setminus\{0\}\\
\lambda, &  & b\in G\setminus N.
\end{array}\right.
\]
\end{defn}
A relative difference set is \emph{semi-regular} if $d=r$. A theorem
has been established that links semi-regular relative $(d,\mu,d,\lambda)$-difference
sets with MU bases.
\begin{thm}
The existence of a semi-regular $(d,\mu,d,\lambda)$-relative difference
set in an Abelian group implies the existence of a set of $\left(\mu+1\right)$
MU bases of $\mathbb{C}^{d}$ 
\end{thm}
By taking a particular finite commutative semifield of order $d$,
relative difference sets with parameters $(d,d,d,1)$ can be constructed.
The characters of the group $G,$ restricted to the set $R$, lead
to a set of $\left(d+1\right)$ MU bases in $\mathbb{C}^{d}$, as
shown in \citep{godsil+09}. A relative $(d,d,d,1)$-difference set
exists only if $d$ is a prime-power \citep{blokhuis+02}, thus, this
method only finds complete sets for $d=p^{n}\in\mathbb{PP}$. 

The relation of relative difference sets to complete sets of MU bases
is also discussed in \citep{belovs+08}, as well as a link to other
combinatorial structures such as planar functions. 

\subsection{Graph-theoretic constructions}

\label{subsec:Graph-states}

Complete sets of MU bases can be built from \emph{graph-states} generated
from a set of graphs, or equivalently, their \emph{adjacency matrices}
\citep{spengler+13}. A set of $p$ graph-states $\ket{G_{b}}\in\mathbb{C}^{p^{n}}$,
$b\in\mathbb{Z}_{p}$, are defined by $p$ different $n$-vertex graphs
(or $n\times n$ adjacency matrices). A graph-state basis is constructed
from a state $\ket{G_{b}}$ by taking as the basis vectors $\ket{G_{b}(m_{1},\ldots,m_{n})}=Z^{m_{1}}\otimes\ldots\otimes Z^{m_{n}}\ket{G_{b}}$
where $m_{i}\in\mathbb{Z}_{p}$.

The adjacency matrices which correspond to the MU bases are the same
symmetric matrices over $\mathbb{Z}_{p}$ which appear in the construction
of a unitary operator basis (see Sec.~\ref{subsec:constructions_on_unitary_operator_bases}).
Therefore, the graph-state bases corresponding to $A_{b}$ and $A_{b'}$
are mutually unbiased if $\det(A_{b}-A_{b'})\neq0\mod p$. To construct
complete sets of MU bases one takes the $p^{n}$ graph-state bases
formed from the graph-states $\ket{G_{b}}\in\mathbb{C}^{p^{n}}$ together
with the computational basis. Thus, $p^{n}$ adjacency matrices $\{A_{1},\ldots A_{p^{n}}\}$
must be found that satisfying the determinant condition. This is exactly
the same requirement given by Bandyopadhyay \emph{et al.} \citep{bandyopadhyay+02},
and the construction given by Wootters and Fields \citep{wootters+89}
can be used.

An alternative method to construct matrices $A_{b}$ using powers
and sums of powers of a single symmetric matrix is presented in \citep{spengler+13}.
In particular, it is possible to find an $n\times n$ symmetric matrix
$Q$ over $\mathbb{Z}_{p}$ such that its characteristic polynomial
is irreducible and primitive; this property implies that the set $S=\{Q^{i}\}^{p^{n}-2}_{i=0}\cup\{O_{n}\}$
is a matrix representation of the field $\mathbb{F}_{p^{n}}$, with
respect to matrix addition and multiplication. Here, $O_{n}$ is the
zero matrix of order $n$.

Since $S$ is a matrix representation of the field $\mathbb{F}_{p^{n}}$,
it contains the required property of the set $\{A_{1},\ldots A_{p^{n}}\}$
that the difference of any two matrices from $S$ is an invertible
matrix, i.e. $\det(A_{b}-A_{b'})\neq0\mod p$. Two methods to construct
the matrix $Q$ are provided in \citep{spengler+13}: a construction
via symmetrised companion matrices which works for any prime-power,
and a construction via tridiagonal matrices which works when $d=2^{n}$.

Other graph-theoretical constructions of MU bases exist involving
clique finding problems in Cayley graphs \citep{van11,alber18}, for
example.

\subsection{Other constructions and inequivalent complete sets}

\label{subsec:other_constructions}

We now point the curious reader to a few more alternative approaches
to construct completes sets of MU bases. Gow \citep{gow07} establishes
a method based on the existence of a unitary matrix whose powers generate
MU bases. Gibbons \emph{et al.} \citep{gibbons+04} start from Wigner
functions defined on discrete phase spaces, which are constructed
using a two-dimensional phase space $(p,q)$, taking the co-ordinates
to be elements of the field $\mathbb{F}_{d}$, with $d\in\mathbb{PP}$
a prime-power. The phase space induces a geometrical way to construct
MU bases, with the technique sharing similarities to an earlier construction
based on generalised Pauli matrices, maximally commuting classes and
finite fields \citep{pittenger+04}. A distinct feature of the Wigner
function approach is that it works for all finite fields, regardless
of $p$ being an even or an odd prime.

In 1997, Calderbank \emph{et al.} \citep{calderbank+97} constructed
complete sets using symplectic spreads and $\mathbb{Z}_{4}-$Kerdock
codes. An interesting feature here is that \textit{inequivalent} complete
sets of MU bases come into play (see Sec.~\ref{subsec:Equivalence-classes}).
In the construction of Calderbank \emph{et al.,} a complete set of
MU bases is associated with a symplectic spread, and two such sets
are equivalent if and only if their symplectic spreads can be mapped
to each other via a symplectic transformation. \emph{Inequivalent}
symplectic spreads, which appear in \citep{kantor82,kantor04}, are
then used to build inequivalent MU bases, e.g., when $d=2^{n}$ and
$n>3$ is odd. Kantor describes this association in \citep{kantor12},
and he presents more details on inequivalent MU bases, including examples
for $d=p^{n}$, with odd $p\in\mathbb{P}$. Other constructions of
MU bases from symplectic spreads have been developed in \citep{thas09,Abdukhalikov2015}.

As pointed out in \citep{godsil+09}, all known complete sets of MU
bases (including those derived here) are covered by the construction
of Calderbank \emph{et al}. in Ref. \citep{calderbank+97}. In particular,
for odd prime-powers, Wootters' complete set, i.e. Eq.~\eqref{eq:MUvectorsd=00003Dpn},
is equivalent to Alltop's complete set in Eqs.~\eqref{alltop_bases}
and the set from Bandyopadhyay \emph{et al.,} in Eq.~(\ref{pauli_group_elements}).
For $d=2^{n}$, the Klappenecker and Rötteler set in Eq.~\eqref{eq:klapp_bases}
is equivalent to those constructed by Wootters and Fields, and Bandyopadhyay
\emph{et al.} The bases from relative $(d,d,d,1)$-difference sets
are equivalent to the sets found by Klappenecker and Rötteler.

Later, Abdukhalikov \citep{Abdukhalikov2015} studied the relations
between various constructions and remarks that ''essentially there
are only three types of constructions up to now'', by which he means
those from symplectic spreads, planar functions over fields of odd
characteristic, and Gow’s construction \citep{gow07} of cyclic MU
bases from a unitary matrix. The latter, he remarks, ''seems to be
isomorphic to the classical one''. A new construction of MU bases
from (pseudo-) planar functions over fields of characteristic two
is also given, which is shown to relate to symplectic spreads by means
of a commutative presemifield.

\section{Complex Hadamard matrices }

\label{sec:Complex-Hadamard-matrices}

First we define the notion of equivalence between complex Hadamard
matrices and compare this to the equivalence of MU pairs. For Hadamard
matrices, the ordering of the columns and their overall phase factors
are not important. Therefore, we can multiply a Hadamard $H$ from
the left by a permutation matrix $P_{1}$ and a unitary diagonal matrix
$D_{1}$, and the resulting matrix $HD_{1}P_{1}$ is regarded as equivalent
to $H$. Similarly, equivalence is also maintained if we multiply
$H$ from the left with permutation and diagonal matrices.
\begin{defn}
\label{def:Hadamard-equivalences}Two Hadamard matrices $H$ and $K$
are \emph{equivalent}, i.e. $H\sim K$, if they satisfy $H=P_{1}D_{1}KD_{2}P_{2}$. 
\end{defn}
It is not known to us whether this mathematical notion of inequivalence
has a simple physical interpretation.

A Hadamard matrix is usually expressed in its dephased form with the
first row and column having elements $H_{i1}=H_{1j}=1/\sqrt{d}$.
Importantly, if two Hadamard matrices $H$ and $K$ are equivalent,
the resulting pairs of MU bases $\{\mathbb{I},H\}$ and $\{\mathbb{I},K\}$
are also equivalent (see Sec.~\ref{subsec:Equivalence-classes} for
the equivalence relations of MU bases). However, two inequivalent
Hadamard matrices \emph{may} form equivalent pairs of MU bases. For
example, the Fourier matrix and its transpose, $F$ and $F^{T}$,
are inequivalent but the MU pairs $\{\mathbb{I},F\}$ and $\{\mathbb{I},F^{T}\}$
are equivalent. 

One challenge is to deduce whether two Hadamard matrices are equivalent.
A useful test involves constructing the Haagerup set $\Lambda(H)$
of the complex Hadamard matrix $H$ defined as
\begin{equation}
\Lambda(H)=\{H_{pq}H^{*}_{qr}H_{rs}H^{*}_{sp}:p,q,r,s=1,\ldots,d\}\,,\label{eq:Haagerup_set}
\end{equation}
where $H^{*}_{ij}$ denotes the complex conjugation of the matrix
element $H_{ij}$ \citep{haagerup97}. The Haagerup set is invariant
under equivalence transformations so that two matrices $H$ and $K$
with \emph{different} Haagerup sets are necessarily \emph{inequivalent}. 

A Hadamard matrix $H$ can be \emph{isolated,} which means that all
Hadamard matrices in its immediate neighbourhood are equivalent to
it \citep{tadej+06}. Otherwise, $H$ is a member of a family of Hadamard
matrices depending on continuous parameters. Different types of parameter
dependence can be distinguished.
\begin{defn}
An \emph{affine} family of Hadamard matrices is a set $H(\mathcal{R})$
stemming from a Hadamard matrix $H$ of order $d$, where 
\begin{equation}
H(\mathcal{R})=\{H\circ\text{EXP}(i\cdot R):R\in\mathcal{R}\}\,,\label{eq: affine Hadamards}
\end{equation}
and $\mathcal{R}$ is a subspace of all real matrices of order $d$
with zeros in the first row and column.
\end{defn}
Here the notation $\circ$ is the entry-wise product of matrices known
as the Hadamard (or Schur) product, and EXP($\cdot$) is the entry-wise
exponential function acting on a matrix. Families not of the form
\eqref{eq: affine Hadamards} are called \emph{non-affine}. 

An upper bound on the number of free parameters for the set of matrices
stemming from $H$ is given by the \emph{defect} $\text{{def}}(H)$
of the matrix, introduced in Ref.~\citep{tadej+06}. It is defined
by multiplying the elements of the \emph{core} of the matrix with
free phase factors and solving the unitary condition on $H$ to first
order; the core consists of all elements different from the first
row and column.
\begin{defn}
\label{def:d (H)}The defect $\text{{def}}(H)$ of a complex Hadamard
matrix $H$ equals the dimension of the solution space of the real
linear system of equations with respect to a matrix variable $R\in\mathbb{R}^{d\times d}$:
\begin{equation}
\sum^{d}_{\ell=1}H_{j\ell}H^{*}_{k\ell}(R_{j\ell}-R_{k\ell})=0\,,\quad1\leq j<k\leq d\,,\label{eq: 1st order defect condition}
\end{equation}
and $R_{11}=R_{1s}=R_{s1}=0$ for $s=2\dots d$.
\end{defn}
Ref.~\citep{bengtsson+07} provides an instructive example. When
$d$ is large a computer program is usually necessary to determine
the value of $\text{{def}}(H)$. The defect is useful to identify
\emph{isolated} Hadamard matrices \citep{tadej+06}. 
\begin{lem}
A dephased Hadamard matrix $H$ is\emph{ isolated} if its defect is
zero. 
\end{lem}
The \emph{span condition} \citep{nicoara06} represents an alternative
method to determine if a matrix is isolated: For a given $d\times d$
Hadamard matrix $H$, the dimension of the vector space span$\{uv-vu:u\in\mathcal{D},v\in H^{*}\mathcal{D}H\}$
must equal $(d-1)^{2}$, where $\mathcal{D}$ is the algebra of diagonal
matrices. It is presently unknown if isolation is equivalent to the
span condition or if there exist matrices with non-zero defect that
are not part of any family.

The defect provides an upper bound on the dimensionality of families
of Hadamard matrices but is not necessarily sharp. Including higher
order equivalents of Def. \ref{def:d (H)} (which is based on the
first order calculation \eqref{eq: 1st order defect condition}) can
lead to sharper upper bounds \citep{barros+12}. As an example, the
Fourier matrix $F_{d}$, with $F_{ij}=\omega^{ij}/\sqrt{d}$ and $\omega=e^{2\pi i/d}$,
has defect
\begin{equation}
\text{def}(F_{d})=\sum^{d-1}_{n=1}(\text{gcd}(n,d)-1)\,.
\end{equation}
This gives an upper bound on any smooth (dephased) family of Hadamard
matrices travelling through $F_{d}$. The bound is saturated when
$d$ is a prime-power. However, the dimension of the largest smooth
family stemming from $F_{d}$ is strictly \emph{less} than the defect
for $6<d\leq100$ when $d$ is not a prime or prime-power \citep{barros+12}.
The $d=6$ case seems to be special: the defect of $F_{6}$ is four
and there exists of a four-parameter family which is likely to contain
$F_{6}$ \citep{skinner+09,barros+12} (cf. Sec.~\ref{sec:pairs_of_MU_bases_C^6}). 

Furthermore, it seems that the bound on the dimension of a smooth
family stemming from the Fourier matrix depends on the prime decomposition
of $d$. It has so far not been possible to find an exact bound for
arbitrary $d$—we have to make do with a conjecture: if $d=p_{1}p^{2}_{2}$
then there is a family of Hadamard matrices stemming from $F_{d}$
which has $(3p_{1}p^{2}_{2}-3p_{1}p_{2}-2p^{2}_{2}+p_{2}+1)$ free
parameters \citep{barros+12}. Since we know that Hadamard matrices
correspond to pairs of complementary bases, this result is another
indication that the geometry of the quantum state space depends heavily
on the number theoretic properties of the dimension $d$. 

\section{MU bases and affine planes}

\label{subsec:affineplanes_mubs_sics}

In Appendix~\ref{sec:affineplanes} we discussed constructions of
MU bases based on affine planes. It is known that affine planes of
order $d$ exist if $d$ is a prime or a prime-power; for certain
composite dimensions suc\.{h} as $d=6$, however, affine planes do
not exist. In fact, the Bruck-Ryser theorem states that no affine
plane of order $d$ exists if $\left(d-1\right)$ or $\left(d-2\right)$
is divisible by four and $d$ is not the sum of two squares \citep{bruck+49}.
Numerical computations also ruled out their existence for $d=10$
\citep{lam+89}.

These results bear a striking resemblance to the MU existence problem
which has led to a conjecture establishing a possible link with projective
planes \citep{saniga+04}:
\begin{conjecture}
The non-existence of a projective plane of order $d$ implies that
there are fewer than $(d+1)$ MU bases in the corresponding Hilbert
space $\mathbb{C}^{d}$, and vice versa. 
\end{conjecture}
A projective rather than an affine plane is used here, but the two
objects are essentially the same in the current context. One can construct
an affine plane from a projective plane by removing a single line
along with all the points it contains. A potential link between MU
bases and affine planes was perhaps first mentioned in \citep{klappenecker+04}.

While no rigorous association between affine planes and MU bases is
known, it has been suggested by Wootters \citep{wootters06} to associate
lines $\lambda$ with projection operators $P_{\lambda}$, projecting
onto orthogonal quantum states. A set of $d$ parallel lines then
corresponds to a basis of $d$ orthogonal projection operators satisfying
$\sum_{\lambda}P_{\lambda}=1$, and two non-parallel lines with associated
projection operators $P_{\lambda}$ and $P_{\nu}$ satisfy $\mbox{Tr}[P_{\lambda}P_{\nu}]=1/d$.
In this way, the $(d+1)$ \emph{striations} of an affine plane correspond
to a set of $(d+1)$ MU bases. An obvious question arises about the
role of the $d^{2}$ points in such a correspondence. In \citep{wootters06},
a point $\alpha$ is chosen to represent a Hermitian operator $A_{\alpha}/d$
such that \emph{(i)} $\mbox{Tr}[A_{\alpha}/d]=1/d$; \emph{(ii)} $\mbox{Tr}[A_{\alpha}A_{\beta}/d^{2}]=\delta_{\alpha\beta}/d$;
and \emph{(iii)} $\sum_{\alpha\in\lambda}(A_{\alpha}/d)=P_{\lambda}$
hold. Unfortunately, in this scheme the existence of $(d+1)$ striations
does not imply the existence of a complete set of MU bases since it
is not known how to construct the operators $A_{\alpha}$.

Caution, however, should be taken with the similarities between affine
planes and MU bases since some unexpected differences appear between
them \citep{weigert+10}. A mismatch arises when considering mutually
unbiased \emph{constellations}, that is, sets of vectors which are
either orthogonal or mutually unbiased (see Sec.~\ref{subsec:nonexistence_of_constellations}).
A numerical search is unable to find a MU constellation consisting
of three MU bases together with four orthogonal states \citep{brierley+08}.
On the other hand, the largest \emph{affine constellation} contains
three striations, each with six lines, and an additional set of four
parallel lines. An affine constellation is defined as a set of points
and lines such that any two lines within a set do not intersect, and
any pair of lines from different sets have one point in common. Thus,
if an affine constellation does not exist, then neither will an affine
plane. If there is a connection between affine planes and MU bases
such that parallel lines correspond to orthonormal bases and intersecting
ones to MU states, as suggested by Wootters \citep{wootters06}, one
would expect the structure of affine and MU constellations to be similar,
if not identical—which seems not to be the case.

Another discrepancy appears when considering the behaviour of MU bases
and affine planes in the limit of ever larger dimensions, $d\rightarrow\infty$.
It was shown by Chowla \emph{et al.} \citep{chowla+60} that the number
of mutually orthogonal Latin squares approaches infinity as $d\rightarrow\infty$,
while only three MU bases (with basis-independent overlaps) have been
found in the infinite-dimensional Hilbert space $L^{2}(\mathbb{R})$ \citep{weigert+08}
(cf. Sec.~\ref{subsec: MUs for CVs}).

\end{document}